\documentclass[phd,twoside]{osudissert96}

\usepackage{graphicx}
\usepackage{dcolumn}
\usepackage{psfrag}
\usepackage{mathrsfs}
\usepackage{mathbbol}
\usepackage{amsmath}
\usepackage{amsfonts}
\usepackage{amssymb}
\usepackage{exscale}
\usepackage{wick}
\usepackage{hyperref}

\renewcommand{\typesetChapterTitle}[1]{\uppercase{#1}}

\numberwithin{equation}{chapter}

\newcolumntype{d}[1]{D{.}{.}{#1}}
\newcolumntype{e}[1]{D{,}{}{#1}}

\DeclareMathOperator{\Trace}{Tr}
\DeclareMathOperator{\trace}{tr}
\DeclareMathOperator{\sTrace}{sTr}
\DeclareMathOperator{\sdet}{sdet}
\DeclareMathOperator{\diag}{diag}
\DeclareMathOperator{\Real}{Re}
\DeclareMathOperator*{\proj}{proj}

\newcommand{\mycaption}[1]{\caption{\small#1}}
\newcommand{\Lag}[1]{\mathscr{L}_\text{#1}}
\newcommand{\fD}[1]{[\mathcal{D}#1]}
\newcommand{\qb}{\bar{q}}
\newcommand{\Qb}{\bar{Q}}
\newcommand{\chib}{\bar{\chi}}
\newcommand{\muh}{\hat{\mu}}
\newcommand{\nuh}{\hat{\nu}}
\newcommand{\g}{\text{\textsl{g}}}
\newcommand{\pad}[1]{{\quad{#1}\quad}}
\newcommand{\Det}[1]{{\det\bigl[#1\bigr]}}

\newcommand{\tr}[1]{{\trace\bigl[#1\bigr]}}
\newcommand{\Tr}[1]{{\Trace\bigl[#1\bigr]}}
\newcommand{\sTr}[1]{{\sTrace\bigl[#1\bigr]}}
\newcommand{\QCD}[1]{\tilde{#1}}
\newcommand{\vect}[1]{\vec{#1}}
\newcommand{\fconfig}[1]{{[#1]}}
\newcommand{\chlog}[1]{{\mathcal{I}_{#1}}}
\newcommand{\halfof}[1]{\tfrac{#1}{2}}
\newcommand{\efrac}[2]{{\mbox{}^{#1}\hspace{-0.25em}/\hspace{-0.13em}\mbox{}^{}_{#2}}}
\newcommand{\bilinear}[2]{{\bigl(\gamma_{#1}\linebreak[0]\otimes\xi_{#2}\bigr)}}
\newcommand{\Bilinear}[2]{{\bigl({#1}\linebreak[0]\otimes{#2}\bigr)}}
\newcommand{\dsub}[2]{{\begin{subarray}{l}{#1}\vphantom{g}\\{#2}\vphantom{l}\end{subarray}}}
\newcommand{\tsub}[2]{{\begin{subarray}{l}{#1}\\{#2}\end{subarray}}}
\newcommand{\prob}[2]{{P_{#1} \mbox{\small$\bigl(#2\bigr)$}}}
\newcommand{\V}[2]{{V^{\vphantom{(\mu)}({#1})}_{\vphantom{\muh}{#2}}}}
\newcommand{\U}[1]{\check{#1}}
\newcommand{\ensemble}[1]{$\mathsf{#1}$}

\newcommand{\redefarrow}{\rightarrowtail}
\newcommand{\symarrow}{\rightarrow}
\newcommand{\maparrow}{\mapsto}
\newcommand{\decayarrow}{\rightarrow}
\newcommand{\cntlmtarrow}{\rightarrow}
\newcommand{\updatearrow}{\rightarrow}
\newcommand{\discarrow}{\rightarrow}
\newcommand{\mshiftarrow}{\rightarrow}
\newcommand{\ring}{\mathring}
\newcommand{\pquad}{{\quad\quad}}
\newcommand{\bra}{\langle}
\newcommand{\ket}{\rangle}
\newcommand{\Op}{\mathcal{O}}
\newcommand{\mS}{m_S}
\newcommand{\mV}{m_V}
\newcommand{\ms}{m_s}
\newcommand{\mv}{m_v}
\newcommand{\mh}{\hat{m}}
\newcommand{\half}{\tfrac{1}{2}}
\newcommand{\m}{M}
\newcommand{\mpn}{\m_{\pi^0}}
\newcommand{\mpp}{\m_{\pi^+}}
\newcommand{\mpm}{\m_{\pi^-}}
\newcommand{\mkn}{\m_{K^0}}
\newcommand{\mkbn}{\m_{\bar{K}^0}}
\newcommand{\mkp}{\m_{K^+}}
\newcommand{\mkm}{\m_{K^-}}

\newcommand{\mpQCD}{\QCD{\m}_\pi}
\newcommand{\mpnQCD}{\QCD{\m}_{\pi^0}}
\newcommand{\mppQCD}{\QCD{\m}_{\pi^+}}
\newcommand{\mpmQCD}{\QCD{\m}_{\pi^-}}
\newcommand{\mkQCD}{\QCD{\m}_K}
\newcommand{\mknQCD}{\QCD{\m}_{K^0}}
\newcommand{\mkbnQCD}{\QCD{\m}_{\bar{K}^0}}
\newcommand{\mkpQCD}{\QCD{\m}_{K^+}}
\newcommand{\mkmQCD}{\QCD{\m}_{K^-}}
\newcommand{\meQCD}{\QCD{\m}_\eta}
\newcommand{\gfconfig}{\fconfig{\phi_a}}
\newcommand{\pathordered}{\mathcal{P}}
\newcommand{\identity}{\Eins}
\newcommand{\discder}{\varDelta}
\newcommand{\CP}{CP}
\newcommand{\qmassm}{\mathcal{M}}
\newcommand{\n}{h}
\newcommand{\avg}{\overline}
\newcommand{\uM}{\U{M}}
\newcommand{\uf}{\U{f}}
\newcommand{\uz}{\U{z}}

\begin{document}

\author{Daniel R. Nelson}
\title{Partially Quenched Chiral Perturbation Theory and a Massless Up Quark:
A Lattice Calculation of the Light-Quark-Mass Ratio}

\authordegrees{Ph.D.}

\unit{Department of Physics}
\graduationyear{2002}

\advisorname{Professor Junko Shigemitsu}
\member{Professor Gregory Kilcup}
\member{Professor Richard Furnstahl}
\member{Professor Richard Hughes}

\date{March 5, 2002}

\maketitle

\begin{abstract}
The nontrivial topological structure of the QCD gauge vacuum generates a $\CP$
breaking term in the QCD Lagrangian.  However, measurements of the neutron
electric dipole moment have demonstrated that the term's coefficient is
unnaturally small, a dilemma known as the strong $\CP$ problem.  A massless up
quark has long been seen as a potential solution, as the term could then be
absorbed through the resulting freedom to perform arbitrary chiral rotations on
the up quark field.

Through the light-quark-mass ratio $m_u / m_d$, leading order Chiral
Perturbation Theory appears to rule this scenario out.  However, the
Kaplan-Manohar ambiguity demonstrates that certain strong next-to-leading order
corrections are indistinguishable from the effects of an up quark mass.  Only a
direct calculation of the Gasser-Leutwyler coefficient combination $2 L_8 -
L_5$ can resolve the issue.

New theoretical insights into partial quenched Chiral Perturbation Theory have
revealed that a calculation of the low-energy constants of the partially
quenched chiral Lagrangian is equivalent to a determination of the physical
Gasser-Leutwyler coefficients.  The coefficient combination in question is
directly accessible through the pion mass's dependence on the valence quark
mass, a dependence ripe for determination via Lattice Quantum Chromodynamics.

We carry out such a partially quenched lattice calculation using $N_f = 3$
staggered fermions and the recently developed smearing technique known as
hypercubic blocking.  Through the use of several ensembles, we make a
quantitative assessment of our systematic error.  We find $2 L_8 - L_5 = \bigl(
0.22 \pm 0.14 \bigr) \times 10^{-3}$, which corresponds to a light-quark-mass
ratio of $m_u / m_d = 0.408 \pm 0.035$.  Thus, our study rules out the
massless-up-quark solution to the strong $\CP$ problem.

This is the first calculation of its type to use a physical number of light
quarks, $N_f = 3$, and the first determination of $2 L_8 - L_5$ to include a
comprehensive study of statistical error.

\end{abstract}

%\begin{externalabstract}
%\input{abstract}
%\end{externalabstract}

\dedication{to Mom and Dad \\ for their unconditional love}

\begin{acknowledgements}
Special thanks go to George Fleming, whose enthusiasm, discipline, and
leadership made this work possible.

I offer thanks to Gregory Kilcup for the use of his matrix inversion code, as
well as for his procurement of the necessary computer resources.

I thank the Ohio Supercomputer Center, which provided a majority of the
computer time consumed by this calculation.

Finally, thanks to Angie Linn for her boundless patience.
\end{acknowledgements}

\tableofcontents
\listoftables
\listoffigures

\chapter{Introduction}
Since its inception, the low-energy dynamics of Quantum Chromodynamics (QCD)
have been poorly understood.  To date, the most successful framework for
building an understanding of low-energy QCD has been Chiral Perturbation
Theory, with the current large uncertainty in its NLO coefficients
quantifying our lackluster understanding.  These low-energy constants, the
Gasser-Leutwyler coefficients, have error bars which range from 10\% to 160\%
of their value.  For many of these coefficients, their uncertainty has never
been reduced below the magnitude determined at the first instance of their
calculation some twenty years ago.

As an effective field theory, light-meson Chiral Perturbation Theory collects
the interactions of QCD into a finite number of meson vertices.  The complex
low-energy dynamics of QCD are boiled down into the coefficients of these
vertices, the Gasser-Leutwyler coefficients.  Thus, while various theoretical
and phenomenological methods have been used to estimate their values, the most
direct determination of the Gasser-Leutwyler coefficients would be a
calculation of these vertices' strength using the fundamental theory, QCD.

While perturbation theory clearly fails in this regard, as the strong coupling
at these energy scales is of order one, lattice techniques have proven
successful.  In fact, this may be a case in which lattice field theory can
provide the physics community with the best results available.

We present here a lattice calculation of a single important combination of the
Gasser-Leutwyler coefficients, $2 L_8 - L_5$.  The motivation behind this
combination choice is twofold.  First, it is the combination whose
calculation on the lattice is most straightforward.  Additionally, this
combination provides insight into the NLO corrections to the light-quark-mass
ratio $m_u / m_d$.  Thus, determining $2 L_8 - L_5$, even with moderate
accuracy, allows one to lay to rest the possibility of a massless up quark.

While we calculate definitively only a single combination of the
Gasser-Leutwyler coefficients, this study is only the first step in a larger
effort by the lattice community to generate results for all accessible
Gasser-Leutwyler coefficients.

Elements of this study have seen earlier publication \cite{Nelson:2001tb}.  A
similar investigation, which uses an unphysical number of light quarks and less
sophisticated analysis techniques, can be found in \cite{Irving:2001vy}.

In Chapter \ref{an:section}, after a brief overview of QCD, the strong $\CP$
problem and its potential solutions are presented, including the directly
relevant solution involving a massless up quark.

In Chapter \ref{n:section} we introduce Chiral Perturbation Theory up to NLO,
focusing on the insight it imparts into the light-quark-mass ratio $m_u / m_d$.
The relationship between the quark-mass ratio and the light mesons is explored,
and the importance of the Gasser-Leutwyler coefficient combination $2 L_8 -
L_5$ in that relationship is presented.  An overview of past phenomenological
and theoretical estimates for the coefficients $L_5$ and $L_8$ is given,
including a discussion of the Kaplan-Manohar ambiguity, which makes an
experimental determination of $2 L_8 - L_5$ impossible and theoretical
estimates challenging.

In Chapter \ref{ao:section} the basics of Lattice Quantum Field Theory are
presented.

In Chapter \ref{o:section} we build on those basics, introducing Lattice
Quantum Chromodynamics.  The lattice techniques used in this study are covered,
including staggered fermions, the conjugate gradient method, the $R$ algorithm,
hypercubic blocking, and the Sommer scale.  We present in explicit detail the
methods used to extract the pion mass and decay constant from staggered
bilinear correlators.

In Chapter \ref{ap:section} we extend the concepts of Chiral Perturbation
Theory to cover the partially quenched case, explaining how partially quenched
Chiral Perturbation Theory allows physical results for the Gasser-Leutwyler
coefficients to be generated from lattice calculations which use the unphysical
partially quenched approximation.

In Chapter \ref{u:section} our methods for data modeling and statistical error
analysis are covered.

In Chapter \ref{aq:section} we detail the lattice ensembles generated for our
study, explaining the motivation behind each ensemble's creation.

In Chapter \ref{q:section} we present a step-by-step analysis of our lattice
data, generating values for the Gasser-Leutwyler coefficient combinations $2
L_8 - L_5$ and $L_5$ for each ensemble studied.  A simultaneous analysis of
several ensembles, which allows us to make a preliminary estimation of the
coefficient combinations $2 L_6 - L_4$ and $L_4$, is also presented.

In Chapter \ref{ac:section} our final result for $2 L_8 - L_5$ is given, along
with an analysis of our study's systematic error.  Using Chiral Perturbation
Theory we produce from our result a prediction for the light-quark-mass ratio.
Secondary results are also given, including values for the coefficients $L_5$,
$L_4$, and $L_6$.  The effects of quenching on results for the Gasser-Leutwyler
coefficients is briefly discussed.

In Chapter \ref{ab:section} we summarize our results and discuss the rich
potential for future work.

\chapter{Quantum Chromodynamics} \label{an:section}
Quarks are the fundamental building blocks of the universe.  Bound together by
gluon exchanges, they are the dominant constituents of hadrons.  It is a
hadron's quarks and the dynamics of the quark-gluon interactions which dictate
that hadron's characteristics and behavior.

The physics of quarks and gluons is dominated by the strong force, Quantum
Chromodynamics (QCD).  Working in Euclidean space, QCD is governed by the
partition function:
\begin{equation}
Z_\text{QCD} = \int  \fD{A_\mu} \fD{q} \fD{\qb} ~ e^{ -\int_x{ \Lag{QCD} }}
\end{equation}
where $A_\mu$ is the gluon field, and $q$ and $\qb$ are the quark and
antiquark fields.  The QCD Euclidean Lagrangian is:
\begin{equation}
\Lag{QCD} = \frac{1}{2} \tr{ F_{\mu\nu} F^{\mu\nu} } + \qb
\bigl( \gamma^\mu D_\mu + \qmassm \bigr) q
\end{equation}

The quark field $q$ is a vector in three spaces:  flavor, color, and spin.  The
flavor index runs from 1 to $N_f$, while the color index runs from 1 to $N_c =
3$.  Finally, quarks are Dirac spinors, four component vectors in spin space.

$D_\mu$ is the covariant derivative and a matrix in color space, while
$\gamma^\mu$ is a Lorentz vector of spin space matrices.  $\qmassm$ is the
quark mass matrix, a diagonal flavor-space matrix with the quark masses along
its diagonal:
\begin{equation}
\qmassm \equiv \diag \bigl( \{ m_i \} \bigr)
\end{equation}

The gluon field $A_\mu$ makes its kinetic appearance in the Lagrangian through
its field strength $F_{\mu\nu}$:
\begin{align}
F_{\mu\nu} & \equiv \frac{i}{\g} \bigl[ D_\mu, D_\nu \bigr] \notag \\
& = \partial_\mu A_\nu - \partial_\nu A_\mu - i \g \bigl[ A_\mu, A_\nu \bigr]
\end{align}
The gluon field is both a
Lorentz vector as well as a vector in adjoint color space.  It has been written
as a matrix in color space, with each component of the adjoint
vector being multiplied by a generator of $SU(N_c)$:
\begin{equation}
A_\mu \equiv A^a_\mu \lambda^a
\end{equation}

The gluon field acts as the parallel transporter of the quarks through color
space, appearing in the covariant derivative:
\begin{equation}
D_\mu \equiv \partial_\mu - i \g A_\mu
\end{equation}
It is through $D_\mu$ that the quark-gluon interaction arises.

\section{Symmetries of Quantum Chromodynamics}
Of the symmetries respected by the QCD Lagrangian, two of them are of
particular interest to us:  color symmetry and flavor symmetry,

\subsection{Color Symmetry}
$SU(N_c)$ color symmetry is an exact and local symmetry of QCD under which
the fields transform as:
\begin{align}
q & \pad{\symarrow} q' = \Omega(x) q \\
\qb & \pad{\symarrow} \qb' = \qb ~ \Omega^\dagger(x) \\
A_\mu & \pad{\symarrow} A'_\mu = \Omega(x) \biggl( A_\mu + \frac{i}{\g}
\partial_\mu \biggr) \Omega^\dagger(x)
\label{cr:equation} \\
& \qquad \Omega(x) = e^{ -i \alpha^a(x) \lambda^a} \in SU(N_c)
\label{a:equation}
\end{align}
where $\alpha^a(x)$ is a set of $N_c^2 - 1$ real functions which parameterize
the transformation.  This transformation of $A_\mu$ results in the gluon field
strength transforming as:
\begin{equation}
F_{\mu\nu} \symarrow F'_{\mu\nu} = \Omega(x) F_{\mu\nu}
\Omega^\dagger(x)
\end{equation}

\subsection{Flavor Symmetry}
In the case of degenerate quarks:
\begin{equation}
\qmassm = m_q \identity
\end{equation}
the Lagrangian respects a global $SU(N_f)$ flavor symmetry, under which the
fields transform as:
\begin{align}
q & \pad{\symarrow} q' = V q \\
\qb & \pad{\symarrow} \qb' = \qb ~ V^\dagger \\
& \qquad V = e^{ -i v^a \tau^a} \in SU(N_f)
\end{align}
where $\tau^a$ are the generators of $SU(N_f)$ flavor and $v^a$ are real
constants which parameterize the transformation.

In the case of massless quarks, $m_q = 0$, known as the chiral limit, the QCD
Lagrangian exhibits an even larger $SU(N_f)_L \otimes SU(N_f)_R$ flavor
symmetry.  The quarks split into left- and right-handed pairs:
\begin{align}
q & = q_L + q_R \\
& \qquad q_L \equiv P_+ q = \tfrac{1}{2} (1 + \gamma_5) q \\
& \qquad q_R \equiv P_- q = \tfrac{1}{2} (1 - \gamma_5) q
\end{align}
\begin{align}
\qb & = \qb_L + \qb_R \\
& \qquad \qb_L \equiv \qb P_- = \tfrac{1}{2} \qb (1 - \gamma_5) \\
& \qquad \qb_R \equiv \qb P_+ = \tfrac{1}{2} \qb (1 + \gamma_5)
\end{align}
each of which rotates independently under flavor symmetry:
\begin{align}
q & \pad{\symarrow} q' = L q_L + R q_R \\
\qb & \pad{\symarrow} \qb' = \qb_L L^\dagger + \qb_R R^\dagger \\
& \qquad L = e^{ -i l^a \tau^a} \in SU(N_f)
\label{f:equation} \\
& \qquad R = e^{ -i r^a \tau^a} \in SU(N_f)
\label{g:equation}
\end{align}
where $l^a$ and $r^a$ are sets of real constants which parameterize the
transformation.

This symmetry can also be expressed in terms of a vector and an axial vector
symmetry, $SU(N_f)_V \otimes SU(N_f)_A$.  Here the vector symmetry corresponds
to the smaller flavor symmetry discussed above, where left- and right-handed
quarks transform equivalently, $L = R$.  In contrast, a pure axial vector
transformation rotates left- and right-handed quarks in opposite directions, $L
= R^\dagger$.  The currents associated with these symmetries are the
non-singlet vector currents:
\begin{equation}
J^a_\mu \equiv \qb \gamma_\mu \tau^a q
\end{equation}
and the non-singlet axial vector currents:
\begin{equation}
J^{5a}_\mu \equiv \qb \gamma_\mu \gamma_5 \tau^a q
\end{equation}
respectively.

In truth all quarks are neither massless nor degenerate.  However, the two
lightest quark flavors, up and down, have masses and a mass splitting which are
quite small relative to the typical baryon mass.  The next lightest quark
flavor, strange, has a somewhat small mass, again relative to the typical
baryon mass.  So, if we restrict ourselves to the three lightest flavors of
quarks, we would expect $SU(N_f = 3)_V \otimes SU(N_f = 3)_A$ flavor symmetry
to be a good approximate symmetry of the strong interactions.

However, while $SU(N_f)_V$ flavor symmetry is manifest in QCD and its
associated currents are conserved, $SU(N_f)_A$ flavor symmetry does not appear
to be respected.  There exist two possibilities when a theory's Lagrangian
contains a symmetry, but the theory itself does not appear to respect that
symmetry.  Either the symmetry is still respected but also hidden via
spontaneous symmetry breaking, or the symmetry is false, ruined by anomalous
quantum corrections to the Lagrangian's classical solution.  The $SU(N_f)$
axial vector flavor symmetry of QCD falls into the first category.

\subsection{Spontaneous Symmetry Breaking}
Spontaneous symmetry breaking occurs when a theory has not one ground state, but
rather a set of ground states which transform into one another via the symmetry
under discussion.  In such a situation the theory's vacuum must choose its
location from among these ground states and thus resides in a position which
is not invariant under the symmetry.  So, while the full theory retains the
given symmetry, the state space in the vicinity of the vacuum does not reflect
that symmetry.  Since this is the region important to low-energy interactions
and the region explored by perturbation theory, the symmetry becomes hidden.

When the set of vacuum states is continuous, local fluctuations of the vacuum
within that set are massless.  Thus, the theory's spectrum will contain
massless particles, known as Goldstone bosons, which correspond to those
fluctuations.  The number of spontaneously broken symmetries corresponds to the
number of orthogonal directions within the set of ground states, and thus
corresponds to the resulting number of Goldstone bosons.

It is the interactions of these Goldstone bosons, with one another and with
the other particles of the theory, which enforces the now hidden symmetry.

\subsection{Chiral Condensate} \label{p:section}
In massless QCD the energy required to create a quark-antiquark pair from the
vacuum is small.  Because such a pair must have zero total linear and angular
momentum, they will contain a net chiral charge.  Thus, the QCD vacuum includes
a chiral condensate characterized by the non-zero vacuum expectation value
of the operator:
\begin{equation}
\bra \qb q \ket = \bra \qb_L q_R \ket + \bra \qb_R q_L \ket \neq 0
\label{cc:equation}
\end{equation}
In truth, the vacuum expectation value includes an arbitrary $SU(N_f)_A$
rotation:
\begin{equation}
\bra \qb_L L^\dagger R q_R \ket + \bra \qb_R R^\dagger L q_L \ket = \bra \qb_L
A^{\dagger2} q_R \ket + \bra \qb_R A^2 q_L \ket \neq 0
\end{equation}
where $L = R^\dagger = A \in SU(N_f)_A$.  The vacuum is forced to choose the
$SU(N_f)_A$ alignment of this expectation value, thus spontaneously breaking
$SU(N_f)_V \otimes SU(N_f)_A$ flavor symmetry down to $SU(N_f)_V$.  Note that
the chiral condensate is invariant under $SU(N_f)_V$:
\begin{align}
\bra \qb q \ket \pad{\symarrow} & \bra \qb_L L^\dagger R q_R \ket + \bra
\qb_R R^\dagger L q_L \ket \notag \\
& \quad = \bra \qb_L V^\dagger V q_R \ket + \bra \qb_R V^\dagger V q_L \ket =
\bra \qb q \ket
\end{align}
where $L = R = V \in SU(N_f)_V$.  Once the vacuum has made its choice for the
condensate's alignment, we in turn can rotate the definition of our quark
fields such that the vacuum expectation value does indeed take the form in
\eqref{cc:equation}.  While the freedom for such a redefinition exists in
massless QCD, the quark masses of true QCD would not allow it.  However in
that case, the vacuum expectation value will naturally align itself with the
quark masses, again taking the form in \eqref{cc:equation}.

Because the set of vacuum states is continuous, connected by elements of
$SU(N_f)_A$ arbitrarily close to identity, we expect to find massless particles
in the spectrum, the $N_f^2 - 1$ Goldstone bosons of the spontaneously broken
symmetry.  The spectrum of QCD does not contain any such massless particles.
However, together the light pseudoscalar mesons --- $\pi^0$, $\pi^+$, $\pi^-$,
$K^0$, $\bar{K}^0$, $K^+$, $K^-$, and $\eta$ --- assume the role of the
Goldstone bosons.  Collected, they form an octet of very light particles, one
for each of the eight generators of the spontaneously broken $SU(N_f = 3)_A$
flavor symmetry.

It is because the spontaneously broken $SU(N_f = 3)_A$ flavor symmetry is only
an approximate symmetry of QCD, not an exact symmetry, that these Goldstone
bosons are not exactly massless.  In fact the squared masses of the light
mesons are proportional to the parameters which break $SU(N_f)_A$ flavor
symmetry:  the quark masses.  Thus, the light pseudoscalar mesons are often
referred to as pseudo-Goldstone bosons.

\section{$U(1)_A$ Problem} \label{ak:section}
In addition to color and flavor symmetry, the QCD Lagrangian is
classically invariant under two additional global $U(1)$ symmetries.  The first
is $U(1)_V$ vector symmetry, under which the quark fields transform by a simple
phase rotation:
\begin{align}
q & \pad{\symarrow} q' = e^{i \alpha} q \\
\qb & \pad{\symarrow} \qb' = \qb e^{-i \alpha}
\end{align}
This symmetry is exact for arbitrary quark masses and corresponds to the
conserved singlet vector current:
\begin{gather}
J_\mu \equiv \qb \gamma_\mu q \\
\partial^\mu J_\mu = 0
\end{gather}
and to baryon number conservation, a phenomenon clearly observed in experiment.
The second symmetry is $U(1)_A$ axial vector symmetry, under which left- and
right-handed quarks undergo opposite phase rotations:
\begin{align}
q & \pad{\symarrow} q' = e^{i \alpha \gamma_5} q \\
\qb & \pad{\symarrow} \qb' = \qb e^{i \alpha \gamma_5}
\end{align}
Similar to the non-singlet flavor symmetries, this symmetry is only exact in
the limit of massless quarks.  The corresponding singlet axial vector current
is:
\begin{equation}
J^5_\mu \equiv \qb \gamma_\mu \gamma_5 q
\end{equation}
Assuming degenerate quarks, naive application of the equations of motion results
in the divergence:
\begin{equation}
\partial^\mu J^5_\mu = 2 m_q \qb \gamma_5 q
\end{equation}
which equals zero in the chiral limit.  Thus, we expect the symmetry to be an
approximate symmetry of QCD.  However, $U(1)_A$ symmetry calls for degenerate
parity doublets which are not even approximately manifest in the QCD particle
spectrum.  The mystery of this missing symmetry is known as the $U(1)_A$
problem.  

If the symmetry were spontaneously broken, the spectrum would contain a
corresponding Goldstone boson.  We could imagine combining the singlet axial
vector symmetry with the non-singlet flavor symmetries:
\begin{equation}
U(N_f)_A = SU(N_f)_A \otimes U(1)_A
\end{equation}
and having them spontaneously break together.  This spontaneous breaking of
$U(N_f = 3)_A$ would result in the eight light mesons mentioned above plus a
new ninth light pseudoscalar meson.  There exists a candidate pseudoscalar
meson to fill the roll of this ninth light meson:  the $\eta'$.  However, it
has been shown that, if the full $U(N_f = 3)_A$ symmetry were spontaneously
broken, it would constrain the $\eta'$ mass:  $\m_{\eta'} < \sqrt{3} \m_\pi$
\cite{Weinberg:1975ui}.  The actual $\eta'$ is much heavier.  $U(1)_A$ has no
Goldstone boson, and thus spontaneous symmetry breaking can not explain the
symmetry's disappearance.  We must look for a second possibility.

\subsection{Adler-Bardeen-Jackiw Anomaly}
When a Lagrangian contains a given symmetry, its classical equations of motion
will respect that symmetry.  However, when the Lagrangian is placed within a
path integral, that symmetry may be lost.  For a given symmetry to survive the
transition to quantum field theory, not only must the action be invariant under
the transformation, but the measure of the path integral must also be
invariant.  When the measure is not invariant, the symmetry is said to be
anomalously broken.  In the context of perturbation theory, this will manifest
as a failure of the regularization of radiative corrections to respect the
symmetry.

The conserved current corresponding to an unbroken symmetry has a divergence of
zero.  For an anomalously broken symmetry, the divergence of the corresponding
current will be non-zero.  This non-zero divergence is referred to as the
anomaly.  The anomaly which breaks the $U(1)_A$ symmetry of QCD is the
Adler-Bardeen-Jackiw (ABJ) anomaly.

Under a $U(1)_A$ rotation, the measure of the QCD path integral transforms as:
\begin{equation}
\fD{A_\mu} \fD{q} \fD{\qb} \pad{\symarrow} \fD{A_\mu} \fD{q} \fD{\qb} e^{
-\alpha \int_x{\Theta}}
\label{d:equation}
\end{equation}
where:
\begin{equation}
\Theta \equiv \frac{N_f \g^2}{8 \pi^2} \tr{ F_{\mu\nu} \tilde{F}^{\mu\nu} }
\end{equation}
and $\tilde{F}_{\mu\nu} \equiv \tfrac{1}{2} \epsilon_{\mu\nu\alpha\beta}
F^{\alpha\beta}$.  This results in the anomalous divergence of the axial vector
current:
\begin{equation}
\partial^\mu J^5_\mu = 2 m_q \qb \gamma_5 q + \Theta
\end{equation}
which no longer equals zero in the massless quark limit.  Thus, the $U(1)_A$
problem appears to be solved.  The $U(1)_A$ symmetry of QCD is missing because
it never existed; it is ruined by the ABJ anomaly.  However, the situation is
not that straightforward.

\subsection{Gauge-Variant Axial Vector Current}
The divergence of the axial vector current can also be written as a divergence
of gauge fields:
\begin{equation}
\Theta = \frac{N_f \g^2}{2 \pi^2} \partial^\mu K_\mu
\label{e:equation}
\end{equation}
where:
\begin{equation}
K_\mu \equiv \frac{1}{2} \epsilon_{\mu\nu\alpha\beta} \tr{ A^\nu
\partial^\alpha A^\beta - \frac{2}{3} i \g A^\nu A^\alpha A^\beta }
\end{equation}
We can then define a new current $\tilde{J}^5_\mu$ which involves both quark
and gauge fields:
\begin{equation}
\tilde{J}^5_\mu \equiv J^5_\mu - \frac{N_f \g^2}{2 \pi^2} K_\mu
\end{equation}
For massless quarks this new current is then conserved:
\begin{equation}
\partial^\mu \tilde{J}^5_\mu = 0
\end{equation}
and once again we find ourselves with a current that appears to be conserved,
but whose associated symmetry is not manifest in the theory and with which
there is no associated Goldstone boson.

The solution to the dilemma centers on the fact that $K_\mu$ is not a gauge
invariant quantity.  Because of this, if the symmetry associated with
$\tilde{J}^5_\mu$ were spontaneously broken, the resulting Goldstone boson
might decouple from the theory's physical states.  When working with a gauge
theory in a covariant gauge, the number of degrees of freedom in the theory is
larger than the number of physical states.  A condition must be applied to the
states of the theory in order to remove unphysical states.  It is during
the application of that condition that the Goldstone bosons of a gauge-variant
symmetry could decouple completely from the physical states.  In fact, it has
been shown that in QCD they do indeed decouple \cite{Kogut:1975kt,
'tHooft:1976fv}.  The modified $U(1)_A$ symmetry is not observed because
it is spontaneously broken, and the Goldstone boson which results decouples from
QCD's physical states.  

To summarize the fate of the chiral symmetries of QCD, we recall that the
massless QCD Lagrangian classically respects the symmetry group $SU(N_f)_V
\otimes SU(N_f)_A \otimes U(1)_V \otimes U(1)_A$, while only the smaller group
of vector symmetries $SU(N_f)_V \otimes U(1)_V$ is manifest in the theory.
The two chiral symmetries have each met a separate fate.  $SU(N_f)_A$ is
spontaneously broken by the chiral condensate, resulting in the light
pseudoscalar mesons as Goldstone bosons of this hidden symmetry.  $U(1)_A$, on
the other hand, is gone entirely, forced to gauge variance by the ABJ anomaly.

\section{$\theta$ Vacua}
Generally a quantum field theory which does not experience spontaneous symmetry
breaking has a single vacuum, the one field configuration with minimum
action.  This is the case when all other field configurations can be smoothly
deformed into the vacuum state.

However, if the field theory contains field configurations which can not be
smoothly deformed into one another, the theory will have multiple vacua.  In
such a case, the space of field states can be grouped into sets of
configurations which are smoothly connected.  In each of these sets, there will
be some configuration with minimal action.  It is these field configurations
which are the multiple vacua of the theory.  They are analogous to multiple
local minima in quantum mechanics, separated by infinitely high barriers.

This phenomenon of multiple vacua occurs for certain gauge theories, including
the $SU(N_c = 3)$ gluon fields of QCD.

It is possible to write down gluon field configurations which have finite
action and localized action density, but where the gauge field does not go to
zero as we approach infinity.  This is because a region of space in which the
gauge field has the form:
\begin{equation}
A_\mu(x) = \frac{i}{\g} \Omega(x) \partial_\mu \Omega^\dagger(x)
\label{cp:equation}
\end{equation}
contains zero action density.  Here $\Omega(x)$ is an arbitrary smooth function
mapping $x$ onto the gauge group.  As evident from \eqref{cr:equation}, the
form in \eqref{cp:equation} is simply the gauge transform of $A_\mu = 0$.  When
constructing such a field configuration where the action density goes to zero
at infinity but the gauge field does not, we must choose $\Omega(x)$ as we
approach infinity in each direction.  In effect we are choosing a smooth
mapping of the sphere at infinity $S_{d - 1}$ onto the gauge group, where $d$
is the dimension of the Euclidean space.

As an example, take the case of a $U(1)$ gauge field in $d = 2$ dimensions.
When we write down a field configuration, we are choosing a mapping of $S_1$,
a simple circle, onto $U(1)$.  Since $U(1)$ is also described by a circle, one
is mapping a circle onto a circle.  Denoting both circles as the phase of a
unit vector in the complex plane, we can write down examples of such mappings:
\begin{equation}
\mathcal{H}: \qquad e^{i \phi} \pad{\maparrow} e^{i \phi'}
\end{equation}
The simplest example is the constant mapping:
\begin{equation}
\mathcal{H}_0: \qquad e^{i \phi} \pad{\maparrow} 1
\label{b:equation}
\end{equation}
The next simplest is the identity mapping:
\begin{equation}
\mathcal{H}_1: \qquad e^{i \phi} \pad{\maparrow} e^{i \phi}
\label{c:equation}
\end{equation}
By visualizing the mappings as vectors located at each point on a circle, we
realize that it is impossible to deform one of the above mappings into the
other using only smooth gauge transformations.  At some point on the circle,
one of the vectors will turn clockwise during the deformation, while its
neighbor will be required to turn counterclockwise.

A set of mappings which can be smoothly deformed into one another is labeled a
homotopy class.  Thus, the two mappings of \eqref{b:equation} and
\eqref{c:equation} are representative members of two separate homotopy
classes.  In fact, we can easily write down an infinite number of mappings,
each of which belongs to a new homotopy class:
\begin{equation}
\mathcal{H}_\nu: \qquad e^{i \phi} \quad \maparrow \quad e^{i \nu \phi}
\end{equation}
where $\nu$ is an integer which enumerates the classes.  For any $U(1)$ gauge
field configuration we write down, there is one and only one mapping among
those above such that the configuration can be smoothly deformed so that its
gauge field orientation at infinity is described by the mapping.

The analog for QCD is mappings of $SU(N_c = 3)$ onto $S_3$.  In this case as
well, the gauge group is rich enough for there to be an infinite number of
homotopy classes.  In each homotopy class there is a field configuration with
the minimum action.  These are the infinite vacuum states of an $SU(N_c = 3)$
gauge theory.  For the homotopy class which contains the constant mapping
$\mathcal{H}_0$, the vacuum state will correspond to the traditional vacuum:
a constant zero gauge field.  However, for all other homotopy classes, the
minimum action configurations will contain local kinks in the gauge field.
These kinks are known as instantons, and each will have a local non-zero action
density associated with it.

The winding number $\nu$ of a field configuration, which is used to label its
homotopy class, can be shown to equal:
\begin{align}
\nu & = \frac{1}{16 \pi^2} \int_x{ \tr{ F_{\mu\nu} \tilde{F}^{\mu\nu} } }
\notag \\
& = \frac{1}{2 N_f \g^2} \int_x{ \Theta }
\end{align}

Thus, we see that the vacuum of QCD is not simple, but rather there is an
infinite number of potential vacua $| \nu \ket$.  The true vacuum $| \theta
\ket$ is then a linear combination of the $| \nu \ket$.  In order for this
vacuum to have the correct behavior under gauge transformations, the
coefficients of the combination must have the form:
\begin{equation}
| \theta \ket = \sum_\nu e^{i \nu \theta} | \nu \ket
\end{equation}
where $\theta$ is a new arbitrary parameter of the theory.  Incorporating this
linear combination of vacua into the path integral of QCD effectively adds an
additional term to the Lagrangian:
\begin{align}
Z_\text{QCD} & = \sum_\nu e^{i \nu \theta} \int{ \fD{A_\mu}_\nu \fD{q} \fD{\qb}
~ e^{ -\int_x{ \Lag{QCD} }}} \\
& = \int{ \fD{A_\mu} \fD{q} \fD{\qb} ~ e^{ -\int_x{ \Lag{QCD eff} }}}
\end{align}
where $\fD{A_\mu}_\nu$ represents a functional integral over field
configurations with winding number $\nu$ and:
\begin{align}
\Lag{QCD eff} & = \Lag{QCD} + \Lag{$\theta$} \notag \\
& = \Lag{QCD} + i \frac{\theta}{16 \pi^2} \tr{ F_{\mu\nu} \tilde{F}^{\mu\nu} }
\end{align}

From \eqref{d:equation} we can see that the ABJ anomaly has the same
form as this new term.  Thus, if we are assuming massless quarks, we have the
freedom to absorb $\theta$ via a $U(1)_A$ rotation of the quark fields.
However, for massive quarks, we lose that freedom.  Because the quark mass
eigenstates of QCD are not the same as those of the full Standard Model, the
actual quark mass matrix $\tilde{\qmassm}$ will include some chiral phase.
In order to remove this chiral phase and put the mass matrix into the form
standard for QCD, we apply a $U(1)_A$ rotation to the quark fields with a
magnitude equal to $\arg \det \tilde{\qmassm}$.  This binds us to that specific
rotation, and restricts us from absorbing $\theta$.  After this rotation, the
physical value $\bar{\theta}$ for the coefficient in $\Lag{$\theta$}$ becomes:
\begin{equation}
\bar{\theta} = \theta + \arg \det \tilde{\qmassm}
\end{equation}

Each choice of value for $\theta$ corresponds to a unique vacuum choice for
QCD.  Such a continuous set of vacuum states is reminiscent of spontaneous
symmetry breaking.  In fact the $\theta$ vacua are the multiple vacua of the
spontaneously broken $\tilde{J}^5_\mu$ symmetry discussed above.  This concurs
with the fact that $U(1)_A$ rotations move one among the $\theta$ vacua.

\section{Strong $\CP$}
As demonstrated by \eqref{e:equation}, $\Lag{$\theta$}$ is a total
divergence and thus can have no effect on perturbation-theory calculations.
However, it still generates non-perturbative symptoms.  In particular,
$\Lag{$\theta$}$ breaks $\CP$ symmetry and leads to $\CP$ violating effects.
The most significant of these violations is a correction to the zero
neutron dipole moment.  However, the neutron dipole moment is strongly bound by
experiment, $d_n < 6.3 \times 10^{-26} ~ e ~ \text{cm}$ \cite{Harris:1999jx}.
This in turn leads to a bound on $\bar{\theta}$ \cite{Nir:2001ge}:
\begin{equation}
\bar{\theta} < 10^{-10}
\end{equation}
This extreme smallness of $\bar{\theta}$ is known as the strong $\CP$ problem.

Were $\bar{\theta}$ the only parameter in the Standard Model to break $\CP$
symmetry, its extreme smallness would be of no particular concern.  In such a
situation, $\bar{\theta}$ can only be multiplicatively renormalized, and thus
can remain small over a broad range of scales.  However, $\CP$ is also broken by
weak interactions.  Thus, even if $\bar{\theta}$ were small at some given
scale, at other scales it would be drastically additively renormalized by
the other $\CP$-breaking elements of the theory.  Therefore, the observed
smallness of $\bar{\theta}$ is deemed unnatural \cite{'tHooft:1994gh}, and
there is likely some as-yet-unknown structure which enforces this smallness.
As such, the strong $\CP$ problem has traditionally been seen as a chink in the
armor of the Standard Model, and as a fertile starting point for
beyond-the-Standard-Model theoretical work.

One proposed solution to the strong $\CP$ problem is through the introduction of
an additional particle:  the axion \cite{Peccei:1977ur}.  This field $\sigma$
would couple to the quarks as a phase factor on the quark mass matrix:
\begin{equation}
\Lag{axion} = \frac{1}{2} \partial_\mu \sigma \partial^\mu \sigma + \qb
\tilde{\qmassm} e^{-i \sigma} q
\end{equation}
In such a situation both $\theta$ and the chiral phase of $\tilde{\qmassm}$
can be absorbed via a field redefinition of the axion.  However, thus far
experimental and astrophysical searches for the axion have been unsuccessful.

A second possible solution is the Nelson-Barr mechanism \cite{Nelson:1984zb,
Barr:1984qx}.  In this scenario, $\CP$ is a symmetry of the fundamental theory
and $\CP$ violations are due to a spontaneous breaking of $\CP$ at the GUT
scale.  Included in the theory are several beyond-the-Standard-Model particles,
including flavors of heavy fermions.  Below the GUT scale, $\bar{\theta}$ gains
a non-zero value proportional to the heavy fermion mass divided by the GUT
scale.  The smallness of this ratio, and therefore $\bar{\theta}$, is no longer
unnatural because, if these fermion flavors were massless, new chiral
symmetries would be introduced into the theory.  In other words, the ratio
obtains only a multiplicative renormalization as we change scales, and can thus
remain small for all scales.

Another potential solution to the strong $\CP$ problem requires that a single
flavor of quark remain massless.  This is in fact the only proposed solution
which does not require physics from beyond the Standard Model.  The obvious
candidate for this massless quark is the lightest flavor:  the up quark.  If
the up quark were massless, we would regain the freedom to apply $U(1)_A$
rotations to it, and through the ABJ anomaly we could absorb the $\CP$
violating term.  Surprisingly, current experimental data does not rule out a
massless up quark.  However, through the lattice calculations presented here,
we show that a massless up quark is unlikely.

It should be noted that a massless up quark would not be the end of the
dilemma.  It would simply shift the focus from an explanation of the smallness
of $\bar{\theta}$ to an explanation of the masslessness of the up quark.
However, several beyond-the-Standard-Model scenarios for the dynamic generation
of quark masses, in which a massless up quark is a natural consequence, have
been proposed \cite{Leurer:1994gy, Banks:1994yg, Kaplan:1998jk}.

\chapter{Chiral Perturbation Theory} \label{n:section}
As a consequence of QCD's non-Abelian nature, it has two important
characteristics:  asymptotic freedom and confinement.  Asymptotic freedom
signifies that at very high energies the coupling of the strong force is
greatly reduced, and quarks behave as if free.  In this regime perturbation
theory is successful.  Conversely, confinement indicates that at low energies
the interactions are strong enough to confine quarks and gluons within bound
states.  It is for this reason that only color singlet states are directly
observed, never free quarks or gluons.  These low energies are beyond the
radius of convergence of perturbation theory, and the method becomes useless.

While it is impossible at low energies to apply perturbation theory to QCD's
fundamental degrees of freedom, we can use clues from QCD to build an effective
quantum field theory of its bound states.  Such an effective theory is known as
Chiral Perturbation Theory (ChPT).  We focus our attention on to the lightest
bound states of QCD, the octet of light pseudoscalar mesons.  However, the
ideas discussed here can also be used to construct an effective theory for the
baryons.

Clues as to the appropriate form for our effective field theory come from QCD
in the form of symmetries.  The Lagrangian of ChPT must respect any symmetries
respected by QCD.  Confinement insures that all bound states are singlets
under color.  Thus, color symmetry does not restrict the form of our chiral
Lagrangian.  Both flavor and Lorentz symmetry, on the other hand, place strong
restrictions on the terms which the Lagrangian may include.  For the moment we
will assume massless quarks, so that $SU(N_f)_V \otimes SU(N_f)_A$ flavor
symmetry is an exact symmetry of QCD.

Because we have no information other than the symmetries of QCD, we are forced
to include in our chiral Lagrangian every possible term which respects both
flavor and Lorentz symmetry.  Of course, there are an infinite number of such
terms.  So, we order the terms by their importance and then ignore those whose
importance is beyond our chosen sensitivity.  The effective Lagrangian is thus
an expansion in some parameter which establishes the importance of each term:
\begin{equation}
\Lag{ChPT} = \Lag{ChPT}^{(2)} + \Lag{ChPT}^{(4)} + \Lag{ChPT}^{(6)} + \dotsb
\end{equation}
Because we are attempting to build a low-energy theory for the light
pseudoscalar mesons, we will take terms with lower powers of meson momentum to
be of greater importance.

We collect the meson fields $\pi^a$ into a flavor-space matrix $\Phi$,
multiplying each meson field by its corresponding broken flavor-symmetry
generator:
\begin{align}
\Phi & \equiv \pi^a \tau^a = \frac{1}{\sqrt{2}} \biggl( \mathfrak{Q}
\bar{\mathfrak{Q}} - \frac{1}{3} \identity \Tr{ \mathfrak{Q} \bar{\mathfrak{Q}}
} \biggr) \notag \\[3mm]
& = \frac{1}{\sqrt{2}} \begin{bmatrix}
\frac{2}{3} u \bar{u} - \frac{1}{3} d \bar{d} - \frac{1}{3} s \bar{s} & u
\bar{d} & u \bar{s} \\[3mm]
d \bar{u} & -\frac{1}{3} u \bar{u} + \frac{2}{3} d \bar{d} - \frac{1}{3} s
\bar{s} & d \bar{s} \\[3mm]
s \bar{u} & s \bar{d} & -\frac{1}{3} u \bar{u} - \frac{1}{3} d \bar{d} +
\frac{2}{3} s \bar{s}
\end{bmatrix}
\end{align}
where:
\begin{align}
\mathfrak{Q} = \begin{bmatrix}
u \\[3mm]
d \\[3mm]
s
\end{bmatrix}
&&
\bar{\mathfrak{Q}} = \begin{bmatrix}
\bar{u} & \bar{d} & \bar{s}
\end{bmatrix}
\end{align}
The singlet piece of $\Phi$ has been subtracted, effectively removing $\eta'$
from the matrix.  The fields $\pi^a$ are known as the Cartesian components of
the mesons and do not correspond to the physical eigenstates of the
light-pseudoscalar-meson fields, which can be identified as:
\begin{equation}
\Phi = \pi^a \tau^a = \frac{1}{\sqrt{2}} \begin{bmatrix}
\frac{1}{\sqrt{2}} \pi^0 + \frac{1}{\sqrt{6}} \eta & \pi^+ & K^+ \\[3mm]
\pi^- & -\frac{1}{\sqrt{2}} \pi^0 + \frac{1}{\sqrt{6}} \eta & K^0 \\[3mm]
K^- & \bar{K}^0 & -\frac{2}{\sqrt{6}} \eta
\end{bmatrix}
\end{equation}
The unitary matrix $\Sigma$ is then built from the meson-field matrix:
\begin{equation}
\Sigma \equiv e^{ 2 i \pi^a \tau^a / f } \in SU(N_f)
\end{equation}

We now define how the meson fields transform under flavor symmetry:
\begin{equation}
\Sigma \pad{\symarrow} \Sigma' = L \Sigma R^\dagger \\
\end{equation}
using $L$ and $R$ from \eqref{f:equation} and \eqref{g:equation}.  With this
definition $\Phi$ transforms linearly under pure vector transformations:
\begin{equation}
\Phi \pad{\symarrow} \Phi' = V \Phi V^\dagger
\end{equation}
where $L = R = V$.  Otherwise, the transformation of the meson fields is
non-linear.

\section{Leading Order Chiral Perturbation Theory}
Considering all terms allowed by symmetry considerations and then expanding in
terms of $p^2 / \Lambda_\chi^2$ --- where $p$ is the meson momentum
and $\Lambda_\chi$ is some scale beyond which the expansion, and ChPT, breaks
down --- we find only one meaningful term at lowest order:
\begin{equation}
\Lag{ChPT}^{\text{LO}'} = \frac{f^2}{4} \Tr{ \partial_\mu \Sigma^\dagger
\partial^\mu \Sigma }
\end{equation}
Note that derivatives of $\Sigma$ correspond to powers of meson momentum.
While the structure of the term is determined by symmetry, its coefficient $f$
is not.  The value of $f$ is set by the behind-the-scenes dynamics of QCD,
which ChPT hides from us.  In the chiral limit, $f$ equals the pion decay
constant:
\begin{equation}
f = f_\pi + O(m_q)
\end{equation}
where we are using the normalization $f_\pi \simeq 92.4 \, \text{MeV}$.
Expanding $\Lag{ChPT}^{\text{LO}'}$ in terms of the meson fields $\pi^a$
reveals a conventional scalar kinetic term for the mesons as well as a tower of
interactions involving an increasing number of meson fields.

Non-zero quark mass breaks the flavor symmetry of QCD, an important building
block of our chiral Lagrangian.  However, there are two critical points which
allow us to include the effects of quark mass in ChPT.  First, we know the form
through which the quark masses break flavor symmetry: the quark mass matrix
$\qmassm$.  Thus, we can correctly break the symmetry in our effective theory
by adding terms to the Lagrangian which break $SU(N_f)_V \otimes SU(N_f)_A$
only via insertions of $\qmassm$.  Secondly, the quark masses are small,
presumably with respect to $\Lambda_\chi$.  So, the most important of such
terms will be those with a low power of $\qmassm$.

While the form of $\qmassm$ is known, it is scaled by an unknown constant $\mu$
with units of mass:
\begin{equation}
\chi \equiv 2 \mu \qmassm = 2 \mu \diag \bigl( \{ m_i \} \bigr)
\end{equation}
In the chiral limit, $\mu$ is directly related to the chiral condensate:
\begin{equation}
\mu = - \frac{\bra \qb q \ket}{f_\pi^2 N_f} + O(m_q)
\end{equation}

Now, considering all terms which respect Lorentz and flavor symmetry, except
via insertions of $\qmassm$, and expanding simultaneously to $O(p^2 /
\Lambda_\chi^2)$ and $O(\mu \qmassm / \Lambda_\chi^2)$, we determine the LO
Euclidean chiral Lagrangian:
\begin{align}
\Lag{ChPT}^\text{LO} & = \Lag{ChPT}^{(2)} \\
& = \frac{f^2}{4} \Tr{ \partial_\mu \Sigma^\dagger \partial^\mu \Sigma } -
\frac{f^2}{4}  \Tr{ \Sigma^\dagger \chi + \chi \Sigma }
\end{align}
The low-energy non-perturbative dynamics of QCD have been boiled down, at this
order, to two unknown constants:  $f$ and $\mu$.

The arrangement of the meson fields in $\Sigma$ seems quite arbitrary.  In fact
there are an infinite number of other arrangements and corresponding
representations of $SU(N_f)_V \otimes SU(N_f)_A$, each of which would result in
a new form for the chiral Lagrangian.  However, any such representation
will result in exactly the same meson physics as the representation presented
here.  This is due to a fundamental characteristic of effective quantum field
theories known as universality.  For a low-energy effective theory, the
symmetries of the theory alone determine the resultant physics, and the details
of the symmetries' representation are unimportant.

\section{Leading Order Quark Mass Ratios}
With the LO chiral Lagrangian in hand, predictions can be made for quantities
in which the constants of the Lagrangian cancel away, such as ratios of
semi-leptonic decay rates.  The meson masses can not be predicted, but their
dependence on the quark masses can.  By inverting those relationships and using
experimental values for the meson masses, we can calculate the light-quark-mass
ratios.

The LO ChPT expressions for the meson masses are straightforward:
\begin{align}
\mpnQCD^2 = \mppQCD^2 = \mpmQCD^2 & = \mu \bigl( m_u + m_d \bigr) \\
\mkpQCD^2 = \mkmQCD^2 & = \mu \bigl( m_u + m_s \bigr) \\
\mknQCD^2 = \mkbnQCD^2 & = \mu \bigl( m_d + m_s \bigr)
\end{align}
These masses $\QCD{\m}_X$ are often referred to as the mesons' QCD masses.
They are related to the physical meson masses by a Quantum Electrodynamics
(QED) correction.  However at lowest order, only the masses of the charged
mesons are corrected and all by the same amount, a result known as Dashen's
theorem \cite{Dashen:1969eg}:
\begin{align}
\delta_{E_\pi} = \delta_{E_K} & = \delta_E \notag \\
& = \mpp^2 - \mpn^2 + O(e^2 m_q)
\end{align}
where $\m_X$ represents a meson's physical mass.  Thus, QED adds only one
additional unknown parameter to the picture:
\begin{align}
\mpn^2 & = \mu \bigl( m_u + m_d \bigr) \\
\mpp^2 = \mpm^2 & = \mu \bigl( m_u + m_d \bigr)  + \delta_E \\
\mkp^2 = \mkm^2 & = \mu \bigl( m_u + m_s \bigr) + \delta_E \\
\mkn^2 = \mkbn^2 & = \mu \bigl( m_d + m_s \bigr)
\end{align}
We are ignoring and will continue to ignore a small correction to $\mpn^2$
due to the mixing of the physical $\pi^0$ and $\eta$ states.  The above
relationships can be inverted to predict the quark-mass ratios
\cite{Weinberg:1977hb}:
\begin{align}
\frac{m_u}{m_d} & = \frac{\mkp^2 - \mkn^2 + 2 \mpn^2 - \mpp^2}{\mkn^2 - \mkp^2 +
\mpp^2} = 0.55 \\
\frac{m_s}{m_d} & = \frac{\mkn^2 + \mkp^2 - \mpp^2}{\mkn^2 - \mkp^2 + \mpp^2} =
20.1
\end{align}
This LO calculation clearly suggests that the up quark mass is non-zero.
However, we will see that the NLO calculation muddy the water.

\section{Next-To-Leading Order Chiral Perturbation Theory}
When we venture beyond tree-level and attempt to calculate one-loop corrections
with our LO chiral Lagrangian, we find that the corrections have new matrix
structures, and thus the divergences can not be absorbed into our LO terms.  As
it turns out, these new terms are the same terms which we would have included
in our Lagrangian had we decided to go out to NLO in meson momentum and quark
mass.  Thus, if we would like to work at one-loop using our LO Lagrangian, we
must also include tree-level effects from NLO terms.  The one-loop corrections
will then renormalize the coefficients of our NLO terms.  This behavior is a
direct consequence of ChPT being an effective field theory and will occur order
by order.

Adding the NLO terms, the Euclidean chiral Lagrangian becomes
\cite{Gasser:1985gg}:
\begin{align}
\Lag{ChPT}^\text{NLO} \pad{=} & \Lag{ChPT}^{(2)} + \Lag{ChPT}^{(4)} \notag \\
\pad{=} & \frac{f^2}{4} \Tr{ \partial_\mu \Sigma^\dagger \partial^\mu \Sigma }
- \frac{f^2}{4}  \Tr{ \Sigma^\dagger \chi + \chi \Sigma } \notag \\
& - L_1 \Bigl( \Tr{ \partial_\mu \Sigma^\dagger \partial^\mu \Sigma } \Bigr)^2
\notag \\
& - L_2 \Bigl( \Tr{ \partial_\mu \Sigma^\dagger \partial_\nu \Sigma } \Tr{
\partial^\mu \Sigma^\dagger \partial^\nu \Sigma } \Bigr) \notag \\
& - L_3 \Bigl( \Tr{ \partial_\mu \Sigma^\dagger \partial^\mu \Sigma \partial_\nu
\Sigma^\dagger \partial^\nu \Sigma } \Bigr) \notag \\
& + L_4 \Bigl( \Tr{ \partial_\mu \Sigma^\dagger \partial^\mu \Sigma } \Tr{
\Sigma^\dagger \chi + \chi \Sigma } \Bigr) \notag \\
& + L_5 \Bigl( \Tr{ \partial_\mu \Sigma^\dagger \partial^\mu \Sigma \bigl(
\Sigma^\dagger \chi + \chi \Sigma \bigr) } \Bigr) \notag \\
& - L_6 \Bigl( \Tr{ \Sigma^\dagger \chi + \chi \Sigma } \Bigr)^2 \notag \\
& - L_7 \Bigl( \Tr{ \Sigma^\dagger \chi - \chi \Sigma } \Bigr)^2 \notag \\
& - L_8 \Bigl( \Tr{ \Sigma^\dagger \chi \Sigma^\dagger \chi } + \Tr{ \chi \Sigma
\chi \Sigma } \Bigr)
\label{j:equation}
\end{align}
where $L_i$ are additional unknown constants known as the Gasser-Leutwyler (GL)
coefficients.  They are not constrained by chiral symmetry.  Rather, they
parameterize our ignorance concerning the low-energy dynamics of QCD.  Terms
which couple the mesons to a background gauge field have been dropped.  Note
that each term in $\Lag{ChPT}^{(4)}$ contains either four derivatives, two
derivatives and one power of $\chi$, or two powers of $\chi$.

The magnitude of the chiral scale $\Lambda_\chi$ is an important element in
determining the validity of ChPT.  By looking at the loop corrections to the
NLO terms in the chiral Lagrangian, we can develop an estimate for
$\Lambda_\chi$.  We find that when the meson momentum in loops is cutoff at
$\Lambda = 4 \pi f$, the radiative corrections from the LO Lagrangian
are on the same order at the contributions from tree-level NLO diagrams.  So,
we can only reasonably expect ChPT to be useful below that scale, and thus:
\begin{equation}
\Lambda_\chi \simeq 4 \pi f
\end{equation}
Since $f = f_\pi$ at lowest order, we find that $\Lambda_\chi \simeq 1.2 ~
\text{GeV}$.  We are now in a position to evaluate the validity of our
expansion in the quark mass.  At lowest order $\mu \qmassm \simeq
\mkn^2$, and thus $(\mu \qmassm / \Lambda_\chi^2) \simeq 0.2$.  The quark
mass expansion is sound.  Therefore, if we restrict ourselves to meson momentum
below $500 \, \text{MeV}$, we can expect ChPT to produce accurate predictions.

\section{Next-To-Leading Order Quark Mass Ratios}
The NLO expressions for the meson masses are calculable from the NLO
Lagrangian.  Using notation similar to the literature, $\mpQCD$ represents the
pion mass without QED corrections, while $\mkQCD$ represents the kaon mass
neglecting the mass difference $m_d - m_u$ and without QED corrections.  These uncorrected masses are related to the physical meson masses via:
\begin{align}
\mpQCD^2 & = \mpnQCD^2 = \mpn^2 \\
2 \mkQCD^2 & = \mknQCD^2 + \mkpQCD^2 = \mkn^2 + \mkp^2 - \bigl( 1 + \Delta_E
\bigr) \bigl( \mpp^2 - \mpn^2 \bigr)
\end{align}
The above equalities are corrected by terms of order $O(e^2 m_q)$ and $O \bigl(
(m_d - m_u)^2 \bigr)$, as well as by NNLO ChPT.  The parameter $\Delta_E$
allows for a difference between the QED contributions to the pion and kaon
masses:
\begin{equation}
\frac{\delta_{E_K}}{\delta_{E_\pi}} = 1 + \Delta_E
\end{equation}
Use of Dashen's theorem, as was done in the LO case, is equivalent to using
$\Delta_E = 0$.  A comprehensive study of experimental data
\cite{Bijnens:1997kk, Amoros:2001cp} suggests a significant deviation from
Dashen's theorem and gives the result:
\begin{equation}
\Delta_E = 0.84 \pm 0.25
\label{w:equation}
\end{equation}
This is the value which will be used in subsequent calculations, unless stated
otherwise.

Using the NLO chiral Lagrangian, the uncorrected NLO meson masses are
calculated \cite{Gasser:1985gg}:
\begin{align}
\mpQCD^2 = \mu \bigl( m_u + m_d \bigr) \biggl\{ 1
& + \chlog{\pi} - \tfrac{1}{3} \chlog{\eta} \notag \\
& + \frac{8}{f^2} \mu \bigl( m_u + m_d \bigr) \bigl( 2 L_8 - L_5 \bigr) \notag
\\
& + \frac{16}{f^2} \mu \bigl( m_u + m_d + m_s \bigr) \bigl( 2 L_6 - L_4 \bigr)
\biggr\}
\label{bg:equation}
\end{align}
\begin{align}
\mkQCD^2 = \mu \bigl( \mh + m_s \bigr) \biggl\{ 1
& + \tfrac{2}{3} \chlog{\eta} \notag \\
& + \frac{8}{f^2} \mu \bigl( \mh + m_s \bigr) \bigl( 2 L_8 - L_5 \bigr) \notag
\\
& + \frac{16}{f^2} \mu \bigl( m_u + m_d + m_s \bigr) \bigl( 2 L_6 - L_4 \bigr)
\biggr\}
\end{align}
where $\mh = \tfrac{1}{2} ( m_u + m_d )$ and $\chlog{X}$ represent chiral logs
which arise from divergent loop corrections.  We choose to cut these loops off
at $\Lambda = \Lambda_\chi = 4 \pi f$:
\begin{equation}
\chlog{X} \equiv \frac{\QCD{\m}_X^2}{2 ( 4 \pi f )^2} \ln \frac{\QCD{\m}_X^2}{(
4 \pi f )^2}
\end{equation}
The additional correction terms involving GL coefficients come from NLO
tree-level interactions.

We can use these expressions to construct two combinations of the meson masses
which experience an equal correction $\Delta_M$ at NLO
\cite{Gasser:1985gg}:
\begin{align}
\frac{\mkQCD^2}{\mpQCD^2} & = \frac{m_s + \mh}{m_d + m_u} \Biggl\{ 1 + \Delta_M
 + O(m_q^2) \Biggr\} \\
\frac{\mknQCD^2 - \mkpQCD^2}{\mkQCD^2 - \mpQCD^2} & = \frac{m_d - m_u}{m_s -
\mh} \Biggl\{ 1 + \Delta_M + O(m_q^2) \Biggr\}
\label{u:equation}
\end{align}
where:
\begin{align}
\Delta_M & = -\chlog{\pi} + \chlog{\eta} + \frac{8}{f^2} \mu \bigl( m_s - \mh
\bigr) \bigl( 2 L_8 - L_5 \bigr) \notag \\
& = -\chlog{\pi} + \chlog{\eta} + \frac{8}{f_\pi^2} \bigl( \mkQCD^2 - \mpQCD^2
\bigr) \bigl( 2 L_8 - L_5 \bigr)
\label{h:equation}
\end{align}
Thus, a ratio independent of NLO corrections can be defined:
\begin{equation}
Q^2 \equiv \frac{m_s^2 - \mh^2}{m_d^2 - m_u^2} = \frac{\mkQCD^2}{\mpQCD^2}
\frac{\mkQCD^2 - \mpQCD^2}{\mknQCD^2 - \mkpQCD^2} + O(m_q^2)
\label{o:equation}
\end{equation}
Replacing the QCD meson masses with physical masses results in:
\begin{multline}
Q^2 = \frac{1}{4 \mpn^2} \frac{ \mkn^2 + \mkp^2 - ( 1 + \Delta_E ) \mpp^2 + ( 1
+ \Delta_E ) \mpn^2 } { \mkn^2 - \mkp^2 + ( 1 + \Delta_E ) \mpp^2 - ( 1 +
\Delta_E ) \mpn^2 } \\
 \times \bigl( \mkn^2 + \mkp^2 - ( 1 + \Delta_E ) \mpp^2 - ( 1 - \Delta_E )
\mpn^2 \bigr) \\
= \bigl( 22.01 \pm 0.57 \bigr)^2
\end{multline}
where experimental error in the meson masses is overwhelmed by uncertainty in
$\Delta_E$.  Use of Dashen's Theorem instead of \eqref{w:equation} results in a
central value of $Q = 24.18$.  The quark-mass ratios can be represented as an
ellipse \cite{Kaplan:1986ru} which conforms to the equation:
\begin{equation}
\biggl( \frac{m_u}{m_d} \biggr)^2 + \frac{1}{Q^2} \biggl( \frac{m_s}{m_d}
\biggr)^2 = 1
\end{equation}
where $\mh$ has been taken to be small relative to $m_s$.

We now can solve for the quark-mass ratios, but only in terms of the unknown
NLO correction $\Delta_M$ \cite{Cohen:1999kk}:
\begin{align}
\frac{m_s}{\mh} & = \frac{2 \mkQCD^2}{\mpQCD^2 \bigl( 1 + \Delta_M \bigr)} - 1
\\
\frac{m_u}{\mh} & = 1 - \frac{\mkQCD^2}{\mpQCD^4} \frac{\mkQCD^2 - \bigl( 1 +
\Delta_M \bigr) \mpQCD^2 }{ Q^2 \bigl( 1 + \Delta_M \bigr) ^2 }
\label{i:equation}
\end{align}
Setting $m_u$ equal to zero and solving \eqref{i:equation} for
$\Delta_M$, we find:
\begin{align}
\Delta_M & = \frac{\mkQCD^2}{4 Q^2 \mpQCD^2} \biggl( -1 \pm \sqrt{ 1 + 4 Q^2 }
\biggr) - 1 \notag \\[2mm]
& = -0.404 ~ \pm ~ 0.002 ~ \pm ~ 0.015 ~ \pm ~ 0.16
\label{t:equation}
\end{align}
where the first uncertainty is due to experimental error in the meson masses,
the second is due to uncertainty in $\Delta_E$, and the third comes from
an assumption that unaccounted for NNLO corrections are on the order of
$\Delta_M^2$.  Use of Dashen's theorem instead of \eqref{w:equation} results in
the value $\Delta_M = -0.455 \pm 0.002 \pm 0.21$, for which no uncertainly due
to $\Delta_E$ is given.  The value for $\Delta_M$ from \eqref{t:equation}
corresponds, via \eqref{h:equation}, to $2 L_8 - L_5 = ( -1.25 \pm 0.77 )
\times 10^{-3}$, or the range:
\begin{equation}
-2.02 \times 10^{-3} \pad{<} 2 L_8 - L_5 \pad{<} -0.48 \times 10^{-3}
\label{ca:equation}
\end{equation}
Using Dashen's theorem corresponds to $2 L_8 - L_5 = ( -1.49 \pm 0.98 ) \times
10^{-3}$.  Sharpe quotes a range similar to \eqref{ca:equation} in
\cite{Sharpe:2000bn}.  The second root of \eqref{i:equation} is not considered
as it requires the corrections to be even more improbably large.

From \eqref{t:equation} we see that if the NLO corrections to ChPT are quite
large, the massless up quark remains a possibility.

At times, an alternative normalization of the GL coefficients is used:
\begin{equation}
\alpha_i = 8 ( 4 \pi )^2  L_i
\end{equation}
This is a more natural normalization in that each $\alpha_i$ can be expected to
be on the order of one.  Under this normalization the range consistent with a
massless up quark is:
\begin{equation}
-2.6 \pad{<} 2 \alpha_8 - \alpha_5 \pad{<} -0.6
\label{cb:equation}
\end{equation}

As discussed above, one-loop graphs generated by the LO chiral Lagrangian
renormalize the GL coefficients.  Thus, the GL coefficients are functions of the
renormalization scale.  That scale dependence has a simple form
\cite{Gasser:1985gg}:
\begin{equation}
L_i (\Lambda_2) = L_i (\Lambda_1) + \frac{\varGamma_i}{(4 \pi)^2} \ln \frac{\Lambda_1}{\Lambda_2}
\end{equation}
where:
\begin{align}
\varGamma_5 = \frac{3}{8}
&&
\varGamma_8 = \frac{5}{48}
\end{align}
$L_7$ is scale invariant, with $\varGamma_7 = 0$.  The full set of scaling
coefficients $\varGamma_i$ can be found in \cite{Bijnens:1994qh}.  Unless
otherwise stated, we report all GL coefficients at $\Lambda = 4 \pi f_\pi$.

\section{Phenomenological Results}
In order to unquestionably rule out a massless up quark, the value of $2 L_8 -
L_5$ must be determined.  Because the GL coefficients encapsulate the low
energy dynamics of QCD, they could in principle be analytically calculated
directly from the QCD Lagrangian.  However, the limitations of perturbation
theory at these low energy scales has stymied such efforts.  Thus, the primary
source of information about these constants is phenomenological studies.

\subsection{Meson Decay Constants}
Calculating NLO expressions for the meson decay constants using the NLO chiral
Lagrangian, we find \cite{Gasser:1985gg}:
\begin{align}
f_\pi = f \biggl\{ 1 & - 2 \chlog{\mu} - \chlog{K} \notag \\
& + \frac{4}{f^2} \mu \bigl( m_u + m_d \bigr) L_5 \notag \\
& + \frac{8}{f^2} \mu \bigl( m_u + m_d + m_s \bigr) L_4 \biggr\}
\label{bh:equation}
\end{align}
\begin{align}
f_K = f \biggl\{ 1 & - \tfrac{3}{4} \chlog{\pi} - \tfrac{3}{2} \chlog{K} -
\tfrac{3}{4} \chlog{\eta} \notag \\
& + \frac{4}{f^2} \mu \bigl( \mh + m_s \bigr) L_5 \notag \\
& + \frac{8}{f^2} \mu \bigl( m_u + m_d + m_s \bigr) L_4 \biggr\}
\end{align}
Their ratio isolates $L_5$:
\begin{equation}
\frac{f_\pi}{f_K} = 1 + \Delta_f + O(m_q^2)
\label{q:equation}
\end{equation}
where:
\begin{align}
\Delta_f & = \tfrac{5}{4} \chlog{\pi} - \tfrac{1}{2} \chlog{K} - \tfrac{3}{4}
\chlog{\eta} + \frac{4}{f^2} \mu \bigl( m_s - \mh \bigr) L_5 \notag \\
& = \tfrac{5}{4} \chlog{\pi} - \tfrac{1}{2} \chlog{K} - \tfrac{3}{4}
\chlog{\eta} + \frac{4}{f_\pi^2} \bigl( \mkQCD^2 - \mpQCD^2 \bigr) L_5
\end{align}
We can then fix $L_5$ using experimental values:
\begin{equation}
L_5 = \bigl( 0.51 \pm 0.47 \bigr) \times 10^{-3}
\label{p:equation}
\end{equation}
where the uncertainty comes from both experimental error and an assumption that
the unknown NNLO corrections to \eqref{q:equation} are on the order of
$\Delta_f^2$.  Use of Dashen's theorem instead of \eqref{w:equation} has very
little effect on the value, resulting in $L_5 = ( 0.50 \pm 0.47 ) \times
10^{-3}$.

\subsection{Gell-Mann-Okubo Relation}
At leading order ChPT confirms the traditional Gell-Mann-Okubo relation:
\begin{equation}
\mpQCD^2 + 3 \meQCD^2 = 4 \mkQCD^2
\end{equation}
At NLO, ChPT predicts a correction which accounts for the meson masses'
observed deviation from the relation:
\begin{equation}
\Delta_\text{GMO} \equiv \frac{ 4 \mkQCD^2 - \mpQCD^2 - 3 \meQCD^2 }{ \meQCD^2
- \mpQCD^2 } = 0.218
\end{equation}
This NLO correction is:
\begin{align}
\Delta_\text{GMO} = & -2 \frac{ 4 \mkQCD^2 \chlog{K} - \mpQCD^2 \chlog{\pi} - 3
\meQCD^2 \chlog{\eta} }{ \meQCD^2 - \mpQCD^2 } \notag \\
& - \frac{6}{f_\pi^2} \bigl( \meQCD^2 - \mpQCD^2 \bigr) \bigl( 12 L_7 + 6 L_8 -
L_5
\bigr)
\end{align}
Using the observed deviation we can bind a linear combination of the GL
coefficients:
\begin{equation}
12 L_7 + 6 L_8 - L_5 = \bigl( -1.08 \pm 0.24 \bigr) \times 10^{-3}
\label{s:equation}
\end{equation}
where uncertainty in $\Delta_E$ and the light meson masses is dwarfed by an
assumption that the unknown NNLO corrections to the Gell-Mann-Okubo relation
are on the order of $\Delta_\text{GMO}^2$.  Use of Dashen's theorem instead of
\eqref{w:equation} results in the value $12 L_7 + 6 L_8 - L_5 = ( -1.10 \pm
0.26 ) \times 10^{-3}$.

With $L_7$ unknown, the combination $2 L_8 - L_5$ remains undetermined.

\section{Kaplan-Manohar Ambiguity}
We will find that it is impossible to fix the value of $2 L_8 - L_5$ using only
expressions from ChPT combined with experimental measurements.  The chiral
Lagrangian contains an ambiguity among its parameters which prevents us from
using the predictions of ChPT to determine $L_6$, $L_7$, $L_8$, and the
quark-mass ratios.  This ambiguity is known as the Kaplan-Manohar (KM)
ambiguity
\cite{Kaplan:1986ru}.

The chiral Lagrangian is invariant under a certain redefinition of its
coefficients.  To see this we shift the quark masses in the following way:
\begin{align}
m_u & \pad{\redefarrow} \ring{m}_u = m_u + \lambda m_d m_s \\
m_d & \pad{\redefarrow} \ring{m}_d = m_d + \lambda m_u m_s \\
m_s & \pad{\redefarrow} \ring{m}_s = m_s + \lambda m_u m_d
\end{align}
where $\lambda$ has units of inverse mass.  Stated in terms of $\chi$, this
redefinition has the form:
\begin{equation}
\chi \pad{\redefarrow} \ring{\chi} = \chi + \hat{\lambda} \chi^{-1} \det \chi
\label{k:equation}
\end{equation}
where:
\begin{equation}
\hat{\lambda} = \frac{\lambda}{2 \mu}
\end{equation}
Using the Cayley-Hamilton theorem for a $3 \times 3$ matrix $\mathcal{B}$:
\begin{equation}
\identity \det \mathcal{B} = \mathcal{B}^3  - \mathcal{B}^2 \Trace
\mathcal{B} - \tfrac{1}{2} \mathcal{B} \Trace \mathcal{B}^2 - \tfrac{1}{2}
\mathcal{B} \bigl( \Trace \mathcal{B} \bigr)^2
\end{equation}
we can write the shift in $\chi$ as:
\begin{gather}
\hat{\lambda} \chi^{-1} \det \chi = \hat{\lambda} \chi^{-1} \Det{ \chi \Sigma }
\notag \\
= \hat{\lambda} \biggl\{ \Sigma \chi \Sigma \chi \Sigma
- \Sigma \chi \Sigma \Tr{ \chi \Sigma }
- \tfrac{1}{2} \Sigma \Tr{ \chi \Sigma \chi \Sigma }
- \tfrac{1}{2} \Sigma \Bigl( \Tr{ \chi \Sigma } \Bigr)^2
\biggr\}
\end{gather}
where $\det \Sigma = 1$ has been used.  From this we find:
\begin{equation}
\Tr{ \Sigma^\dagger \ring{\chi} } = \Tr{ \Sigma^\dagger \chi } -
\frac{\hat{\lambda}}{2} \biggl\{ \Tr{ \chi \Sigma \chi \Sigma } - \Bigl( \Tr{
\chi \Sigma } \Bigr)^2 \biggr\}
\end{equation}
Thus, the mass term in the chiral Lagrangian becomes:
\begin{align}
\Tr{ \Sigma^\dagger \ring{\chi} + \ring{\chi} \Sigma } = & ~ \Tr{ \Sigma^\dagger
\chi + \chi \Sigma } \notag \\
& + \frac{\hat{\lambda}}{4} \Bigl( \Tr{ \Sigma^\dagger \chi + \chi \Sigma }
\Bigr)^2 \notag \\
& + \frac{\hat{\lambda}}{4} \Bigl( \Tr{ \Sigma^\dagger \chi - \chi \Sigma }
\Bigr)^2 \notag \\
& - \frac{\hat{\lambda}}{2} \Bigl( \Tr{ \Sigma^\dagger \chi \Sigma^\dagger \chi + \chi \Sigma \chi \Sigma } \Bigr)
\end{align}
From \eqref{j:equation} we see that a redefinition of three GL coefficients
will absorb these additional terms:
\begin{align}
L_6 & \pad{\redefarrow} \ring{L}_6 = L_6 - \tilde{\lambda}
\label{l:equation} \\
L_7 & \pad{\redefarrow} \ring{L}_7 = L_7 - \tilde{\lambda}
\label{m:equation} \\
L_8 & \pad{\redefarrow} \ring{L}_8 = L_8 + 2 \tilde{\lambda}
\label{n:equation}
\end{align}
where:
\begin{equation}
\tilde{\lambda} = \frac{f^2}{32 \mu} \lambda
\end{equation}

Stated explicitly, the KM ambiguity is the invariance in form of the chiral
Lagrangian under the redefinitions \eqref{k:equation}, \eqref{l:equation},
\eqref{m:equation}, and \eqref{n:equation}.  While we had believed that there
was only one well-defined Lagrangian for ChPT, we now discover that there
exists a family of Lagrangians, all equally valid and connected via the above
transformations.  

The KM ambiguity implies that all expressions for physical quantities obtained
from the chiral Lagrangian will be invariant under the parameter redefinitions.
Thus, combining these expressions with experimental results can never allow us
to distinguish between one coefficient set $( \chi, L_6, L_7, L_8 )$ and
another $( \ring{\chi}, \ring{L}_6, \ring{L}_7, \ring{L}_8 )$.  For example,
the combinations of GL coefficients $L_5$ and $12 L_7 + 6 L_8 - L_5$, which we
were able to determine using expressions from ChPT, are both clearly invariant
under the redefinitions.  The ratio $Q^2$ defined in \eqref{o:equation}, which
was also fixed using ChPT, is invariant to this order under the redefinitions.
This becomes clear when we note that the squares of the quark masses transform
as:
\begin{equation}
\ring{m}_i^2 = m_i^2 + 2 \lambda m_u m_d m_s + O(m_q^4)
\end{equation}
Even the meson mass expressions are invariant to the order at which we are
working:
\begin{align}
\mpQCD^2 & = \mu \bigl( \ring{m}_u + \ring{m}_d \bigr) \biggl\{ 1
+ \chlog{\pi} - \tfrac{1}{3} \chlog{\eta} \notag \\
& \phantom{= \mu \bigl( \ring{m}_u + \ring{m}_d \bigr) \biggl\{ 1}
~ + \frac{8}{f^2} \mu \bigl( \ring{m}_u + \ring{m}_d \bigr) \bigl( 2 \ring{L}_8
- L_5 \bigr) \notag \\
& \phantom{= \mu \bigl( \ring{m}_u + \ring{m}_d \bigr) \biggl\{ 1}
~ + \frac{16}{f^2} \mu \bigl( \ring{m}_u + \ring{m}_d + \ring{m}_s \bigr)
\bigl( 2 \ring{L}_6 - L_4 \bigr) \biggr\} \notag \\
& = \mpQCD^2 + 2 \lambda \mu m_s \mh + 2 \mu \mh \biggl\{ \frac{8}{f^2} \mu \mh
\bigl( 4 \tilde{\lambda} \bigr) + \frac{16}{f^2} \mu \bigl( 2 \mh + m_s \bigr)
\bigl( -2 \tilde{\lambda} \bigl) \biggr\} + \dotsb \notag \\
& = \mpQCD^2  + 2 \lambda \mu m_s \mh - \tilde{\lambda} \frac{64 \mu}{f^2} \mu
m_s \mh + \dotsb \notag \\
& = \mpQCD^2 + \dotsb
\end{align}
Note that while some measurable quantities appear to break invariance at
higher order, they in fact continue to be invariant order by order.

The quark-mass ratios, as well as the quantity $2 L_8 - L_5$, are not invariant
under the redefinition.  Thus, using ChPT alone, they can not be fixed.  We can
only hope to determine a one-parameter family of allowed values.

The KM ambiguity does not represent a symmetry of either QCD or ChPT.  It is
nothing more than an ambiguity in the effective theory's couplings.  Thus, true
values for the quark-mass ratios and the GL coefficients do exist.  We have
simply found that determining those quantities requires theoretical input from
outside ChPT.

ChPT does place one constraint on $L_6$, $L_7$, and $L_8$.  The GL coefficients
must be of natural order.  If they were not, it would imply that ChPT is based
on a poor expansion, and ChPT would have proven useless.  Yet, ChPT's accurate
predictions in other contexts imply otherwise.  We are curious what sort of
bounds this naturalness places on the up quark mass.  If we assume that the
maximum reasonable shift in an $\alpha_i$ is on the order of one:
\begin{equation}
\ring{\alpha}_i = \alpha_i + \frac{( 4 \pi f )^2}{4 \mu} \lambda = \alpha_i +
O(1)
\end{equation}
we find an estimate for the largest reasonable value for $\lambda$:
\begin{equation}
\lambda \approx \frac{4 \mu}{( 4 \pi f )^2} \approx 4 \frac{\mkn^2}{m_s}
\frac{1}{\Lambda_\chi^2} \approx \frac{1}{m_s}
\end{equation}
This $\lambda$ corresponds to a shift in $m_u$ on the order of the down quark
mass:
\begin{equation}
\ring{m}_u - m_u = \lambda m_d m_s \approx m_d
\end{equation}
Thus, we find that a massless up quark is within the natural range and is not
shown by this argument to be beyond the scope of possibility.

In light of the KM ambiguity, we see that we were somewhat naive in our
construction of the chiral Lagrangian.  We had stated previously that we knew
exactly the form of the chiral symmetry breaking structure, and that we could
account for that breaking via insertions of $\qmassm$.  Yet, we now find
that there exists a continuous set of matrices:
\begin{equation}
\ring{\qmassm} ( \lambda ) = \qmassm + \lambda \qmassm^{-1} \det
\qmassm
\end{equation}
each of which is an equally valid choice for breaking the chiral symmetry of our
Lagrangian.  If fact, even if the up quark were massless, the form of the
chiral symmetry breaking term could naively be mistaken for a non-zero up quark
mass:
\begin{equation}
\Bigl. \ring{\qmassm} ( \lambda ) \Bigr\rvert_{m_u = 0} = \begin{bmatrix}
\lambda m_d m_s & & \\
& m_d & \\
& & m_s
\end{bmatrix}
\end{equation}
It is impossible for ChPT to distinguish between the effects of a non-zero up
quark mass and certain large NLO corrections.

\section{Theoretical Estimates}
Various theoretical approximations can be made in an attempt to estimate
$\Delta_M$ and the Gasser-Leutwyler coefficients.

\subsection{Resonance Saturation}
While each interaction term in an effective field theory such as ChPT occurs at
a point, in the full theory they correspond to short-distance interactions,
each encompassing a tower of graphs involving heavier particles.  These
particles are heavy in the sense that they are more massive than the scale of
the effective theory.  In the case of our ChPT, these heavy particles include
all bound states of QCD heavier than the light mesons.  When the transition is
made from the full theory to an effective theory, the heavy states are
integrated out.  This integration shrinks the short-distance interaction to a
point and condenses the effects of the heavy particle exchanges into the
coefficient of the effective theory's corresponding coupling term.

Thus, if we can identifying the most significant of the heavy particle
exchanges corresponding to a given coupling term, we can estimate the
full integration process by merely integrating over the identified important
exchanges.  This leads to a rough estimate of the coupling's coefficient in
the effective theory.

For certain simple coupling terms, the most significant of the
contributing interactions is the exchange of a single heavy particle, the
lightest of the heavy particles with the correct quantum numbers to mediate the
coupling.  In such a case the resulting estimate for the coupling constant is
proportional to the inverse square of the heavy particle's mass.

Both $L_5$ and $L_7$ are examples of coupling constants whose value can be
estimated as described.  We can model their corresponding vertices as being
saturated by exchange of the members of the scalar octet and $\eta'$
respectively \cite{Ecker:1989te, Leutwyler:1996et}.  Adding factors which arise
from the integration, the estimates become:
\begin{align}
L_5 \simeq \frac{f_\pi^2}{4 \m_S^2} \simeq 2.3 \times 10^{-3}
&&
L_7 \simeq -\frac{f_\pi^2}{48 \m_{\eta'}^2} = -0.2 \times 10^{-3}
\end{align}
with $\m_S \simeq 980 \, \text{MeV}$.  Comparing with the determined value of
$L_5$ \eqref{p:equation}, we can see that the estimate is off but is of the
correct order of magnitude.  This allows us to approximate the uncertainty of
the $L_7$ estimate:
\begin{equation}
L_7 = \bigl( -0.2 \pm 2.5 \bigr) \times 10^{-3}
\label{r:equation}
\end{equation}
Combined with \eqref{p:equation} and \eqref{s:equation}, \eqref{r:equation}
leads to estimates for $2 L_8 - L_5$, $\Delta_M$, and the light-quark-mass
ratio:
\begin{equation}
\frac{m_u}{m_d} = 0.4 \pm 2.1
\end{equation}
We can see that the uncertainly in the resonance saturation estimate is too
far significant to rule out a massless up quark.

\subsection{Large $N_c$}
In the large-$N_c$ limit the ABJ anomaly is suppressed and massless QCD has a
full $U(N_f = 3)_V \otimes U(N_f = 3)_A$ flavor symmetry.  If we assume that
this symmetry spontaneously breaks to $U(N_f = 3)_V$, it would result in nine
Goldstone bosons as discussed in Section \ref{ak:section}.  A now light $\eta'$
would join the ranks of the light mesons and take on the role of the ninth
Goldstone boson.  At NLO in $1 / N_c$, the anomaly and the $\eta'$ mass return.

The perspective of large $N_c$ can be incorporated into ChPT by constructing a
Lagrangian which is not only an expansion in $p^2 / \Lambda_\chi^2$ and $\mu
\qmassm / \Lambda_\chi^2$, but also in $1 / N_c$ \cite{Witten:1980sp}.
Additionally, the now light $\eta'$ can no longer be excluded from the
low-energy theory.  Instead, we include it as the trace of our meson field
matrix $\Phi$, making $\Sigma$ an element of $U(N_f)$.  This procedure is
complicated by the fact that, while the ABJ anomaly is suppressed, it is not
absent.  Correspondingly, the $\eta'$ is light, yet the symmetry group which
must be respected by our Lagrangian is the anomalously-broken chiral symmetry,
$U(N_f)_V \otimes SU(N_f)_A$.  Under this reduced group, $\Tr{ \ln \Sigma }
\propto \Trace \Phi$ is invariant.  Thus, the symmetry leaves the Lagrangian's
dependence on $\Tr{ \ln \Sigma }$ unconstrained, and the Lagrangian must
incorporate arbitrary functions of the $\eta'$ field.  In the end, however, we
are saved from these arbitrary functions by the fact that their Taylor
expansion is an expansion in $1 / N_c$.  Thus, large $N_c$ truncates the
functions to simple forms.  

An analysis of the $\eta$ and $\eta'$ masses in large-$N_c$ ChPT out to order
$O(N_c p^2 / \Lambda_\chi^2)$, $O(N_c \mu \qmassm / \Lambda_\chi^2)$, and
$O(N_c^0 = 1)$ implies a constraint on $\Delta_M$ \cite{Leutwyler:1996sa}:
\begin{equation}
\Delta_M > - \frac{4 \mkQCD^2 - 3 \meQCD^2 - \mpQCD^2}{4 \mkQCD^2 - 4 \mpQCD^2}
= -0.07
\end{equation}
Comparing the constraint to \eqref{t:equation}, we can see that large-$N_c$
considerations suggest that a massless up quark is unlikely.

\section{Indirect Phenomenological Results}
To continue our attempt to determine $\Delta_M$ via phenomenological results,
we are forced to draw from evidence beyond the light-meson sector.  These
results are often introduced to light-meson ChPT in the form of the quantity
$R$, the ratio of the strength of $SU(N_f = 3)$ breaking over the strength of
isospin breaking:
\begin{equation}
R \equiv \frac{m_s - \mh}{m_d - m_u}
\end{equation}
Via \eqref{u:equation}, knowledge of $R$ allows for a determination of
$\Delta_M$.

Results from other systems tend to give somewhat consistent values for $R$ at
leading order.  However, consistency in LO results does not preclude large NLO
corrections.  Additionally, spread in the LO results is significant enough to
suggest somewhat sizable NLO corrections, perhaps sizable enough to allow for a
massless up quark.  Unfortunately, each of the phenomenological results
available requires theoretical assumptions in order to tackle their own NLO
corrections.  Thus, none of them provide a model-free determination of the
light-meson sector's NLO corrections.

\subsection{$\psi'$ Branching Ratios}
In the limit of degenerate quarks, the decays:
\begin{align}
\psi' & \decayarrow J/\psi + \pi^0 \\
\psi' & \decayarrow J/\psi + \eta
\end{align}
vanish.  The first decay is allowed by isospin breaking, while the second is
allowed by $SU(N_f = 3)$ breaking.  Thus, at leading order the ratio of their
amplitudes allows for a measure of $R$ \cite{Ioffe:1980mx}:
\begin{equation}
\frac{ T \bigl( \psi' \decayarrow J/\psi + \pi^0 \bigr) }{ T \bigl( \psi'
\decayarrow J/\psi + \eta \bigr) } = \frac{3 \sqrt{3}}{4 R} \bigl( 1 +
\Delta_{\psi'} \bigr)
\end{equation}
where $\Delta_{\psi'}$ represents NLO corrections.  Attempts have been made to
account for the NLO corrections \cite{Donoghue:1992ac}, resulting in the value:
\begin{equation}
R_{\psi'} = 30 \pm 4
\end{equation}
However, the methods are not direct and various theoretical assumptions are
required.  In fact doubts have been raised concerning one of the primary
assumptions \cite{Luty:1993xf}.

\subsection{$\rho^0$-$\omega$ Mixing}
In the limit of perfect isospin symmetry, the $\omega$ would be a pure isospin
singlet.  However, isospin is broken and thus the $\rho^0$ and $\omega$ mix.
The strength of this mixing allows for a measurement of $R$ based on the decay:
\begin{equation}
\omega \decayarrow \pi^+ \pi^-
\end{equation}
This results in the value \cite{Urech:1995ry}:
\begin{equation}
R_\omega = 41 \pm 4
\end{equation}
which does not account for NLO corrections to the vector meson masses.

\subsection{Baryon Masses}
At leading order the mass splitting of the baryon octet leads to three
independent measurements of $R$.  Corrections out to order $O(m_q^2)$ have been
accounted for \cite{Gasser:1982ap}, although theoretical assumptions were
required.  The nucleon, $\Sigma$, and $\Xi$ splittings result in the values:
\begin{align}
R_N & = 51 \pm 10 \\
R_\Sigma & = 43 \pm 4 \\
R_\Xi & = 42 \pm 6
\end{align}
respectively.

\subsection{Accepted Values}
$L_8$ has a generally accepted value, often quoted in reviews and first
presented in \cite{Gasser:1985gg}.  This value is based on the $R$ obtained by
averaging the baryon mass splitting predictions and the $\rho^0$-$\omega$
mixing prediction:
\begin{equation}
R = 42.6 \pm 2.5
\end{equation}
where the value used for $R_{\omega}$ is more recent than that used in
\cite{Gasser:1985gg}.  This value for $R$ results in:
\begin{gather}
\Delta_M = 0.179 \pm 0.097
\label{cl:equation} \\
2 L_8 - L_5 = \bigl( 1.50 \pm 0.46 \bigr) \times 10^{-3} \\
\begin{align}
L_7 = \bigl( -0.55 \pm 0.14 \bigr) \times 10^{-3}
&&
L_8 = \bigl( 1.00 \pm 0.33 \bigr) \times 10^{-3}
\end{align} \\
\frac{m_u}{m_d} = 0.608 \pm 0.056
\label{cm:equation}
\end{gather}
where equations \eqref{u:equation}, \eqref{h:equation}, \eqref{i:equation},
\eqref{p:equation}, and \eqref{s:equation} have been used.  The uncertainty in
$\Delta_M$ due to NNLO corrections has been assumed to be on the order of
$\Delta_M^2$.  Because \eqref{u:equation} is sensitive to the value of
$\Delta_E$, use of Dashen's theorem instead of \eqref{w:equation} results in
significantly different values:
\begin{gather}
\Delta_M = -0.022 \pm 0.057 \\
2 L_8 - L_5 = \bigl( 0.55 \pm 0.27 \bigr) \times 10^{-3} \\
\begin{align}
L_7 = \bigl( -0.31 \pm 0.12 \bigr) \times 10^{-3}
&&
L_8 = \bigl( 0.53 \pm 0.31 \bigr) \times 10^{-3}
\end{align} \\
\frac{m_u}{m_d} = 0.539 \pm 0.043
\end{gather}
Any attempts to calculate $m_u / m_d$ through a determination of $R$ will
encounter this sensitivity to the poorly understood QED mass contributions to
the light meson masses.

A recent comprehensive study of the relevant experimental data can be found in
\cite{Amoros:2001cp}.  The analysis culminates in the result:
\begin{equation}
\frac{m_u}{m_d} = 0.46 \pm 0.09
\label{cn:equation}
\end{equation}
Although the analysis of the data is sophisticated, it still requires the use
of indirect phenomenological results and various theoretical assumptions in
order to determine the strength of $SU(N_f)$ breaking.

All of these results clearly suggest that the up quark is massive.  However,
the value for $R$ was determined using model-dependent assumptions about NLO
corrections to quantities outside the light-meson sector.  We then used this
value to calculate the NLO correction to the light meson masses.  The validity
of this process is somewhat questionable.

\section{First Principles Calculation}
The strong consequences of a massless up quark and the importance of the strong
$\CP$ problem make a first principles calculation of $2 L_8 - L_5$, free of
model-dependent assumptions, very desirable.  Currently the only context in
which such a calculation can be attempted is Lattice Quantum Chromodynamics.
Lattice QCD allows for a direct and non-perturbative measurement of the
Gasser-Leutwyler coefficients, numerically evaluating the underlying QCD
dynamics from which they obtain their values.

\section{Degenerate Quark Masses}
As will be discussed in Chapter \ref{o:section}, in the context of a Lattice
QCD calculation, we have the freedom to choose the masses of our quarks.  For
our study, we have chosen to use $N_f = 3$ degenerate light quarks.  Thus we
present here, assuming $m_q \equiv m_u = m_d = m_s$ and cutting off loops at
$\Lambda = 4 \pi f$, ChPT's NLO expressions for the chiral pseudo-Goldstone
boson mass:
\begin{align}
\mpQCD^2 & = z m_q (4 \pi f)^2 \biggl\{ 1 + \frac{z m_q}{N_f} \ln z m_q \notag
\\
& ~ \phantom{= z m_q (4 \pi f)^2 \biggl\{ 1} + z m_q \bigl( 2 \alpha_8 -
\alpha_5 \bigr) + z m_q N_f \bigl( 2 \alpha_6 - \alpha_4 \bigr) \biggr\}
\label{bi:equation} \\
\intertext{and decay constant:}
f_\pi & = f \biggl\{ 1 + \frac{z m_q N_f}{2} \ln z m_q + z m_q
\frac{\alpha_5}{2} + z m_q N_f \frac{\alpha_4}{2} \biggr\}
\label{bj:equation}
\end{align}
where we introduce:
\begin{equation}
z \equiv \frac{2 \mu}{(4 \pi f)^2}
\label{bp:equation}
\end{equation}
and we have made use of the alternative normalization of the GL coefficients.
Note that, other than the fact that we have now left $N_f$ unspecified, these
expressions follow directly from \eqref{bg:equation} and \eqref{bh:equation}.

\chapter{Lattice Quantum Field Theory} \label{ao:section}
Lattice Quantum Field Theory (LQFT) is a first-principles non-perturbative
numerical approach to Euclidean-space quantum field theory.

\section{Discretization}
It begins with the Euclidean-space partition function of a field theory:
\begin{equation}
Z = \int{ \Biggl( \prod_a \fD{\phi_a} \Biggr) ~ e^{ -S \gfconfig } }
\end{equation}
where the theory contains some set of fields $\phi_a$, and $S \gfconfig$ is the
Euclidean action for a given field configuration $\gfconfig$.  Any physical
observable of the theory can be expressed as the expectation value of an
operator:  the value of the operator evaluated under the distribution defined
by the partition function:
\begin{equation}
\bra \Op \ket = Z^{-1} \int{ \Biggl( \prod_a \fD{\phi_a} \Biggr) ~ \Op
\gfconfig ~ e^{ -S \gfconfig } }
\end{equation}
where the operator $\Op \gfconfig$ is some mapping of field configurations into
real numbers.

In order to manage the theory numerically, continuous space is replaced by a
discrete lattice of points, and the infinite extent of space is made finite and
compact.  The number of lattice sites along a given direction $\mu$ is denoted
by $L_\mu$, which in all cases we will take to be even, while the distance
between lattice sites is denoted by $a$.  The theory's fields are attributed
values only on the discrete set of locations $x_i$:
\begin{align}
\phi_{a;n_i} & \equiv \phi_a(x_i) \notag \\
& n_{i;\mu} \equiv a^{-1} x_{i;\mu} \in \mathbb{Z}^4
\end{align}
where each component of $n_i$ is confined to integers in the range $[ 0, L_\mu
- 1 ]$.  Derivatives within $S \gfconfig$ must also be discretized,
\eqref{cs:equation} and \eqref{ab:equation}.  A single field configuration
now contains a finite number of degrees of freedom, and the functional integral
is replaced by a finite product of standard integrals:
\begin{equation}
\bra \Op \ket = Z^{-1} \prod_a \prod_i \Biggl( \int^{\infty}_{-\infty}
d \phi_{a;n_i} \Biggr) ~ \Op \gfconfig ~ e^{ -S \gfconfig }
\label{v:equation}
\end{equation}
where $Z$ has been redefined accordingly.  In the limit of a small spacing
between the lattice sites, $a \cntlmtarrow 0$, and a large lattice extent,
$L_\mu \cntlmtarrow \infty$, the results of the lattice theory will coincide
with the continuum theory.

While our express motivation for discretization is to allow for a numerical
approach to quantum field theory, it is worth noting that the discretization
procedure is also a valid regularization scheme.  The finite lattice spacing
results in a Lorentz-variant ultraviolet momentum cutoff at $p_\mu = \pi / a$,
removing infinities which arise in a perturbative formulation due to loop
corrections.  Renormalized physical quantities become functions of the lattice
spacing and remain finite in the continuum limit.  However, because the
regulator breaks Lorentz invariance, it is cumbersome to work with and has a
narrow range of practical applications.

While a single field configuration contains a finite number of degrees of
freedom, the integral in \eqref{v:equation} runs over an infinite number of
such configurations.  Thus, in order to attempt the integral numerically, we
must apply Monte Carlo techniques.  A finite sampling of the infinite
field-configuration space is generated, where the probability that a given
field configuration $\gfconfig$ is included in the sample is:
\begin{align}
P \bigl( \gfconfig \bigr) & \equiv Z^{-1} W \gfconfig \notag \\
& = Z^{-1} e^{-S \gfconfig}
\end{align}
where $W \gfconfig$ is the Boltzmann weight of the configuration.  Such a set
of field configurations is referred to as an ensemble.  In the limit of having
a large number of configurations in an ensemble, the expectation value of an
operator is simply the average of its value evaluated on each configuration in
the ensemble:
\begin{equation}
\bra \Op \ket = \frac{1}{N} \sum_n \bra \Op \ket_{\gfconfig_n}
\end{equation}
where the ensemble contains $N$ configurations and $\bra \Op \ket_{\gfconfig_n}
\equiv \Op \gfconfig_n$ denotes the evaluation of the operator on the fixed
field configuration $\gfconfig_n$, the $n$-th field configuration of the
ensemble.

A given quantum field theory contains some number of fundamental constants.  In
order to fix these constants in the context of LQFT, an equal number of
observables must be calculated.  The results of these calculations are then set
to experimentally measured values, binding the fundamental constants.  From
that point on, LQFT is predictive.  Any additional calculated observables must
match experiment.  This procedure is analogous to choosing the
renormalization conditions when implementing a continuum regularization scheme.

\section{Ensemble Creation}
Much of the computation time required to numerically implement LQFT is consumed
generating the weighted ensemble of field configurations.  Thus, identifying
efficient algorithm is of paramount importance.

Any algorithm for generating an ensemble with the proper field-configuration
distribution will involve the iteration of a two-step update process.  This
process consists of proposing the addition of a configuration to the ensemble,
and then accepting or rejecting that proposal based on an appropriate
probability.  Such an algorithm will generate the correct ensemble if we insure
that the probability of proposing a given configuration times the probability
of accepting that configuration results in the correct probability for that
configuration's inclusion in the ensemble.

The most straightforward update process involves proposing field configurations
generated randomly, with a distribution that is flat relative the measure of
the partition function.  A configuration $\gfconfig$ is then accepted into the
ensemble with a probability based directly on its Boltzmann weight $W \gfconfig
= e^{-S \gfconfig}$.  If a random number in the range $[0, 1]$ is less than $W
\gfconfig$, the configuration is added to the ensemble.

However, the volume of field-configuration space is quite large, and $W
\gfconfig$ tends to be very sharply peaked at a specific set of
field configurations.  Thus, while this procedure generates the correct
configuration distribution in a straightforward manner, it is very inefficient.
Much of the computation time will be spent generating configurations which
are subsequently rejected by the acceptance step.

\subsection{Markov Chains}
We require an algorithm which allows us to primarily propose configurations
which are in the vicinity of the peak, but which does not skew the probability
of each configuration's inclusion in the ensemble.

In order to propose configurations near the peak in $W \gfconfig$, we will
generate our proposal based on the last configuration accepted into the
ensemble.  The proposal configuration is produced by introducing some change in
the previous configuration.  The magnitude of that change must be small enough
that we do not stray too far from the peak, but large enough that we can hope
to sample a substantial volume of configuration space with a reasonably sized
ensemble.  An ensemble generated using such a chain of configurations, each one
spawned from the previous, is known as a Markov chain.

Most algorithms for generating Markov chains satisfy two conditions:
ergodicity and detailed balance.  While ergodicity is required for any Markov
algorithm, detailed balance is sufficient, but not necessary, to insure the
creation of the correct configuration distribution.

\subsection{Ergodicity}
As we step along our Markov chain, we must insure that the update algorithm
does not restrict us to some subset of field-configuration space.  Rather,
whatever process we develop for changing the previous configuration and
generating a proposal must, given an arbitrary number of iterations, span all
of configuration space.  This condition on the update process is known as
ergodicity.

\subsection{Detailed Balance} \label{d:section}
To correctly construct our Markov chain, we must determine the
proposal-acceptance probability which leads to our desired distribution.  We
begin by noting that, once an ensemble has the correct distribution, the
configuration densities must be in equilibrium.  That is, the correct
distribution is a fixed point of the update process.  Therefore, for such an
ensemble, the probability of adding field configuration $\gfconfig_B$ to the
ensemble, given that configuration $\gfconfig_A$ was the previous configuration
accepted, must be equal to the probability of adding field configuration
$\gfconfig_A$ to the ensemble, given that $\gfconfig_B$ was the previous
configuration accepted:
\begin{equation}
\prob{}{\gfconfig_A \updatearrow \gfconfig_B} = \prob{}{\gfconfig_B
\updatearrow \gfconfig_A}
\end{equation}
This condition is known as detailed balance.

Three distinct factors combine to determine $\prob{}{\gfconfig_A \updatearrow
\gfconfig_B}$:  the probability of $\gfconfig_A$ being the previous
configuration accepted into the ensemble $\prob{W}{\gfconfig_A}$; the
probability of proposing $\gfconfig_B$, given that $\gfconfig_A$ was the
previous configuration $\prob{P}{\gfconfig_A \updatearrow \gfconfig_B}$; and
the probability of accepting $\gfconfig_B$, given that $\gfconfig_A$ was the
previous configuration $\prob{A}{\gfconfig_A \updatearrow \gfconfig_B}$:
\begin{equation}
\prob{}{\gfconfig_A \updatearrow \gfconfig_B} = \prob{W}{\gfconfig_A}
\prob{P}{\gfconfig_A \updatearrow \gfconfig_B} \prob{A}{\gfconfig_A
\updatearrow \gfconfig_B}
\end{equation}
Equilibrium dictates that:
\begin{multline}
\prob{W}{\gfconfig_A} \prob{P}{\gfconfig_A \updatearrow \gfconfig_B}
\prob{A}{\gfconfig_A \updatearrow \gfconfig_B} \\
= \prob{W}{\gfconfig_B} \prob{P}{\gfconfig_B \updatearrow \gfconfig_A}
\prob{A}{\gfconfig_B \updatearrow \gfconfig_A}
\end{multline}

If we require that the equilibrium point corresponds to an ensemble with our
desired configuration distribution, then the probability of a configuration
being the latest configuration accepted is proportional to its Boltzmann weight:
\begin{align}
\prob{W}{\gfconfig_A} = Z^{-1} W \gfconfig_A
&&
\prob{W}{\gfconfig_B} = Z^{-1} W \gfconfig_B
\end{align}

We have said very little about the process which generates a proposal
configuration from the previous configuration.  However, placing a simple
condition on it will allow us to determine the acceptance probability.  We
require that the probability of making any given change is equal to the
probability of making the reverse change:
\begin{align}
\prob{P}{\gfconfig_A \updatearrow \gfconfig_B} = \prob{P}{\gfconfig_B
\updatearrow \gfconfig_A}
\end{align}

Bound by such a condition, we can easily see that an appropriate acceptance
probability would be:
\begin{align}
\prob{A}{\gfconfig_A \updatearrow \gfconfig_B} & = \min \biggl\{ 1, \frac{W
\gfconfig_B}{W \gfconfig_A} \biggr\} \notag \\
& = \min \Bigl\{ 1, e^{S \gfconfig_A - S \gfconfig_B} \Bigr\}
\end{align}
If the proposed configuration has a larger Boltzmann weight than the previous
configuration, it is always accepted into the ensemble.  If it has a smaller
weight than the previous configuration, it is accepted with a probability based
on the change in the action.

Such a Markov process, which accepts or rejects a proposal configuration with
a probability based on its difference in weight from the previous
configuration, is known as a Metropolis algorithm.

\subsection{Autocorrelation Length} \label{i:section}
Because each new configuration is generated by making changes to the
previously accepted configuration, any given configuration in a Markov chain
will have similarities to the configurations which come before it.  This
correlation between ensembles along a Markov chain is known as autocorrelation.

Due to autocorrelation, a single update step does not sample configuration
space with the correct weight.  Instead, a number of update steps must be
performed in order to generate only a single independent and correctly weighted
sample of configuration space.  The length that must be moved along a Markov
chain in order to go from one independent configuration to the next is known as
the autocorrelation length.  

We can estimate the autocorrelation length by choosing some observable and
watching its evolution as we evaluate it on individual configurations along the
Markov chain.  The observable will experience fluctuations whose frequency can
be taken to suggest the order of the autocorrelation length.  Clearly, the
perceived autocorrelation length will depend heavily on the observable chosen.
Because two configurations are only truly decorrelated after an infinite number
of update steps, any estimation of autocorrelation length will always involve a
somewhat arbitrary cutoff decision.

Reducing the autocorrelation length is desirable, as it allows us to sample a
larger region of configuration space using a shorter Markov chain.  The
autocorrelation length can be reduced by increasing the magnitude of the
changes made when generating a proposal configuration.  However, such gains can
be offset by a reduction in the acceptance rate.

\subsection{Acceptance Rate}
In order to minimize the computation time spent generating configurations which
ultimately go unused, the acceptance rate of a Markov chain process must be
kept reasonably large.  This can be done by reducing the magnitude of the
changes which are made when generating a proposal configuration, and thus
reducing the chances of straying significantly from the peak in $W \gfconfig$.
However, as discussed above, such a reduction causes a corresponding increase
in the autocorrelation length.  Thus, the optimal magnitude of the update
step can be elusive, but is generally one which results in an acceptance rate
of approximately 50\%.

In order to retain a reasonable acceptance rate, but still allow large changes
in the proposal configuration, update methods which move through configuration
space along, or nearly along, lines of constant action can be used.  Minimizing
the change in action increases the acceptance rate without ruining detailed
balance, and conversely allows for larger update steps.  By interspersing such
constant-action update steps with standard update steps, ergodicity is
retained.

\subsection{Thermalization} \label{r:section}
When beginning a Markov chain, one must choose an initial field configuration.
Because this choice for the head of the chain is arbitrary, and not chosen by
the Markov process itself, that first configuration, and any configurations
which are correlated with it, will not have the correct distribution.  Thus,
the ensemble will only obtain the correct distribution in the limit of an
infinitely large ensemble, when the effects of the earliest configurations have
washed out.

This process of decorrelation from the initial configuration is known as
thermalization.  The number of update steps required by thermalization is
obviously closely related to an ensemble's autocorrelation length.  Since we
can only work with finite ensembles, configurations generated before the Markov
chain is thermalized will be dropped from the ensemble.  We refer to the point
on the Markov chain at which we first begin to retain configurations as the
thermalization point, with $N_T$ denoting the number of configurations dropped.
Unfortunately, an estimation of the thermalization point involves the same
uncertainties and arbitrariness as an estimation of the autocorrelation length.

\section{Two-Point Correlation Functions} \label{a:section}
A particularly useful observable in LQFT is the two-point correlation function:
\begin{equation}
C(x) \equiv \bra \Op(x) \Op(0) \ket
\end{equation}
where the operator $\Op(x)$ is some function of the fields local to $x$ and
corresponds to a set of values for the theory's quantum numbers.  The
Euclidean-space two-point correlation function is analogous to the two-point
Green's function of Minkowski space.  $\Op(x)$ creates or annihilates at $x$
every eigenstate of the theory with matching quantum numbers, each with an
amplitude that is dependent upon the operator's exact form.  Thus, the
correlation function gives the amplitude for creating that tower of states at
the origin, having them propagate to $x$, and then annihilating them.

The Euclidean-space states of a quantum field theory are defined to be the
eigenstates of the time component of the translation operator, also known as
the transfer matrix.  These states decay exponentially with time, each picking
up a factor equal to its eigenvalue $e^{-E t}$ after propagating a distance $t$
in time, where $E$ is the state's energy.  This corresponds directly to the
phase oscillation of a Minkowski-space state with a frequency proportional to
its energy.  Knowing this factor, the tower of states created by $\Op(x)$ can
be made explicit:
\begin{equation}
C ( \vect{x} = \vec{0}, t ) = \sum_n{ \frac{1}{2 E_n V} \bra 0 | \Op | n \ket
\bra n | \Op | 0 \ket ~ e^{-E_n t} }
\end{equation}
where the sum is over all states with appropriate quantum numbers, both
single- and multi-particle, and $\bra n | \Op | 0 \ket$ represents the
amplitude for the operator $\Op(x)$ creating the state $| n \ket$ from the
vacuum.  The factor $(2 E_n V)^{-1}$ comes from a relativistic normalization of
the states, where $V$ is the spatial volume of a time slice.

Because $\Op(x)$ is local in space, the tower of created states includes states
of all momenta.  We can restrict the states created to those with a specific
momentum $\vect{p}$ by Fourier transforming the annihilation operator:
\begin{equation}
\Op ( \vect{p}, t ) = \sum_{\vect{x}}{ e^{ -i \vect{p} \cdot \vect{x} } \Op (
\vect{x}, t ) }
\end{equation}
Choosing only states with zero total momentum, we simply sum over all spatial
positions of a time slice:
\begin{equation}
\Op ( \vect{p} = \vec{0}, t ) = \sum_{\vect{x}}{ \Op ( \vect{x}, t ) }
\end{equation}
At zero momentum the energy $E_n$ of a state is reduced to its mass $M_n$.
Thus, using a zero-momentum annihilation operator, the correlation function
becomes:
\begin{align}
C ( t ) & \equiv C ( \vect{p} = \vec{0}, t ) = \Bigl\bra
\sum_{\vec{x}}
\Op ( \vec{x}, t )
\Op ( \vec{0}, 0 ) \Bigr\ket \notag \\
& = \sum_n{ \frac{1}{2 M_n} \lvert \bra n | \Op | 0 \ket \rvert^2 e^{-M_n
t} }
\end{align}
where the sum is now only over zero-momentum states, and the sum over $\vect{x}$
has canceled with the factor of $V^{-1}$.  Note that restricting the
annihilation operator to zero momentum is sufficient.  There is no need to
Fourier transform the creation operator.

If the creation and annihilation points are well separated in time, and
assuming that there is an energy gap between the lightest and second-lightest
states, all states will be exponentially damped relative to the lowest energy
state $| 1 \ket$.  Thus:
\begin{equation}
\lim_{t \rightarrow \infty} C ( t ) = \frac{1}{2 M_1} \lvert \bra 1 | \Op | 0
\ket \rvert^2 e^{-M_1 t}
\label{ar:equation}
\end{equation}
At this point the correlation function has been expressed in terms of only two
unknowns:  the operator's overlap with the lightest state $\bra 1 | \Op | 0
\ket$ and the mass of that state $M_1$.

Using the techniques of LQFT, we can numerically calculate, for a given set of
quantum numbers, the two-point correlation function $C(t)$ at large time
separations.  Then, analyzing the $t$ dependence of our result, we can extract
the creation amplitude and mass of the lowest energy state.  By repeating this
process for all appropriate combinations of quantum numbers, LQFT allows us to
non-perturbatively calculate from first principles the low-energy spectrum of a
quantum field theory.

\chapter{Lattice Quantum Chromodynamics} \label{o:section}
The study of QCD using the techniques of LQFT is known as Lattice Quantum
Chromodynamics (LQCD).

\section{Gluon Fields}
We first describe the discretization of gluon fields.

\subsection{Gauge Discretization}
Many of the symmetries of a quantum field theory, the most significant of which
is Poincar\'e symmetry, are lost during discretization and are only regained
in the continuum limit.  Because local $SU(N_c)$ gauge symmetry is the defining
symmetry of QCD, the discretization process for the gluon fields will focus on
preserving this symmetry at non-zero lattice spacing.

The gluon fields act as a parallel transporter, describing the color
transformation of the quarks as they move through space.  A color vector moving
through some path $C$ picks up a unitary rotation $U_C$:
\begin{equation}
U_C = \pathordered \exp \biggl( i \g \int_C{ dx A_\mu(x) } \biggr)
\label{x:equation}
\end{equation}
where $\pathordered$ denotes path-ordering within the exponentiated integral.
In order to retain gauge symmetry at non-zero lattice spacing, and to preserve
the parallel-transporting nature of the gluon fields, we choose to work, not
with the gluon fields $A_\mu$ themselves, but instead with unitary
color-transformation matrices which we assign to the links connecting
neighboring lattice sites.  These matrices transport a color vector from one
lattice site to the next, and thus are defined using \eqref{x:equation}, where
the path $C$ is now the straight line linking neighboring sites:
\begin{equation}
U_{\mu; n} \equiv \pathordered \exp \biggl( i \g \int_x^{x + a \muh}{
dx' A_\mu(x') } \biggr)
\end{equation}
where $U_{\mu; n}$ denotes the matrix which transports a vector from the site
at $x$ to the site at $x + a \muh$, and $\muh$ denotes a unit vector in the
direction $\mu$.  Correspondingly, $U^\dagger_{\mu; n - \muh}$ is the matrix
which transports a vector from $x$ to $x - a \muh$.  Recall that the
four-vector $n$ has been defined such that its elements are integers at the
lattice sites:
\begin{equation}
n_\mu \equiv a^{-1} x_\mu
\end{equation}

At lowest order in $a$, $U_{\mu; n}$ can be expressed in terms of only
the parallel component of the gauge field at a point halfway along the
relevant link:
\begin{equation}
U_{\mu; n} = \exp \biggl( i a \g A_\mu ( x + \tfrac{a}{2} \muh ) +
\dotsb \biggr)
\end{equation}

Using link matrices $U_{\mu; n}$ to describe our gauge fields allows for
the preservation of a discretized local color symmetry.  The associated gauge
transformation is:
\begin{align}
U_{\mu; n} & \pad{\symarrow} U'_{\mu; n} = \Omega^{}_{n + \muh}
U^{}_{\mu; n} \Omega^\dagger_n
\label{y:equation} \\
& \qquad \Omega_n \in SU(N_c)
\end{align}
where $\Omega_n$ is ascribed values only on the lattice sites.

\subsection{Gauge Action}
It is clear that, under the gauge transformation described in
\eqref{y:equation}, any closed loop of link matrices will be invariant.  The
simplest of such loops is a square with sides one link in length.  This loop
in known as the plaquette:
\begin{equation}
P^{}_{\mu\nu; n} \equiv W^{1 \times 1}_{\mu\nu; n} = U^\dagger_{\nu; n}
U^\dagger_{\smash{\mu; n + \hat{\nu}}} U^{}_{\nu; n + \muh} U^{}_{\mu; n}
\end{equation}
Expressing the plaquette in terms of the gluon field and expanding around its
center:
\begin{align}
P_{\mu\nu; x} = & ~ e^{ -a \mathcal{A}_\nu ( x + \halfof{\bar{\nu}} ) }
e^{ -a \mathcal{A}_\mu ( x + \bar{\nu} + \halfof{\bar{\mu}} ) } e^{ a
\mathcal{A}_\nu ( x + \bar{\mu} + \halfof{\bar{\nu}} ) } e^{ a \mathcal{A}_\mu
( x + \halfof{\bar{\mu}} ) } + \dotsb \notag \\
= & ~ e^{ -a \mathcal{A}_\nu ( x + \halfof{\bar{\nu}} ) } e^{ -a
\mathcal{A}_\mu ( x + \halfof{\bar{\mu}} ) - a^2 \discder_\nu \mathcal{A}_\mu (
x + \halfof{\bar{\mu}} + \halfof{\bar{\nu}} ) } \notag \\
& \quad \times e^{ a \mathcal{A}_\nu ( x + \halfof{\bar{\nu}} ) + a^2
\discder_\mu \mathcal{A}_\nu ( x + \halfof{\bar{\nu}} + \halfof{\bar{\mu}} ) }
e^{ a \mathcal{A}_\mu ( x + \halfof{\bar{\mu}} ) } + \dotsb \notag \\
= & ~ e^{ -a \mathcal{A}_\nu ( x + \halfof{\bar{\nu}} ) - a \mathcal{A}_\mu ( x
+ \halfof{\bar{\mu}} ) - a^2 \discder_\nu \mathcal{A}_\mu ( x +
\halfof{\bar{\mu}} + \halfof{\bar{\nu}} ) + \tfrac{a^2}{2} \bigl[
\mathcal{A}_\nu ( x + \halfof{\bar{\nu}} ) , \mathcal{A}_\mu ( x +
\halfof{\bar{\mu}} ) \bigr] } \notag \\
& \quad \times e^{ a \mathcal{A}_\nu ( x + \halfof{\bar{\nu}} ) + a^2
\discder_\mu \mathcal{A}_\nu ( x + \halfof{\bar{\nu}} + \halfof{\bar{\mu}} ) +
a \mathcal{A}_\mu ( x + \halfof{\bar{\mu}} ) + \tfrac{a^2}{2} \bigl[
\mathcal{A}_\nu ( x + \halfof{\bar{\nu}} ), \mathcal{A}_\mu ( x +
\halfof{\bar{\mu}} ) \bigr] } + \dotsb \notag \\
= & ~ e^{ a^2 \bigl( \discder_\mu \mathcal{A}_\nu ( x + \halfof{\bar{\mu}} +
\halfof{\bar{\nu}} ) - \discder_\nu \mathcal{A}_\mu ( x + \halfof{\bar{\mu}} +
\halfof{\bar{\nu}} ) - \bigl[ \mathcal{A}_\mu ( x + \halfof{\bar{\nu}} ),
\mathcal{A}_\nu ( x + \halfof{\bar{\mu}} ) \bigr] \bigr) } + \dotsb \notag \\
= & ~ e^{ a^2 \bigl( \discder_\mu \mathcal{A}_\nu ( x + \halfof{\bar{\mu}} +
\halfof{\bar{\nu}} ) - \discder_\nu \mathcal{A}_\mu ( x + \halfof{\bar{\mu}} +
\halfof{\bar{\nu}} ) - \bigl[ \mathcal{A}_\mu ( x + \halfof{\bar{\mu}} +
\halfof{\bar{\nu}} ), \mathcal{A}_\nu ( x + \halfof{\bar{\mu}} +
\halfof{\bar{\nu}} ) \bigr] \bigr) } + \dotsb \notag \\
= & ~ e^{ i a^2 \g F_{\mu\nu} ( x + \halfof{\bar{\mu}} + \halfof{\bar{\nu}} ) }
+ \dotsb \notag \\
= & ~ \identity + i a^2 \g F_{\mu\nu} ( x + \tfrac{a}{2} \muh +
\tfrac{a}{2} \hat{\nu} ) - \frac{a^4 \g^2}{2} \bigl( F_{\mu\nu} ( x +
\tfrac{a}{2} \muh + \tfrac{a}{2} \hat{\nu} ) \bigr)^2 + \dotsb
\end{align}
where $\mathcal{A}_\mu(x) \equiv i \g A_\mu(x)$, $\bar{\mu} \equiv a \muh$, the
matrix identity:
\begin{equation}
e^\mathcal{A} e^\mathcal{B} = e^{ \mathcal{A} + \mathcal{B} + \half [
\mathcal{A}, \mathcal{B} ] } + \dotsb
\end{equation}
has been used, and terms higher order in $a$ have been dropped at each step.
Thus, at lowest order in $a$, the real trace of the plaquette gives us access
to the gauge field strength at its center:
\begin{equation}
\Real \tr{ \identity - P_{\mu\nu; n} }
= \frac{a^4 \g^2}{2} \trace \Bigl[ \bigl( F_{\mu\nu} ( x + \tfrac{a}{2} \muh
+ \tfrac{a}{2} \hat{\nu} ) \bigr)^2 \Bigr] + O(a^6)
\end{equation}

We are now able to assemble an action which is dependent only on the degrees of
freedom which remain in our gluon field after discretization, and equals the
continuum action at lowest order in $a$.  Using $\int_x \discarrow a^4 \sum_n$:
\begin{align}
S_g \fconfig{U} = \int_x \frac{1}{2} & \tr{ F_{\mu\nu}(x) F^{\mu\nu}(x) }
\pad{\discarrow} \notag \\
& a^4 \sum_n \frac{1}{2} \tr{ F_{\mu\nu} ( an + \tfrac{a}{2} \muh +
\tfrac{a}{2} \hat{\nu} ) F^{\mu\nu} ( an + \tfrac{a}{2} \muh + \tfrac{a}{2}
\hat{\nu} ) } \notag \\
& = a^4 \sum_n \sum_\mu \sum_{\nu < \mu} \trace \Bigl[ \bigl( F_{\mu\nu}
( an + \tfrac{a}{2} \muh + \tfrac{a}{2} \hat{\nu} ) \bigr)^2 \Bigr] \notag \\
& = \frac{2}{\g^2} \sum_n \sum_\mu \sum_{\nu < \mu} \Real \tr{ \identity -
P_{\mu\nu; n} } + O(a^2)
\end{align}
Finally:
\begin{gather}
S_g \fconfig{U} = \beta \sum_n \sum_\mu \sum_{\nu < \mu} \frac{1}{N_c} \Real
\tr{ \identity - P_{\mu\nu; n} }
\label{z:equation} \\
\beta \equiv \frac{2 N_c}{\g^2}
\end{gather}
Because the action is constructed only of closed gauge-link loops, we can be
sure that gauge symmetry has been preserved.

\subsection{Gauge Coupling} \label{e:section}
It is evident from \eqref{z:equation} that, for any lattice calculation which
includes only gauge fields, there is only one free parameter in the action, the
strong coupling constant $\g$.  We might expect the lattice space $a$ to also
appear in the action, but in fact it does not.  This is because the lattice
spacing is not a physical parameter, but rather must be thought of as a
momentum cutoff for our quantum field theory.  Thus, the action should depend
on our cutoff $a$ only through the renormalization-scale dependence of the true
parameters of the action.

We see then that the strong coupling constant and the lattice spacing are not
independently free parameters.  Once $\beta$ is chosen for a pure-gauge
lattice calculation, both $\g$ and $a$ are fixed, related by the strong
coupling's renormalization group equation.  It is this dependence of the strong
coupling on the lattice spacing which allows us to conceptualize a continuum
limit, and which distinguishes a LQFT calculation from a standard statistical
mechanics system.

In more concrete terms, when doing such a lattice calculation, we first choose
a value for $\beta$.  Then we calculate some dimensionful physical observable
whose value has been determined experimentally.  The units in our lattice
calculation of the observable will appear as powers of $a$.  Thus, by setting
our calculated value equal to the experimentally measured value, the lattice
spacing is fixed.  This procedure corresponds directly to the application of a
renormalization condition in continuum field theory.

\subsection{Gauge Updates}
The gauge action \eqref{z:equation} has the very convenient property of
locality.  That is, any given field degree of freedom couples directly only to
other degrees of freedom in close spatial vicinity.  Thus, calculating the
shift in action due to some change in the field variables requires only
information local to that change, and Markov-chain update steps can be made
with relatively little computational effort.

In the case of a pure gauge calculation, the update step would involve making a
change to an individual link in such a manner as to preserve detailed balance.
Only the plaquettes that contain the changed link need to be recalculated in
order to determine the shift in the action.  The update is then accepted or
rejected based on the size of that shift.  By doing a number of these
individual updates to each link in the lattice, we can quickly generate a new
configuration to add to the Markov chain.  

Unfortunately, the introduction of quarks to our lattice calculation will
destroy locality, significantly increasing the computation time required for an
update.

\section{Quark Fields}
We will now bring quarks into our lattice theory.  The process will be
complicated significantly by the fact that they are Dirac spinors, and that the
Dirac action is linear in momentum.

\subsection{Fermion Discretization}
First we define a discretized quark field, attributing it values only on the lattice sites:
\begin{align}
q_n \equiv a^\efrac{3}{2} q(x)
&&
\qb_n \equiv a^\efrac{3}{2} \qb(x)
\end{align}
The powers of $a$ which appear in the definition cancel units contained by
the continuum quark field, making the discretized field suitable for use in
numerical calculations.

To construct a lattice quark action, we must first discretize the covariant
derivative:
\begin{align}
D_\mu q(x) = & ~ \partial_\mu q(x) - i \g A_\mu(x) q(x) \notag \\
= & ~ \frac{1}{2 a} \Bigl[ q ( x + \muh ) - q ( x - \muh ) \Bigr]
\notag \\
& \pquad - \frac{i \g}{2} \Bigl[ A_\mu ( x + \halfof{\bar{\mu}} ) q ( x +
\muh ) + A_\mu ( x - \halfof{\bar{\mu}} ) q ( x - \muh ) \Bigr] +
\dotsb \notag \\
= & ~ \frac{1}{2 a} \biggl\{ \Bigl[ 1 - i \g a A_\mu ( x + \halfof{\bar{\mu}} )
\Bigr] q ( x + \muh ) \notag \\
& \pquad - \Bigl[ 1 + i \g a A_\mu ( x - \halfof{\bar{\mu}} ) \Bigr] q ( x -
\muh ) \biggr\} + \dotsb \notag \\
= & ~ \frac{1}{2 a^\efrac5{2}} \Bigl[ U^\dagger_{\mu; n} q^{}_{n + \muh} -
U^{}_{\mu; n - \muh} q^{}_{n - \muh} \Bigr] + \dotsb
\end{align}
Note that in order to calculate the finite difference at site $n$, the value of
the quark field at neighboring sites is first parallel transported to $n$ via
the appropriate gauge link matrices.  In this way the quark-gluon interaction
manifests in our lattice theory.

\subsection{Fermion Action}
The quark action can now be discretized:
\begin{align}
S_q^N \fconfig{U, q, \qb} & = \int_x \qb(x) \bigl(
\gamma^\mu D_\mu + \qmassm \bigr) q(x) \pad{\discarrow} \notag \\
& a^4 \sum_n \frac{1}{a^4} \biggl\{ \frac{1}{2} \sum_\mu \qb_n
\gamma_\mu \Bigl[ U^\dagger_{\mu; n} q^{}_{n + \muh} -
U^{}_{\mu; n - \muh} q^{}_{n - \muh} \Bigr] + a
\qb_n \qmassm q_n \biggr\} \notag \\
& = \sum_n \sum_m \qb_n M^N_{n,m} \fconfig{U} q_m
\end{align}
where, if we assume degenerate quarks, the interaction matrix $M \fconfig{U} =
M^N \fconfig{U}$ is:
\begin{align}
M^N_{n,m} \fconfig{U} & \equiv \frac{1}{2} \sum_\mu \gamma_\mu \Bigl[
U^\dagger_{\mu; n} \delta^{}_{n, m - \muh} - U^{}_{\mu; n - \muh} \delta^{}_{n,
m + \muh} \Bigr] + m_Q \delta_{n,m} \notag \\
& = D^N_{n,m} \fconfig{U} + m_Q \delta_{n,m}
\end{align}
Note that the action includes an implied sum over color, spin, and flavor
indices and that $M_{n,m} \fconfig{U}$ is a matrix in those three indices, as
well as in position.  We represent the kinetic term of $M_{n,m} \fconfig{U}$
using $D_{n,m} \fconfig{U}$.  The lattice quark mass $m_Q$ absorbs one power
of $a$ and becomes a unitless parameter, suitable for use in a numerical
calculation.  It is related to the dimensionful quark mass by:
\begin{align}
m_Q = \U{m}_q \equiv a m_q
\end{align}

Now including both quarks and gluons in our LQCD action, $S'_\text{LQCD}
\fconfig{U, q, \qb} = S_g \fconfig{U} + S_q \fconfig{U, q, \qb}$, our partition
function is:
\begin{equation}
Z_\text{LQCD} = \int \fD{U} \fD{q} \fD{\qb} ~ e^{ -S_g \fconfig{U} } ~ e^{
-\sum_{nm} \qb_n M_{n,m} \fconfig{U} q_m }
\end{equation}
As a fermion, the quark field is expressed in terms of Grassmann numbers.
Because there is no simple way to numerically account for the anticommutating
property of Grassmann numbers, we will integrate out the quark degrees of
freedom analytically.  The result is the determinant of the interaction matrix:
\begin{align}
Z_\text{LQCD} & = \int \fD{U} ~ e^{ -S_g \fconfig{U} } ~ \det M \fconfig{U}
\notag \\
& = \int \fD{U} ~ e^{ -S_\text{LQCD} \fconfig{U} }
\end{align}
where
\begin{align}
S_\text{LQCD} \fconfig{U} & = S_g \fconfig{U} - \ln \det M \fconfig{U} \notag
\\
& = S_g \fconfig{U} - \Trace \ln M \fconfig{U} 
\end{align}
This determinant, or trace, is over all indices of $M \fconfig{U}$.   In the
case of $M^N \fconfig{U}$, it is over color, spin, flavor, and position
indices.  Note that if the interaction matrix is diagonal in flavor space,
as is true for $M^N \fconfig{U}$, the determinant can be factored into a
product of distinct flavor determinants.

\subsection{Color Symmetry}
The full LQCD action continues to have local color symmetry at non-zero lattice
spacing:
\begin{align}
q_n & \pad{\symarrow} q'_n = \Omega_n q_n \\
\qb_n & \pad{\symarrow} \qb'_n = \qb^{}_n \Omega^\dagger_n \\
U_{\mu; n} & \pad{\symarrow} U'_{\mu; n} = \Omega^{}_{n + \muh} U^{}_{\mu; n}
\Omega^\dagger_n \\
& \qquad \Omega_n \in SU(N_c)
\end{align}

\subsection{Quark Propagators}
In order to calculate a quark-antiquark correlator, we again must first
analytically integrate over the fermion degrees of freedom.  Via the
integration, the quark and antiquark insertions generate an inverse of the
interaction matrix:
\begin{align}
\bra \qb_{a \alpha i; n} q_{b \beta j; m} \ket & =
Z_\text{LQCD}^{-1} \int \fD{U} \fD{q} \fD{\qb} ~ e^{ -S'_\text{LQCD}
\fconfig{U, q, \qb} } ~ \qb_{a \alpha i; n} ~ q_{b \beta j; m}
\notag \\
& = Z_\text{LQCD}^{-1} \int \fD{U} ~ e^{ -S_\text{LQCD} \fconfig{U} } ~ M
\fconfig{U}^{-1}_{a \alpha i n, b \beta j m} \notag \\
& = Z_\text{LQCD}^{-1} \int \fD{U} ~ e^{ -S_g \fconfig{U} } ~ \det M
\fconfig{U} ~ M \fconfig{U}^{-1}_{a \alpha i n, b \beta j m}
\end{align}
where the correlator's color, spin, flavor, and position indices respectively
are shown explicitly.  We see that $M \fconfig{U}^{-1}$ is acting as the quark
propagator.

Because the interaction matrix is constructed to take quarks into antiquarks,
its transpose appears as the antiquark propagator:
\begin{align}
\bra q_{a \alpha i; n} & \qb_{b \beta j; m} \ket \notag \\
& = Z_\text{LQCD}^{-1} \int \fD{U} ~ e^{ -S_g \fconfig{U} } ~ \det M
\fconfig{U} ~ ( M \fconfig{U}^T )^{-1}_{a \alpha i n, b \beta
j m}
\end{align}
Inspection of the adjoint of the interaction matrix allows us to express the
antiquark propagator in terms of the quark propagator:
\begin{align}
M^N \fconfig{U}^\dagger_{n,m} & = \frac{1}{2} \sum_\mu \gamma_\mu \Bigl[
U^{}_{\mu; m} \delta^{}_{m, n - \muh} - U^\dagger_{\mu; m - \muh} \delta^{}_{m,
n + \muh} \Bigr] + m_Q \delta_{m,n} \notag \\
& = \frac{1}{2} \sum_\mu \bigl( -\gamma_\mu \bigr) \Bigl[ U^{\dagger}_{\mu; m
- \muh} \delta^{}_{m, n + \muh} - U^{}_{\mu;m} \delta^{}_{m, n - \muh} \Bigr]
+ m_Q \delta_{n,m} \notag \\
& = \gamma_5 \biggl\{ \frac{1}{2} \sum_\mu \gamma_\mu \Bigl[ U^\dagger_{\mu;
n} \delta^{}_{n, m - \muh} - U^{}_{\mu; n - \muh} \delta^{}_{n, m + \muh}
\Bigr] + m_Q \delta_{n,m} \biggr\} \gamma_5 \notag \\
& = \gamma_5 M^N_{n,m} \fconfig{U} \gamma_5
\label{aq:equation}
\end{align}
Thus, for this interaction matrix:
\begin{align}
\bra q_{a \alpha i; n} & \qb_{b \beta j; m} \ket \notag \\
& = Z_\text{LQCD}^{-1} \int \fD{U} ~ e^{ -S_g \fconfig{U} } ~ \det M
\fconfig{U} ~ \bigl( \gamma_5 M^N \fconfig{U}^{-1} \gamma_5 \bigr)^*_{a
\alpha i n, b \beta j m}
\end{align}
Note that $\gamma_5$ is its own inverse.

While it is instructive to consider the calculation of a quark-antiquark
correlator due to its simplicity, it is worth noting that, because it is not a
gauge-invariant quantity, its expectation value on each configuration includes
a random phase and thus equals zero after the ensemble average.  Quark
operators with a non-zero expectation value are more complex than the simple
two-point quark correlator.  Using Wick contractions, operators with four or
more quark insertions can be reduced to a sum of products of quark and
antiquark propagators.  If the operator is such that these products of
propagators form closed loops, their expectation value with be gauge invariant
and potentially non-zero.

\subsection{Quenched Approximation}
The quark fields have introduced the determinant of their interaction matrix
into the Boltzmann weight.  From the perspective of the gauge fields, this
determinant acts as a non-local interaction, coupling each link to every other
link in the lattice.  While it is still possible to account for this
complicated Boltzmann weight in our Markov process, the computation time
required will be significantly increased.

In order to avoid this increase, and to return to the simplicity of the pure
gauge action, an approximation is often made within the partition function
known as the quenched approximation.  The quenched approximation assumes that:
\begin{equation}
\det M \fconfig{U} \propto e^{ -S_g }
\end{equation}
Thus, the determinant can be accounted for by a simple shift in $\beta$, and
the partition function reverts to that for pure gauge.  Calculation of a
quark-antiquark correlator under the quenched approximation amounts to the
expression:
\begin{equation}
\bra \qb_{a \alpha i; n} q_{b \beta j; m} \ket = Z_\text{qQCD}^{-1}
\int \fD{U} ~ e^{ -S_g \fconfig{U} } ~ M \fconfig{U}^{-1}_{a
\alpha i n, b \beta j m}
\end{equation}
where $Z_\text{qQCD}$ is appropriately defined.

While the quenched partition function does not correspond to any well-defined
unitary field theory, the resulting system of interactions is often referred to
as quenched QCD (qQCD).

From the perspective of perturbation theory, removal of the interaction matrix
determinant from the QCD partition function corresponds to the removal of quark
loops from Feynman diagrams.

\subsection{Partially Quenched Approximation} \label{m:section}
Even when the interaction-matrix determinant is included in the Boltzmann
weight during a Markov process, a second approximation known as the partially
quenched approximation is often made.

The partially quenched approximation arises from the fact that the quark
interaction matrix is a function of the quark mass, $M_{(m_Q)} \fconfig{U}$.
Thus, if we wish to calculate the expectation value of an operator at a variety
of quark masses, we must generate a separate ensemble for each quark mass.
However, the creation of a Markov chain is computationally expensive, much more
so than the calculation of a quark propagator under that Markov chain.

In the partially quenched approximation we generate only a single Markov chain
using a single quark mass known as the sea, or dynamical, quark mass $m_S$.
Observables, such as a quark propagator, are then calculated under that
ensemble using a number of different quark mass values.  The quark mass used in
the observable is known as the valence quark mass $m_V$.  Partial quenching is
thus an approximation of the interaction-matrix determinant at the valence
quark mass by that determinant evaluated at the dynamical quark mass.

Calculation of a quark-antiquark correlator under the partially quenched
approximation amounts to the expression:
\begin{align}
\bra \qb_{a \alpha i; n} & q_{b \beta j; m} \ket \notag \\
& = Z_\text{pqQCD}^{-1} \int \fD{U} ~ e^{ -S_g \fconfig{U} } ~ \det M_{(m_S)}
\fconfig{U} ~ M_{(m_V)} \fconfig{U}^{-1}_{a \alpha i n, b \beta j m}
\label{bs:equation}
\end{align}
where $Z_\text{pqQCD}$ is appropriately defined.

In a similar fashion to quenching, partial quenching does not result in a
unitary field theory.  Nonetheless, the system of interactions which results
from partial quenching is referred to as partially quenched QCD (pqQCD).

From the perspective of partial quenching, qQCD is the special case of pqQCD in
which the dynamical quark mass is taken to infinity.  The mass term in the
quark action $S_q \fconfig{U, q, \qb}$ dominates, and the dynamical quarks
decouple from the gauge fields, affecting the partition function only as an
irrelevant constant factor.

Here we have presented pqQCD as a convenient approximation of full QCD.
However, we will find that extensions to the concepts behind ChPT will take
partial quenching beyond a mere approximation, allowing us to calculate
physical results incalculable within unquenched QCD.

\subsection{Fermion Doubling Problem}
In order to determine the particles contained in our discretized quark
action, we must identify the zeros of the momentum-space action.

Introducing the momentum-space quark fields:
\begin{align}
q_n = \int_k e^{i a k \cdot n} q(k)
&&
\qb_n = \int_k e^{i a k \cdot n} \qb(k)
\end{align}
allows for a Fourier transformation of the free-field action:
\begin{align}
S_q^N \fconfig{\identity, q, \qb} & = \sum_n \biggl\{ \frac{1}{2} \sum_\mu
\qb_n \gamma_\mu \Bigl[ q_{n + \muh} - q_{n - \muh} \Bigr] + m_Q \qb_n
q_n \biggr\} \notag \\
& = \int_k \biggl\{ \frac{1}{2} \sum_\mu \qb(k) \gamma_\mu \Bigl[ e^{i a k_\mu}
- e^{-i a k_\mu} \Bigr] q(k) + m_Q \qb(k) q(k) \biggr\} \notag \\
& = \int_k \qb(k) M^N_{(k)} \fconfig{\identity} q(k)
\end{align}
where:
\begin{equation}
M^N_{(k)} \fconfig{\identity} = i \sum_\mu \gamma_\mu \sin a k_\mu + m_Q
\end{equation}
As in the continuum case, the free-field action is diagonal in momentum space.

For the remainder of this section, we will assume $m_Q = 0$.

Using an infinitely large lattice volume has resulted in a continuous momentum
space.  However, our discretization of position space has resulted in an action
which is periodic with respect to momentum.  Additionally, momentum space
itself is periodic, with each component confined to a Brillouin zone of length
$2 \pi a^{-1}$.  We choose the range of allowed momenta to be $ -\efrac{\pi}{2
a} < k_\mu \leq \efrac{3 \pi}{2 a}$, such that the maxima of the action occur
on the limits of the Brillouin zone.  

Each zero of the momentum-space action corresponds to a pole in the propagator,
and thus a particle of the theory.  We expect one particle, the particle
corresponding to the zero at $k = 0$.  However, the periodicity of the action
results in fifteen additional zeros, generating fifteen additional species of
quarks.  This unexpected proliferation of quark species is known as the fermion
doubling problem.

The fermion doubling problem is a direct consequence of the linear nature of
the Dirac action.  Had the action been quadratic, as is the case for scalar
fields, the additional zeros would have been pushed beyond the Brillouin zone.

Any of these additional quark species can be shifted to the origin via the
redefinition:
\begin{equation}
k_\mu \pad{\mshiftarrow} k'_\mu = k_\mu + A_\mu \frac{\pi}{a}
\end{equation}
where $A$ is one of sixteen possible binary four-vectors, each of which
corresponds to a single quark species.  We define a binary vector to be a
vector whose components take on only the values zero and one.  Under this
redefinition of momentum:
\begin{align}
M^N_{(k)} \fconfig{\identity} \pad{\mshiftarrow} & i \sum_\mu (-)^{A_\mu}
\gamma_\mu \sin a k_\mu \notag \\
& = i \sum_\mu \gamma^A_\mu \sin a k_\mu
\end{align}
where, for each quark species, the Dirac matrices have been redefined in order
to return $M^N_{(k)} \fconfig{\identity}$ to its proper form:
\begin{align}
\gamma^A_\mu & \equiv (-)^{A_\mu} \gamma_\mu \\
\gamma^A_5 & \equiv \gamma^A_1 \gamma^A_2 \gamma^A_3 \gamma^A_4 = \gamma_5
\prod_\mu (-)^{A_\mu}
\end{align}
For half of the quark species, $\gamma_5$ has switched sign, resulting in a
corresponding switch in the definition of chirality.  Thus, not only does the
fermion doubling problem introduce unexpected quark species to the theory, but
it also hopelessly entangles the theory's left- and right-handed degrees of
freedom, making impossible their independent rotation.  Thus, the definition of
a discretized version of chiral symmetry becomes impossible.

The inevitability of fermion doubling for any naive action is demonstrated by
the Nielsen-Ninomiya theorem and is detailed in \cite{Karsten:1981wd,
Nielsen:1981rz, Nielsen:1981xu}

\section{Staggered Fermions}
There are several ways of handling the fermion doubling problem, each of which
has its own advantages and disadvantages.  The original and perhaps most
commonly used method is known as Wilson fermions \cite{Wilson:1975id}.  A term
is added to the fermion action which grants the extra quark species a large
mass, effectively decoupling them from the theory.  However, this term also
destroys all chiral symmetry.  Because of the importance of chiral symmetry to
our study, such a trampling of the symmetry is unacceptable.  Thus, we make use
of a second fermion formulation: Kogut-Susskind, or staggered, fermions
\cite{Kogut:1975ag, Banks:1976gq, Susskind:1977jm}, a formulation in which a
subset of chiral symmetry is retained at non-zero lattice spacing.

\subsection{Staggered Action}
The staggered fermion formulation begins with a realization that, through a
redefinition of the quark fields, we can diagonalize the fermion action with
respect to spin.  Once each spin component is independent, we drop three of the
four components, reducing the degrees of freedom within the quark species by a
factor of four.  Finally, the resulting sixteen single-spin-component quark
species are collected together to form four flavors of Dirac quarks.

The formulation of staggered fermions makes use of several phase functions,
which we define now:
\begin{gather}
\begin{align}
\eta_{\mu; n} \equiv \prod_{\nu < \mu} (-)^{n_\nu}
&&
\zeta_{\mu; n} \equiv \prod_{\nu > \mu} (-)^{n_\nu}
\end{align} \\
\epsilon_n \equiv \prod_\mu (-)^{n_\mu}
\end{gather}

The staggered fermion fields are defined by a redefinition of the quark fields:
\begin{align}
\chi_n \equiv \Gamma_n q_n
&&
\chib_n \equiv \qb^{}_n \Gamma^\dagger_n
\end{align}
where $\Gamma_n$ is a spin-space matrix, defined by:
\begin{equation}
\Gamma_n \equiv \prod_{\mu} \gamma^{n_\mu}_\mu = \gamma^{n_1}_1 \gamma^{n_2}_2
\gamma^{n_3}_3 \gamma^{n_4}_4
\end{equation}
Recall that the components of $n$ are integers and that $\gamma_\mu \gamma_\mu
= \identity$.  Thus, $\Gamma_n$ traverses only sixteen possible values.

Substituting $\chi$ and $\chib$ into the single-flavor fermion action:
\begin{align}
& S_q^N \fconfig{U, q, \qb} = \sum_n \biggl\{ \frac{1}{2} \sum_\mu \qb_n
\gamma_\mu \Bigl[ U^\dagger_{\mu; n} q^{}_{n + \muh} - U^{}_{\mu; n - \muh}
q^{}_{n - \muh} \Bigr] + m_Q \qb_n q_n \biggr\} \notag \\
& \hspace{1ex} = \sum_n \biggl\{ \frac{1}{2} \sum_\mu \chib^{}_n
\Gamma^\dagger_n \gamma_\mu \Bigl[ U^\dagger_{\mu; n} \Gamma^{}_{n + \muh}
\chi^{}_{n + \muh} - U^{}_{\mu; n - \muh} \Gamma^{}_{n - \muh} \chi^{}_{n -
\muh} \Bigr] + m_Q \chib_n \Gamma^\dagger_n \Gamma^{}_n \chi_n \biggr\} \notag
\\
& \hspace{1ex} = \sum_n \biggl\{ \frac{1}{2} \sum_\mu \eta_{\mu; n} \chib^{}_n
\Gamma^\dagger_n \gamma_\mu \gamma_\mu \Gamma_n \Bigl[ U^\dagger_{\mu; n}
\chi^{}_{n + \muh} - U^{}_{\mu; n - \muh} \chi^{}_{n - \muh} \Bigr] + m_Q
\chib_n \chi_n \biggr\} \notag \\
& \hspace{1ex} = \sum_n \biggl\{ \frac{1}{2} \sum_\mu \eta_{\mu; n} \chib_n
\Bigl[ U^\dagger_{\mu; n} \chi^{}_{n + \muh} - U^{}_{\mu; n - \muh} \chi^{}_{n
- \muh} \Bigr] + m_Q \chib_n \chi_n \biggr\}
\end{align}
where the following identities have been used:
\begin{align}
\Gamma^\dagger_n \Gamma^{}_n & = \identity \\
\Gamma_{n + \muh} = \Gamma_{n - \muh} & = \eta_{\mu; n} \gamma_\mu \Gamma_n =
\zeta_{\mu; n} \Gamma_n \gamma_\mu
\end{align}
The Dirac spin matrix $\gamma_\mu$ in the fermion action has been replaced by a
simple scalar phase function $\eta_{\mu; n}$.  Thus, the action is now diagonal
with respect to spin, and the spin components of the redefined fields have
decoupled.  We now drop three of the four spin components, leaving $\chi_n$ a
scalar in spin space and a vector only in color space.  The result is the
staggered fermion action:
\begin{align}
S_q^S \fconfig{U, \chi, \chib} & = \sum_n \sum_m \chib_n M^S_{n,m} \fconfig{U}
\chi_m
\label{bb:equation} \\
M^S_{n,m} \fconfig{U} & = \frac{1}{2} \sum_\mu \eta_{\mu; n}
\Bigl[ U^\dagger_{\mu; n} \delta^{}_{n, m - \muh} - U^{}_{\mu; n - \muh}
\delta^{}_{n, m + \muh} \Bigr] + m_Q \delta_{n,m} \notag \\
& = D^S_{n,m} \fconfig{U} + m_Q \delta_{n,m}
\end{align}
where $M^S \fconfig{U}$ is now a matrix in only color and position.

\subsection{Shift Symmetry}
In general, a lattice field theory will have a discrete translation symmetry in
which the fields are translated multiples of the lattice spacing along
directions flush with the lattice.  In the continuum limit this discrete
symmetry becomes the continuum's continuous translation symmetry.

In the case of staggered fermions, however, the appearance of the phase factor
$\eta_{\mu;n}$ in the action, whose value varies between lattice sites, does
not allow for a straightforward definition of translation symmetry.  Instead,
the action has the more complex shift symmetry, sometimes referred to as
$\zeta$-shift symmetry, defined by the transformation:
\begin{align}
\chi_n & \pad{\symarrow} \chi_n' = \zeta_{\nu;n} \chi_{n + \hat{\nu}} \\
\chib_n & \pad{\symarrow} \chib_n' = \zeta_{\nu;n} \chib_{n + \hat{\nu}}
\end{align}
where $\hat{\nu}$ points in the direction of the shift.  It is easy to
verify that this is a symmetry of the staggered action through use of the
identity:
\begin{equation}
\eta_{\mu;\hat{\nu}} = \zeta_{\nu;\muh}
\end{equation}

Noting $\zeta_{\nu;n} \zeta_{\nu;n + \hat{\nu}} = \zeta_{\nu;\hat{\nu}} =
1$ reveals that the standard translation symmetries can still be defined, yet
only for translation lengths which are a multiple of two lattice spacings.

The transfer matrix, which propagates states through time and whose eigenstates
define the states of a theory, is equivalent to the time component of the
translation operator.  Thus, a staggered transfer matrix which propagates
states forward a single lattice step can not be defined.  Rather, the transfer
matrix can only be constructed such that it propagates states forward two time
steps per application.

\subsection{Even-Odd Symmetry} \label{al:section}
We define even sites to be those for which summing the components of $n$
results in an even number.  Conversely, odd sites are defined as those for
which the sum is odd.  Inspection of $M^S \fconfig{U}$ reveals that, for $m_Q =
0$, $\chi$ on even sites couples only to $\chib$ on odd sites.  Similarly,
$\chi$ on odd sites couples only to $\chib$ on even sites.  Thus, the staggered
fermion action has a $U(1)_e \otimes U(1)_o$ symmetry, defined by the
transformation:
\begin{align}
\chi_n & \pad{\symarrow} \chi_n' = \left\{
\begin{aligned}
& e^{i \alpha_e } \chi_n \phantom{e^{-i \alpha_o}} \text{even $n$} \\
& e^{i \alpha_o } \chi_n \phantom{e^{-i \alpha_e}} \text{odd $n$}
\end{aligned}
\right. \\[4mm]
\chib_n & \pad{\symarrow} \chib_n' = \left\{
\begin{aligned}
& \chib_n e^{-i \alpha_o } \phantom{e^{i \alpha_e}} \text{even $n$} \\
& \chib_n e^{-i \alpha_e } \phantom{e^{i \alpha_o}} \text{odd $n$}
\end{aligned}
\right.
\end{align}
For $m_Q \neq 0$, only a subset of the symmetry remains, that for which
$\alpha_e = \alpha_o$.

The symmetry can also be expressed in terms of a $U(1)_1 \otimes U(1)_\epsilon$
symmetry:
\begin{align}
\chi_n & \pad{\symarrow} \chi_n' = e^{i ( \alpha_1 + \epsilon_n
\alpha_\epsilon ) } \chi_n
\label{ae:equation} \\
\chib_n & \pad{\symarrow} \chib_n' = \chib_n e^{-i ( \alpha_1 + \epsilon_n
\alpha_\epsilon ) }
\label{af:equation} 
\end{align}
where the connection to even-odd symmetry is made via $\alpha_1 = \half (
\alpha_e + \alpha_o )$ and $\alpha_\epsilon = \half ( \alpha_e - \alpha_o )$.
For $m_Q \neq 0$, the $U(1)_\epsilon$ symmetry is broken, and only $U(1)_1$
remains.  This $U(1)_\epsilon$ symmetry is the one component of chiral symmetry
which survives the discretization process and is the reason we use the
staggered fermion formulation.

One other consequence of even-odd symmetry is that the square of the
interaction matrix $M^S \fconfig{U}^\dagger M^S \fconfig{U} $ couples only even
sites to even sites and odd sites to odd sites.  This becomes useful in
simplifying certain calculations.

\subsection{Quarks}
The staggered fermion action leaves us with sixteen species of particles, each
of which has a single spin and flavor component.  Collecting these degrees of
freedom, we define our quark field such that the result is four flavors of
Dirac spinors.

One option is to make the definition in momentum space
\cite{Sharatchandra:1981si, vandenDoel:1983mf, Golterman:1984cy}, assigning
each of the zeros of the momentum-space action a distinct spin and flavor
index.  However, such a procedure leads to a highly non-local definition of the
quark field.  Instead, we chose to define the quark field in position space,
constructing the quark field $Q$ at a single point by gathering together nearby
degrees of freedom in $\chi$.  

We divide the lattice into hypercubes.  A subset of our full lattice, to which
each hypercube contributes its lowest member site, makes up a lattice with a
spacing of $2 a$.  We label the sites on this lattice with the index $\n$, such
that $\n / 2 \in \mathbb{Z}^4$, and define our quark field $Q$ such that it
takes on values only at the sites of this coarser lattice.  Each spin- and
flavor-component combination of $Q_\n$ is built from a unique linear
combination of $\chi$ at the sixteen corners of the hypercube $\n$.  The quark
field $(Q_\n)_{\alpha i}$ is constructed as a $4 \times 4$ matrix which mixes
spin and flavor space, with one spin index $\alpha$ and one flavor index $i$:
\begin{align}
(Q_\n)_{\alpha i} \equiv \frac{1}{2} \sum_A (\Gamma_A)_{\alpha i} \chi_{\n + A}
&&
(\Qb_\n)_{i \alpha} \equiv \frac{1}{2} \sum_A \chib_{\n + A}
(\Gamma^\dagger_A)_{i \alpha}
\end{align}
where the summation is over all possible binary four-vectors, and visiting each
corner of the hypercube $\n$ in turn.  Note that the normalization of this
definition varies in the literature.

We will find that this spatial separation between the various spin and flavor
degrees of freedom in our quark field breaks a majority of spin and flavor
symmetry.  Since the different spin and flavor degrees of freedom of the quark
field couple to different gauge links, they each experience a distinct gauge
environment.  Only in the continuum limit, where the lattice spacing goes to
zero and the spatial separation is removed, will the full spin and flavor
symmetries of our theory return.

\subsection{Quark Bilinears}
With our quark field defined, we can construct quark bilinears, operators
consisting of local quark-antiquark pairs with a specific spin and flavor
structure.  It is through bilinears that we will access the spin and flavor
structure of the staggered formalism, using them as the creation and
annihilation operators of two-quark bound states, our lattice mesons.

In the continuum we represented bilinears with the notation
$\mathcal{J}_\Gamma^a = \qb \Gamma \tau^a q$, where $\Gamma$ was some set of
spin-space Dirac matrices which determined the bilinear's spin structure and
$\tau^a$ was some generator of flavor-space rotations which determined the
bilinear's flavor structure.

In the context of staggered quarks, bilinears are represented using the
notation:
\begin{align}
\mathcal{J}_{\mathcal{S}, \mathcal{F}; \n} & = \Qb_\n
\bilinear{\mathcal{S}}{\mathcal{F}} Q_\n \notag \\
& \equiv (\Qb_\n)_{i \alpha} (\gamma_\mathcal{S})_{\alpha \beta}
(\xi_\mathcal{F})_{i j} (Q_\n)_{\beta j} \notag \\
& = \Tr{ \Qb_\n \gamma_\mathcal{S} Q_\n \xi^T_\mathcal{F} }
\end{align}
where $\gamma_\mathcal{S}$ is a matrix which contracts the quark field's spin
indices and represents the spin structure of the bilinear, and
$\xi_\mathcal{F}$ is a matrix which contracts the quark field's flavor indices
and represents the flavor structure of the bilinear.

Because the number of quark flavors equals the number of spin components,
similar bases can be used for the bilinears' spin and flavor matrices.  Each
matrix takes on the value of one of the sixteen members of a Clifford algebra:
\begin{align}
\gamma_\mathcal{S} \pad{\in}
\left\{
\begin{gathered}
\identity \\
\gamma_\mu \\
\gamma_\mu \gamma_\nu \\
\gamma_\mu \gamma_5 \\
\gamma_5
\end{gathered}
\right.
&&
\xi_\mathcal{F} \pad{\in}
\left\{
\begin{gathered}
\identity \\
\gamma^*_\mu \\
\gamma^*_\mu \gamma^*_\nu \\
\gamma^*_\mu \gamma^*_5 \\
\gamma^*_5
\end{gathered}
\right.
\end{align}

The members of a Clifford algebra can be expressed in a manner which parallels
the definition of $\Gamma_n$.  Treating $\mathcal{S}$ and $\mathcal{F}$ as
binary four-vectors which encode a bilinear's spin and flavor structure:
\begin{align}
\gamma_\mathcal{S} & \equiv \Gamma_\mathcal{S} = \prod_\mu
\gamma^{\mathcal{S}_\mu}_\mu = \gamma^{\mathcal{S}_1}_1
\gamma^{\mathcal{S}_2}_2 \gamma^{\mathcal{S}_3}_3 \gamma^{\mathcal{S}_4}_4
\label{ax:equation} \\
\xi^*_\mathcal{F} & \equiv \gamma_\mathcal{F} = \Gamma_\mathcal{F} = \prod_\mu
\gamma^{\mathcal{F}_\mu}_\mu = \gamma^{\mathcal{F}_1}_1
\gamma^{\mathcal{F}_2}_2 \gamma^{\mathcal{F}_3}_3 \gamma^{\mathcal{F}_4}_4
\end{align}

With this notation in place, we can express bilinears in terms of the $\chi$
and $\chib$ fields:
\begin{align}
\Qb_\n \bilinear{\mathcal{S}}{\mathcal{F}} Q_\n & = \Tr{ \Qb_\n
\gamma_\mathcal{S} Q_\n \xi^T_\mathcal{F} } \notag \\
& = \frac{1}{4} \sum_A \sum_B \Tr{ \Gamma^\dagger_A \gamma_\mathcal{S}
\Gamma_B \gamma^\dagger_\mathcal{F} } \chib_{\n + A} \chi_{\n + B}
\label{aa:equation}
\end{align}
Thus, a bilinear at a hypercube consists of a linear combination of all
possible contractions of $\chi$ and $\chib$ within that hypercube, where the
coefficients of the linear combination depend on the spin and flavor structure
of the bilinear.

In truth \eqref{aa:equation} is somewhat oversimplified.  The definition of the
quark bilinears is complicated by the fact that link matrices live in the
hypercube over which the bilinear is defined.  Thus, in order for the quark
bilinear definition to have the correct color structure, link matrices must be
included between non-local contractions of $\chi$ and $\chib$:
\begin{equation}
\Qb_\n \bilinear{\mathcal{S}}{\xi_\mathcal{F}} Q_\n = \frac{1}{4} \sum_A
\sum_B \Tr{ \Gamma^\dagger_A \gamma_\mathcal{S} \Gamma_B
\gamma^\dagger_\mathcal{F} } \chib_{\n + A} \mathcal{U}_{A,B;\n} \chi_{\n + B}
\label{ak:equation}
\end{equation}
where $\mathcal{U}_{A,B;\n}$ represents an equal weighting of all possible
shortest gauge link chains connecting the corner $B$ to the corner $A$ of the
hypercube $\n$.  Note that because $\mathcal{U}_{A,B;h}$ is in general a sum of
unitary matrices, it is not itself unitary.

Returning again to the free-field case, we can solve for the coefficients of
the linear combination:
\begin{align}
\Qb_\n \bilinear{\mathcal{S}}{\mathcal{F}} Q_\n & = \frac{1}{4} \sum_A \sum_B
\Tr{ \Gamma^\dagger_A \gamma_\mathcal{S} \Gamma_B \gamma^\dagger_\mathcal{F} }
\chib_{\n + A} \chi_{\n + B} \notag \\
& = \sum_A \sum_B \delta_{A + \mathcal{S}, B + \mathcal{F}} (-)^{A \cdot
\bar{\eta}_{A + \mathcal{S}}} (-)^{B \cdot \bar{\zeta}_{B + \mathcal{F}}}
\chib_{h + A} \chi_{h + B} \notag \\
& = \sum_A (-)^{\varphi_{\mathcal{S}, \mathcal{F}; A}} \chib_{h + A} \chi_{h +
A + \mathcal{S} + \mathcal{F}}
\label{ag:equation}
\end{align}
where:
\begin{equation}
\varphi_{\mathcal{S}, \mathcal{F}; A} \equiv A \cdot \bigl(
\bar{\zeta}_\mathcal{S} + \bar{\eta}_\mathcal{F} \bigr) + \mathcal{S} \cdot
\bar{\eta}_{\mathcal{S} + \mathcal{F}}
\end{equation}
and we have used the following definitions:
\begin{align}
\eta_{\mu;A} = (-)^{{\bar{\eta}}_{\mu;A}}
&&
\bar{\eta}_{\mu;A} \equiv \sum_{\nu < \mu} A_\nu
\label{cf:equation} \\
\zeta_{\mu;A} = (-)^{{\bar{\zeta}}_{\mu;A}}
&&
\bar{\zeta}_{\mu;A} \equiv \sum_{\nu > \mu} A_\nu
\label{cd:equation}
\end{align}
and identities:
\begin{gather}
\begin{align}
A \cdot \bar{\eta}_B & = B \cdot \bar{\zeta}_A \\
\Tr{ \Gamma^\dagger_A \Gamma^{}_B } & = 4 \delta_{A,B}
\end{align} \\
\begin{align}
\Gamma^\dagger_A \Gamma^{}_B = (-)^{A \cdot \bar{\eta}_{A + B}} \Gamma^{}_{A
+ B}
&&
\Gamma^{}_A \Gamma^\dagger_B = (-)^{A \cdot \bar{\zeta}_{A + B}}
\Gamma^\dagger_{A + B}
\end{align}
\end{gather}
We define the addition of binary vectors modulo two, such that the
sum of two binary vectors is itself a binary vector.  When adding binary and
standard vectors together, as per the case $h + A + \mathcal{S} + \mathcal{F}$,
the addition of the binary vectors is carried out first.

From \eqref{ag:equation} we see that, for any given $\mathcal{S}$ and
$\mathcal{F}$ combination, only sixteen of the sum's coefficients are non-zero.
Also, those which are non-zero equal a simple sign factor.

The second term in $\varphi_{\mathcal{S}, \mathcal{F}; A}$ can be factored out
of the hypercubic sum, and contributes only an overall sign:
\begin{equation}
\Qb_\n \bilinear{\mathcal{S}}{\mathcal{F}} Q_\n = (-)^{\mathcal{S} \cdot
\bar{\eta}_{\mathcal{S} + \mathcal{F}}} \sum_A (-)^{\varphi'_{\mathcal{S},
\mathcal{F}; A}} \chib_{h + A} \chi_{h + A + \mathcal{S} + \mathcal{F}}
\end{equation}
where:
\begin{equation}
\varphi'_{\mathcal{S}, \mathcal{F}; A} \equiv A \cdot \bigl(
\bar{\zeta}_\mathcal{S} + \bar{\eta}_\mathcal{F} \bigr)
\end{equation}

For a given bilinear $\mathcal{J}_{\mathcal{S}, \mathcal{F}}$ we define the
binary four-vector $\mathcal{D}_{\mathcal{S}, \mathcal{F}} = \mathcal{S} +
\mathcal{F}$, and we define the distance of that bilinear to be the number of
non-zero elements of $\mathcal{D}_{\mathcal{S}, \mathcal{F}}$.  Inspection of
\eqref{ag:equation} reveals that the non-zero contractions within a bilinear
are always between corners of the hypercube separated by an offset equal to
$\mathcal{D}_{\mathcal{S}, \mathcal{F}}$, and thus the number of links
separating those corners always equals the distance of the bilinear.

The simplest bilinears are those in which $\mathcal{S} = \mathcal{F}$.  They
are known as local bilinears as they have a distance of zero, and thus all of
their contractions are local:
\begin{equation}
\Qb_\n \bilinear{\mathcal{S}}{\mathcal{S}} Q_\n = \sum_A (-)^{A \cdot (
\bar{\zeta}_\mathcal{S} + \bar{\eta}_\mathcal{S} ) } \chib_{\n + A} \chi_{\n +
A}
\label{ai:equation}
\end{equation}

In the more specific cases of $\gamma_\mathcal{S} = \xi^*_\mathcal{F} =
\identity$ and $\gamma_\mathcal{S} = \xi^*_\mathcal{F} = \gamma_5$, the phase
factors become quite simple:
\begin{align}
\Qb_\n \bigl( \identity \otimes \identity \bigr) Q_\n & = \sum_A \chib_{\n + A}
\chi_{\n + A} \\
\Qb_\n \bilinear55 Q_\n & = \sum_A \epsilon_{A} \chib_{\n + A} \chi_{\n + A}
\end{align}

We can generalize our perspective on the structure
$\bilinear{\mathcal{S}}{\mathcal{F}}$ beyond its use in the notation of
bilinears by viewing it as a position- and color-space matrix which operates
on $\chi$ and $\chib$ field configurations, also referred to as fermion field
vectors.  For the local case the matrix is diagonal, and simply applies a sign
factor to the $\chi$ and $\chib$ field at each lattice site.  In the non-local
case the matrix also swaps the fermion degrees of freedom around within each
hypercube, shifting the $\chi$ field at corner $A$ on each hypercube to corner
$A + \mathcal{S} + \mathcal{F}$ and applying gauge link matrices as
appropriate.  Stated explicitly as a matrix operating on $\chi$ and $\chib$
field vectors, $\bilinear{\mathcal{S}}{\mathcal{F}}$ has the form:
\begin{equation}
\bilinear{\mathcal{S}}{\mathcal{F}}_{n,m} = (-)^{\mathcal{S} \cdot
\bar{\eta}_{\mathcal{S} + \mathcal{F}}} (-)^{A \cdot ( \bar{\zeta}_\mathcal{S}
+ \bar{\eta}_\mathcal{F} )} \delta_{h + A + \mathcal{S} + \mathcal{F}, m}
\mathcal{U}_{A, A + \mathcal{S} + \mathcal{F}; h}
\label{ah:equation}
\end{equation}
where $h$ denotes the hypercube containing $n$, and $A$ denotes its position
within that hypercube:
\begin{align}
h = 2 \Bigl\lfloor \frac{n}{2} \Bigr\rfloor
&&
A = n - h
\end{align}
The first two terms on the right-hand side of \eqref{ah:equation} apply a sign
factor to $\chi$, the second of which is position dependent, the third term
swaps the $\chi$ degrees of freedom around within each hypercube, and the
fourth term applies the appropriate color matrices for that movement.  From this
perspective, a bilinear is seen as the application of such a matrix to a
fermion field vector which is non-zero only at all corners of a single
hypercube, and then the contraction of the result with the adjoint of the
original vector.

\subsection{Meson States} \label{ai:section}
Each bilinear corresponds to a set of values for the spin and flavor quantum
numbers of our staggered lattice theory and thus create and annihilate all
states with like quantum numbers.  The lightest state created by a bilinear
will be a two-quark bound state with appropriate quantum numbers, a staggered
lattice meson.  By determining a correspondence between lattice and continuum
quantum numbers, we can identify what continuum mesons our lattice mesons
become as we move to the continuum limit.

We choose to label each staggered lattice meson using the name of the lightest
Standard Model meson which has corresponding spin quantum numbers.  Flavor
structure takes a backseat to spin structure when labeling our staggered states
because our staggered theory contains four degenerate light quarks, while the
Standard Model contains three non-degenerate light quarks.  Thus, there is no
direct correspondence between the flavor quantum numbers of our staggered
mesons and the mesons of the Standard Model.

The staggered meson states, each listed with a bilinear having matching quantum
numbers, include:
\begin{alignat}{2}
\text{pion:} & & \qquad & \Qb \bilinear5{\mathcal{F}} Q \\
\text{rho:} & & & \Qb \bilinear{i}{\mathcal{F}} Q \\
\text{scalar:} & & & \Qb \Bilinear{\identity}{\xi_\mathcal{F}} Q
\end{alignat}
where $i \in \{ 1, 2, 3 \}$ chooses the polarization of the rho and the
flavor structure of the states has been left unspecified.

The flavor structure $\xi_\mathcal{F}$ of each listed bilinear takes on one of
sixteen possible values.  In the continuum limit the degeneracy of the four
staggered quark flavors causes these sixteen states to be themselves
degenerate.  However, at non-zero lattice spacing, all but one of the flavor
structures becomes non-local, flavor symmetry is broken, and the masses of the
sixteen states split.

For each spin structure, one of the flavor structure choices will correspond to
a local bilinear.  In the case of the staggered pion, that state is:
\begin{alignat}{2}
\text{local pion:} & & \qquad & \Qb \bilinear55 Q
\end{alignat}
At non-zero lattice spacing only this local pion remains light, while the other
fifteen non-local pions gain a mass contribution dependent upon the lattice
spacing.

Thus far, we have ignored two complications which are inherent in the
correspondence between the bilinear operators and the staggered meson states.

The first complication relates to the time component of a state's spin.  Recall
that the spin of a particle is defined relative to its momentum four-vector.
In the case of states with zero spatial momentum, which we create by summing
over a time slice as described in Section \ref{a:section}, the state's momentum
four-vector points along the time direction.  Thus, the time component of our
state's spin is undefined, and there exist more bilinears than there are states
to create.  As a consequence, the bilinears $\Qb
\bilinear{\mathcal{S}}{\mathcal{F}} Q$ and $\Qb \Bilinear{\gamma_4
\gamma_\mathcal{S}}{\xi_\mathcal{F}} Q$ both couple to the same state.

The second complication arises because states decay exponentially as they
propagate through time, and thus the use of creation operators, or sources,
which are distributed in time, such as bilinears, is not straightforward.  We
will find that each bilinear couples to states with two sets of quantum
numbers, one set which was naively unexpected.

Looking at the two bilinears $\Qb \bilinear{\mathcal{S}}{\mathcal{F}} Q$ and
$\Qb \Bilinear{\gamma_4 \gamma_5 \gamma_\mathcal{S}}{\xi_4 \xi_5
\xi_\mathcal{F}} Q$ and comparing the sign factors of their $\chi$ with $\chib$
contractions, we find the factors the same on a given time slice.  The
difference between their sign factors only arises when we compare neighboring
time slices.  The sign factors of one bilinear will be constant between time
slices, while the other's factors will flip in sign each time step.

By defining the binary four-vector $\mathcal{K}$ such that $\gamma_\mathcal{K}
= \gamma_5 \gamma_4$, we can state the above explicitly:
\begin{equation}
\varphi'_{\mathcal{S} + \mathcal{K}, \mathcal{F} + \mathcal{K}; A} =
\varphi'_{\mathcal{S}, \mathcal{F}; A} + A_4
\end{equation}
The only distinction between $\varphi'_{\mathcal{S} + \mathcal{K}, \mathcal{F}
+ \mathcal{K}; A}$ and $\varphi'_{\mathcal{S}, \mathcal{F}; A}$ is an
oscillation along the time direction represented by the $A_4$ term.

If we choose to use a creation source for our mesons which is limited to a
single time slice, we will always in effect be choosing a source which is an
equal linear combination of the bilinears $\Qb
\bilinear{\mathcal{S}}{\mathcal{F}} Q$ and $\Qb \Bilinear{\gamma_4 \gamma_5
\gamma_\mathcal{S}}{\xi_4 \xi_5 \xi_\mathcal{F}} Q$.  The alternating in time
of one of the bilinear's sign factors will cause their combined contribution to
cancel on one of the two time slices of the hypercube.  The result of this
cancellation is a single-time-slice source.

In an effort to generate only one set of quantum numbers, we might consider
adding a second time slice to our source.  This allows us to use exactly the
bilinear $\Qb \bilinear{\mathcal{S}}{\mathcal{F}} Q$ as our source.  However,
doing so only suppresses, but does not eliminate, the coupling to states having
the quantum numbers of the bilinear $\Qb \Bilinear{\gamma_4 \gamma_5
\gamma_\mathcal{S}}{\xi_4 \xi_5 \xi_\mathcal{F}} Q$.  This continued coupling
is because states decay exponentially as they propagate in time.  Thus, in
order to overlap only with states of one quantum number set or another, the
exponential decay of states which occurs within our source, which is distribute
in time, must be accounted for.  Yet, such an accounting can only be done if we
have complete knowledge of the spectrum for the two sets of quantum numbers in
question.  Thus, in practice it is impossible.

The quantum number sets of the bilinears $\Qb
\bilinear{\mathcal{S}}{\mathcal{F}} Q$ and $\Qb \Bilinear{\gamma_4 \gamma_5
\gamma_\mathcal{S}}{\xi_4 \xi_5 \xi_\mathcal{F}} \linebreak[0] Q$ each have an
associated lowest energy state, their respective staggered mesons.  These two
mesons are often referred to as parity partners, as they have the same quantum
numbers up to opposite parity.  If, instead of attempting to project out the
entire tower of states of one of the quantum number sets, we wish merely to
project out its associated meson state, only knowledge of the time evolution of
those two lowest-energy states is required.  That is, we must know the meson
masses ahead of time.

If no attempt is made to account for the time evolution of the meson states,
and instead the exact two-time-slice bilinear $\Qb
\bilinear{\mathcal{S}}{\mathcal{F}} Q$ is used as our source, the amplitude of
the parity-partner meson is suppressed by a factor of $\tanh \half a M$, where
the two mesons are assumed to have similar masses and $M$ is the average of
those masses.  Note that in the continuum limit, $a \cntlmtarrow 0$, the
suppression is complete.  This is not surprising as the operators are then
local in time.

For any two parity partners, the positive parity state will propagate through
time in a straightforward fashion.  Its negative parity partner, however,
will alternate in sign each time step.  This unorthodox behavior is a direct
consequence of the staggered transfer matrix being defined only for time
translations of two lattice steps.  If we consider the states only on every
other time step, the sign oscillation of the negative-parity state is hidden,
and both states behave as expected.

To summarize the core issue, when the bilinear $\Qb
\bilinear{\mathcal{S}}{\mathcal{F}} Q$ is used to create states, it will
invariably couple both to states with its own quantum numbers and to states
with quantum numbers associated with the bilinear $\Qb \Bilinear{\gamma_4
\gamma_5 \gamma_\mathcal{S}}{\xi_4 \xi_5 \xi_\mathcal{F}} Q$.  Thus, the direct
connection between bilinears and meson states is blurred.

The local pion is free from this second complication, as it has no parity
partner.  The bilinear $\Qb \bilinear{5}{5} Q$ creates only states with the
quantum numbers of the local pion, and creates no negative-parity states.

\subsection{Quark Action} \label{b:section}
Using our definition of quark bilinears, we can express the free-field
staggered fermion action in terms of $Q$ and $\Qb$:
\begin{align}
& S^S \fconfig{\identity, \chi, \chib} = \sum_n \biggl\{ \frac{1}{2} \sum_\mu
\eta_{\mu; n} \chib_n \Bigl[ \chi_{n + \muh} - \chi_{n - \muh} \Bigr] +
m_Q \chib_n \chi_n \biggr\} \notag \\
& \hspace{1ex} = \sum_\n \sum_A
\biggl\{ \frac{1}{2} \sum_\mu \eta_{\mu; A} \chib_{\n + A} \Bigl[ \chi_{\n +
A + \muh} - \chi_{\n + A - \muh} \Bigr] + m_Q \chib_{\n + A} \chi_{\n + A}
\biggr\} \notag \\
& \hspace{1ex} = \sum_\n \biggl\{ \frac{1}{4} \sum_\mu \Bigl[ \Qb_\n \bigl(
\gamma_\mu \otimes \identity \bigr) \bigl( Q_{\n + 2 \muh} - Q_{\n - 2 \muh}
\bigr) \notag \\
& \hspace{1ex} \pquad + \Qb_\n \bigl( \gamma_5 \otimes \xi_\mu \xi_5 \bigr)
\bigr( Q_{\n + 2 \muh} - 2 Q_\n + Q_{\n - 2 \muh} \bigl) \Bigr] + m_Q \Qb_\n
\bigl( \identity \otimes \identity \bigr) Q_\n \biggr\}
\label{ac:equation}
\end{align}
where we have used the following identities:
\begin{gather}
\chib_{\n + A} = \Tr{ \Qb_\n \Gamma_A } \\
\chi_{\n + A + \muh} = \Trace \Big[ \big( \delta_{0, A_\mu} Q_\n +
\delta_{1, A_\mu} Q_{\n + 2 \muh} \big) \Gamma^\dagger_{A + \muh} \Big] \\
\chi_{\n + A - \muh} = \Trace \Big[ \big( \delta_{0, A_\mu} Q_{\n - 2 \muh} +
\delta_{1, A_\mu} Q_\n \big) \Gamma^\dagger_{A - \muh} \Big] \\[1mm]
\sum_A \Tr{ \Qb_\n \Gamma_A } \Tr{ Q_\n \Gamma^\dagger_\mathcal{F}
\Gamma^\dagger_A \Gamma_\mathcal{S} } = 4 \Tr{ \Qb_\n \Gamma_\mathcal{S} Q_\n
\Gamma^\dagger_\mathcal{F} } \\
\begin{align}
\epsilon_A \Gamma_A & = \gamma_5 \Gamma_A \gamma_5
&
\delta_{0, A_\mu} & = \half ( 1 + \epsilon_A \eta_{\mu; A} \zeta_{\mu; A} ) \\
\eta_{\mu; A} \zeta_{\mu; A} \Gamma_A & = \gamma_\mu \Gamma_A \gamma_\mu
&
\delta_{1, A_\mu} & = \half ( 1 - \epsilon_A \eta_{\mu; A} \zeta_{\mu; A} )
\end{align}
\end{gather}

The first term in \eqref{ac:equation} is expected.  It is a discretization of
the standard flavor-diagonal kinetic term of the continuum quark action.
However, the second term corresponds to the lattice spacing $a$ times a
discretization of the second derivative of the quark field \eqref{ab:equation}.
Thus, this second term is a lattice artifact.  In the continuum limit, $a
\cntlmtarrow 0$, it disappears and the staggered quark action becomes the
standard continuum action.

Yet at non-zero lattice spacing, this term remains, and as the term is not
diagonal in flavor space, it demonstrates explicitly the flavor symmetry
breaking present within the staggered formalism.

\subsection{Flavor Symmetry}
A continuum four-flavor massless QCD theory classically has a $U(4)_V \otimes
U(4)_A$ flavor symmetry.  However, in our staggered discretization of that
four-flavor theory, the second kinetic term in the staggered quark action
\eqref{ac:equation} breaks much of the flavor symmetry, and at finite lattice
spacing only a remnant is preserved.

For massive quarks, the staggered quark action has a $U(1)_\identity$ symmetry,
defined by the transformation:
\begin{align}
Q & \pad{\symarrow} Q' = e^{i \alpha_1 ( \identity \otimes \identity ) } Q \\
\Qb & \pad{\symarrow} \Qb' = \Qb e^{-i \alpha_1 ( \identity \otimes \identity )
}
\end{align}
For massless quarks, the symmetry expands to $U(1)_\identity \otimes U(1)_{\gamma_5}$, where $U(1)_{\gamma_5}$ is defined by the transformation:
\begin{align}
Q & \pad{\symarrow} Q' = e^{i \alpha_\epsilon ( \gamma_5 \otimes \xi_5 ) } Q \\
\Qb & \pad{\symarrow} \Qb' = \Qb e^{i \alpha_\epsilon ( \gamma_5 \otimes \xi_5
) }
\end{align}
This $U(1)_\identity \otimes U(1)_{\gamma_5}$ symmetry is equivalent to the
$U(1)_1 \otimes U(1)_\epsilon$ symmetry described by \eqref{ae:equation} and
\eqref{af:equation}, simply reexpressed in terms of the staggered quark fields.  
In the continuum limit, the $U(1)_1$ symmetry becomes the flavor-singlet vector
symmetry $U(1)_V$.  The $U(1)_\epsilon$ symmetry becomes a single $U(1)$
subgroup of the continuum's flavor non-singlet axial vector symmetry.
Following its role as a non-singlet axial vector symmetry, $U(1)_\epsilon$ is
spontaneously broken, with the resulting Goldstone boson being a two-particle
bound state created and annihilated by the bilinear $\Qb \bigl( \gamma_5
\otimes \xi_5 \bigr) Q$.  When we move away from the chiral limit,
$U(1)_\epsilon$ becomes an approximate symmetry, and the bound state becomes a
pseudo-Goldstone boson with a squared mass proportional to the quark mass.

The form of the creation bilinear reveals that the Goldstone boson of the
spontaneously broken $U(1)_\epsilon$ symmetry is the local staggered pion.  It
is clear now that the local pion remains light at non-zero lattice spacing
because a remnant of the continuum's non-singlet axial symmetry is preserved.
It is also not surprising that the fifteen other non-local pions gain a mass, as
the staggered formulation breaks the flavor symmetries for which they would
have been Goldstone bosons.

It is the existence of this robust Goldstone boson, still present an non-zero
lattice spacing, which will prove to make the staggered fermion formulation a
valuable tool in our study.  Because the local pion is a Goldstone boson,
Chiral Perturbation Theory allows us to calculate its mass and decay constant
in terms of certain GL coefficients.  By evaluating these quantities using a
lattice calculation, we can determine the value of those GL coefficients.

\subsection{$N_f \neq 4$} \label{aj:section}
One remaining stumbling block is that the number of quark flavors is restricted
by the staggered fermion formulation to a multiple of four.  Yet, the Standard
Model contains three light flavors.  We resolve this issue by taking the
interaction-matrix determinant to a fractional power.

As discussed in Section \ref{b:section}, the staggered interaction matrix $M^S
\fconfig{U}$ becomes flavor-space diagonal in the continuum limit.  Thus in
that limit, its determinant can be factored into the product of four equivalent
determinants, one associated with each quark flavor:
\begin{equation}
\det M^S
\fconfig{U} = \Bigl( \sqrt[4]{\det M^S \fconfig{U}} ~ \Bigr)^4
\end{equation}
If we desire some number of quark flavors $N_f$ other than four, we need only
to include $N_f$ powers of the fourth root of the interaction-matrix
determinant in our partition function.  In terms of the effective gauge action,
this choice of flavors becomes a simple factor:
\begin{align}
S^S_\text{LQCD} \fconfig{U} & = S_g \fconfig{U} - \ln \biggl[ \Bigl(
\sqrt[4]{\det M^S \fconfig{U}} ~ \Bigr)^{N_f} \biggr] \notag \\
& = S_g \fconfig{U} - \frac{N_f}{4} \Trace \ln M^S \fconfig{U}
\end{align}

Of course at finite lattice spacing, the staggered action is not flavor
diagonal and its determinant will not factor.  Thus, this procedure for
choosing $N_f$ clearly becomes invalid.  Yet, the inclusion of the $N_f / 4$
factor in a numerical calculation proves to be straightforward and, whatever
effect that factor has at non-zero lattice spacing, as $a$ becomes small that
effect will become an improving approximation of $N_f$ flavors.  Thus, in our
lattice calculations we will use this method to approximate $N_f = 3$ quark
flavors, the same number of light flavors as is present in the Standard Model.

\section{Conjugate Gradient}
At their core most lattice calculations involve the inverse of the fermion
interaction matrix.  In the case of our study, the inverse is required during
both the generation of our ensemble configurations and during the calculation
of our meson propagators.

The staggered fermion interaction matrix $M^S \fconfig{U}$ is very large, $L^4
N_c \times L^4 N_c$, but since it only connects each site to its neighbors, it
is also quite sparse, with only $L^4 N_c ( 4 N_c + 1)$ non-zero elements.  In
both cases, we use $L^4$ to represent $L_1 L_2 L_3 L_4$.  Because of this
sparseness, the memory required to store the interaction matrix grows only
linearly with the lattice volume.  Its inverse $M^S \fconfig{U}^{-1}$, however,
is not sparse.  Thus, we make no attempt to calculate and store the inverse
interaction matrix directly.  Rather, we use a numerical algorithm which allows
us to apply the inverse matrix to a vector via repeated application of the
original matrix.  This algorithm, which finds use in a broad range of fields,
is known as the Conjugate Gradient (CG) method \cite{hest:52}.  For a clear yet
complete explanation of CG, we refer the reader to \cite{shewchuk}.

We use CG to apply the inverse interaction matrix to some field vector $W$,
calculating $X$, where $X$ has the form:
\begin{equation}
X = M^S \fconfig{U}^{-1} W
\label{at:equation}
\end{equation}
In practice however, we can not use CG to apply $M^S \fconfig{U}^{-1}$ to a
vector directly because $M^S \fconfig{U}$ is not positive definite, a
characteristic required by CG.  Yet the square of the interaction matrix $M^S
\fconfig{U}^\dagger M^S \fconfig{U}$ is positive definite.  Thus, we are able
to calculate $X$ by first applying $M^S \fconfig{U}^\dagger$ to $W$, and then
using CG to apply the inverse of $M^S \fconfig{U}^\dagger M^S \fconfig{U}$:
\begin{equation}
X = \bigl( M^S \fconfig{U}^\dagger M^S \fconfig{U} \bigr)^{-1} M^S
\fconfig{U}^\dagger W
\end{equation}

Recalling Section \ref{al:section}, the calculation is simplified by the fact
that the square of the interaction matrix, and thus its inverse, couples only
even sites to even sites and odd sites to odd sites.  The even components of
$X$ can be calculated by inverting a matrix which consists of only the
even-to-even elements of $M^S \fconfig{U}^\dagger M^S \fconfig{U}$, a smaller
matrix which is thus easier to invert:
\begin{align}
X_e & = \bigl( M^S \fconfig{U}^\dagger M^S \fconfig{U} \bigr)^{-1}_{e,e} \bigl(
M^S \fconfig{U}^\dagger W \bigr)_e \notag \\
& = \bigl( M^S \fconfig{U}^\dagger M^S \fconfig{U} \bigr)^{-1}_{e,e} \bigl(
-D^S \fconfig{U}_{e,o} W_o + m_Q W_e \bigr)
\end{align}
where the subscript $e$ ($o$) denotes the even (odd) half of the lattice sites.
We have used $D \fconfig{U}^\dagger = -D \fconfig{U}$.  From
\eqref{at:equation} we know that:
\begin{align}
W_o & = \bigl( M^S \fconfig{U} X \bigr)_o \notag \\
& = D^S \fconfig{U}_{o,e} X_e + m_Q X_o
\end{align}
Thus, the odd elements of $X$ can be put in terms of its even elements:
\begin{equation}
X_o = \frac{1}{m_Q} \bigl( W_o - D^S \fconfig{U}_{o,e} X_e \bigr)
\end{equation}
We see that both the even and odd components of our resulting field
vector $X$ can be computed by inverting only the even half of the interaction
matrix.  This procedure is known as preconditioning and will significantly
speed up our numerical calculations.

\section{$R$ Algorithm}
The advent of fermion fields in our lattice field theory results in a non-local
effective gauge action:
\begin{align}
Z^S_\text{LQCD} & = \int \fD{U} ~ e^{ -S^S_\text{LQCD} \fconfig{U} } \\
& S^S_\text{LQCD} \fconfig{U} = S_g \fconfig{U} + \frac{N_f}{4} \Trace \ln M^S
\fconfig{U}
\end{align}
where the $\Trace \ln M^S \fconfig{U}$ term mixes all gauge degrees of freedom.
This non-locality significantly increases the computational effort required to
generate an ensemble.  When producing a Markov chain using the local pure-gauge
action, determining the change in action due to some local modification
requires only information in the neighborhood of the modification.  With our
non-local action, however, determining the change in action due to any
modification requires a full recalculation of the configuration's action.  As
such, we can no longer afford to grow our Markov chain via a large number of
small localized changes.

Instead, we require a procedure which permits large steps through configuration
space during which all gauge field degrees of freedom are updated
simultaneously.  These update steps must change the equilibrium ensemble
distribution either not at all or very slightly, such that there is either no
chance, or only a very small chance, of rejection.  Additionally, the algorithm
must permit simulation of $N_f = 3$ flavors of dynamical fermions.

The $R$ algorithm \cite{Gottlieb:1987mq}, also known as Hybrid Molecular
Dynamics (HMD), fits all of these criteria.  The strong point of HMD is that it
chooses the next configuration in our Markov chain with the correct probability
distribution.  No acceptance step is required.  Additionally, the simulation of
any number of dynamical flavors is straightforward.  Unfortunately, the
effective probability distribution used by the $R$ algorithm is only accurate
to within a certain uncorrectable error.

\subsection{Hybrid Molecular Dynamics}
The first conceptual step towards HMD is the introduction of an auxiliary field
$H_{\mu;n}$ to the theory.  This field consists of traceless Hermitian color
matrices, four of which live on each lattice site.  They are parameterized as:
\begin{equation}
H_{\mu;n} = \sum_a \lambda^a h^a_{\mu;n}
\end{equation}
where $\lambda^a$ are the generators of $SU(3)$ color.  Being an auxiliary
field, $H_{\mu;n}$ has a trivial action:
\begin{align}
Z^S_\text{HMD} & = \int \fD{H} \fD{U} ~ e^{ -S^S_\text{HMD} \fconfig{H, U} } \\
& S^S_\text{HMD} \fconfig{H, U} = S_H \fconfig{H} +
S^S_\text{LQCD} \fconfig{U} \\
& S_H \fconfig{H} = \frac{1}{2} \sum_n \sum_\mu \trace H^2_{\mu;n}
\end{align}
As there are no interaction terms between the auxiliary field $H_{\mu;n}$ and
the gauge field $U_{\mu;n}$, its introduction into the theory does not affect
the expectation value of gauge or fermion field operators.  The quark and gluon
physics of the HMD partition function is the same as that of the LQCD partition
function.

With its simple action, updating the auxiliary field is effortless.  A
heat-bath update can be used, in which the next $H_{\mu;u}$ field configuration
is simply chosen using the proper probability distribution.  This field
configuration choice is made by setting each $h^a_{\mu;n}$ to a Gaussian random
number with appropriate normalization.  No acceptance step is required.

Given that $\fconfig{U}_A$ was the last gauge configuration added to our Markov
chain, after including a heat-bath updated auxiliary field $\fconfig{H}_A$, we
have the HMD configuration $\fconfig{H, U}_A$.  What we now require is a
procedure which will take us from $\fconfig{H, U}_A$ to a new configuration
$\fconfig{H, U}_B$ with equal HMD action $S^S_\text{HMD} \fconfig{H, U}$ in a
deterministic and reversible manor.  Once $\fconfig{H, U}_B$ is found,
$\fconfig{U}_B$ can be added to our ensemble with no acceptance step, as there
has been no change in the HMD action.  The procedure used by HMD to generate
$\fconfig{H, U}_B$ from $\fconfig{H, U}_A$ is based on classical molecular
dynamics.  This is from where HMD draws its name.

In the context of HMD, the gauge field configuration $\fconfig{U}$ is thought
of as denoting the position of a classical particle in the
very-large-dimensional configuration space.  The LQCD action $S^S_\text{LQCD}
\fconfig{U}$ is taken to be a static potential through which the particle
moves.  Finally, the auxiliary field configuration $\fconfig{H}$ acts as the
particle's canonical momentum.  With these identifications, the HMD action
$S^S_\text{HMD} \fconfig{H, U}$ has the same form as the one-particle system's
Hamiltonian $\mathscr{H}$.

The configuration $\fconfig{H, U}_A$ is treated as the initial condition for a
trajectory through configuration space.  Using the Hamiltonian's classical
equations of motion, we move the configuration along the trajectory for some
interval in artificial HMD time.  By the time we reach the endpoint of the
trajectory $\fconfig{H, U}_B$, the gauge field degrees of freedom have changed
significantly.  We have taken a large directed step through configuration
space.  Additionally, $\fconfig{H, U}_B$ has the same HMD action as our
starting point $\fconfig{H, U}_A$.  This is because, along a classical
trajectory through a time-independent potential, a system's Hamiltonian is
conserved.  Thus, we can safely add the gauge configuration $\fconfig{U}_B$ to
our ensemble with no acceptance step.  In a sense, when a heat-bath update is
applied to the auxiliary field, HMD is choosing, with the proper distribution,
our ensemble's next gauge configuration.  It is then simply a matter of
calculating, via the classical equations of motion, which configuration the $R$
algorithm has chosen.

\subsection{Ensemble Equilibrium}
To see clearly that this sort of update process is valid, we can investigate
the ensemble equilibrium condition discussed in Section \ref{d:section}.  That
is, an ensemble with the desired configuration distribution must sit in
equilibrium during the update process:
\begin{equation}
\prob{}{\fconfig{U}_A \updatearrow \fconfig{U}_B} = \prob{}{\fconfig{U}_B
\updatearrow \fconfig{U}_A}
\end{equation}
As before, $\prob{}{\fconfig{U}_A \updatearrow \fconfig{U}_B}$ has three
factors:
\begin{equation}
\prob{}{\fconfig{U}_A \updatearrow \fconfig{U}_B} = \prob{W}{\fconfig{U}_A}
\prob{P}{\fconfig{U}_A \updatearrow \fconfig{U}_B} \prob{A}{\fconfig{U}_A
\updatearrow \fconfig{U}_B}
\end{equation}
where:
\begin{equation}
\prob{W}{\fconfig{U}_A} = Z_\text{LQCD}^{-1} e^{-S^S_\text{LQCD} \fconfig{U}_A}
\end{equation}
and as we have asserted that no acceptance step is required:
\begin{equation}
\prob{A}{\fconfig{U}_A \updatearrow \fconfig{U}_B} = 1
\end{equation}

The remaining factor
$\prob{P}{\fconfig{U}_A \updatearrow \fconfig{U}_B}$
is essentially the probability that the heat-bath update of the auxiliary field
will choose a certain trajectory, the trajectory which takes us from the
position $\fconfig{U}_A$ in configuration space to the position
$\fconfig{U}_B$.  Conversely, $\prob{P}{\fconfig{U}_B \updatearrow
\fconfig{U}_A}$ is the probability of choosing that same trajectory in the
reverse direction.  Associated with the trajectory is a value for the conserved
Hamiltonian $\mathscr{H}$ which is independent of the direction the trajectory
is traversed.

The probability of choosing a given trajectory is based solely on the kinetic
energy $S_H \fconfig{H}$ required to take the trajectory.  The kinetic energy
must make up the difference between the trajectory's Hamiltonian $\mathscr{H}$
and the static potential at the trajectory's starting point $S^S_\text{LQCD}
\fconfig{U}_A$:
\begin{equation}
S_H \fconfig{H} = \mathscr{H} - S^S_\text{LQCD} \fconfig{U}_A
\end{equation}
Thus, the probability of choosing to take the trajectory in the forward
direction over choosing to take that same trajectory in the reverse direction
is:
\begin{equation}
\frac{\prob{P}{\fconfig{U}_A \updatearrow
\fconfig{U}_B}}{\prob{P}{\fconfig{U}_B \updatearrow \fconfig{U}_A}} =
\frac{e^{-\mathscr{H} + S^S_\text{LQCD} \fconfig{U}_A}}{e^{-\mathscr{H} +
S^S_\text{LQCD} \fconfig{U}_B}} = e^{S^S_\text{LQCD} \fconfig{U}_A -
S^S_\text{LQCD} \fconfig{U}_B}
\end{equation}
This is exactly the probability required to generate the desired configuration
distribution:
\begin{align}
\prob{}{\fconfig{U}_A \updatearrow \fconfig{U}_B} & = \prob{W}{\fconfig{U}_A}
\prob{P}{\fconfig{U}_A \updatearrow \fconfig{U}_B} \notag \\
& = Z_\text{LQCD}^{-1} e^{-S^S_\text{LQCD} \fconfig{U}_A}
\prob{P}{\fconfig{U}_B \updatearrow \fconfig{U}_A} e^{S^S_\text{LQCD}
\fconfig{U}_A - S^S_\text{LQCD} \fconfig{U}_B} \notag \\
& = Z_\text{LQCD}^{-1} e^{-S^S_\text{LQCD} \fconfig{U}_B}
\prob{P}{\fconfig{U}_B \updatearrow \fconfig{U}_A} \notag \\
& = \prob{W}{\fconfig{U}_B} \prob{P}{\fconfig{U}_B \updatearrow \fconfig{U}_A}
\notag \\
& = \prob{}{\fconfig{U}_B \updatearrow \fconfig{U}_A}
\end{align}

\subsection{Finite Step-Size Error}
We numerically evolve an auxiliary and gauge field configuration along its
trajectory using finite steps in HMD time of some chosen size.  Preceding each
step, the gradient of the static potential $S^S_\text{LQCD} \fconfig{U}$ is
calculated in order to determine the magnitude and direction of the step.  Each
such calculation requires an application of the inverse interaction matrix, the
most numerically demanding aspect of the algorithm.  As such, it may be
tempting to use a large time step in the interests of efficiency.  However, the
finite size of these steps introduces error into the $R$ algorithm's ensemble
distribution.  The finite steps cause the numerical evolution to stray from the
true trajectory.  Thus, the end-point configuration will not be exactly the
configuration chosen by the heat-bath update of the auxiliary field.  We must
insure that the step size used is sufficiently small such that our ensemble's
configuration distribution is not significantly skewed.

\subsection{$N_f \neq 4$}
When simulating $N_f = 4$ dynamical staggered quarks, the $\Phi$ algorithm
\cite{Gottlieb:1987mq}, also known as Hybrid Monte Carlo (HMC), can be used.
This algorithm uses molecular dynamics in the same fashion as the $R$
algorithm.  The difference between the two lies in how they handle the
interaction-matrix determinant in the effective gauge action $S^S_\text{LQCD}
\fconfig{U}$.

In the case of the $\Phi$ algorithm, the determinant is replaced by a set of
pseudo-fermion fields whose interaction matrix is the inverse of the staggered
quark interaction matrix.  Such a pseudo-fermion interaction matrix generates
the appropriate determinant factor in the ensemble average.  As the
pseudo-fermions are not Grassmann fields, we can work with them numerically,
and as their action is straightforward, they are easily updated at the start of
each trajectory using a heat-bath method similar to that used for the auxiliary
field.

A particularly noteworthy feature of the $\Phi$ algorithm is that the use of
the pseudo-fermion fields allows the HMC action to be known exactly.  As a
result, the error introduced by finite step size can be accounted for with an
acceptance step at the end of each trajectory.  While the numerical evolution
will stray from the correct trajectory before it reaches its endpoint, we can
make up for this error by accepting or rejecting the entire trajectory based on
the difference between the the HMC action at the beginning and end points of
the trajectory

In the case of the $R$ algorithm, because we are taking the fermion
interaction-matrix determinant to a fractional power in order to simulate at
$N_f \neq 4$, the determinant can not be replaced with a functional integral
over pseudo-fermionic degrees of freedom.  Instead, the determinant is
evaluated using a noisy estimator.  Such a procedure does not allow for an
exact determination of the HMD action, and thus the error introduced by finite
step size can not be accounted for via an acceptance step.

In summary, the $R$ algorithm allows us to take large directed steps through
configuration space with no chance of losing numerical effort due to an
acceptance test.  Additionally, it allows us to operate at $N_f \neq 4$.
Its primary disadvantage is that the algorithm is inexact, as the finite
step-size errors can not be accounted for.

For the specifics of the $R$ algorithm, including the exact form of the
classical equations of motion, the reader is referred to \cite{Gottlieb:1987mq}.

\section{Hypercubic Blocking} \label{h:section}
In order to gain access to the GL coefficients, we are studying the local pion
mass's dependence on the quark mass.  At NLO in ChPT, the local pion propagator
includes graphs with non-local pion loops.  Thus, its mass gains a dependence
on the masses of the non-local pions, and through them a NLO dependence on the
strength of the staggered fermion formulation's flavor symmetry breaking.
Because this is the same order at which the GL coefficients appear, flavor
symmetry breaking has the potential to introduce significant systematic error
in the values we observe for the GL coefficients.

If the lattice spacing of our ensembles were small enough, flavor symmetry
breaking would become negligible.  However, as is always the case in lattice
calculations, due to limited computer resources our lattice spacing is not as
small as we would like.  Thus, some other method for controlling flavor
symmetry breaking is required.

An oft-used method for reducing lattice artifacts is the smearing of gauge
links.  Each of a configuration's gauge links is replaced by a linear
combination of itself and other local gauge paths which connect the same
endpoints.  The result is a smeared, or fat, link configuration.  Such a
process does not affect long distance behavior, but reduces lattice artifacts
by smoothing out short-distance gauge fluctuations.  A reduction in statistical
error, especially in quantities sensitive to short-distance fluctuations, is
often seen, as well as an improvement in rotational symmetry.

In the context of staggered fermions, flavor symmetry breaking is the result of
the various corners of a hypercube experiencing distinct gauge backgrounds.
Smearing the gauge links reduces these differences, and thus reduces flavor
symmetry breaking.  This has been demonstrated in both quenched
\cite{Orginos:1999cr} and full LQCD \cite{Orginos:1998ue}, and is also
supported by perturbation theory \cite{Lepage:1998vj}.  It is not surprising
that a procedure which in general improves rotational symmetry would, in the
context of staggered fermions, improve flavor symmetry as well, as the Lorentz
and flavor symmetries have been tightly entangled.

Often multiple iterations, or levels, of a smearing process are applied to an
ensemble, as the effects of smearing become more pronounced with the number of
levels applied.  However, a large number of smearing levels results in a highly
non-local definition of the fat link and significantly alters short distance
behavior.  Thus, there is a trade-off between using a large number of smearing
levels to greatly reduce short-distance gauge noise, and using a moderate
number of smearing levels to preserve short-distance physics.

From the perspective of reducing flavor symmetry breaking in staggered
fermions, the optimal point in this trade-off is a smearing procedure which
maximally smears a gauge link without reaching beyond that link's hypercubes.
A procedure developed for this exact purpose is hypercubic blocking
\cite{Hasenfratz:2001hp, Alexandru:2002jr}.  It constructs a fat link by
smearing it only with links contained by the eight hypercubes of which that
link is an edge. 

In the application of a standard smearing level, a link is replaced by a
projected linear combination of itself and the six three-link gauge paths,
known as staples, which connect the same endpoints.  With hypercubic blocking,
three levels similar to this manner of smearing are applied.  The defining
feature is that staples which reach beyond the hypercube of the resulting fat
link are not used.  As such, the staples used in a given smearing level must be
orthogonal to the fat link and to the staples used in all previous levels.
With only three directions orthogonal to the original link, hypercubic blocking
stops after three levels.

Taking $U_{\mu;n}$ to be the original thin links of a configuration and
$V_{\mu;n}$ to be the resulting fat links of the hypercubic-blocked
configuration, we layout the definition of hypercubic blocking explicitly:
\begin{align}
V_{\mu;n} \equiv \V{1}{\mu;n} & \equiv \proj_{SU(3)} \Biggl[ \bigl( 1 -
\alpha_1 \bigr) U_{\mu;n} + \frac{\alpha_1}{6} \sum_{\nu \neq \mu} \sum_\pm
\V{2; \mu}{\mp \nu; n + \muh \pm \nuh} \V{2; \nu}{\mu; n \pm \nuh} \V{2;
\mu}{\pm \nu; n} \Biggr] \displaybreak[0] \\
\V{2; \nu}{\mu;n} & \equiv \proj_{SU(3)} \Biggl[ \bigl( 1 - \alpha_2 \bigr)
U_{\mu;n} + \frac{\alpha_2}{4} \sum_{\substack{\rho \neq \mu \\ \rho \neq \nu}}
\sum_\pm \V{3; \mu \nu}{\mp \rho; n + \muh \pm \hat{\rho}} \V{3; \nu \rho}{\mu;
n \pm \hat{\rho}} \V{3; \mu \nu}{\pm \rho; n} \Biggr] \displaybreak[0] \\
\V{3; \nu \rho}{\mu;n} & \equiv \proj_{SU(3)} \Biggl[ \bigl( 1 - \alpha_3
\bigr) U_{\mu;n} + \frac{\alpha_3}{2} \sum_{\substack{\sigma \neq \mu \\ \sigma
\neq \nu \\ \sigma \neq \rho}} \sum_\pm U_{\mp \sigma; n + \muh \pm
\hat{\sigma}} U_{\mu; n \pm \hat{\sigma}} U_{\pm \sigma; n} \Biggr]
\end{align}
where $X^{}_{-\nu; n} \equiv X^{\smash{\dagger}}_{\nu; n - \nuh}$.  The
parameters $\alpha_1$, $\alpha_2$, and $\alpha_3$ are the adjustable parameters
of the procedure.  Throughout our work we use values for these parameters
suggested by early hypercubic blocking literature \cite{Hasenfratz:2001hp}:
\begin{align}
\alpha_1 = 0.75
&&
\alpha_2 = 0.6
&&
\alpha_3 = 0.3
\end{align}

Because the sum of a set of unitary matrices is not itself unitary, the result
of each level of smearing must be projected back into the space of unitary
matrices.  This projection process must result in a unitary matrix which is
as close as possible to the original non-unitary matrix so that the resulting
fat link characterizes as well as possible the sum over local paths.  It is not
obvious, however, what metric this measure of closeness should use.  As such,
we simply define a reasonable method for identifying the closest unitary
matrix, and do not trouble ourselves with the exact form of the resulting
metric.  For the definition of the $SU(3)$ projection and details of the
procedure used to implement that definition, see Appendix \ref{l:section}.

It has been demonstrated that hypercubic blocking reduces the mass splitting
between the local and lightest non-local staggered pions
\cite{Hasenfratz:2001hp} and thus clearly reduces flavor symmetry breaking.
As such, we use hypercubic blocking to both estimate and reduce the
effects of flavor symmetry breaking on our results.

We generate our ensembles using the standard thin-link staggered fermion action
\eqref{bb:equation}.  We then apply hypercubic blocking to our ensembles and
proceed with our analysis using both the thin-link and hypercubic-blocked
ensembles.  We have a greater trust in results arising from hypercubic-blocked
ensembles, but by comparing those results to the thin-link results, we can
estimate the magnitude of the systematic error that flavor symmetry breaking
introduces.

By using hypercubic blocking in this way, we are in effect using a different
interaction matrix for our dynamical and valence quarks.  Calculation of a
partially quenched quark-antiquark correlator after hypercubic blocking amounts
to the expression:
\begin{align}
\bra \qb_{a \alpha i; n} & q_{b \beta j; m} \ket \notag \\
& = Z_\text{pqQCD}^{-1} \int \fD{U} ~ e^{ -S_g \fconfig{U} } ~ \det M_{(m_S)}
\fconfig{U} ~ M_{(m_V)} \fconfig{V}^{-1}_{a \alpha i n, b \beta j m}
\end{align}
While we admit that it is neither clear that this procedure has a clean
continuum limit, nor that it corresponds to a well-defined field theory, it
should at least prove useful in estimating the magnitude of
flavor-symmetry-breaking effects.

It should be noted that, while we do not use hypercubic blocked links in our
dynamical interaction matrix, an algorithm for doing so has since been proposed
\cite{Alexandru:2002jr, Alexandru:2002sw, Hasenfratz:2002pt}.

\section{Sommer Scale} \label{t:section}
When the creation of an ensemble begins, a set of values for the parameters in
the lattice Lagrangian must be chosen.  In our case these parameters are the
gauge coupling, via $\beta$, and the unitless quark mass $m_Q$.  As discussed
in Section \ref{e:section}, the lattice spacing $a$ is not chosen directly, but
rather is related to the other parameters by their renormalization group
equations.  In order to determine $a$, we use the ensemble to calculate some
dimensionful quantity and then match the result to a physical measurement.

A quantity which proves to be particularly useful for this purpose is the
Sommer scale \cite{Sommer:1994ce, Guagnelli:1998ud}, also known as $r_0$.  Its
popularity as a scale-setting quantity arises from the fact that it can be
calculated on the lattice easily and with small statistical error, its physical
value is well determined, and its definition is equally valid in both quenched
and unquenched simulations.  No other quantity combines all of these
advantages.

\subsection{Static Quark Potential}
The definition of the Sommer scale is based on the static quark potential
$V(r)$, the potential energy function between two static quarks.  The Sommer
scale $r_0$ is defined to be the separation distance $r$ between two static
quarks at which:
\begin{equation}
r^2 F(r) = c = 1.65
\label{av:equation}
\end{equation}
where $F(r)$ is the force between the two quarks:
\begin{equation}
F(r) = \frac{d}{dr} V(r)
\end{equation}
The dimensionless parameter $c = 1.65$ is chosen so as to correspond to an
inter-quark distance which is both convenient for calculation on the lattice
and well explored phenomenologically.  A second common choice is $c = 1$ which
defines an alternative scale known as $r_1$.

The static quark potential dominates the internal physics of heavy quark
mesons.  Thus, its form can be gleaned from the spectra of both $J / \psi$ and
$\varUpsilon$.  All successful phenomenological potentials are in agreement as
to the behavior of the static quark potential near the Sommer scale, giving:
\begin{equation}
r_0 = 0.49 \, \text{fm}
\label{au:equation}
\end{equation}

\subsection{Wilson Loops}
On the lattice we calculate the static quark potential, and from it $r_0$,
using the expectation value of rectangular Wilson loops $W^{s \times t}$.
In general a Wilson loop is any closed path of gauge-link matrices.  In the
case of rectangular Wilson loops, we define $W^{s \times t}_{\mu\nu;n}$ to be
a loop with its lowest corner at site $n$ and with a length of $s$ gauge links
in the direction $\mu$ and a breadth of $t$ gauge links in the direction $\nu$:
\begin{align}
W^{s \times t}_{\mu\nu;n} \equiv \Biggl( \prod^t_{\ell = 1} & U^\dagger_{\nu; n
+ (\ell - 1) \nuh} \Biggr) \Biggl( \prod^s_{\ell = 1} U^\dagger_{\mu; n + (\ell
- 1) \muh + t \nuh} \Biggr) \notag \\
& \times \Biggl( \prod^1_{\ell = t} U^{}_{\nu; n + s \muh + (\ell - 1) \nuh}
\Biggr) \Biggl( \prod^1_{\ell = s} U^{}_{\mu; n + (\ell - 1) \muh} \Biggr)
\end{align}

The action associated with a heavy point-like color charge is simply the
path-ordered integral of the gauge field along its trajectory.  From this
perspective the real trace of the Wilson loop, $\Real \tr{ W^{s \times t}_{i 4;
0} }$, is thought of as the action associated with creating a quark-antiquark
pair at the origin and instantaneously separating them a distance $a s$,
having the resulting states propagate an interval of time $a t$, and finally
instantaneously reuniting the pair.  If we go to the limit of large time
separations, the state which dominates the expectation value of the Wilson loop
is two static quarks separated by a distance $a s$.  By its very definition,
the static quark potential $V(r = a s)$ corresponds to the energy of this
state:
\begin{equation}
\lim_{t \rightarrow \infty} \bigl\bra \Real \tr{ W^{s \times t} } \bigr\ket =
\mathscr{A} e^{-a V(a s) t}
\end{equation}
where $\mathscr{A}$ is the amplitude for creating and annihilating the state.
In practice, in order to maximize the information extracted from each
configuration and thus minimize statistical error, we calculate the Wilson-loop
expectation value by summing over all possible positions and allowed
orientations of the loop:
\begin{equation}
W^{s \times t} \equiv \sum_n \sum_{\mu \neq 4} W^{s \times
t}_{\mu4;n}
\end{equation}

\subsection{Corrected Cornell Potential}
A standard ansatz for the continuum static quark potential is the Cornell
potential:
\begin{equation}
a V_C ( a s ) = v_0 + v_1 s + v_2 \frac{1}{s}
\end{equation}
This ansatz is based on two of the defining features of QCD: confinement and
asymptotic freedom.  The second term is a string-tension term, which dominates
at large $s$ and causes the potential to be confining.  The third term is a
Coulomb term, which dominates at small $s$ and results in asymptotic freedom.
In our case we have expressed the potential in terms of the unitless parameters
$v_i$ in order to facilitate our numerical analysis.

The form for the continuum's Coulomb potential follows directly from the
continuum's gluon propagator $G_C ( \vec{k} )$:
\begin{equation}
\Bigl[ \frac{1}{s} \Bigr]_C \equiv \frac{1}{s} = 4 \pi \int^\infty_{-\infty}
\frac{d^3 k}{(2 \pi)^3} G_C ( \vec{k} ) \cos s k_\mu
\end{equation}
where:
\begin{equation}
G_C ( \vec{k} ) = \frac{1}{\vec{k} \cdot \vec{k}}
\end{equation}
and the momentum integration variable $\vec{k}$ is unitless.  Recall that we
have defined the displacement $s$ to be in the direction $\mu$.

On the lattice, finite lattice spacing and the loss of rotational invariance
alter the form of the gluon propagator at tree level, and thus modify the
Coulomb potential.  We denote the tree-level Coulomb potential for a system as:
\begin{equation}
\Bigl[ \frac{1}{s} \Bigr]_X \equiv 4 \pi \int^\pi_{-\pi} \frac{d^3 k}{(2
\pi)^3} G_X ( \vec{k} ) \cos s k_\mu
\label{aw:equation}
\end{equation}
where $X$ becomes $L$ for the thin-link lattice and $H$ for the
hypercubic-blocked lattice.  The corresponding tree-level gluon propagators are
\cite{Hasenfratz:2001tw}:
\begin{align}
G_L ( \vec{k} ) = \frac{1}{\vec{p} \cdot \vec{p}}
&&
G_H ( \vec{k} ) = \frac{\mathfrak{H}(\vec{k})}{\vec{p} \cdot \vec{p}}
\end{align}
where:
\begin{equation}
p_i \equiv 2 \sin \frac{k_i}{2}
\end{equation}
The factor $\mathfrak{H}(\vec{k})$ accounts for the effects of the hypercubic
blocking:
\begin{equation}
\mathfrak{H}(\vec{k}) \equiv \Bigl( 1 - \frac{\alpha_1}{6} \vec{p} \cdot
\vec{\omega} \Bigr)^2
\end{equation}
where:
\begin{equation}
\omega_i \equiv p_i \Biggl[ 1 + \alpha_2 \bigl( 1 + \alpha_3 \bigr) -
\frac{\alpha_2}{4} \bigl( 1 + 2 \alpha_3 \bigr) \bigl( \vec{p} \cdot \vec{p} -
p_i^2 \bigr) + \frac{\alpha_2 \alpha_3}{4} \prod_{j \neq i} p_j^2 \Biggr]
\end{equation}
and $\alpha_i$ are the hypercubic-blocking coefficients.

The values used for the corrected Coulomb potentials are shown in Table
\ref{a:table}.  They were determined using a simple Monte Carlo integration of
the momenta integrals \eqref{aw:equation}.

\begin{table}
\centering
\begin{tabular}{|c||e{2.6}|d{2.9}|d{2.9}|}
\hline
$\pad{s}$ & \multicolumn{3}{c|}{\parbox{1cm}{
\begin{equation*}
\Bigl[ \frac{1}{s} \Bigr]_X
\end{equation*}
}} \\
\cline{2-4}
& \multicolumn{1}{c|}{$C$} & \multicolumn{1}{c|}{$L$} &
\multicolumn{1}{c|}{$H$}\\
\hline
\hline
1   & 1,       & 1.0817(12)  & 0.7330(12)  \\
2   & 0,.5     & 0.53793(73) & 0.48464(73) \\
3   & 0,.33333 & 0.34572(54) & 0.33359(53) \\
4   & 0,.25    & 0.25513(68) & 0.25187(68) \\
\hline
5   & 0,.2     & 0.20176(43) & 0.20071(43) \\
6   & 0,.16667 & 0.16868(71) & 0.16827(71) \\
7   & 0,.14286 & 0.14337(46) & 0.14319(46) \\
8   & 0,.125   & 0.12597(43) & 0.12586(43) \\
\hline
9   & 0,.11111 & 0.11157(38) & 0.11523(38) \\
10  & 0,.1     & 0.10009(46) & 0.10006(46) \\
11  & 0,.09091 & 0.09139(61) & 0.09137(61) \\
12  & 0,.08333 & 0.08355(66) & 0.08355(66) \\
\hline
\end{tabular}
\mycaption{Tree-level Coulomb potential for the continuum $C$, thin-link
lattice $L$, and hypercubic-blocked lattice $H$.  The error listed is the
statistical error of the Monte Carlo integration through which the values were
determined.  Because the same set of sample points was used for the Monte Carlo
integration of the thin-link and hypercubic-blocked propagators, their error
is highly correlated.}
\label{a:table}
\end{table}

We can now introduce a corrected potential:
\begin{equation}
a V_X(as) = v_0 + v_1 s + v_2 \frac{1}{s} + \tilde{v}_2 \biggl( \Bigl[
\frac{1}{s} \Bigr]_X - \frac{1}{s} \biggr)
\label{ct:equation}
\end{equation}
where the parameter $\tilde{v}_2$ allows for a correction of the continuum's
Coulomb term.

Once an ensemble's Wilson-loop expectation values have been calculated, we fit
that two-dimensional data to the form:
\begin{equation}
\bigl\bra \Real \tr{ W^{s \times t} } \bigr\ket = \mathscr{A}_s \exp \Biggl\{
-\Biggl[ v_0 + v_1 s + v_2 \frac{1}{s} + \tilde{v}_2 \biggl( \Bigl[ \frac{1}{s}
\Bigr]_X - \frac{1}{s} \biggr) \Biggr] t \Biggr\}
\label{bf:equation}
\end{equation}
The range of $s$ and $t$ used in the fit must be chosen carefully in order
that $t$ remains large enough that the static-quark state dominates, while both
$s$ and $t$ remain small enough that the finite extent of the lattice does not
come into play.  The results of the fit are values for its free parameters:
$v_0$, $v_1$, $v_2$, $\tilde{v}_2$, and an amplitude $\mathscr{A}_s$ for each
value of $s$ included in the fit range.

With these values in hand, the ensemble's lattice spacing can be determined via
the Sommer scale.  The Sommer scale $r_0$ is defined to be the distance at
which the continuum static quark potential meets the criteria of
\eqref{av:equation}.  Thus after fitting, we drop the correction term from our
potential \cite{Edwards:1998xf}, determining $r_0$ to be:
\begin{equation}
r_0 = a \sqrt{\frac{1.65 + v_2}{v_1}}
\label{bt:equation}
\end{equation}
Matching with phenomenological results \eqref{au:equation}, we find:
\begin{align}
a & = \Biggl( 0.49 \sqrt{\frac{v_1}{1.65 + v_2}} \Biggr) \, \text{fm} \\
\intertext{or:}
a^{-1} & = \Biggl( 395 \sqrt{\frac{1.65 + v_2}{v_1}} \Biggr) \, \text{MeV}
\end{align}

It is worth noting that, because the GL coefficients are unitless parameters,
their calculation does not involve the lattice spacing directly.
However, knowledge of the lattice spacing is still critical in that it allows
access to such important values as the physical extent of an ensemble's
lattice, the physical mass of the pseudo-Goldstone boson, and the expected
strength of finite-lattice-spacing errors.

\section{Meson Propagators} \label{c:section}
The lightest states produced by each quark bilinear are that bilinear's
corresponding staggered meson and its parity partner.  A two-point correlation
function which uses a bilinear as its creation and annihilation operator will
thus, at large time separations, be dominated by the propagator of these two
states.  Therefore, in the manner outlined in Section \ref{a:section}, the
bilinear correlators allow us access to the masses of the staggered mesons, and
in the case of the local pion, the decay constant as well.

\subsection{Bilinear Correlators} \label{f:section}
In order to analyze meson propagators using lattice techniques, we must first
express the bilinear correlators in a form which leaves them vulnerable to
lattice investigation.

We discuss here only local-bilinear correlators, $\mathcal{S} = \mathcal{F}$,
as they are significantly simpler than more general non-local correlators and
are of primary importance in our study.  For a similar discussion which
encompasses non-local bilinears, see Appendix \ref{j:section}.

We begin with a simple correlator of the form presented in Section
\ref{a:section}.  For our creation operator, or source, we use the bilinear
with the quantum numbers of whatever meson we have chosen for study.  For our
annihilation operator, or sink, we use a wall of the same bilinear, summing
over all positions on a time slice in order to restrict the annihilated states
to those with zero spatial momentum:
\begin{equation}
\Bigl\bra \sum_{\substack{\vec{g} \\ g_4 = t}} \mathcal{J}_{\mathcal{S},
\mathcal{S}; g} \mathcal{J}_{\mathcal{S}, \mathcal{S}; 0} \Bigr\ket
\end{equation}
The large time behavior of this correlator will be dominated by the propagator
of the corresponding staggered meson and its parity partner.  We denote the
time separation of the correlator with the unitless parameter $t$, which takes
on the integer values $[ 0, L_4 - 1 ]$

We replace the single bilinear in our source with a linear combination of all
local bilinears:
\begin{equation}
\frac{1}{16} \Bigl\bra \sum_{\substack{\vec{g} \\ g_4 = t}}
\mathcal{J}_{\mathcal{S}, \mathcal{S}; g} \sum_\mathcal{R}
\mathcal{J}_{\mathcal{R}, \mathcal{R}; 0} \Bigr\ket
\end{equation}
Our source now overlaps with significantly more states, including all local
mesons.  However, with the bilinear form of the sink unchanged, only our
desired states are annihilated, and thus the correlator continues to measure
only the states we desire.  Furthermore, we will find that this change allows
us to calculate all local correlators simultaneously, minimizing our
computational effort.

For our source we switch from a single bilinear to a wall of bilinears at the
appropriate time slice.  This reduces the statistical error of our calculation
without significantly increasing the computation time.  The resulting factor
of $V \equiv L_1 L_2 L_3$ is accounted for in Appendix \ref{g:section}.
\begin{equation}
C_{\mathcal{S}, \mathcal{S}; t} = \frac{1}{16} \Bigl\bra
\sum_{\substack{\vec{g} \\ g_4 = t}} \mathcal{J}_{\mathcal{S}, \mathcal{S}; g}
\sum_{\substack{\vec{h} \\ h_4 = 0}} \sum_\mathcal{R} \mathcal{J}_{\mathcal{R},
\mathcal{R}; h} \Bigr\ket
\end{equation}

The sum of all local bilinears has a simple form in terms of $\chi$ and
$\chib$, as demonstrated in Appendix \ref{k:section}.  Only a single
contraction over the hypercube is non-zero, that at the lowest corner of the
hypercube:
\begin{equation}
\sum_\mathcal{R} \mathcal{J}_{\mathcal{R}, \mathcal{R}; h} = 16 \chib_h \chi_h
\label{aj:equation}
\end{equation}
Using \eqref{ai:equation} and \eqref{aj:equation}, we put the correlator in
terms of the fundamental lattice fermion degrees of freedom, $\chi$ and $\chib$:
\begin{equation}
C_{\mathcal{S}, \mathcal{S}; t} = \Bigl\bra \sum_{\substack{\vec{g} \\ g_4 =
t}} \sum_B (-)^{\varphi'_{\mathcal{S}, \mathcal{S}; B}} \chib_{g + B} \chi_{g +
B} \sum_{\substack{\vec{h} \\ h_4 = 0}} \chib_{h} \chi_{h} \Bigr\ket
\label{as:equation}
\end{equation}

The situation is complicated slightly by the fact that the fermion field at
each site is not a scalar but rather a color vector.  Within each bilinear
these color vectors are contracted with an implied sum over color.  That is,
taking $\chi_\tsub{n}{a}$ to be the fermion degree of freedom with color $a$ at
site $n$:
\begin{equation}
C_{\mathcal{S}, \mathcal{S}; t} = \Bigl\bra \sum_{\substack{\vec{g} \\ g_4 =
t}} \sum_B (-)^{\varphi'_{\mathcal{S}, \mathcal{S}; B}} \sum_a \chib_\dsub{g +
B}{a} \chi_\dsub{g + B}{a} \sum_{\substack{\vec{h} \\ h_4 = 0}} \sum_c
\chib_\dsub{h}{c} \chi_\dsub{h}{c} \Bigr\ket
\end{equation}

Via Wick contractions we compute the Grassmann integral over the fermion fields
analytically, putting our correlator in terms of only gauge degrees of freedom.
We do not allow the bilinears to self-contract by asserting that we only wish
to investigate flavor-non-singlet mesons.  The contracted $\chi$ and $\chib$
fields are transformed by the integral into the inverse of the fermion
interaction matrix:
% all this gymnastics just to get wick.sty working :(
\newsavebox{\layoutchib}
\savebox{\layoutchib}{$\chib$}
\newsavebox{\layoutdsubh}
\savebox{\layoutdsubh}{$\dsub{h}{c}$}
\newsavebox{\layoutdsubg}
\savebox{\layoutdsubg}{$\dsub{g + B}{a}$}
\newsavebox{\layoutwick}
\savebox{\layoutwick}{$\overwick{2}{<+\chi_{\usebox{\layoutdsubg}}
>+{\usebox{\layoutchib}}_{\usebox{\layoutdsubh}}}$}
\begin{align}
C_{\mathcal{S}, \mathcal{S}; t} & = \Bigl\bra \sum_{a,c}
\sum_{\substack{\vec{g} \\ g_4 = t}} \sum_B (-)^{\varphi'_{\mathcal{S},
\mathcal{S}; B}} \sum_{\substack{\vec{h} \\ h_4 = 0}}
\overwick{1}{<+{\usebox{\layoutchib}}_{\usebox{\layoutdsubg}}
\usebox{\layoutwick} >+\chi_{\usebox{\layoutdsubh}}} \Bigr\ket \notag \\
& = \Bigl\bra \sum_{a,c} \sum_{\substack{\vec{g} \\ g_4 = t}} \sum_B
(-)^{\varphi'_{\mathcal{S}, \mathcal{S}; B}} \sum_{\substack{\vec{h} \\ h_4 =
0}} M^S \fconfig{U}^{-1}_\dsub{g + B, h}{a,c} ( M^S \fconfig{U}^T)^{-1}_\dsub{g
+ B, h}{a,c} \Bigr\ket \notag \\
& = \Bigl\bra \sum_{a,c} \sum_{\substack{\vec{g} \\ g_4 = t}} \sum_B
(-)^{\varphi'_{\mathcal{S}, \mathcal{S}; B}} \sum_{\substack{\vec{h} \\ h_4 =
0}} \epsilon_{g + B} \epsilon_h \bigl\lvert M^S \fconfig{U}^{-1}_\dsub{g + B,
h}{a,c} \bigr\rvert^2 \Bigr\ket \notag \\
& = \Bigl\bra \sum_{a,c} \sum_{\substack{\vec{g} \\ g_4 = t}} \sum_B
(-)^{\varphi'_{\mathcal{S} + 5, \mathcal{S} + 5; B}} \sum_{\substack{\vec{h} \\
h_4 = 0}} \bigl\lvert M^S \fconfig{U}^{-1}_\dsub{g + B, h}{a,c} \bigr\rvert^2
\Bigr\ket
\end{align}
where $( M^S \fconfig{U}^T )^{-1}$ has been put in terms of $M^S
\fconfig{U}^{-1}$ as described in \eqref{an:equation}, and 5 is taken to be the
binary four-vector with all elements set to one, such that $\gamma_5$ has the
proper value under the definition \eqref{ax:equation}.

The bilinear correlator is now in terms of quantities which are calculable on
the lattice:  elements of the inverse of the fermion interaction matrix.
Because we do not have access to the inverse interaction matrix itself, but
rather use CG to apply the inverse to a fermion field vector, calculating the
individual elements of the inverse matrix, as appears to be required by the
correlator, requires a prohibitive number $N_c V$ of applications of the
inverse interaction matrix.

In order to reduce the number of inversions required, we calculate the field
vectors $X^{(c)\ell}$:
\begin{equation}
X^{(c)\ell}_\dsub{n}{a} = \sum_{\substack{\vec{h} \\ h_4 = 0}} M^S
\fconfig{U}^{-1}_\dsub{n,h}{a,c} \eta^\ell_h
\end{equation}
each of which requires only a single application of the inverse interaction
matrix.

$\eta^\ell$ is a set of noise field vectors for which each element equals a
random unit-length phase.  The introduction of this random-phase vector
eliminates unwanted cross terms from the square of $X^{(c)\ell}$.  In the limit
of an infinite number of noise vectors $N_\ell$:
\begin{equation}
\lim_{N_\ell \rightarrow \infty} \frac{1}{N_\ell} \sum_\ell^{N_\ell}
\eta^{\ell*}_n \eta^\ell_m = \delta_{n,m}
\end{equation}
The sum for the off-diagonal elements is over an infinite number of random
phases and washes out to zero.  Thus:
\begin{align}
\sum_\ell^{N_\ell} \bigl\lvert X^{(c)\ell}_\dsub{n}{a} \bigr\rvert^2 & =
\sum_\ell^{N_\ell} \sum_{\substack{\vec{h} \\ h_4 = 0}} M^S
\fconfig{U}^{-1*}_\dsub{n,h}{a,c} \eta^{\ell*}_h \sum_{\substack{\vec{f} \\ f_4
= 0}} M^S \fconfig{U}^{-1}_\dsub{n,f}{a,c} \eta^\ell_f \notag \\
& = \sum_{\substack{\vec{h} \\ h_4 = 0}} \bigl\lvert M^S
\fconfig{U}^{-1}_\dsub{n,h}{a,c} \bigr\rvert^2
\end{align}
The use of the random-phase vector has contracted $M^S \fconfig{U}^{-1}$ with
itself, allowing us to avoid the calculation of its individual elements.

In practice, we use only a single noise vector, $N_\ell = 1$.  This is allowed
as the random phases between the cross terms within a single element of the sum
over noise vectors is enough to wash out the position-off-diagonal terms in the
square of $X^{(c)}$.  This single noise vector is effectively provided by the
random phase which is automatically present at each lattice site due to the
gauge link's local gauge freedom.

Our correlator becomes:
\begin{equation}
C_{\mathcal{S}, \mathcal{S}; t} = \Bigl\bra \sum_c \sum_{\substack{\vec{g} \\
g_4 = t}} \sum_B (-)^{\varphi'_{\mathcal{S} + 5, \mathcal{S} + 5; B}} \sum_a
\bigl\lvert X^{(c)}_\dsub{g + B}{a} \bigr\rvert^2 \Bigr\ket
\end{equation}
where we have one field vector $X^{(c)}$ for each color:
\begin{align}
X^{(c)}_\dsub{n}{a} & = \sum_{\substack{\vec{h} \\ h_4 = 0}} M^S
\fconfig{U}^{-1}_\dsub{n,h}{a,c} \notag \\
& = \bigl( M^S \fconfig{U}^{-1} W^{(c)} \bigr)_\dsub{n}{a}
\end{align}
These $N_c$ field vectors $X^{(c)}$ each require an application of the inverse
of the interaction matrix to calculate.  To do so, we construct the field
vector $W^{(c)}$, which equals one only at color $c$ in the lowest corner of
each hypercube on the time slice $n_4 = 0$ and zero elsewhere.  The result of
applying the inverse interaction matrix to $W^{(c)}$ is $X^{(c)}$.

From inspection it is clear that the calculation of $X^{(c)}$ is independent of
both $\mathcal{S}$ and $t$.  Thus, due to our use of the sum of all local
bilinears as our source, we are able to calculate the correlator between all
local bilinears at all time separations using only $N_c$ inversions per
configuration.

In the case of the local pion, $\mathcal{S} = 5$ and $\varphi'_{\mathcal{S} +
5, \mathcal{S} + 5; B} = \varphi'_{0, 0; B} = 0$ for all $B$.  The correlator
becomes quite simple:
\begin{equation}
C_{5, 5; t} = \Bigl\bra \sum_c \sum_{\substack{\vec{g} \\ g_4 = t}} \sum_B
\sum_a \bigl\lvert X^{(c)}_\dsub{g + B}{a} \bigr\rvert^2 \Bigr\ket
\label{bd:equation}
\end{equation}

In practice we calculate the correlator multiple times, shifting our origin's
time component and averaging over the results.  For our study we repeated the
calculation every four lattice steps in time.  This allows us to extract the
maximum amount of information from each configuration, minimizing our
statistical error.

\subsection{Meson Masses}
For large time separations the bilinear correlator will be dominated by the
propagator of the lightest states it creates.  In our case those are the
appropriate staggered meson and its parity partner.  For the moment we will
ignore the parity partner.

For a bilinear correlator with its source contained on the time slice $n_4 = 0$
and its sink contained on the time slice $n_4 = t$, we see from
\eqref{ar:equation} that the large $t$ behavior will be:
\begin{equation}
\mathscr{A} e^{-a M t}
\end{equation}
where $M$ is the mass of the staggered meson and $\mathscr{A}$ is the amplitude
for creating and annihilating the meson state.  Note that we can define a
unitless mass $\uM = aM$ by absorbing into $M$ the factor of $a$.  Such
unit-free quantities prove useful, as numerical procedures can only be used to
determine dimensionless values.  Throughout, we use the notation $\U{X}$ to
indicate some quantity $X$ after the appropriate powers of lattice spacing
have been absorbed so as to make $\U{X}$ dimensionless.

The lattice's periodic boundary conditions allow states to propagate from
source to sink along both time directions.  Accounting for this in our
propagator, we find:
\begin{equation}
\mathscr{A} \bigl( e^{-a M t} + e^{-a M ( L_4 - t) } \bigr)
\end{equation}

As discussed in Section \ref{f:section}, we do not use a single-time-slice
sink, but rather a full-bilinear sink which lives both on time slice $n_4 = t$
and $n_4 = t + 1$.  Thus, in addition to this approximately doubling the
magnitude of our correlator, the states must propagate different distances when
being annihilated at the two time slices of our sink:
\begin{gather}
\mathscr{A} \Bigl[ \bigl( e^{-a M t} + e^{-a M ( L_4 - t ) } \bigr) + \bigl(
e^{-a M ( t + 1 ) } + e^{-a M ( L_4 - 1 - t ) } \bigr) \Bigr] \notag \\
= \mathscr{A} \bigl( 1 + e^{-a M} \bigr) \bigl( e^{-a M t} + e^{-a M ( L_4 - 1
- t) } \bigr)
\end{gather}

Accounting now for the negative-parity state, whose propagator oscillates in
sign each time step, the final form for our large-time bilinear correlator is:
\begin{align}
C_{\mathcal{S}, \mathcal{F}; t} & = \mathscr{A}_+ \bigl( 1 + e^{-a M_+} \bigr)
\bigl( e^{-a M_+ t} + e^{-a M_+ ( L_4 - 1 - t) } \bigr) \notag \\
& \quad + \mathscr{A}_- (-)^t \bigl( 1 - e^{-a M_-} \bigr)  \bigl( e^{-a M_- t}
- e^{-a M_- ( L_4 - 1 - t ) } \bigr)
\end{align}
where $\mathscr{A}_+$ and $\mathscr{A}_-$ are the amplitude for the creation
and annihilation of the positive and negative parity states, and $M_+$ and
$M_-$ are those states' masses.  Recall that $L_4$ has been defined to be even.

In the case of the local pion, there is no overlap with the negative parity
state, $\mathscr{A}_- = 0$:
\begin{equation}
C_{5, 5; t} = \mathscr{A}_{\pi_5} \bigl( 1 + e^{-a M_{\pi_5}} \bigr) \bigl(
e^{-a M_{\pi_5} t} + e^{-a M_{\pi_5} ( L_4 - 1 - t) } \bigr)
\label{ap:equation}
\end{equation}
where $\mathscr{A}_{\pi_5}$ is the amplitude for creating and
annihilating the local pion state using our source and sink operators, and
$M_{\pi_5}$ is the mass of the local pion.

Thus, after completing a lattice calculation of $C_{5, 5; t}$, we can fit the
results to the form described by \eqref{ap:equation} and extract the mass of
the local pion.

\subsection{Meson Decay Constants} \label{g:section}
While the local pion mass can be extracted without any consideration for the
correlator's amplitude $\mathscr{A}_{\pi_5}$, the local pion decay constant
$f_{\pi_5}$ can not.  Thus, careful attention must be given to the
normalization of the source and sink operators.

Recalling \eqref{ar:equation}, we put our large-time correlator's amplitude in
terms of our source and sink operator's overlap with the local pion state:
\begin{equation}
C_{5, 5; t} = \frac{1}{2 \uM_{\pi_5} V} \bra 0 \vert \Op_\text{sink} \vert
\pi_5 \ket \bra \pi_5 \vert \Op_\text{source} \vert 0 \ket \mathfrak{C}_{5, 5;
t}
\end{equation}
where $\vert \pi_5 \ket$ is the dimensionless zero-momentum staggered
local-pion state and:
\begin{equation}
\mathfrak{C}_{5, 5; t} \equiv \bigl( 1 + e^{-\uM_{\pi_5}} \bigr) \bigl(
e^{-\uM_{\pi_5} t} + e^{-\uM_{\pi_5} ( L_4 - 1 - t) } \bigr)
\end{equation}

From \eqref{as:equation} we identify our source and sink operators as:
\begin{align}
\Op_\text{source} & = \sum_{\substack{\vec{h} \\ h_4 = 0}} \chib_h \chi_h =
\sum_{\substack{\vec{n} ~ \text{even} \vphantom{y} \\ n_4 = 0}} \chib_n \chi_n \notag \\
& = \sum_{\substack{\vec{n} ~ \text{even} \vphantom{y} \\ n_4 = 0}} \epsilon_n \chib_n
\chi_n \\
\intertext{and:}
\Op_\text{sink} & = \sum_{\substack{\vphantom{\vec{B}} \vec{g} \\
\vphantom{B_4} g_4 = t}} \sum_{\substack{\vphantom{\vec{g}} \vec{B} \\
\vphantom{g_4} B_4 = 0}} (-)^{\varphi'_{5, 5; B}} \chib_{g + B} \chi_{g + B}
\notag \\
& = \sum_{\substack{\vec{n} \\ n_4 = t}} \epsilon_n \chib_n \chi_n
\end{align}
Our addition of an $\epsilon_n$ factor to the source is allowed as the source
is only non-zero at even sites.  We consider only a single time slice, $B_4 =
0$, of our two-time-slice sink operator, as the form of $\mathfrak{C}_{5, 5;
t}$ already accounts for a second time slice.

We can now put our correlator in terms of $\bra 0 \vert \epsilon_n \chib_n
\chi_n \vert \pi_5 \ket$, the overlap between the local pion state and a
single-site operator with the local pion's quantum numbers.  We find:
\begin{align}
\bra 0 \vert \Op_\text{source} \vert \pi_5 \ket & = \frac{V}{2^3} \bra 0 \vert
\epsilon_n \chib_n \chi_n \vert \pi_5 \ket \\
\bra 0 \vert \Op_\text{sink} \vert \pi_5 \ket & = V \bra 0 \vert \epsilon_n
\chib_n \chi_n \vert \pi_5 \ket
\end{align}
and, thus:
\begin{equation}
C_{5, 5; t} = \frac{V}{16 \uM_{\pi_5}} \lvert \bra 0 \vert \epsilon_n \chib_n
\chi_n \vert \pi_5 \ket \rvert^2 \mathfrak{C}_{5, 5; t}
\label{ay:equation}
\end{equation}

In the continuum, a meson's decay constant can be defined by the meson state's
overlap with the appropriate weak interaction bilinear.  In the case of the
continuum pion, this definition is:
\begin{equation}
\sqrt{2} f_\pi M^2_\pi = \bigl( m_u + m_d \bigr) \bra 0 \vert \bar{u} \gamma_5
d \vert \pi^+ \ket
\label{az:equation}
\end{equation}
where $\vert \pi^+ \ket$ is the continuum's zero-momentum $\pi^+$ state.
Recall that we are using the pion decay constant normalization in which $f_\pi
\simeq 92.4 \, \text{MeV}$.  In order to put our correlator in terms of the
pion decay constant, we use the discretization correspondence
\cite{Kilcup:1987dg}:
\begin{equation}
\bra 0 \vert \bar{u} \gamma_5 d \vert \pi^+ \ket \pad{\discarrow}
\frac{1}{a^2 \sqrt{4}} \bra 0 \vert \epsilon_n \chib_n \chi_n \vert \pi_5 \ket
\label{ba:equation}
\end{equation}
Thus, combining \eqref{ay:equation}, \eqref{az:equation}, and
\eqref{ba:equation}, our correlator becomes:
\begin{align}
C_{5, 5; t} & = \frac{f_{\pi_5}^2 M_{\pi_5}^3 (a^3 V)}{8 m_q^2} \mathfrak{C}_{5,
5; t} \notag \\
& = \frac{\uf_{\pi_5}^2 \uM_{\pi_5}^3 V}{8 m_Q^2} \mathfrak{C}_{5, 5; t}
\label{be:equation}
\end{align}
where $f_{\pi_5}$ is the local pion decay constant and $\uf_{\pi_5}$ is the
corresponding dimensionless value.

Knowing the normalization of our correlator, we now have the ability to extract
both the local pion mass and decay constant.

The local pion decay constant is protected by the staggered formulation's
even-odd symmetry.  This protection does not occur for any of the other fifteen
staggered pions, as the staggered action breaks their corresponding axial
vector symmetries.  Thus, making a connection between the lattice propagator of
one of these pions and its continuum decay constant requires the calculation of
a renormalization factor.  The process is thus much more involved than for the
local pion case.

\chapter{Partially Quenched Chiral Perturbation Theory} \label{ap:section}
In Chapter \ref{n:section} we presented ChPT's prediction for the quark mass
dependence of the chiral pseudo-Goldstone boson's mass and decay constant.  In
Chapter \ref{o:section} we detailed lattice techniques for calculating that
mass and decay constant.  However, performing a set of lattice calculations
using different values for the quark mass is extremely computationally
cumbersome.  As we are constrained by finite computer resources, we find it
easier to vary only the valence quark mass.  That is, we make use of the
partially quenched approximation of Section \ref{m:section}.

ChPT is the low-energy effective field theory for the light bound states of
unquenched QCD.  Thus, its predictions are not valid in the context of our
partially quenched calculations.  In order to bridge this gap between ChPT and
partial quenching, we must make quantitative sense of the effects of partial
quenching.  The culmination of this process is partially quenched Chiral
Perturbation Theory (pqChPT), a low-energy effective field theory for the light
bound states of pqQCD \cite{Bernard:1992mk, Bernard:1994sv}.

Note that throughout this chapter, as in Chapter \ref{n:section}, we work in an
infinite volume continuum.

\section{Partially Quenched Quantum Chromodynamics}
As discussed in Section \ref{m:section}, pqQCD is not a well-behaved unitary
field theory.  Nonetheless, we are not prevented from proposing a quark content
for pqQCD, nor from discussing pqQCD in a Lagrangian context.

Partially quenched QCD clearly contains two types of fermionic quark flavors:
dynamical quarks and valence quarks.  As is illustrated by \eqref{bs:equation},
the dynamical quark field appears in the partition function's Boltzmann weight
but does not appear in the operators whose expectation value we calculate.  The
valence quark field, on the other hand, appears only in these operators and
does not contribute to the Boltzmann weight.  The challenge before us is to
construct a quark content for pqQCD in which these abnormal attributes are a
natural consequence.

Eliminating dynamical quarks from external operators is a simple matter of
restricting ourselves to calculating only the expectation value of operators
which involve valence quarks.  By definition, this was our intent from the
beginning.

Eliminating the valence quarks' natural contribution to the theory's Boltzmann
weight is more involved.  We introduce a set of ghost quark flavors, scalar
quark fields with incorrect spin statistics, with an interaction matrix
equivalent to the valence quarks'.  As the ghost quarks are bosons, their
functional integral results in the inverse of the interaction-matrix
determinant.  If, for each valence quark flavor, we include a ghost quark
flavor of equal mass, the two contributions to the theory's Boltzmann weight
will cancel.  The valence quark interaction-matrix determinant, which would
naturally appear in the Boltzmann weight, is effectively eliminated.

Thus, pqQCD can be thought of as having three types of quark flavors: dynamical
quarks, valence quarks, and ghost quarks.  The first two are standard fermionic
quarks and the last is bosonic.  In our study, we use dynamical and valence
quark flavors which are independently degenerate.  The mass of the ghost quarks
is then set equal to the valence quark mass.  We represent the number of
dynamical quarks as $N_f$ and the number of valence quarks, and therefore ghost
quarks as well, as $N_v$.  From the perspective of this quark content, the
pqQCD partition function, before the integration over quark degrees of freedom,
is seen to be:
\begin{equation}
Z_\text{pqQCD} = \int \fD{A_\mu} \fD{q_v} \fD{\qb_v} \fD{q_s} \fD{\qb_s}
\fD{\tilde{q}_g} \fD{\bar{\tilde{q}}_g} ~ e^{-\int_x \mathscr{L}_\text{pqQCD}}
\end{equation}
where:
\begin{align}
& \mathscr{L}_\text{pqQCD} = \frac{1}{2} \tr{ F_{\mu\nu} F^{\mu\nu} } \notag \\
& \quad
+ \qb_v
\bigl( \gamma^\mu D_\mu + \mv \bigr)
q_v
+ \qb_s
\bigl( \gamma^\mu D_\mu + \ms \bigr)
q_s
+ \bar{\tilde{q}}_g
\bigl( \gamma^\mu D_\mu + \mv \bigr)
\tilde{q}_g
\end{align}
We use $q_v$ to represent the vector of $N_v$ fermionic valence quark flavors
with mass $\mv$, $q_s$ to represent the vector of $N_f$ fermionic dynamical
quark flavors with mass $\ms$, and $\tilde{q}_g$ to represent the vector of
$N_v$ bosonic ghost quark flavors with mass $\mv$.  Our use of $\ms$ here to
denote the dynamical quark mass, and elsewhere to denote the Standard Model's
strange quark mass, should remain clear via context.

Performing the functional integral over quark degrees of freedom results in the
cancellation discussed above:
\begin{align}
Z_\text{pqQCD} & = \int \fD{A_\mu} ~ e^{-\int_x \mathscr{L}_g} ~
\frac{\Det{\gamma^\mu D_\mu + \mv}}{\Det{\gamma^\mu D_\mu + \mv}} ~
\Det{\gamma^\mu D_\mu + \ms} \notag \\
& = \int \fD{A_\mu} ~ e^{-\int_x \mathscr{L}_g} ~ \Det{\gamma^\mu D_\mu + \ms}
\label{cq:equation}
\end{align}
We can see by comparison to \eqref{bs:equation} that the pqQCD partition
function \eqref{cq:equation} has the appropriate form to correspond to the
calculation of partially quenched expectation values.  

The above construction of the quark content of pqQCD can also be understood in
the context of perturbation theory.  From this perspective, dynamical quarks
appear only in closed quark loops, and not in the external legs of Feynman
diagrams.  Valence quarks, on the other hand, appear only in external legs and
never in loops.  To insure that dynamical quarks appear only in loops, we
restrict ourselves to calculating matrix elements between valence-quark
external states.  To remove valence-quark loops, we again introduce ghost
quarks.  For every Feynman diagram which includes a valence-quark loop, there
will be an identical diagram in which that loop is replaced by a ghost-quark
loop.  As bosons, the ghost-quark loops will appear with opposite sign.  Thus,
all diagrams which include valence-quark loops are canceled.

It should be noted that, although we have conceived of a quark content and
constructed a Lagrangian for pqQCD, the theory remains pathological.  The
advent of scalar quarks, with incorrect spin statistics, spoils unitarity.
Furthermore, we are artificially restricting the physical states of the theory,
removing dynamical- and ghost-quark external states.  This effectively sets
elements of the $S$ matrix to zero, further ruining unitary.  However, because
it is the partially quenched approximation itself which is non-unitary, we can
not hope to construct a unitary field theory with the necessary partition
function.

\section{Partially Quenched Flavor Symmetry}
Just as the flavor symmetry group of QCD dictates the form of ChPT, so will the
flavor symmetry group of pqQCD dictate the form of pqChPT.

The flavor symmetry group of massless QCD, $SU(N_f)_L \otimes SU(N_f)_R$,
follows directly from its quark content.  Massless pqQCD has an analogous
flavor symmetry group which also follows from the quark content proposed above.
First, we collect the standard and ghost quarks into a single vector:
\begin{align}
q = \begin{bmatrix}
q_v \\[3mm]
q_s \\[3mm]
\tilde{q}_g
\end{bmatrix}
&&
\qb = \begin{bmatrix}
\qb_v & \qb_s & \bar{\tilde{q}}_g
\end{bmatrix}
\end{align}
This mixture of fermions and bosons causes the flavor symmetry group to be a
graded group, $SU(N_v + N_f \vert N_v)_L \otimes SU(N_v + N_f \vert N_v)_R$.
In truth, there are subtleties which arise in the identification of pqQCD's
flavor symmetry group.  However, Sharpe demonstrates in \cite{Sharpe:2001fh}
that the above graded group generates the correct Ward identities for pqQCD.
Thus, while it is not the exact flavor symmetry group of pqQCD, it is still
appropriate to use in constructing pqChPT.

\section{Graded Groups}
Matrices which are elements of a graded group have the following distinguishing
properties.  The graded group element $\Sigma \in SU(N \vert M)$ can be written
in block form:
\begin{equation}
\Sigma = \begin{bmatrix} A & C \\[2mm] D & B \end{bmatrix}
\end{equation}
where $A$ and $B$ are respectively $N \times N$ and $M \times M$ matrices
of standard commuting numbers, while $C$ and $D$ are respectively $N \times M$
and $M \times N$ matrices of anticommuting numbers.  In order to preserve the
correct behavior of the adjoint, the complex conjugate of a product of these
anticommuting numbers is defined to switch their order:
\begin{equation}
(a b)^* \equiv b^* a^*
\end{equation}
where $a$ and $b$ are elements of $C$ or $D$.  Finally, the super trace of a
graded group element is defined as:
\begin{equation}
\sTrace \Sigma \equiv \Trace A - \Trace B
\end{equation}
from which the definition of the super determinant is based:
\begin{equation}
\sdet \Sigma \equiv \exp \Bigl\{ \sTr{\ln \Sigma} \Bigr\} = \frac{\Det{A - C
B^{-1} D}}{\det B}
\end{equation}

Using these properties for graded group elements, the graded group
$SU(N_v + N_f \vert N_v)_L \otimes SU(N_v + N_f \vert N_v)_R$ has the
appropriate behavior to act as the flavor symmetry group for pqQCD's mixture of
fermionic and bosonic quarks.

\section{Partially Quenched Chiral Lagrangian} \label{v:section}
With the flavor symmetry group for pqQCD in hand, we are now in a position to
develop a low-energy effective field theory for the light bound states of
pqQCD, mirroring closely the process in Chapter \ref{n:section}.

We begin with an assumption that the full flavor symmetry of pqQCD is
spontaneously broken by a chiral condensate, from $SU(N_v + N_f \vert N_v)_L
\otimes SU(N_v + N_f \vert N_v)_R$ down to $SU(N_v + N_f \vert N_v)_V$, just as
occurs in full QCD as discussed in Section \ref{p:section}.  The result is
set of Goldstone particles, pqQCD's light mesons.   Note that we use the term
mesons loosely, as several of these particles are fermions.  The fermionic
subset of Goldstone particles are those which correspond to broken symmetries
whose generator's non-zero elements are contained by the sub-matrices $C$ and
$D$.

Partially quenched ChPT will be a low-energy effective field theory for these
Goldstone particles.  We collect the partially quenched mesons into a
flavor-space matrix $\Phi$, multiplying each meson field $\pi^a$ by its
corresponding broken flavor symmetry generator:
\begin{equation}
\Phi \equiv \pi^a \tau^a = \begin{bmatrix}
\phi_{q \qb} & \phi_{q \bar{\tilde{q}}} \\[2mm]
\phi_{\tilde{q} \qb} & \phi_{\tilde{q} \bar{\tilde{q}}}
\end{bmatrix}
\end{equation}
where $\phi_{q \qb}$ is an $N_v + N_f \times N_v + N_f$ matrix containing
bosonic bound states of ordinary quark-antiquark pairs, $\phi_{\tilde{q}
\bar{\tilde{q}}}$ is an $N_v \times N_v$ matrix containing bosonic bound states
of ghost quark-antiquark pairs, and $\phi_{q \bar{\tilde{q}}}$ and
$\phi_{\tilde{q} \qb}$ are respectively $N_v + N_f \times N_v$ and
$N_v \times N_v + N_f$ matrices containing fermionic bound states of mixed
quark-antiquark pairs.  We further subdivide the matrix $\phi_{q \qb}$, making
its dynamical- and valence-quark blocks explicit:
\begin{equation}
\phi_{q \qb} = \begin{bmatrix}
\phi_{v \bar{v}} & \phi_{v \bar{s}} \\
\phi_{s \bar{v}} & \phi_{s \bar{s}}
\end{bmatrix}
\end{equation}
where the $N_v \times N_v$ matrix $\phi_{v \bar{v}}$ holds the mesons which
contain a valence quark-antiquark pair.  As only valence-quark fields are
used in the operators whose expectation values we calculate, it is only this
block of mesons whose masses we measure directly in our partially quenched
lattice calculations.  From the perspective of pqChPT, all other mesons appear
only at one-loop order and higher.  From now on we will refer to a valence
quark-antiquark meson as a pion, as it coincides with the state in our
partially quenched staggered calculation to which we have designated that name.  
We define the unitary field matrix $\Sigma$ as:
\begin{equation}
\Sigma \equiv e^{ 2 i \pi^a \tau^a / f } \in SU(N_v + N_f \vert N_v)
\end{equation}

The quark masses break pqQCD's flavor symmetry explicitly, granting the
Goldstone particles a light mass.  We incorporate this symmetry breaking into
pqChPT using a factor whose matrix structure is equivalent to the quark mass
matrix:
\begin{equation}
\chi \equiv 2 \mu \qmassm = 2 \mu \diag ( \underbrace{\mv, \ldots}_{N_v},
\underbrace{\ms, \ldots}_{N_f}, \underbrace{\mv, \ldots}_{N_v} )
\end{equation}
As stated above, we use $N_v$ valence quarks of mass $m_v$, $N_f$ dynamical
quarks of mass $m_s$, and $N_v$ ghost quarks of mass $m_v$.

Considering all terms which respect both Lorentz and pqQCD's graded flavor
symmetry, except via insertion of $\qmassm$, and expanding simultaneously in
both meson momentum and meson mass, we construct the NLO partially quenched
chiral Euclidean Lagrangian:
\begin{align}
\Lag{pqChPT}^\text{NLO} \pad{=} & \frac{f^2}{4} \sTr{ \partial_\mu
\Sigma^\dagger \partial^\mu \Sigma } - \frac{f^2}{4}  \sTr{ \Sigma^\dagger \chi
+ \chi \Sigma } \notag \\
& - L_1 \Bigl( \sTr{ \partial_\mu \Sigma^\dagger \partial^\mu \Sigma } \Bigr)^2
\notag \\
& - L_2 \Bigl( \sTr{ \partial_\mu \Sigma^\dagger \partial_\nu \Sigma } \sTr{
\partial^\mu \Sigma^\dagger \partial^\nu \Sigma } \Bigr) \notag \\
& - L_3 \Bigl( \sTr{ \partial_\mu \Sigma^\dagger \partial^\mu \Sigma
\partial_\nu\Sigma^\dagger \partial^\nu \Sigma } \Bigr) \notag \\
& + L_4 \Bigl( \sTr{ \partial_\mu \Sigma^\dagger \partial^\mu \Sigma } \sTr{
\Sigma^\dagger \chi + \chi \Sigma } \Bigr) \notag \\
& + L_5 \Bigl( \sTr{ \partial_\mu \Sigma^\dagger \partial^\mu \Sigma \bigl(
\Sigma^\dagger \chi + \chi \Sigma \bigr) } \Bigr) \notag \\
& - L_6 \Bigl( \sTr{ \Sigma^\dagger \chi + \chi \Sigma } \Bigr)^2 \notag \\
& - L_7 \Bigl( \sTr{ \Sigma^\dagger \chi - \chi \Sigma } \Bigr)^2 \notag \\
& - L_8 \Bigl( \sTr{ \Sigma^\dagger \chi \Sigma^\dagger \chi } + \sTr{ \chi
\Sigma\chi \Sigma } \Bigr)
\end{align}
Note that this Lagrangian has the same form as that for standard ChPT
\eqref{j:equation}, with the only differences arising from our use of a graded
flavor symmetry group.

As long as we restrict our choice for the dynamical and valence quark masses
such that they remain within the radius of convergence of pqChPT, the partially
quenched chiral Lagrangian will accurately describe the two-dimensional quark
mass dependence of the mesons' characteristics and behavior.

\section{Next-to-Leading Order Meson Properties}
Using the NLO partially quenched chiral Lagrangian, we determine NLO expressions
for the mass \cite{Sharpe:1997by}:
\begin{align}
\m^2_{vv} & = z \mv (4 \pi f)^2 \biggl\{ 1 + \frac{z}{N_f} \bigl( 2 \mv
- \ms \bigr) \ln z \mv + \frac{z}{N_f} \bigl( \mv - \ms \bigr) \notag \\
& ~ \phantom{= z \mv (4 \pi f)^2 \biggl\{ 1} + z \mv \bigl( 2 \alpha_8 -
\alpha_5 \bigr) + z \ms N_f \bigl( 2 \alpha_6 - \alpha_4 \bigr) \biggr\}
\label{bk:equation} \\
\intertext{and decay constant:}
f_{vv} & = f \biggl\{ 1 + \frac{z N_f}{4} \bigl( \mv + \ms \bigr) \ln
\frac{z}{2} \bigl( \mv + \ms \bigr) + z \mv \frac{\alpha_5}{2} + z \ms N_f
\frac{\alpha_4}{2} \biggr\}
\label{bl:equation}
\end{align}
of the partially quenched chiral pseudo-Goldstone boson containing two
degenerate valence quarks, our pion.  The quantity $z$ is defined in
\eqref{bp:equation}.  Note that in the unquenched case, $m_q = m_v = m_s$,
these expressions correspond exactly to those from ChPT:  \eqref{bi:equation}
and \eqref{bj:equation}.

For the pqChPT expressions for the pion mass and decay constant in the case of
arbitrary quark masses, the reader is referred to \cite{Sharpe:2000bc}.

Note that in the limit of zero valence quark mass, $m_v \rightarrow 0$, the
logarithmic term in $\m^2_{vv}$ diverges.  Such divergent log terms in pqChPT
expressions are often referred to as quenched chiral logs, or sometimes simply
as quenched logs.  They are directly analogous to the chiral logs of standard
ChPT.  The appearance of quenched chiral logs makes it clear that the true
chiral limit of pqQCD is reached only by taking the dynamical and valence quark
masses to zero simultaneously.  In that case, the quenched log terms remain
finite.

Quenched logs also appear in expressions derived from the low-energy chiral
theory of fully quenched QCD, quenched Chiral Perturbation Theory (qChPT).  We
will not discuss qChPT in detail, noting only that such logs are divergent when
the only quark mass available for adjustment, the valence quark mass, is taken
to zero.  Thus, doubt is thrown onto the very existence of a chiral limit under
the quenched approximation.

\section{Physical Results from Partially Quenched Calculations}
\label{ag:section}
There are two critical points which allow pqChPT to act as a bridge between our
unphysical partially quenched calculations and the physical GL coefficients.

\begin{figure}
\centering
\psfrag{ms}{$m_s$}
\psfrag{mv}{$m_v$}
\includegraphics[width=0.75\textwidth,clip=]{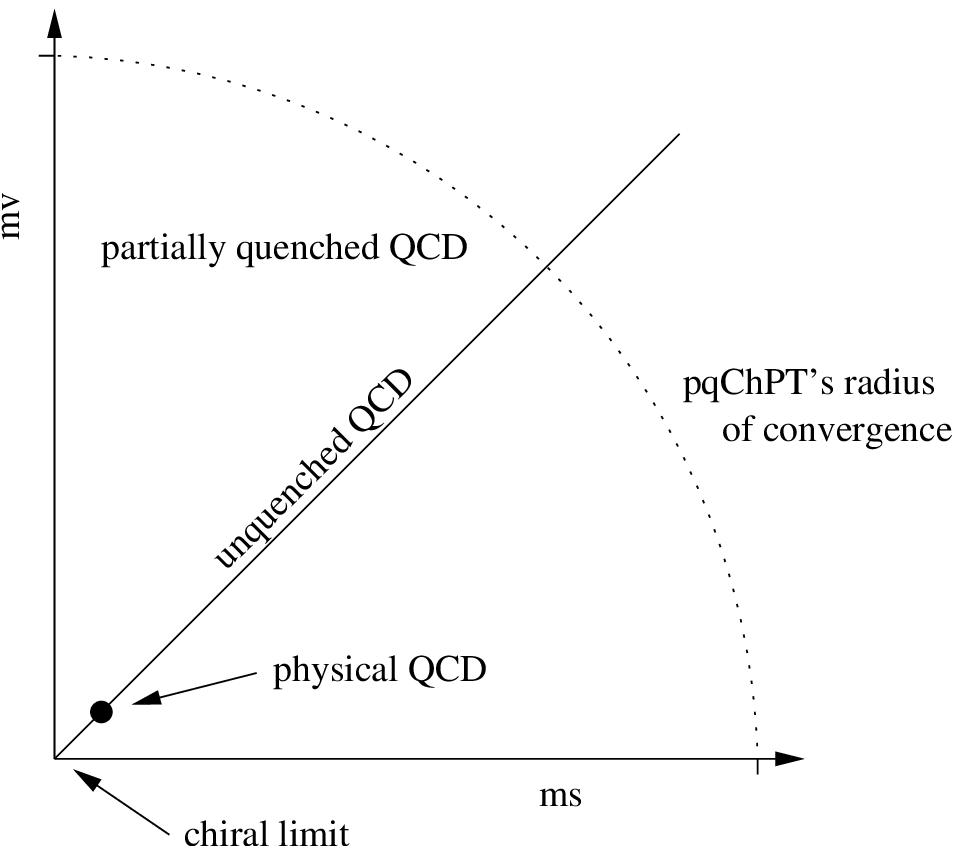}
\mycaption{The quark-mass plane of pqQCD.}
\label{a:figure}
\end{figure}

The first is that unquenched QCD is a subset of pqQCD.  If we visualize all
possible dynamical- and valence-quark-mass choices as defining the plane of
pqQCD, as shown in Figure \ref{a:figure}, unquenched QCD is seen to lie on the
diagonal line at which $m_v = m_s$.

The second point is that the two-dimensional quark mass dependence of pqChPT's
Lagrangian is known and explicitly included up to whatever order we have chosen
to work.  Thus, the GL coefficients are not functions of quark mass, but
rather are constants throughout the quark-mass plane.  Most importantly, they
are constants as we move across the line of unquenched QCD.

Thus, the GL coefficients of pqChPT and standard ChPT are the same.  If we
perform a partially quenched calculation and, via pqChPT, determine from the
calculation values for the GL coefficients, we have in fact generated valid
results for the GL coefficients of the physical world.

Additionally, in a unquenched calculation, we are restricted to exploring quark
mass dependence only along the unquenched line.  In the case of the pion mass,
it is clear from \eqref{bi:equation} that such an exploration would grant us
access only to the GL coefficient combination $2 \alpha_8 - \alpha_5 + 2 N_f
\alpha_6 - N_f \alpha_4$.  On the other hand, the use of partial quenching
grants us additional moment arms with which to extract information from our
calculated quantities.  In the case of the pion mass, varying the dynamical and
valence quark masses independently allows for the determination of two
coefficient combinations:  $2 \alpha_8 - \alpha_5$ and $2 \alpha_6 - \alpha_4$.

It should be noted that the dependence of the partially quenched chiral
Lagrangian on the number of dynamical quark flavors $N_f$ is not known.  Thus,
the GL coefficients are functions of $N_f$.  As a consequence, in order for our
partially quenched results to be considered valid results for the physical
coefficients, we are forced to use a physical number of dynamical light quarks,
$N_f = 3$, in our calculations.  We in fact do just that.

In standard QCD, the $\eta'$ can be thought of as obtaining its large mass via
the summation over quark-loop chains which appear in its propagator.  From this
perspective it is evident that the $\eta'$ of fully quenched QCD will be
lighter than expected, as any such quark loops have been removed.  Thus, the
$\eta'$ must be included in any low-energy effective theory for qQCD such as
qChPT.  This inclusion introduces additional couplings to qChPT and destroys
any correspondence of its couplings to the couplings of physical ChPT.  While
it was not initially known whether partially quenched QCD suffers from this
same flaw, Sharpe demonstrates in \cite{Sharpe:2001fh} that the $\eta'$ of
pqQCD is in fact heavy, and can be safely left out of pqChPT.  Thus, the
correspondence between the GL coefficients of pqChPT and ChPT is retained.

\section{Calculation of the Gasser-Leutwyler Coefficients}
A basic outline of our procedure for the calculation of the GL coefficients is
as follows.   We calculate the local pion's bilinear correlator, using $N_f =
3$ staggered lattice techniques as detailed in Chapter \ref{o:section}, for a
range of dynamical and valence quark masses.  We then fit those results,
using the techniques described in Chapter \ref{u:section} and over a fit range
contained by the radius of convergence of pqChPT, to pqChPT's corresponding
predictions for the correlator's quark-mass dependence.

In order to facilitate a numerical fit of the lattice results to the predicted
forms from pqChPT, those forms must be expressed in terms of dimensionless
parameters.  Making use of \eqref{be:equation}, we build pqChPT's prediction
for the correlator's form:
\begin{equation}
C_{5, 5;t} = \frac{\uf^2_{\pi_5} \uM^3_{\pi_5} V}{8 m^2_V} \bigl( 1 +
e^{-\uM_{\pi_5}} \bigr) \bigl( e^{-\uM_{\pi_5} t} + e^{-\uM_{\pi_5} ( L_4 - 1 -
t) } \bigr)
\label{bm:equation}
\end{equation}
where, using \eqref{bk:equation}:
\begin{align}
\uM^2_{\pi_5} & = \uz \mV (4 \pi \uf)^2 \biggl\{ 1 + \frac{\uz}{N_f} \bigl( 2
\mV - \mS \bigr) \ln \uz \mV + \frac{\uz}{N_f} \bigl( \mV - \mS \bigr) \notag
\\
& ~ \phantom{= \uz \mV (4 \pi \uf)^2 \biggl\{ 1} + \uz \mV \bigl( 2 \alpha_8 -
\alpha_5 \bigr) + \uz \mS N_f \bigl( 2 \alpha_6 - \alpha_4 \bigr) \biggr\}
\label{bq:equation} \\
\intertext{using \eqref{bl:equation}:}
\uf_{\pi_5} & = \uf \biggl\{ 1 + \frac{\uz N_f}{4} \bigl( \mV + \mS \bigr) \ln
\frac{\uz}{2} \bigl( \mV + \mS \bigr) + \uz \mV \frac{\alpha_5}{2} + \uz \mS
N_f \frac{\alpha_4}{2} \biggr\}
\label{br:equation}
\end{align}
and $\uz$ and $\uf$ are dimensionless parameters related to $z$ and $f$ by
appropriate powers of the lattice spacing.  The unitless dynamical and valence
quark masses are denoted by $\mS$ and $\mV$.

The products of this fit are values for the physical GL coefficients of ChPT.

\section{Constant Dynamical Quark Mass}
As discussed in Section \ref{m:section}, changing the valence quark mass is
significantly less computationally demanding than changing the dynamical quark
mass.  Thus, for those ensembles which have the smallest expected systematic
error, and thus those which are the most computationally demanding, we use a
single dynamical-quark-mass value, and vary only the valence quark mass.
Fortunately, variation of the valence quark mass grants us access to the
GL coefficient combination we are most interested in, the combination $2
\alpha_8 - \alpha_5$.

The GL coefficient combination $2 \alpha_8 - \alpha_5$ originally became of
interest because it is the unknown term in $\Delta_M$ \eqref{h:equation}, where
$\Delta_M$ is the difference in the NLO contributions to the physical pion and
kaon masses \eqref{u:equation}.  Clearly, any such differences in the NLO terms
arise due to the contrasting valence-quark content of the pion and kaon.  Thus,
it is of no surprise that variation of only the valence quark mass grants
access to the coefficient combination critical in the determination of
$\Delta_M$.

Without variation of the dynamical quark mass, the terms in
\eqref{bq:equation} and \eqref{br:equation} which depend on $\mS$ can not be
accounted for.  Thus, in cases where we have calculated the local-bilinear
correlator only along a line of constant dynamical quark mass, we absorb the
unknown NLO $\mS$ dependence into the LO coefficients $\uz$ and $\uf$.  This
change results in a deviation from the original expressions only when $\uz$ and
$\uf$ appear in NLO terms, and thus the error due to this change occurs at
NNLO.  The resulting expression for the pion mass is:
\begin{align}
\uM^2_{\pi_5} & = \uz \mV (4 \pi \uf)^2 \biggl\{ 1 + \frac{\uz}{N_f} \bigl( 2
\mV - \mS \bigr) \ln \uz \mV + \frac{\uz}{N_f} \bigl( \mV - \mS \bigr) \notag
\\
& ~ \phantom{= \uz \mV (4 \pi \uf)^2 \biggl\{ 1} + \uz \mV \bigl( 2 \alpha_8 -
\alpha_5 \bigr) \biggr\}
\label{bn:equation} \\
\intertext{and for the pion decay constant is:}
\uf_{\pi_5} & = \uf \biggl\{ 1 + \frac{\uz N_f}{4} \bigl( \mV + \mS \bigr) \ln
\frac{\uz}{2} \bigl( \mV + \mS \bigr) + \uz \mV \frac{\alpha_5}{2} \biggr\}
\label{bo:equation}
\end{align}
These forms are then used in \eqref{bm:equation} in place of
\eqref{bq:equation} and \eqref{br:equation}.

\chapter{Data Modeling} \label{u:section}
Throughout our study, in order to extract information from our lattice
calculations, we model the results to theoretical predictions.  We present
here details of that process.

\section{Lattice Measurements}
The product of a lattice calculation is the value of some operator evaluated
under each gauge configuration of an ensemble.  We refer to such an operator
evaluation as a measurement, using the term in the loosest sense, as no actual
experimental measurement is taking place.

An operator's form will contain some number of adjustable parameters $x_n$,
which we label as measure parameters.  For each measurement a value is be
chosen for each of these measure parameters.  We codify the results of the
measurements as:
\begin{equation}
y_a ( x_1, x_2, x_3, \ldots) \equiv \bra \Op ( x_1, x_2, x_3, \ldots )
\ket_{\fconfig{U}_a}
\end{equation}
where $a$ denotes the configuration under which the operator was evaluated, and
runs from 1 to $N$, where $N$ is the number of configurations in the ensemble,
and the various $x_n$ designate the location in the operator's
adjustable-parameter space at which the measurement is made.  Each measure
parameter $x_n$ takes on a discrete set of values $x_{n;i}$, where $i$
enumerates the values for that measure parameter at which we have chosen to
make measurements, running from 1 to $N_{M;n}$.  As an example, in the
evaluation of a single bilinear correlator, there is one measure parameter, the
bilinears' time separation, which takes on integer values in the range $[ 0,
L_4 - 1]$.

We generally use the more compact notation $y_{a;x}$, where the index $x$
enumerates all possible combinations of allowed values for the measure
parameters.  It runs from 1 to $N_M$, where:
\begin{equation}
N_M = \prod_n N_{M;n}
\end{equation}
representing the total number of measurements made on each configuration.

\section{Data Blocking} \label{s:section}
As a result of autocorrelation, as described in Section \ref{i:section},
measurements on neighboring configurations of a Markov chain are correlated.
However, to correctly account for the statistical error in our data, we require
that each measurement be independent.  In order to meet this requirement, we
divide the Markov chain into $N_K$ blocks, each of which contains $N_B = N /
N_K$ neighboring configurations.  We choose an $N_B$ larger than the Markov
chain's autocorrelation length.  Thus, the average measurement within each
block can be treated as a single uncorrelated measurement, independent of the
results from other blocks.

From now on we work only with this blocked data:
\begin{equation}
y_{B;x} = \frac{1}{N_B} \sum_{a \in B} y_{a;x}
\end{equation}

It is worth noting that, for Gaussian distributed data, the standard error of
our result is independent of the value chosen for $N_B$, assuming $N_B$ is
larger than the chain's autocorrelation length but is still small enough that
$N_K$ remains large.  Increasing $N_B$ reduces the variance
between blocks but also reduces our sample size.  For Gaussian distributed
data, the two effects cancel exactly.

\section{Data Correlation}
We visualize all of the measurements on a single block as defining the
location of a single point in a large-dimensional measurement space, where each
possible combination of measure-parameter values represents a dimension of the
space.  From this perspective, the index $x$ of each measurement $y_{B;x}$
is seen as indexing the components of the point's position.

The full set of measurements for an ensemble is seen as a cloud of points in
this measurement space.  We expect the data points to cluster around some
central value and their density to die off as a Gaussian in all directions.
The shape of this cloud, in addition to the rate of its falloff along the
various directions, characterize the statistical error of our measurements.

For each measurement there is an ensemble-average value:
\begin{equation}
\avg{y}_x = \bra y_x \ket = \frac{1}{N_K} \sum_B y_{B;x}
\end{equation}
which corresponds to the expectation value of the operator $\Op_x$.

Additionally, each measurement has a variance:
\begin{align}
\sigma^2_x & = \Bigl\bra \bigl( y_x - \avg{y}_x \bigr)^2 \Bigr\ket \notag
\\
& = \frac{1}{N_K} \sum_B \bigl( y_{B;x} - \avg{y}_x \bigr)^2
\end{align}
which expresses the width of the cloud along the measurement-space coordinate
$x$.  However, $\sigma^2_x$ only effectively characterizes the measurements'
error if the cloud is an ellipse with axes that are flush with the coordinate
directions of the measurement space.  In such a case the width of the cloud
along each coordinate direction is all that is needed to characterize the shape
of the cloud.  Yet, if the measurements are correlated, we will find that the
axes of the ellipse do not fall along the coordinate directions.

Invariably, measurements are correlated.  That is, the measurements vary
coherently between blocks.  If, when a given measurement on a given block is
found to be above the ensemble average, another nearby measurement tends to
also be above the ensemble average, then those two measurements are said to be
correlated.  If the second measurement tends to be on the opposite side of the
ensemble average as the first, the two measurements are anticorrelated.  If the
two measurements move above and below the ensemble average in an independent
fashion, they are uncorrelated.

The correlations between the $N_M$ measurements are quantified by the
covariance matrix $C$:
\begin{align}
C_{x,y} & = \Bigl\bra \bigl( y_x - \avg{y}_x \bigr) \bigl( y_y
- \avg{y}_y \bigr) \Bigr\ket \notag \\
& = \frac{1}{N_K} \sum_B \bigl( y_{B;x} - \avg{y}_x \bigr) \bigl( y_{B;y}
- \avg{y}_y \bigr)
\end{align}
Along the diagonal, the covariance matrix corresponds to the standard variance
of a measurement, $C_{x,x} = \sigma^2_x$, while the off-diagonal elements
represent correlations between measurements.  A large positive off-diagonal
value corresponds to a strong correlation, while a large negative off-diagonal
value corresponds to a strong anticorrelation.  Completely uncorrelated data
will have a diagonal covariance matrix.

Often literature refers to a correlation matrix $\rho$ rather than a covariance
matrix.  The correlation matrix is constructed to specify only the correlation
between measurements, with all information concerning their error removed:
\begin{equation}
\rho_{x,y} = \frac{1}{\sigma_x \sigma_y} C_{x,y}
\end{equation}

If two measurements are correlated, their value will tend to vary coherently.
From the perspective of the data cloud, measurement points will tend to fall
along a line diagonal to the coordinate directions of the two measurements.
Thus, the data will form an ellipse with diagonal axes.

The eigenvectors of the covariance matrix correspond to the off-diagonal axes
of this ellipse.  Additionally, the eigenvalues of the matrix correspond to the
width of the cloud along those directions, effectively giving the data's
standard error measured along the ellipse's axes.

It is possible in certain pathological situations that data could form shapes
more complex than a simple ellipse.  The most common example is a
boomerang-like shape, where the cloud may actually miss the ensemble average
completely.  Correct statistical error analysis of such data requires the use
of moments higher than the second, and is beyond the scope of this discussion.

\section{Theoretical Models}
For each set of measurements taken, we have some theoretical form which we
expect the data to match.  Included in this theory function are some number of
fit parameters $c_\ell$ whose value we hope to glean from the data:
\begin{equation}
f ( c_1, c_2, c_3, \ldots; x_1, x_2, x_3, \ldots )
\end{equation}
We also make use of the more compact notation $f_x ( c_1, c_2, c_3, \ldots )$
or $f_x(c)$.

Given a set of values for the fit parameters, the theory function $f_x(c)$
specifies a point in measurement space.  We would like to find the set of fit
parameters which result in that point being as close as possible to the data's
ensemble average.  Additionally, we would like to use a metric for this measure
of closeness which accounts for the variation in magnitude of the data's
statistical error with the direction of the distance in question.

The correct metric for this purpose is the inverse of the covariance matrix.
It weights distance in a given direction with a strength inversely proportional
to the variance of the data along that direction.  Using this metric the
squared distance between our ensemble average and the theory function for a
given set of fit parameters is:
\begin{equation}
\chi^2 = \sum_x \sum_y \bigl( \avg{y}_x - f_x(c) \bigr) C^{-1}_{x,y} \bigl(
\avg{y}_y - f_y(c) \bigr)
\end{equation}
We are now left to minimize $\chi^2$ with respect to the fit parameters, and
thus determine the optimal set of fit parameter values.

By the nature of its definition, the covariance matrix is symmetric, and must
be positive definite in the limit of a large number of configurations, $N_K
\rightarrow \infty$.  Thus, Cholesky decomposition \cite{Numerical-Recipes} can
be used to determine a triangular matrix $L$ such that:
\begin{align}
L L^T & = C
\intertext{or equivalently:}
( L^{-1} )^T L^{-1} & = C^{-1}
\end{align}
By applying $L^{-1}$ to our vector of correlated distances, we generate a
vector of uncorrelated distances weighted by the inverse of their uncertainty:
\begin{equation}
r_x = L^{-1}_{x,y} \bigl( \avg{y}_y - f_y(c) \bigr)
\end{equation}
In terms of $r_x$, $\chi^2$ is now simply:
\begin{equation}
\chi^2 = \sum_x r^2_x
\end{equation}

In cases where the number of data blocks $N_K$ is not large relative to the
number of measurements $N_M$, there is no guarantee that the covariance matrix
will be positive definite.  When it is not, we have no choice other than to set
the matrix's off-diagonal elements to zero, effectively disregarding
measurement correlation in the context of the fit.  We find that this is often
required in our study, as many of the fits have two, or even three, measurement
parameters.  Additionally, increasing our available data requires lengthening
our Markov chains, a process which is exceedingly computationally intensive.
These two factors often lead us to a very large $N_M$ relative to our available
$N_K$.

Minimizing $\chi^2$ can be done using a nonlinear least-squares fitting
technique.   We begin each fit with the downhill simplex method
\cite{Numerical-Recipes, Nelder:1965:SMF}, also known as the amoeba method, and
complete it using the Levenberg-Marquardt method \cite{Numerical-Recipes,
Marquardt:1963:ALS}.

\section{Jackknife Error}
We determine the statistical error of our fit parameters via a jackknife
procedure.  This involves repeating the entire fit process $N_K$ times, each
time leaving out a different block of data from the data set.  The result is a
vector of values $c_{\ell;j}$, where $j$ runs from 1 to $N_K$, for each fit
parameter $c_\ell$.  The standard error for each fit parameter is then:
\begin{equation}
\sigma^2_{c_\ell} = \frac{N_K - 1}{N_K} \sum_j \bigl( c_{\ell;j} - \avg{c}_\ell
\bigr)^2
\label{bc:equation}
\end{equation}
where:
\begin{equation}
\avg{c}_\ell = \frac{1}{N_K} \sum_j c_{\ell;j}
\end{equation}

When reporting the value of a fit parameter, we quote the statistical error
using the form:
\begin{equation}
c_\ell \pm \sqrt{\sigma^2_{c_\ell}}
\end{equation}
where $c_\ell$ is the optimal fit parameter value found in the full fit and
$\sigma^2_{c_\ell}$ is given by \eqref{bc:equation}.

When our desired result is some function of the fit parameters $g ( c_1, c_2,
c_3, \ldots )$, the variation in the fit parameters may be correlated such that
the subsequent variation in the result is larger or smaller than what would be
expected.  In such a case we apply the jackknife procedure to the result
itself:
\begin{equation}
\sigma^2_g = \frac{N_K - 1}{N_K} \sum_j \bigl( g ( c_{1;j}, c_{2;j}, c_{3;j},
\ldots ) - \avg{g} \bigr)^2
\end{equation}
where:
\begin{equation}
\avg{g} = \frac{1}{N_K} \sum_j g ( c_{1;j}, c_{2;j}, c_{3;j}, \ldots )
\end{equation}
The result is then reported in the form:
\begin{equation}
g ( c_1, c_2, c_3, \ldots ) \pm \sqrt{\sigma^2_g}
\end{equation}

\section{Specific Applications}
We applied the data modeling techniques described above in six aspects of our
study.  For clarity we present explicitly the measure parameters $x_n$, fit
parameters $c_\ell$, measurement operator $\Op_x$, and theory function $f_x(c)$
used in each fit.  

Because we are only able to manipulate dimensionless quantities in our
numerical fit procedure, measure and fit parameters must be unitless.  Often we
absorb into dimensionful quantities some number of powers of the lattice
spacing $a$ in order to construct a fit's required unitless quantities.  We
remind the reader of our use of $\U{X}$ to denote the unitless version of some
quantity $X$.

\subsection{Static Quark Potential}
We determine the static quark potential at a single spatial separation by
fitting the rectangular Wilson-loop expectation value to an exponential:
\begin{align}
x_n & : t
&
\Op_x & : \eqref{bu:equation} \notag \\
c_\ell & : \mathscr{A}, a V ( a s )
&
f_x(c) & : \eqref{bu:equation} \notag
\end{align}

\subsection{Spatial Dependence of Static Quark Potential}
We extracted the form of the static quark potential by fitting the rectangular
Wilson loop expectation value to its predicted form:
\begin{align}
x_n & : s, t
&
\Op_x & : \eqref{bf:equation} \notag \\
c_\ell & : \mathscr{A}_s, v_0, v_1, v_2, \tilde{v}_2
&
f_x(c) & : \eqref{bf:equation} \notag
\end{align}
Note that a unique fit parameter $\mathscr{A}_s$ is used for each value of $s$
in the fit range.

\subsection{Single Correlator}
In order to determine the mass and decay constant of the local pion for a
single value of the valence quark mass, we fit the local-bilinear correlator to
an exponential:
\begin{align}
x_n & : t
&
\Op_x & : \eqref{bd:equation} \notag \\
c_\ell & : \uf_{\pi_5}, \uM_{\pi_5}
&
f_x(c) & : \eqref{be:equation} \notag
\end{align}

\subsection{Quadratic Valence-Quark-Mass Dependence}
To determine the valence quark mass value at which the local pion mass equals
the physical kaon mass, we fit the local-bilinear correlator at various valence
quark masses to a phenomenologically motivated quadratic form for the local
pion mass:
\begin{align}
x_n & : t, m_V
&
\Op_x & : \eqref{bd:equation} \notag \\
c_\ell & : \mathscr{A}_{m_V}, a_0, a_1, a_2
&
f_x(c) & : \eqref{bw:equation}, \eqref{bv:equation} \notag
\end{align}
Note that an independent fit parameter $\mathscr{A}_{m_V}$ is used for each
value of $m_V$ studied.

For illustrative purposes, in one case we also fit the correlator to a cubic
form, using the fit characteristics described above, but replacing
\eqref{bv:equation} with \eqref{by:equation}.

\subsection{Chiral Valence-Quark-Mass Dependence}
To determine the GL coefficient combinations $2 \alpha_8 - \alpha_5$ and
$\alpha_5$, we fit the local-bilinear correlator at various valence quark
masses to the form for the local pion mass and decay constant predicted by
pqChPT:
\begin{align}
x_n & : t, m_V
&
\Op_x & : \eqref{bd:equation} \notag \\
c_\ell & : \uz, \uf, 2 \alpha_8 - \alpha_5, \alpha_5
&
f_x(c) & :
\eqref{bm:equation}, \eqref{bn:equation}, \eqref{bo:equation} \notag
\end{align}

For small volume ensembles, we are forced to add a constant term to the
pion-mass form in the above fit.  For these ensembles we use the fit
characteristics described above, but replace \eqref{bn:equation} with
\eqref{bz:equation}.

\subsection{Chiral Dynamical- and Valence-Quark-Mass Dependence}
In order to determine the GL coefficient combinations $2 \alpha_8 - \alpha_5$,
$2 \alpha_6 - \alpha_4$, $\alpha_5$, and $\alpha_4$, we fit the local-bilinear
correlator at various dynamical and valence quark masses to the form for the
local pion mass and decay constant predicted by pqChPT:
\begin{align}
x_n & : t, m_V, m_S
&
\Op_x & : \eqref{bd:equation} \notag \\
c_\ell & : \uz, \uf, 2 \alpha_8 - \alpha_5, 2 \alpha_6 - \alpha_4, \alpha_5,
\alpha_4
&
f_x(c) & : \eqref{bm:equation}, \eqref{bq:equation}, \eqref{br:equation} \notag
\end{align}
In this case the measurement operator is not a function of one of the measure
parameters, the dynamical quark mass $m_S$.  Instead, $m_S$ is a parameter of
the ensemble itself.  Thus, measurements at different values of $m_S$ come from
different ensembles, and there can be no correlation between them.  The
elements of the covariance matrix which correspond to measurements at
different dynamical quark masses are simply zero.

\chapter{Ensemble Details} \label{aq:section}
We present here the ensembles used in our study, revealing the motivation
behind and the parameters used in the generation of each ensemble.  Table
\ref{d:table} displays a summary of this information.

\begin{table}
\centering
\begin{tabular}{|c||cccc|d{1.3}d{1.2}c|ccccc|}
\hline
& $L_1$ & $L_2$ & $L_3$ & $L_4$ & \multicolumn{1}{c}{$\beta$} &
\multicolumn{1}{c}{$m_S$} & $N_f$ & start & $N$ & $N_T$ & $N_B$ & $N_K$ \\
\hline
\hline
\ensemble{A} & 16 & 16 & 16 & 32 & 5.3    & 0.01   & 3 & O &  2250 & 250 & 200 &
 20 \\
             &    &    &    &    &        &        &   & D &  2250 & 250 &     &
    \\
\ensemble{B} & 12 & 12 & 16 & 32 & 5.3    & 0.01   & 3 & O &  2250 & 250 & 200 &
 10 \\
\ensemble{C} &  8 &  8 &  8 & 32 & 5.3    & 0.01   & 3 & O & 10050 & 250 & 200 &
 49 \\
\hline
\ensemble{W} &  8 &  8 &  8 & 32 & 5.115  & 0.015  & 3 & O & 10300 & 300 & 100 &
100 \\
\ensemble{X} &  8 &  8 &  8 & 32 & 5.1235 & 0.02   & 3 & O & 10300 & 300 & 100 &
100 \\
\ensemble{Y} &  8 &  8 &  8 & 32 & 5.132  & 0.025  & 3 & T & 10000 &   0 & 100 &
100 \\
\ensemble{Z} &  8 &  8 &  8 & 32 & 5.151  & 0.035  & 3 & T & 10000 &   0 & 100 &
100 \\
\hline
\ensemble{Q} & 16 & 16 & 16 & 32 & 5.8    & \multicolumn{1}{c}{-} & 0 &
\multicolumn{5}{c|}{144 configurations} \\
\hline
\end{tabular}
\mycaption{The lattice parameters for the ensembles used in our study.  The
starting conditions are denoted as O for an ordered start, D for a disordered
start, and T for a thermal start.  The values for Markov chain length $N$,
thermalization point $N_T$, and block length $N_B$ are given as trajectory
counts.  $N_K$ corresponds to the number of blocks available in an ensemble.}
\label{d:table}
\end{table}

For all dynamical ensembles, a configuration was added to the ensemble every
ten HMD trajectories along the Markov chain, with the first configuration in
each ensemble being ten trajectories from the ensemble's starting condition.

\section{Primary Ensemble}
We refer to our primary ensemble as ensemble \ensemble{A}, or after hypercubic
blocking as ensemble \ensemble{A ~ hyp}.  This ensemble has our largest lattice
extent, $16^3 \times 32$, and reasonable lattice spacing.  As such, it is from
\ensemble{A ~ hyp} that we will generate our quoted results.  This is our only
ensemble to include two Markov chains, one of which has an ordered starting
condition, and the second of which has a disordered starting condition.  Such a
set of Markov chains is very helpful in determining the thermalization point,
as in a sense the chains begin on opposite sides of the desired equilibrium
ensemble.

\section{Finite-Volume Ensembles}
In order to study the magnitude of finite-volume error in our results, we use
two ensembles with smaller lattice extent than our primary ensemble, leaving
all other lattice parameters unchanged.  We hope that results from the $12^2
\times 16 \times 32$ ensemble, or ensemble \ensemble{B}, deviate only slightly
from those of our primary ensemble, demonstrating that the finite-volume error
in ensemble \ensemble{A} is under control.  We expect that results from the
$8^3 \times 32$ ensemble, or ensemble \ensemble{C}, will deviate significantly
from our primary ensemble's results, as its physical volume is thought to be
too small for our study.  This should, however, make clear the effects of
finite volume.  Both of these finite-volume ensembles have ordered starting
conditions.  

\section{Varying-Dynamical-Quark-Mass Ensembles}
To determine the GL coefficient combination $2 \alpha_6 - \alpha_4$, we require
a set of ensembles between which the dynamical quark mass varies, but all
other ensemble parameters are constant.  In order to vary the dynamical quark
mass, but leave the lattice spacing unchanged, we have made use of results from
the Columbia group \cite{c:url:cite}.  They have mapped out the $N_f = 3$, $L_4
= 4$ finite temperature transition, determining several critical $\beta$ and
$m_Q$ value pairs \cite{Liao:2001en}.  We have generated four ensembles using
these values, and refer to them as ensembles \ensemble{W}, \ensemble{X},
\ensemble{Y}, and \ensemble{Z}.  Ensembles \ensemble{W} and \ensemble{X} have
ordered starting conditions, while ensembles \ensemble{Y} and \ensemble{Z} have
thermalized starting conditions, using initial configurations from the
ensembles of \cite{Liao:2001en}.

Because of the significant computational effort required to generate four
distinct ensembles, we have used a small lattice extent.  A reasonable physical
volume is still obtained, as the lattice spacing corresponding to the ensemble
parameters used is rather large.

We also make use of ensemble \ensemble{W} to study the magnitude of
finite-lattice-spacing effects on our results.  This ensemble has approximately
twice the lattice spacing of our primary ensemble, yet has approximately equal
physical volume.  Thus, the deviation of its results from those of our primary
ensemble give an indication of the effects of finite lattice spacing.

\section{Quenched Ensemble}
Quenched ChPT predicts a different functional form than pqChPT for the pion
mass and decay constant's dependence on the valence quark mass
\cite{Bernard:1992mk}.  Thus, it may be possible to distinguish a quenched
ensemble from a partially quenched ensemble via a study of that dependence.  To
this end, we include in our study a fully quenched ensemble with the same
lattice extent as our primary ensemble and approximately equal lattice spacing.
We refer to this ensemble as ensemble \ensemble{Q ~ hyp}, studying it only
after hypercubic blocking.  We do not make use of the predictions of qChPT, but
instead treat the quenched ensemble as though it were partially quenched, with
$N_f = 3$ and $m_Q = 0.01$.  If the effects of quenching are strong, the
results from ensemble \ensemble{Q ~ hyp} should deviate significantly from
those of ensemble \ensemble{A ~ hyp}.

\chapter{Analysis} \label{q:section}
We present here the analysis of our ensembles and the processes which lead to
our results.

\section{Thermalization and Block Length}
Figures \ref{b:figure} through \ref{c:figure} show the local pion's bilinear
correlator $C_{5,5;t}$ calculated on the individual configurations of our
ensembles.  Such a plot makes obvious the effects of both thermalization and
autocorrelation.  For each ensemble we present the bilinear correlator at two
time separations: $t = 0$ and $t = 15$.

The vertical dotted line in each plot demonstrates the value used for that
ensemble's thermalization point $N_T$ as described in Section \ref{r:section}.
The vertical range represents the block length $N_B$ used for that ensemble as
described in Section \ref{s:section}.

In the case of ensemble \ensemble{A}, with its pair of Markov chains, we
overlay the evolution of the two chains on a single plot.  This makes the
appropriate thermalization point quite clear.

\section{Sommer Scale}
As described in Section \ref{t:section}, the lattice spacing of each ensemble
is set via the Sommer scale.  We present here our calculation of the lattice
spacings, as well as a study of the effects of our fit-range choices on the
results.  We calculate the lattice spacing of the thin-link and
hypercubic-blocked versions of each ensemble independently.

In all cases a value of $s_\text{min} = 1$ is used as the lower spatial bound
of our fit ranges.

\subsection{Effective Potential}
In order to determine an appropriate minimum time separation $t_\text{min}$ to
use for our fit range, we calculate the effective potential $V_\text{eff} (s)$
for a range of time separations.  The value of the effective potential between
$t$ and $t + 1$ is the exponential decay constant which describes the falloff
of the Wilson-loop expectation value considering only those two time
separations.  Explicitly, the effective potential at $t + \half$ is defined as:
\begin{equation}
a V_\text{eff} (s) \Big\vert_{t + \half} \equiv \ln \frac{\bigl\bra \Real \tr{
W^{s \times t} } \bigr\ket}{\bigl\bra \Real \tr{ W^{s \times t + 1} }
\bigr\ket}
\end{equation}
The effective potential is expected to vary rapidly for small $t$, where the
asymptotic static quark state is infected by higher-energy states.  Beyond some
larger $t$ the value is expected to plateau at the static quark potential.  We
chose $t_\text{min}$ such that our fit range spans only the region in which
the static quark potential is uncontaminated.  

Figures \ref{d:figure} through \ref{e:figure} show the effective potential over
a range of time separations $t$ and at two spatial separations $s$ for each
ensemble.  The chosen spatial separations bracket, or nearly bracket, that
ensemble's result for $r_0 / a$.  The dotted vertical line in each plot
represents the value of $t_\text{min}$ chosen for the final fit.  The error
bars displayed are the result of a jackknife analysis.  Those points whose
error bars span their plot's vertical range have been dropped.

For ensembles \ensemble{W} through \ensemble{Z}, and their hypercubic blocked
counterparts, our choice of $t_\text{min} = 2$ may not seem optimal in light of
the effective potential plots.  However, $t_\text{min}$ must be low enough so
that the statistical errors of the static quark potential are under control out
to a spatial separation of $s = 4$.  This requirement arises from the fact that
we are using a four-parameter form for the $s$ dependence of the Wilson-loop
expectation value \eqref{ct:equation}.  Thus, without four well-determined
values for the static quark potential along the $s$ direction, the fit is
under-determined.  Choosing $t_\text{min} = 3$, which would perhaps seem
advisable based on the effective potential's behavior, causes the statistical
error of the static quark potential at $s = 4$ to be very large.  Thus, we
choose $t_\text{min} = 2$.  In Section \ref{as:section} we present the
dependence of the resulting value of $r_0 / a$ on our choice of $t_\text{min}$.
This dependence demonstrates that our choice of $t_\text{min} = 2$ for
ensembles \ensemble{W} through \ensemble{Z} in not unreasonable.  

If we were not using a corrected ansatz for the static quark potential, an
under-determined spatial dependence would not be as significant an issue.  When
calculating jackknife error for such an under-determined potential, the results
for the fit parameters will vary wildly.  However, as long as the statistical
error in the vicinity of the Sommer scale is under control, these variations
will be correlated such that the corresponding variation in $r_0 / a$ will
remain small.  For our corrected potential, we do not have this luxury.  In
this case, we ignore a term in the static quark potential's fit form when
determining the Sommer scale.  It is critical that the coefficient for that
term $\tilde{v}_2$ be well determined by the fit, as the correlation of its
error with that of the other fit parameters is discarded, and thus is not
available to stabilize $r_0 / a$.

It follows from the definition of the rectangular Wilson loop that its
expectation value has a symmetric dependence on $s$ and $t$.  Based on our
ansatz for the static quark potential \eqref{ct:equation}, we can only expect
the expectation value to fall off in time as a single exponential for time
separations $t$ larger than some value $s_\text{str}$, where $s_\text{str}$ is
the spatial separation beyond which the static quark potential is dominated by
its string-tension term.  The Sommer scale $r_0$ is specifically defined so
that it falls in the transition region between a Coulomb-like and a string-like
quark potential.  Thus, it is clear that one should choose a value for
$t_\text{min}$ which is just greater than $r_0 / a$.  In all cases we choose
$t_\text{min}$ to be just greater than the value determined for $r_0 / a$ on
the corresponding hypercubic blocked ensemble.

\subsection{Static Quark Potential}
For each ensemble the Wilson-loop expectation value is fit to our ansatz
\eqref{bf:equation} over the range of spatial and time separations bound by
$s_\text{min}$, $s_\text{max}$, $t_\text{min}$, and $t_\text{max}$.  The value
of $r_0 / a$ is then determined via \eqref{bt:equation}.

In Figures \ref{f:figure} through \ref{g:figure} we display the results of
these fits, plotting the determined corrected static quark potential, defined
as:
\begin{equation}
a V_\text{corr} ( a s ) \equiv v_0 + v_1 s + v_2 \frac{1}{s}
\end{equation}
The parameters $v_0$, $v_1$, and $v_2$ are determined by the fit.  The
parameter $\tilde{v}_2$ is also determined by the fit, but its term is dropped
when calculating the corrected static quark potential.

Included in the plots are the results of independent fits of the Wilson-loop
expectation value along lines of constant $s$.  The Wilson-loop expectation
value is fit to the form:
\begin{equation}
\bigl\bra \Real \tr{  W^{s \times t} } \bigr\ket = \mathscr{A} e^{-a V ( a s )
t}
\label{bu:equation}
\end{equation}
for a single spatial separation $s$ and over the range of time separations $t$
from $t_\text{min}$ to $t_\text{max}$.  The result is a value for the static
quark potential at a single spatial separation $a V ( a s )$.  These values are
represented on the plots by $\times$'s.  The plots' diamonds correspond to the
corrected static quark potential, which in this case is defined as:
\begin{equation}
a V_\text{corr} ( a s ) \equiv a V ( a s ) - \tilde{v}_2 \biggl( \Bigl[
\frac{1}{s} \Bigr]_X - \frac{1}{s} \biggr)
\end{equation}
where $\tilde{v}_2$ is determined by the full fit as discussed above.

It should be stressed that the corrected static-quark-potential curve is not
the result of a fit to the corrected static-quark-potential points.  The curve
and points are only related in that they are both the result of fits to the
same expectation-value data.  Agreement between the corrected
static-quark-potential curve and points demonstrates that our ansatz for the
$s$ dependence of the Wilson-loop expectation value is appropriate.

The error bars are determined via a jackknife analysis.  For clarity, error
bars are only shown on the corrected potential points, although equivalent
error bars on the corresponding uncorrected potential points would be
appropriate.  Each fit curve's one-sigma range is bound by dotted lines.  This
range is the result of a jackknife analysis of the curve's value at each point
along the horizontal axis.  Due to the large number of Wilson-loop expectation
values used for the fit, the data's covariance matrix fails to be positive
definite for all ensembles.  Thus, in all cases we use a diagonal covariance
matrix.

On each plot the vertical dashed line corresponds to the ensemble's
determined value of $r_0 / a$, while the shaded region corresponds that
result's jackknife error bars.

\subsection{Dependence on $t_\text{min}$} \label{as:section}
In order to directly study the $t_\text{min}$ dependence of our results, we
repeat the fits using a range of values for $t_\text{min}$.  All other fit
range parameters are left unchanged.  The fruits of this analysis are presented
in Figures \ref{j:figure} through \ref{k:figure}.  In each plot the
$t_\text{min}$ used in the final fit is denoted by a filled diamond, which thus
corresponds to that ensemble's determined value for $r_0 / a$.  For several
ensembles, especially the thin-link ensembles, very few points are given.  For
those values of $t_\text{min}$ at which a point is not given, the fit
either fails entirely, results in an imaginary value for $r_0 / a$, or produces
statistical error bars which span the plot's vertical axis.

\subsection{Dependence on $t_\text{max}$}
In order to directly study the $t_\text{max}$ dependence of our results, we
repeat the fits using a range of values for $t_\text{max}$.  All other fit
range parameters are left unchanged.  The products of this analysis are
presented in Figures \ref{l:figure} through \ref{m:figure}.  In each plot the
$t_\text{max}$ used in the final fit is denoted by a filled diamond, which thus
corresponds to that ensemble's determined value for $r_0 / a$.  Due to
increasingly large statistical error in the Wilson-loop expectation value for
large time separations, the fit results have a weak dependence on
$t_\text{max}$.  Ensemble \ensemble{C ~ hyp} shows the strongest dependence,
most likely due to significant finite-volume effects.

\subsection{Dependence on $s_\text{max}$}
In order to directly study the $s_\text{max}$ dependence of our results, we
repeat the fits using a range of values for $s_\text{max}$.  All other fit
range parameters are left unchanged.  The outcome of this analysis is presented
in Figures \ref{n:figure} through \ref{o:figure}.  In each plot the
$s_\text{max}$ used in the final fit is denoted by a filled diamond, which thus
corresponds to that ensemble's determined value for $r_0 / a$.  Due to
increasingly large statistical error in the Wilson-loop expectation value for
large spatial separations, the fit results have a weak dependence on
$s_\text{max}$.  Again, ensemble \ensemble{C ~ hyp} shows the strongest
dependence.

\subsection{Results}
\begin{table}
\centering
\begin{tabular}{|l||ccc|d{2.7}d{1.8}d{2.7}d{2.6}|}
\hline
& $s_\text{max}$ & $t_\text{min}$ & $t_\text{max}$ &
\multicolumn{1}{c}{$v_0$} & \multicolumn{1}{c}{$v_1$} &
\multicolumn{1}{c}{$v_2$} & \multicolumn{1}{c|}{$\tilde{v}_2$} \\
\hline
\hline
\ensemble{A ~ hyp} & 7 & 4 & 10 & 0.243(16)  & 0.1031(31) & -0.336(21) &
-0.352(31) \\
\ensemble{B ~ hyp} & 5 & 4 & 10 & 0.293(24)  & 0.0924(41) & -0.396(33) &
-0.432(48) \\
\ensemble{C ~ hyp} & 4 & 4 & 10 & 0.374(15)  & 0.0649(33) & -0.479(19) &
-0.523(26) \\
\hline
\ensemble{A}       & 7 & 4 & 10 & 0.77(13)   & 0.097(45)  & -0.16(11)  &
-2.4(24)   \\
\ensemble{B}       & 5 & 4 & 10 & 0.81(16)   & 0.081(55)  & -0.30(17)  &
-1.0(33)   \\
\ensemble{C}       & 4 & 4 & 10 & 0.81(10)   & 0.074(36)  & -0.33(13)  &
-0.6(23)   \\
\hline
\ensemble{W ~ hyp} & 4 & 2 & 10 & -0.239(50) & 0.4715(99) & -0.026(64) &
-0.119(89) \\
\ensemble{X ~ hyp} & 4 & 2 & 10 & -0.129(40) & 0.4393(76) & -0.152(52) &
-0.278(74) \\
\ensemble{Y ~ hyp} & 4 & 2 & 10 & -0.144(47) & 0.4322(90) & -0.114(61) &
-0.205(86) \\
\ensemble{Z ~ hyp} & 4 & 2 & 10 & -0.128(43) & 0.4190(79) & -0.129(57) &
-0.233(82) \\
\hline
\ensemble{W}       & 4 & 2 & 10 & 0.65(21)   & 0.384(72)  & 0.05(18)   &
-4.9(38)   \\
\ensemble{X}       & 4 & 2 & 10 & 0.33(19)   & 0.487(67)  & -0.10(17)  &
-0.4(36)   \\
\ensemble{Y}       & 4 & 2 & 10 & -0.05(20)  & 0.613(69)  & -0.48(17)  &
7.4(36)    \\
\ensemble{Z}       & 4 & 2 & 10 &  0.11(20)  & 0.548(72)  & -0.39(16)  &
5.1(37)    \\
\hline
\ensemble{Q ~ hyp} & 7 & 4 & 10 & 0.169(12)  & 0.1043(22) & -0.261(15) &
-0.290(22) \\
\hline
\end{tabular}
\mycaption{Fit-range limits and the results for a subset of the fit parameters
from the corrected static-quark-potential fits.}
\label{c:table}
\end{table}

Tables \ref{c:table} and \ref{b:table} compile the results of the corrected
static-quark-potential fits.  Table \ref{c:table} presents the fit ranges used,
and displays each ensemble's determined values for a relevant subset of the fit
parameters.  Table \ref{b:table} presents the corresponding values for $r_0 /
a$ and the inverse lattice spacing.  The quoted uncertainties are from a
jackknife analysis of the statistical error.

In order to gauge the effects of using a corrected potential, we repeat the
fits using an uncorrected static quark potential, fixing the correction term's
parameter to zero, $\tilde{v}_2 = 0$.  The results of these uncorrected fits
are given in Table \ref{b:table}.

Note that between the thin-link ensembles \ensemble{W} through \ensemble{Z},
the results of the corrected fit show uncontrolled variation.  This is most
obvious in the values obtained for the correction term's parameter
$\tilde{v}_2$, and can be seen clearly in Figures \ref{h:figure} through
\ref{i:figure} and in Table \ref{c:table}.  This is the result of a poorly
determined $s$ dependence for the static quark potential, resulting in an
extremely flat minimum for $\chi^2$ along a path through fit-parameter space.
This leads to large, but highly correlated, variations in the fit parameters.
Because we ignore the correction term when determining $r_0 / a$, these
correlated errors do not have a chance to counteract one another.  The end
result for $r_0 / a$ is large statistical errors within each ensemble and
large variations in the value between ensembles.  For the corresponding
uncorrected static-quark-potential fits, the situation is improved twofold.
The ansatz for the potential contains fewer terms, and thus the resulting fit
parameters are better determined.  Additionally, all terms are retained when
determining $r_0 / a$, and thus any correlated error has an opportunity to
cancel in the final result for $r_0 / a$.  This improved situation is clear
from the relative consistency between the uncorrected results for the Sommer
scale of ensembles \ensemble{W} through \ensemble{Z}, as shown in Table
\ref{b:table}.

\begin{table}
\centering
\begin{tabular}{|l||d{1.7}d{1.7}|e{4.5}e{4.5}|}
\hline
& 
\multicolumn{2}{c|}{$r_0 / a$} &
\multicolumn{2}{c|}{$a^{-1} \, \text{(MeV)}$} \\
& \multicolumn{1}{c}{uncorr} & \multicolumn{1}{c|}{corr} &
\multicolumn{1}{c}{uncorr} & \multicolumn{1}{c|}{corr} \\
\hline
\hline
\ensemble{A ~ hyp} & 3.507(23)  & 3.570(27) & 1384,.1(90) & 1409,(10)   \\
\ensemble{B ~ hyp} & 3.581(68)  & 3.684(44) & 1414,(27)   & 1454,(17)   \\
\ensemble{C ~ hyp} & 3.757(28)  & 4.248(78) & 1483,(11)   & 1677,(31)   \\
\hline
\ensemble{A}       & 3.17(12)   & 3.9(10)   & 1251,(49)   & 1550,(400)  \\
\ensemble{B}       & 3.69(24)   & 4.1(17)   & 1457,(95)   & 1610,(670)  \\
\ensemble{C}       & 3.94(29)   & 4.2(12)   & 1560,(110)  & 1670,(480)  \\
\hline
\ensemble{W ~ hyp} & 1.8809(37) & 1.856(19) & 742,.4(15)  & 732,.5(73)  \\
\ensemble{X ~ hyp} & 1.9076(33) & 1.847(17) & 752,.9(13)  & 728,.9(69)  \\
\ensemble{Y ~ hyp} & 1.9293(35) & 1.885(19) & 761,.5(14)  & 744,.2(75)  \\
\ensemble{Z ~ hyp} & 1.9547(40) & 1.905(19) & 771,.5(16)  & 751,.9(75)  \\
\hline
\ensemble{W}       & 1.766(12)  & 2.10(30)  & 697,.2(47)  & 830,(120)   \\
\ensemble{X}       & 1.763(11)  & 1.78(21)  & 696,.0(44)  & 704,(84)    \\
\ensemble{Y}       & 1.7809(93) & 1.38(17)  & 702,.9(37)  & 545,(68)    \\
\ensemble{Z}       & 1.812(10)  & 1.51(19)  & 715,.1(41)  & 598,(76)    \\
\hline
\ensemble{Q ~ hyp} & 3.570(12)  & 3.650(19) & 1409,.2(47) & 1440,.5(76) \\
\hline
\end{tabular}
\mycaption{The final results for the Sommer scale and lattice spacing from the
corrected and uncorrected static-quark-potential fits.}
\label{b:table}
\end{table}

One advantage of hypercubic blocking, a reduction in statistical error, is
evident in our results for the Sommer scale.  As such, we have a greater trust
in our hypercubic-blocked results.  In following sections we use an ensemble's
hypercubic-blocked lattice spacing for both the thin-link and
hypercubic-blocked versions of the ensemble.

We suspect the variation in the determined lattice spacing for ensembles
\ensemble{A} through \ensemble{C} is due not to true changes in the lattice
spacing, but rather to our limited ability to determine the lattice spacing on
the smaller-volume ensembles.  Thus, in order to clarify the effects of finite
volume in other quantities, we use, in the following sections, the lattice
spacing determined for ensemble \ensemble{A ~ hyp} for ensembles \ensemble{B}
and \ensemble{C} as well.

\clearpage

\section{Effective Mass}
We can extract the value of the local pion mass and decay constant from the
bilinear correlator $C_{5, 5; t}$ because, for large time separations $t$, it
contains only a single state, the static local pion.  Thus, we must determine
some time separation $t_\text{min}$ beyond which we will assume the pion state
is uncontaminated.

In order to make this choice for each ensemble, we calculate the effective mass
of the local pion over a range of time separations.  The effective mass
$M_\text{eff}$ at $t + \half$ is the mass which describes the exponential
falloff of the correlator, considering only the correlator's value at the time
separations $t$ and $t + 1$.  In practice we determine the effective mass by
fitting the correlator at a single valence quark mass to the form in
\eqref{be:equation}, using the value of the correlator only at four time
separations: $t$, $t + 1$, $L_4 - 2 - t$, and $L_4 - 1 - t$.

The resulting values for the effective mass are presented in Figures
\ref{r:figure} through \ref{s:figure}.  The error bars shown are the result of
a jackknife analysis.  In each case the effective mass is high for small $t$,
where the correlator still contains higher-energy states, and then plateaus for
large $t$.  The value of $t_\text{min}$ chosen for each ensemble is
represented in the plots by a vertical dotted line.  Those values can also be
found in Table \ref{e:table}.  All other fits of the correlator in this study
use these values for $t_\text{min}$, fitting the correlator only over the range
from $t_\text{min}$ to $L_4 - 1 - t_\text{min}$.

\section{Pion Mass} \label{w:section}
Once we have determined an appropriate $t_\text{min}$, the local pion mass at a
given valence quark mass can be ascertained by fitting the bilinear correlator
$C_{5, 5; t}$ to its expected exponential falloff \eqref{be:equation} in $t$.
Table \ref{e:table} shows the results of these fits using the
valence-quark-mass values $m_V = 0.01$ and $m_V = m_S$.  For those ensembles in
which $m_S = 0.01$, the results are not repeated.  The uncertainties quoted are
the result of a jackknife analysis of the statistical error.  In the case of
the dimensionful result, the statistical error of the inverse lattice spacing
is added in quadrature.  We use a diagonal correlation matrix for these fits.

\begin{table}
\centering
\begin{tabular}{|l||c|d{1.8}e{3.4}|d{1.8}d{3.4}|}
\hline
& & \multicolumn{2}{c|}{$m_V = 0.01$} & \multicolumn{2}{c|}{$m_V = m_S$} \\
& $t_\text{min}$ & \multicolumn{1}{c}{$a M_{\pi_5}$} &
\multicolumn{1}{c|}{$M_{\pi_5} \, \text{(MeV)}$} & \multicolumn{1}{c}{$a
M_{\pi_5}$} & \multicolumn{1}{c|}{$M_{\pi_5} \, \text{(MeV)}$} \\
\hline
\hline
\ensemble{A ~ hyp} & 7 & 0.19769(57)  & 278,.5(21) & & \\
\ensemble{B ~ hyp} & 7 & 0.20235(93)  & 285,.1(24) & & \\
\ensemble{C ~ hyp} & 7 & 0.335(11)    & 471,(16)   & & \\
\hline
\ensemble{A}       & 7 & 0.29704(31)  & 418,.5(30) & & \\
\ensemble{B}       & 7 & 0.29792(37)  & 419,.8(30) & & \\
\ensemble{C}       & 7 & 0.3483(32)   & 490,.8(57) & & \\
\hline
\ensemble{W ~ hyp} & 4 & 0.25189(19)  & 184,.5(18) & 0.30659(19)  & 224.6(22) \\
\ensemble{X ~ hyp} & 4 & 0.25143(25)  & 183,.3(17) & 0.35134(24)  & 256.1(24) \\
\ensemble{Y ~ hyp} & 4 & 0.25098(23)  & 186,.8(19) & 0.39029(25)  & 290.5(29) \\
\ensemble{Z ~ hyp} & 4 & 0.25091(26)  & 188,.7(19) & 0.45804(26)  & 344.4(34) \\
\hline
\ensemble{W}       & 4 & 0.258565(96) & 189,.4(19) & 0.315645(98) & 231.2(23) \\
\ensemble{X}       & 4 & 0.25955(11)  & 189,.2(18) & 0.36470(11)  & 265.8(25) \\
\ensemble{Y}       & 4 & 0.260230(99) & 193,.7(20) & 0.40742(11)  & 303.2(31) \\
\ensemble{Z}       & 4 & 0.26140(12)  & 196,.5(20) & 0.48119(12)  & 361.8(36) \\
\hline
\ensemble{Q ~ hyp} & 7 & 0.19472(70)  & 280,.5(18) & \multicolumn{1}{c}{-} &
\multicolumn{1}{c|}{-} \\
\hline
\end{tabular}
\mycaption{Minimum time separation and local pion mass at $m_V = 0.01$ and $m_V
= m_S$.  For those ensembles in which $m_S = 0.01$, the results are not
repeated.}
\label{e:table}
\end{table}

Ensembles \ensemble{A} and \ensemble{B} generate very similar pion masses,
indicating that the local pion mass is relatively free of finite-volume
effects.  The pion mass of ensemble \ensemble{C} is significantly larger.  This
indicates that the small volume of the ensemble is constricting the pion,
preventing it from relaxing to its lowest energy state.

As discussed in Section \ref{v:section}, pqChPT is only valid for quark masses
within some radius of convergence from the chiral limit.  In order to gain some
insight into the physical magnitude of our quark masses, we compare our
calculated pion masses to the experimental mass of the the Standard Model's
light mesons.  The experimental mass of the Standard Model's pion is
$M_{\pi^0} = 135.0 \, \text{MeV}$, while the kaon mass is $M_{K^+} = 493.7 \,
\text{MeV}$ \cite{PDBook}.  From Table \ref{e:table}, it is clear that the
quark masses we use are larger than those of the up and down quarks.  Yet, the
ensembles' pion masses remain below the kaon mass.  In the case of ensemble
\ensemble{A ~ hyp}, the pion mass is well below the kaon mass.  This leaves us
hopeful that pqChPT is valid within the quark-mass range of our study.  It
should be noted that our coarse lattice spacings, especially in ensembles
\ensemble{W} through \ensemble{Z}, are a large part of why our pion masses
remain low.

\clearpage

\section{Kaon Quark-Mass Threshold} \label{ad:section}
For staggered lattice fermions the quark mass is multiplicatively renormalized.
While we do not wish to calculate the renormalization factors exactly, we do
require some method for comparing the quark mass between ensembles.  Thus, for
each ensemble we calculate what we identify as the kaon quark-mass threshold
$m_{Q_K}$, the valence quark mass at which the local pion mass equals the
Standard Model's kaon mass.  At that point, the physical valence quark mass
should be on the order of half the strange quark mass.

In order to determine $m_{Q_K}$, we require an ansatz for the pion mass's
valence-quark-mass dependence which is accurate up to large valence quark mass.
While pqChPT predicts a form for the pion mass, it is not appropriate to use it
in this case, as the fit includes valence quark masses which are beyond the
expected radius of convergence of pqChPT.  Instead, we simply fit the squared
pion mass to a quadratic polynomial, a form which it follows very well up to
the largest valence quark mass we investigate.

The phenomenological fit form we use for this purpose alone is:
\begin{equation}
C_{5, 5;t} = \mathscr{A}_{m_V} \bigl( e^{-\uM_{\pi_5} t} + e^{-\uM_{\pi_5} (
L_4 - 1 - t) } \bigr)
\label{bw:equation}
\end{equation}
where:
\begin{equation}
\uM^2_{\pi_5} = a_0 + a_1 m_V + a_2 m^2_V
\label{bv:equation}
\end{equation}
and $\mathscr{A}_{m_V}$ is a set of independent fit parameters, one for each
valence-quark-mass value studied.  We should stress that this is clearly not
the form predicted by pqChPT.  We use it here only because we require a good
fit of the local pion mass up to large valence quark mass.  The results of
these fits are used for nothing other than an interpolation to determine the
point at which the local pion mass crosses the Standard Model's physical kaon
mass:
\begin{equation}
m_{Q_K} = \frac{-a_1 + \sqrt{a_1^2 - 4 a_2 \bigl( a_0 - ( a \times 493.7 \,
\text{MeV})^2 \bigr)}}{2 a_2}
\end{equation}
where $a$ is the lattice spacing used for the ensemble.

\begin{table}
\centering
\begin{tabular}{|l||d{1.9}d{1.7}d{3.6}|d{1.9}|}
\hline
& \multicolumn{1}{c}{$a_0$} & \multicolumn{1}{c}{$a_1$} &
\multicolumn{1}{c|}{$a_2$} & \multicolumn{1}{c|}{$m_{Q_K}$} \\
\hline
\hline
\ensemble{A ~ hyp} & 0.00223(11)  & 3.684(20)  & -1.99(22)  & 0.03331(13)  \\
\ensemble{B ~ hyp} & 0.00401(29)  & 3.694(37)  & -2.15(54)  & 0.03276(20)  \\
\ensemble{C ~ hyp} & 0.0671(76)   & 4.59(21)   & -10.1(25)  & 0.0125(17)   \\
\hline
\ensemble{A}       & 0.00435(18)  & 8.558(15)  & -21.99(24) & 0.014365(21) \\
\ensemble{B}       & 0.00503(20)  & 8.547(15)  & -21.85(24) & 0.014295(27) \\
\ensemble{C}       & 0.0359(25)   & 8.744(49)  & -27.49(52) & 0.01027(27)  \\
\hline
\ensemble{W ~ hyp} & 0.00378(12)  & 6.0518(72) & -2.778(47) & 0.077151(55) \\
\ensemble{X ~ hyp} & 0.00395(13)  & 6.0084(71) & -2.654(45) & 0.078381(63) \\
\ensemble{Y ~ hyp} & 0.00406(12)  & 5.9758(73) & -2.617(49) & 0.075492(63) \\
\ensemble{Z ~ hyp} & 0.00434(15)  & 5.9523(81) & -2.669(51) & 0.074160(64) \\
\hline
\ensemble{W}       & 0.001263(59) & 6.6238(40) & -4.911(28) & 0.072242(28) \\
\ensemble{X}       & 0.001388(55) & 6.6679(40) & -5.100(30) & 0.072598(25) \\
\ensemble{Y}       & 0.001509(56) & 6.6935(40) & -5.204(28) & 0.069284(26) \\
\ensemble{Z}       & 0.001612(72) & 6.7502(43) & -5.456(29) & 0.067282(27) \\
\hline
\ensemble{Q ~ hyp} & 0.00269(22)  & 3.524(21)  & -2.25(27)  & 0.03327(12)  \\
\hline
\end{tabular}
\mycaption{Kaon quark-mass threshold and results for a subset of the fit
parameters from the quadratic fits of the pion mass's valence-quark-mass
dependence.}
\label{f:table}
\end{table}

Figures \ref{p:figure} through \ref{q:figure} display, and Table \ref{f:table}
summarizes, the results of these fits.  The curve in each plot demonstrates the
valence-quark-mass dependence determined by the fit, and is generated by using
\eqref{bv:equation} and the resultant values for the fit parameters.  The
diamonds are the result of independent fits of the correlator at separate
valence quark masses, using the same method as was used in Section
\ref{w:section}.  It should be stressed that the full quadratic fit of the
valence-quark-mass dependence is not a fit to these points.  The points are
related to the fit only in that they are both derived from the same correlator
data.  Agreement between the fit curve and the diamonds demonstrates that our
phenomenological ansatz for the local pion mass's valence-quark-mass dependence
is appropriate.  The valence-quark-mass values at which pion-mass points are
given correspond to the values used in the quadratic fit.

The error bars were determined via a jackknife analysis.   Each fit curve's
one-sigma range is bound by dotted lines.  This range is the result of a
jackknife analysis of the curve's value at each point along the horizontal
axis.  In most cases this region is thinner than the curve's line, and thus is
not easily visible.  Uncertainty in the lattice spacing is not taken into
account.  Instead, the plots' vertical axes are simply rescaled using the
lattice spacing's central value.  We use a diagonal correlation matrix in the
fit, as the full correlation matrix is not positive definite for several of the
ensembles.  

The dashed vertical line appearing in the plots corresponds to that ensemble's
determined value for $m_{Q_K}$, while the shaded region corresponds to the
uncertainty in that result, as determined by a jackknife analysis of the
statistical error.  In many cases this region is thiner than the line itself,
and is not readily visible.  This error, which is repeated in Table
\ref{f:table}, does not take into account uncertainty in the ensembles' lattice
spacings.

Because finite-volume effects inflate the pion mass in ensembles \ensemble{C ~
hyp} and \ensemble{C}, the kaon quark-mass threshold for these ensembles is
unexpectedly small.  Thus, in order to clarify finite-volume effects in
other quantities, we use the value of $m_{Q_K}$ determined for \ensemble{A ~
hyp} for both ensembles \ensemble{B ~ hyp} and \ensemble{C ~ hyp}.  Similarly,
\ensemble{A}'s value for $m_{Q_K}$ is used for ensembles \ensemble{B} and
\ensemble{C}.

\section{$2 \alpha_8 - \alpha_5$} \label{ae:section}
In order to determine the value of the GL coefficient combination $2 \alpha_8 -
\alpha_5$, we fit the correlator $C_{5, 5; t}$ over a range of time
separations $t$ and valence quark masses $m_V$ to the form predicted by pqChPT:
\eqref{bm:equation}, \eqref{bn:equation}, and \eqref{bo:equation}.  The result
is values for the fit's free parameters, one of which is $2 \alpha_8 -
\alpha_5$.

\subsection{Valence-Quark-Mass Cutoff}
pqChPT is expected to accurately model the behavior of the local pion only
within some radius of convergence of the chiral limit.  Thus, we should only
fit the correlator to pqChPT's predictions at and below some cutoff in the
valence quark mass $\Lambda_{m_V}$.  While we expect the appropriate cutoff to
be below the kaon quark-mass threshold, there is no clear a priori value.  As
such, we turn to the data to determine our cutoff.

Figure \ref{t:figure} plots $\chi^2$ per degree of freedom for fits of
ensemble \ensemble{A ~ hyp}'s correlator data to the predictions of pqChPT
against various choices for the cutoff $\Lambda_{m_V}$.  A diagonal correlation
matrix was used in these fits.  The error bars are the result of a jackknife
analysis.  We choose as our value for $\Lambda_{m_V}$ the cutoff at which
$\chi^2$ per degree of freedom is closest to one:  $\Lambda_{m_V} = 0.025$.  In
Figure \ref{t:figure} the chosen cutoff is denoted by a filled diamond.  This
cutoff is then used for ensembles \ensemble{A ~ hyp} through \ensemble{C ~
hyp}, as well as for ensemble \ensemble{Q ~ hyp}.

We know from our study of the kaon quark-mass threshold $m_{Q_K}$ that the
quark-mass renormalization factor of the thin-link ensembles is significantly
different than that of the hypercubic-blocked ensembles.  Thus, it would not be
appropriate to use the same valence-quark-mass cutoff for the thin-link
ensembles.  Instead we determine a cutoff which is consistent between the
thin-link and hypercubic-blocked ensembles using $m_{Q_K}$.

The ratio of the valence-quark-mass cutoff over the kaon quark-mass threshold
for ensemble \ensemble{A ~ hyp} is $\Lambda_{m_V} / m_{Q_K} = 0.751$.  We
choose the cutoff for ensemble \ensemble{A} to be that which gives the closest
ratio beyond this value.  We choose $\Lambda_{m_V} = 0.0125$, which corresponds
to a ratio of $\Lambda_{m_V} / m_{Q_K} = 0.870$.  This cutoff is then used for
ensembles \ensemble{A} through \ensemble{C}.  We select the cutoff for
ensembles \ensemble{W ~ hyp} through \ensemble{Z ~ hyp} similarly, choosing
$\Lambda_{m_V} = 0.06$.  Between the four ensembles, this gives an average
ratio of $\Lambda_{m_V} / \avg{m}_{Q_K} = 0.0763$.  For ensembles \ensemble{W}
through \ensemble{Z}, we use the cutoff $\Lambda_{m_V} = 0.055$, which between
the four ensembles gives an average ratio of $\Lambda_{m_V} / \avg{m}_{Q_K} =
0.782$.

Given the evident correlation present in the data, and that our calculations of
$\chi^2$ do not take that correlation into account, it could be argued that,
while clearly minimization of this $\chi^2$ is appropriate, its actual value is
meaningless.  As such, there is no compelling reason to believe that a choice
of valence-quark-mass cutoff which results in a $\chi^2$ per degree of freedom
of one is appropriate.  However, as we are left with no other quantitative
evidence as to an appropriate cutoff choice, we feel that it is important to at
least use a consistent and systematic method for resolving the choice.

In section \ref{y:section} we study the effect of these cutoff choices on the
resulting value of $2 \alpha_8 - \alpha_5$ and find that it is a significant
source of systematic uncertainty.

\subsection{Pion Mass and Decay Constant} \label{z:section}
Figures \ref{u:figure} and \ref{v:figure} present results from the fit of
ensemble \ensemble{A ~ hyp}'s correlator to the predictions of pqChPT.  These
plots of the dependence of the pion mass and decay constant on the
valence quark mass were generated by inserting the fit's determined values for
its free parameters into \eqref{bn:equation} and \eqref{bo:equation}.  Recall
that we use the pion-decay-constant normalization in which $f_\pi \simeq 92.4
\, \text{MeV}$.

The diamonds display the result of a set of independent fits of the correlator
at separate valence quark masses, using the same method as was used in Section
\ref{w:section}.  The result is a value for the pion mass and decay constant at
each valence quark mass at which the correlator was calculated.  The
valence-quark-mass values at which a filled diamond appears correspond to the
set of values used in the full pqChPT fit; that is, those within the
valence-quark-mass cutoff.  Open diamonds correspond to valence-quark-mass
values beyond the cutoff.  We stress that the full pqChPT fit is not a fit to
the filled diamonds.  Rather, the fit curve and points are related only in that
they are derived from the same correlator data.  Their agreement, or lack
thereof, demonstrates the correlator data's tendency to match the predictions
of pqChPT.  

Note that while two plots are used to present the results of the pqChPT fit,
they are both the product of a single fit, encompassing the valence-quark-mass
dependence of both the pion mass and decay constant, as well as the
correlator's time dependence.

Error bars were determined via a jackknife analysis of the statistical error.
The fit curve's one-sigma range appears in the plots bound by dotted lines.
This range was determined by individual jackknife analyses of the curve's value
at each point along the horizontal axis.  Uncertainty in the lattice spacing
and kaon quark-mass threshold are not taken into account.  Instead, the plots'
axes have simply been rescaled using their central value,  so as not to obscure
statistical error in the quantities of interest.

A diagonal correlation matrix was used in the fit, as the full matrix proved to
not be positive definite in all cases.

Because the statistical error in Figure \ref{u:figure} is too small to be
visible, we have included, inset in the plot, close-ups at three values of the
valence quark mass.  These inset plots do not share a single scale.  However,
they do have a horizontal-to-vertical scale ratio which matches that of the
main plot.

The fit's determined values for its free parameters appear in Table
\ref{g:table}.

\subsection{$R_M$} \label{aa:section}
As is clear from Figure \ref{u:figure}, the dependence of the squared pion mass
on the valence quark mass is very nearly linear.  In such plots this strong
linearity obscures the presence of higher-order effects.  Thus, in order to
accentuate non-linear terms, we include plots of a ratio first suggested in
\cite{Heitger:2000ay}:
\begin{equation}
R_M \equiv \frac{m_V M^2_{\pi_5} ( m_S )}{m_S M^2_{\pi_5} ( m_V )}
\label{bx:equation}
\end{equation}
where $M_{\pi_5} ( m_Q )$ represents the mass of the local pion containing
valence quarks of mass $m_Q$.

The ratio $R_M$ is designed to allow a visual assessment of the strength and
nature of the squared pion mass's non-linearity.  Were the dependence of
$M^2_{\pi_5}$ on $m_V$ linear, with a massless pion in the chiral limit, a plot
of $R_M$ would be flat at $R_M = 1$.  If the dependence were quadratic with no
constant term, and did not include any higher-order or non-polynomial terms, an
$R_M$ plot would be linear.  In both cases a small non-zero pion mass at the
chiral limit introduces a sharp downturn in $R_M$ for small $m_V$.  For larger
$m_V$ the behavior described above is unaffected.

Taking the points and curve from Figure \ref{u:figure} and transforming them
based on the definition of $R_M$ results in Figure \ref{w:figure}.

In order not to cloud the statistical error of the fit results, a jackknife
analysis of the quantity $R_M$ has not been performed.  As such, correlation
between $M^2_{\pi_5}$ at a given valence quark mass and at the dynamical quark
mass is not accounted for, nor is uncertainty in $M^2_{\pi_5}$ at the
dynamical quark mass.  Instead, the plot has simply been transformed using the
central value for $M^2_{\pi_5} ( m_S )$ as determined in Section
\ref{w:section}.  We feel that this gives a more appropriate representation of
the statistical error present in the pqChPT fit.  The goal of presenting an
$R_M$ plot is not to accurately determine $R_M$ and its uncertainty, but rather
to compare the non-linearity of the data and the resulting fit curve.  If we
were to account for correlation with and uncertainty in $M^2_{\pi_5} ( m_S )$,
it would obfuscate the points' error bars and the fit curve's one-sigma range.
Furthermore, because the independent points and the fit curve do not
necessarily agree on the value of $M^2_{\pi_5} ( m_S )$, a jackknife analysis
would generate a misleading relative shift between their values for $R_M$.

Note that we use $R_M$ for plotting purposes only.  For the pqChPT fit of the
correlator data, the expressions for pion mass and decay constant,
\eqref{bn:equation} and \eqref{bo:equation}, are used directly.  Additionally,
we do not use the simplification of $R_M$ suggested in \cite{Heitger:2000ay}.

\subsection{Results} \label{af:section}
Figures \ref{y:figure} through \ref{z:figure} display the results of each
ensemble's pqChPT fit.  For each ensemble, three plots are presented:  an $R_M$
plot, a pion-mass plot, and a pion-decay-constant plot.  The creation and
presentation of these plots mirrors Figures \ref{u:figure} through
\ref{w:figure}, as discussed in Sections \ref{z:section} and \ref{aa:section}.
For completeness, the plots for ensemble \ensemble{A ~ hyp} are repeated.  The
values determined for the fits' free parameters are compiled in Table
\ref{g:table}.  

\begin{table}
\centering
\begin{tabular}{|l||d{1.4}|d{1.7}d{3.7}d{1.8}d{2.6}d{2.7}|}
\hline
\raisebox{0pt}[0.5cm][0cm]{} & 
\multicolumn{1}{c|}{$\Lambda_{m_V}$} &
\multicolumn{1}{c}{$a_0$} &
\multicolumn{1}{c}{$\uz$} &
\multicolumn{1}{c}{$\uf$} &
\multicolumn{1}{c}{$\alpha_5$} &
\multicolumn{1}{c|}{$2 \alpha_8 - \alpha_5$} \\
\hline
\hline
\ensemble{A ~ hyp} & 0.025  & \multicolumn{1}{c}{-} & 9.35(17)   & 0.05237(35)
& 0.240(33)  & 0.275(17)  \\
\ensemble{B ~ hyp} & 0.025  & \multicolumn{1}{c}{-} & 11.95(74)  & 0.0471(11)
& 0.857(58)  & 0.211(53)  \\
\ensemble{C ~ hyp} & 0.025  & 0.0979(77)            & 149.1(85)  & 0.00693(49)
& 5.02(14)   & -0.28(11)  \\
\hline
\ensemble{A}       & 0.0125 & \multicolumn{1}{c}{-} & 9.60(15)   & 0.07778(50)
& 2.065(20)  & 0.360(17)  \\
\ensemble{B}       & 0.0125 & \multicolumn{1}{c}{-} & 10.17(32)  & 0.0758(11)
& 2.187(49)  & 0.356(41)  \\
\ensemble{C}       & 0.0125 & 0.0411(41)            & 43.7(20)   & 0.0317(14)
& 4.82(25)   & 0.55(13)   \\
\hline
\ensemble{W ~ hyp} & 0.06   & \multicolumn{1}{c}{-} & 2.867(15)  & 0.12036(26)
& -0.298(15) & 0.2472(66) \\
\ensemble{X ~ hyp} & 0.06   & \multicolumn{1}{c}{-} & 3.104(19)  & 0.11481(29)
& -0.066(21) & 0.2749(82) \\
\ensemble{Y ~ hyp} & 0.06   & \multicolumn{1}{c}{-} & 3.281(23)  & 0.11086(31)
& 0.111(16)  & 0.3004(90) \\
\ensemble{Z ~ hyp} & 0.06   & \multicolumn{1}{c}{-} & 3.778(29)  & 0.10221(32)
& 0.491(17)  & 0.3271(89) \\
\hline
\ensemble{W}       & 0.055  & \multicolumn{1}{c}{-} & 0.7649(20) & 0.23635(26)
& -0.527(26) & 0.278(14)  \\
\ensemble{X}       & 0.055  & \multicolumn{1}{c}{-} & 0.8069(23) & 0.23049(27)
& -0.351(27) & 0.304(13)  \\
\ensemble{Y}       & 0.055  & \multicolumn{1}{c}{-} & 0.8430(26) & 0.22548(28)
& -0.089(23) & 0.348(11)  \\
\ensemble{Z}       & 0.055  & \multicolumn{1}{c}{-} & 0.9176(29) & 0.21596(28)
& 0.254(25)  & 0.4385(99) \\
\hline
\ensemble{Q ~ hyp} & 0.025  & \multicolumn{1}{c}{-} & 9.05(21)  & 0.05250(46)
& -0.014(53) & 0.223(30) \\
\hline
\end{tabular}
\mycaption{Results from the pqChPT correlator fits and corresponding
valence-quark-mass cutoff values.}
\label{g:table}
\end{table}

Comparison of the pqChPT-predicted curve to the individual points in each
ensemble's $R_M$ plot demonstrates that the correlator data is systematically
missing the predictions of pqChPT.  This is most evident in the results for
ensembles \ensemble{W ~ hyp} through \ensemble{Z ~ hyp}, and their thin-link
counterparts.  It is telling that these are the same ensembles which have the
coarsest lattice spacing, and thus the largest expected flavor-symmetry
breaking.  It is likely that a majority of the data's systematic deviation from
the predictions of pqChPT is due to our failure to account for the effects of
the staggered formulation's inherent flavor symmetry breaking.  As will be
discussed in Section \ref{am:section}, any future work in this area will
require a more robust handling of flavor-symmetry-breaking effects.

Our results for the GL coefficient combination $2 \alpha_8 - \alpha_5$ are
relatively stable between ensembles.  Comparing ensembles \ensemble{A ~ hyp},
\ensemble{A}, \ensemble{B ~ hyp}, and \ensemble{W ~ hyp} demonstrates that,
while the corresponding systematic effects are strong, they are not beyond
control.  A detailed analysis of the systematic error present in our
calculation of $2 \alpha_8 - \alpha_5$ is presented in Section
\ref{ah:section}.

Results for $\alpha_5$, which controls the polynomial NLO term in the pion
decay constant, are much less consistent.  This is interesting when we note
that the data generally follow pqChPT's predictions for the form of the pion
decay constant better than its predictions for the form of the pion mass.

The determined values for $2 \alpha_8 - \alpha_5$ from ensembles \ensemble{W ~
hyp} through \ensemble{Z ~ hyp} show a definite trend, demonstrating an
apparent dependence on the dynamical quark mass.  This trend can be explained
if we recall the $m_S$-dependent term which was dropped between the true
predictions of pqChPT \eqref{bq:equation} and the form used in our
single-dynamical-quark-mass fits \eqref{bn:equation}.  This term is
accounted for in our simultaneous fit over these four ensembles, which is
presented in Section \ref{x:section}.

\subsection{Polynomial Fits}
For comparison we include the $R_M$ plot for both a quadratic and a cubic
polynomial fit of the squared pion mass of ensemble \ensemble{A ~ hyp}.  These
can be found in Figures \ref{ab:figure} and \ref{ac:figure}.  The quadratic fit
is taken directly from Section \ref{ad:section}, while the cubic fit uses the
same methods as Section \ref{ad:section}, replacing the fit's form for the
squared pion mass \eqref{bv:equation} with:
\begin{equation}
\uM^2_{\pi_5} = a_0 + a_1 m_V + a_2 m^2_V + a_3 m^3_V
\label{by:equation}
\end{equation}
The resulting values for the fit parameters of the cubic fit are given in Table
\ref{h:table}.  The agreement between the independent mass points and the fit
curve, even in the quadratic case, is strikingly good, especially when compared
to the results of the corresponding pqChPT fit, Figure \ref{w:figure}.

\begin{table}
\centering
\begin{tabular}{|l||d{1.9}d{1.6}d{2.5}d{2.4}|}
\hline
& \multicolumn{1}{c}{$a_0$} & \multicolumn{1}{c}{$a_1$} &
\multicolumn{1}{c}{$a_2$} & \multicolumn{1}{c|}{$a_3$} \\
\hline
\hline
\ensemble{A ~ hyp} & 0.001576(82) & 3.815(28) & -8.12(81) & 78.0(81) \\
\hline
\end{tabular}
\mycaption{Results for a subset of the fit parameters from the cubic fit of the
pion mass's valence-quark-mass dependence.}
\label{h:table}
\end{table}

\subsection{Finite Volume}
The fit results of ensembles \ensemble{C ~ hyp} and \ensemble{C} are given
last, as their analyses required special attention.  The small volume of these
ensembles significantly affects their correlator data, a fact which is made
most clear by the large non-zero value obtained for the constant term $a_0$ in
the calculation of their kaon quark-mass thresholds.  The reader is directed to
Table \ref{f:table} and Figure \ref{p:figure}.  These finite-volume effects
overwhelm the forms predicted by infinite-volume pqChPT, such that attempts to
fit these ensembles' correlator data to pqChPT's forms either fail completely
or generate nonsensical results.

In order to produce an even moderately reasonable fit, we added a constant
term $a_0$ to pqChPT's predictions for the form of the squared pion mass.
Thus, in these fits, the standard pqChPT form \eqref{bn:equation} was replaced
with:
\begin{align}
\uM^2_{\pi_5} & = a_0 + \uz \mV (4 \pi \uf)^2 \biggl\{ 1 + \frac{\uz}{N_f}
\bigl( 2 \mV - \mS \bigr) \ln \uz \mV + \frac{\uz}{N_f} \bigl( \mV - \mS \bigr)
\notag
\\
& ~ \phantom{= a_0 + \uz \mV (4 \pi \uf)^2 \biggl\{ 1} + \uz \mV \bigl( 2
\alpha_8 - \alpha_5 \bigr) \biggr\}
\label{bz:equation}
\end{align}
where $a_0$ is an additional fit parameter.  Comparing Table \ref{f:table} and
\ref{g:table}, it is interesting to note the similarity between the values
obtained for the constant term $a_0$ in the quadratic and pqChPT-predicted
forms.

Even after the addition of a constant term, the results of the fits, Figures
\ref{ad:figure} and \ref{ae:figure}, remain questionable, especially in the
case of ensemble \ensemble{C ~ hyp}.  Luckily, our study of these ensembles has
no purpose other than to elucidate the effects of finite volume.

\subsection{Dependence on $\Lambda_{m_V}$} \label{y:section}
In order to study the dependence of our results for $2 \alpha_8 - \alpha_5$ on
the choice of valence-quark-mass cutoff $\Lambda_{m_V}$, the pqChPT fits were
repeated using a range of cutoff values.  The results of this investigation are
presented in Figures \ref{af:figure} through \ref{ag:figure}.  Figure
\ref{af:figure} displays the $\Lambda_{m_V}$ dependence of all four parameters
of the fit of ensemble \ensemble{A ~ hyp}, while Figures \ref{ah:figure}
through \ref{ag:figure} display only the dependence of the parameter $2
\alpha_8 - \alpha_5$ for all ensembles.

These plots clearly demonstrate that the cutoff choice is a significant source
of systematic error.  The strong dependence of $2 \alpha_8 - \alpha_5$ on
$\Lambda_{m_V}$ is a direct result of the failure of the theoretical forms to
closely match the data.  As the mass cutoff is increased, and additional
valence-quark-mass values are added to the fit range, the additional correlator
data consistently fails to match the values predicted for it by previous
lower-cutoff fits.  As such, each addition of correlator data significantly
alters the fit results.

We expect this sort of behavior for large cutoffs which fall beyond the range
of pqChPT.  However, below some threshold value for $\Lambda_{m_V}$, we expect
the value of $2 \alpha_8 - \alpha_5$ to level off.  This threshold would
indicate the outer limit of pqChPT's domain.  Yet, in our data we see no clear
plateau.  Instead, we observe a strong dependence on $\Lambda_{m_V}$ down to
the smallest valence quark masses studied.

Two possible reasons for this behavior are readily available.  First, it is
possible that the valence quark masses under study are too large for pqChPT to
generate accurate predictions.  That is, our entire study lies beyond the
threshold.  However, as the mass of our local pion reaches values well below
the physical kaon mass, we feel that this possibility is unlikely.  Second, the
behavior of our correlator data may be significantly skewed by
flavor-symmetry-breaking effects, rendering the predictions of continuum pqChPT
inaccurate.

While the choice of valence-quark-mass cutoff is clearly a significant source
of systematic error, the variation of $2 \alpha_8 - \alpha_5$ is not so great
as to render our results meaningless.  In Section \ref{ah:section} we
incorporate this quantitative analysis of this source of systematic error into
an estimate for the full systematic uncertainty of our result.

While the utility of an analysis of the $\Lambda_{m_V}$ dependence of the
results of ensembles \ensemble{C ~ hyp} and \ensemble{C} is not clear, it has
been included for completeness.  Because, for these ensembles, we are using a
pion mass form with three free parameters, the smallest cutoff which generates
sensible results is one which leaves us with three valence-quark-mass values
within the cutoff.  As such, the range of $\Lambda_{m_V}$ studied for these
ensembles does not reach as low as for other ensembles.

\section{$2 \alpha_6 - \alpha_4$} \label{x:section}
Four of our ensembles, ensembles \ensemble{W} through \ensemble{Z}, are
constructed to have the same lattice spacing and volume, such that the only
variation between them is their dynamical quark mass.  Through these ensembles
we are granted the opportunity to fit lattice data to the predictions of pqChPT
over a block of pqQCD's two-dimensional quark-mass plane.

We fit the correlator $C_{5, 5; t}$ over a range of time separations $t$,
valence quark masses $m_V$, and dynamical quark masses $m_S$ to the form
predicted by pqChPT: \eqref{bm:equation}, \eqref{bq:equation}, and
\eqref{br:equation}.  Because we are working over a range of dynamical quark
masses, we do not need to drop the unknown $m_S$ dependence from these forms,
as was done in Section \ref{ae:section}.  The result is values for the fit's
free parameters, which include the GL coefficient combinations $2 \alpha_8 -
\alpha_5$, $2 \alpha_6 - \alpha_4$, $\alpha_5$, and $\alpha_4$.

\begin{table}
\centering
\begin{tabular}{|l||d{1.7}d{1.8}d{1.7}d{2.6}d{2.7}d{2.6}|}
\hline
\raisebox{0pt}[0.5cm][0cm]{} & 
\multicolumn{1}{c}{$\uz$} &
\multicolumn{1}{c}{$\uf$} &
\multicolumn{1}{c}{$2 \alpha_8 - \alpha_5$} &
\multicolumn{1}{c}{$\alpha_5$} &
\multicolumn{1}{c}{$2 \alpha_6 - \alpha_4$} &
\multicolumn{1}{c|}{$\alpha_4$} \\
\hline
\hline
\ensemble{hyp}  & 2.500(16)  & 0.12993(41) & 0.3860(63) & -0.119(19) &
-0.2703(74) & -1.446(33) \\
\ensemble{thin} & 0.6970(26) & 0.24780(39) & 0.2292(88) & -0.093(29) &
0.006(14)   & -3.207(59) \\
\hline
\end{tabular}
\mycaption{Results from the simultaneous pqChPT fits of the correlators of
ensemble sets \ensemble{hyp}, which includes ensembles \ensemble{W ~ hyp}
through \ensemble{Z ~ hyp}, and \ensemble{thin}, which includes ensembles
\ensemble{W} through \ensemble{Z}.}
\label{i:table}
\end{table}

The fit was carried out twice, once for the hypercubic-blocked ensemble set,
and once for the corresponding thin-link set.  The resulting fit parameter
values can be found in Table \ref{i:table}, with the ensemble set
\ensemble{hyp} containing ensembles \ensemble{W ~ hyp} through \ensemble{Z ~
hyp} and the ensemble set \ensemble{thin} containing ensembles \ensemble{W}
through \ensemble{Z}.  The corresponding fit curves are displayed in Figures
\ref{ai:figure} through \ref{aj:figure}.  These curves are generated using the
fits' resulting parameter values in equations \eqref{bq:equation} and
\eqref{br:equation} and constructing the $R_M$ curves as described in Section
\ref{aa:section}.  Each plot includes four cross sections through the
quark-mass plane, each along a different line of constant dynamical quark mass.
Note that while eight fit curves are shown for each ensemble set, four in each
plot, they together represent the results of a single fit.

The plots' diamonds represent the results of a set of independent fits of the
correlator data at separate valence- and dynamical-quark-mass values, using the
same method as was used in Section \ref{w:section}.  From this we obtain a
value for the pion mass and decay constant at each valence and dynamical quark
mass at which the correlator was calculated.  The quark-mass values at which a
filled diamond appears correspond to the set of values used in the full pqChPT
fit.  Open diamonds correspond to quark-mass values  beyond the
valence-quark-mass cutoff.  The full pqChPT fit of an ensemble set is not a fit
to the filled diamonds.  Rather, the fit curve and diamonds are related only in
that they are derived from the same correlator data.  Their agreement, or lack
thereof, demonstrates the correlator's tendency to match the predictions of
pqChPT.

Error bars were determined via a jackknife analysis of the statistical error.
The fit curves' one-sigma range appears in the plots bound by dotted lines.
This range was determined by individual jackknife analyses at each point along
the horizontal axis.  Uncertainty in the lattice spacing and kaon quark-mass
threshold are not taken into account.  Instead, the plots' axes have simply
been rescaled using the average central value between the ensembles.

A diagonal correlation matrix was used in the fit, as the large number of
correlator values generated a matrix which was far from positive definite.

For illustrative purposes we have compiled the results of the independent pqChPT
fits of ensembles \ensemble{W ~ hyp} through \ensemble{Z ~ hyp} and
\ensemble{W} through \ensemble{Z} from Section \ref{ae:section} into plots
whose formats mirror Figures \ref{ai:figure} through \ref{aj:figure}.  These
compiled plots of the independent fits can be found in Figures \ref{ap:figure}
through \ref{aq:figure}.  Comparison between these plots makes clear the
difference in results between the independent and simultaneous correlator fits.

We also present the results of the simultaneous fits using a cross section
through the quark-mass plane along the unquenched line, $m_Q \equiv m_V = m_S$.
The plots are found in Figures \ref{ak:figure} and \ref{al:figure}.  Other
than the choice of cross section, these plots were generated in the same
fashion as described above.  Because the definition of $R_M$ is meaningful only
along lines of constant dynamical quark mass, we introduce a new ratio:
\begin{equation}
R'_M \equiv \frac{m_Q M^2_{\pi_5} ( m_R )}{m_R M^2_{\pi_5} ( m_Q )}
\end{equation}
For our plots we use a reference quark mass of $m_R = 0.025$.

From all the plots presented, we can clearly see that the correlator data
systematically misses the predictions of pqChPT.  As discussed in Section
\ref{af:section}, this is most likely due to strong flavor-symmetry-breaking
effects, a consequence of these ensembles' coarse lattice spacing.  Because of
the computational expense involved in the generation of a set of four
reasonably long partially quenched Markov chains, we were forced to use small
lattice extents.  Thus, a very coarse lattice spacing was required in order to
maintain a reasonable lattice volume.  The spacing is coarser than what is
generally deemed acceptable by the community.  As such, this calculation of $2
\alpha_6 - \alpha_4$ can only be taken as a preliminary study.  Yet, as ours is
the first attempt at such a calculation, such a preliminary study is not
without value.

The preliminary nature of this study is further emphasized by the fact that
only a single ensemble set is available to us.  As such, we do not have the
ability to generate estimates for the magnitude of the various systematic
errors we know are present in our result.

\subsection{Dependence on $\Lambda_{m_V}$}
In order to study the dependence of our results on the choice of
valence-quark-mass cutoff $\Lambda_{m_V}$, we repeated the fits using a range
of cutoff values.  The results of this investigation are shown in Figures
\ref{am:figure} and \ref{an:figure}.  Just as was seen in the independent
fits of Section \ref{ae:section}, there is a strong dependence on
$\Lambda_{m_V}$ down to small quark mass.  We would hope that in a more
complete study using a smaller lattice spacing, this dependence would disappear
for small quark mass.

\chapter{Results} \label{ac:section}
The primary result of our study is a value for the Gasser-Leutwyler coefficient
combination $2 L_8 - L_5$ \eqref{ch:equation}, along with its corresponding
value for the light-quark-mass ratio $m_u / m_d$ \eqref{ci:equation}.  The
secondary results include a value for the the GL coefficient $L_5$ with very
large systematic errors \eqref{cj:equation} and values for $L_4$ and $L_6$ with
large and unestimated systematic errors, \eqref{co:equation} and
\eqref{ck:equation}.

A subset of these results were presented in \cite{Nelson:2001tb}.  Their
account here is significantly more comprehensive and incorporates several
improvements in our analysis techniques.

\section{Primary Results}
We take the central value of our quoted result for the GL coefficient
combination $2 \alpha_8 - \alpha_5$ from the correlator data of our primary
ensemble, ensemble \ensemble{A ~ hyp}.  This produces the value $2 \alpha_8 -
\alpha_5 = 0.275 \pm 0.017$, where the given uncertainty accounts only for
statistical error.

\subsection{Systematic Error} \label{ah:section}
An important aspect of our study is the ability to make a quantitative estimate
of our systematic error.  Past theoretical estimates for the GL coefficient
combination $2 \alpha_8 - \alpha_5$ have suffered from an inability to quantify
the systematic error resulting from the approximations they require.  Our
first-principles approach, however, allows for such an investigation.  All of
the systematics which separate our calculation's result from the true value of
$2 \alpha_8 - \alpha_5$ are known and open to investigation.

To determine our systematic error, we note the variation in $2 \alpha_8 -
\alpha_5$ due to four changes in our calculation:  reducing the lattice volume,
hypercubic blocking, doubling the lattice spacing, and shifting the
valence-quark-mass cutoff.  While it is likely that several of the variations
are correlated --- for example, both the removal of hypercubic blocking and
doubling the lattice spacing increase a single systematic-error source:  flavor
and Lorentz symmetry breaking --- in order to generate a generous estimate of
our error, we will add the variations in quadrature, as if uncorrelated.

Our study of ensemble \ensemble{B ~ hyp} grants us insight into the effects of
finite volume on our result.  Ensemble \ensemble{B ~ hyp} produces values for
the fit parameters similar to those of ensemble \ensemble{A ~ hyp}, indicating
that finite-volume effects in our primary ensemble are well under control.  We
estimate the uncertainty due to finite volume using the shift in value of $2
\alpha_8 - \alpha_5$ between the volumes:  $\pm 0.064$.

Hypercubic blocking has been shown to reduce the flavor symmetry breaking
inherent in staggered calculations.  In order to estimate the strength of
flavor-symmetry-breaking effects on our result, we compare our result to the
value obtained from the thin-link version of our primary ensemble, ensemble
\ensemble{A}.  We take the resulting shift in value as our estimate of
flavor-symmetry-breaking error:  $\pm 0.085$.

The lattice spacing has a direct impact on the accuracy of our lattice
calculations, controlling the strength of unwanted Lorentz- and
flavor-symmetry-breaking terms in our action.  To estimate the strength of
these finite-lattice-spacing effects, we compare our value for $2 \alpha_8 -
\alpha_5$ to that obtained from an ensemble with approximately double the
lattice spacing, ensemble \ensemble{W ~ hyp}.  This shift gives us a
qualitative estimate for the corresponding systematic error: $\pm 0.028$.  

\begin{figure}
\centering
\psfrag{2a8a5}[t][t]{\Large $2 \alpha_8 - \alpha_5$}
\psfrag{Final}[l][l]{final result}
\psfrag{Ah}[l][l]{\ensemble{A ~ hyp}}
\psfrag{Cut}[l][l]{$\Lambda_{m_V}$ variation}
\psfrag{Bh}[l][l]{\ensemble{B ~ hyp}}
\psfrag{A}[l][l]{\ensemble{A}}
\psfrag{B}[l][l]{\ensemble{B}}
\psfrag{Wh}[l][l]{\ensemble{W ~ hyp}}
\psfrag{Xh}[l][l]{\ensemble{X ~ hyp}}
\psfrag{Yh}[l][l]{\ensemble{Y ~ hyp}}
\psfrag{Zh}[l][l]{\ensemble{Z ~ hyp}}
\psfrag{WWWWWW}[l][l]{\ensemble{W}}
\psfrag{X}[l][l]{\ensemble{X}}
\psfrag{Y}[l][l]{\ensemble{Y}}
\psfrag{Z}[l][l]{\ensemble{Z}}
\psfrag{H}[l][l]{set \ensemble{hyp}}
\psfrag{T}[l][l]{set \ensemble{thin}}
\psfrag{Mu0}[l][l]{$m_u = 0$}
\includegraphics[width=\textwidth,clip=]{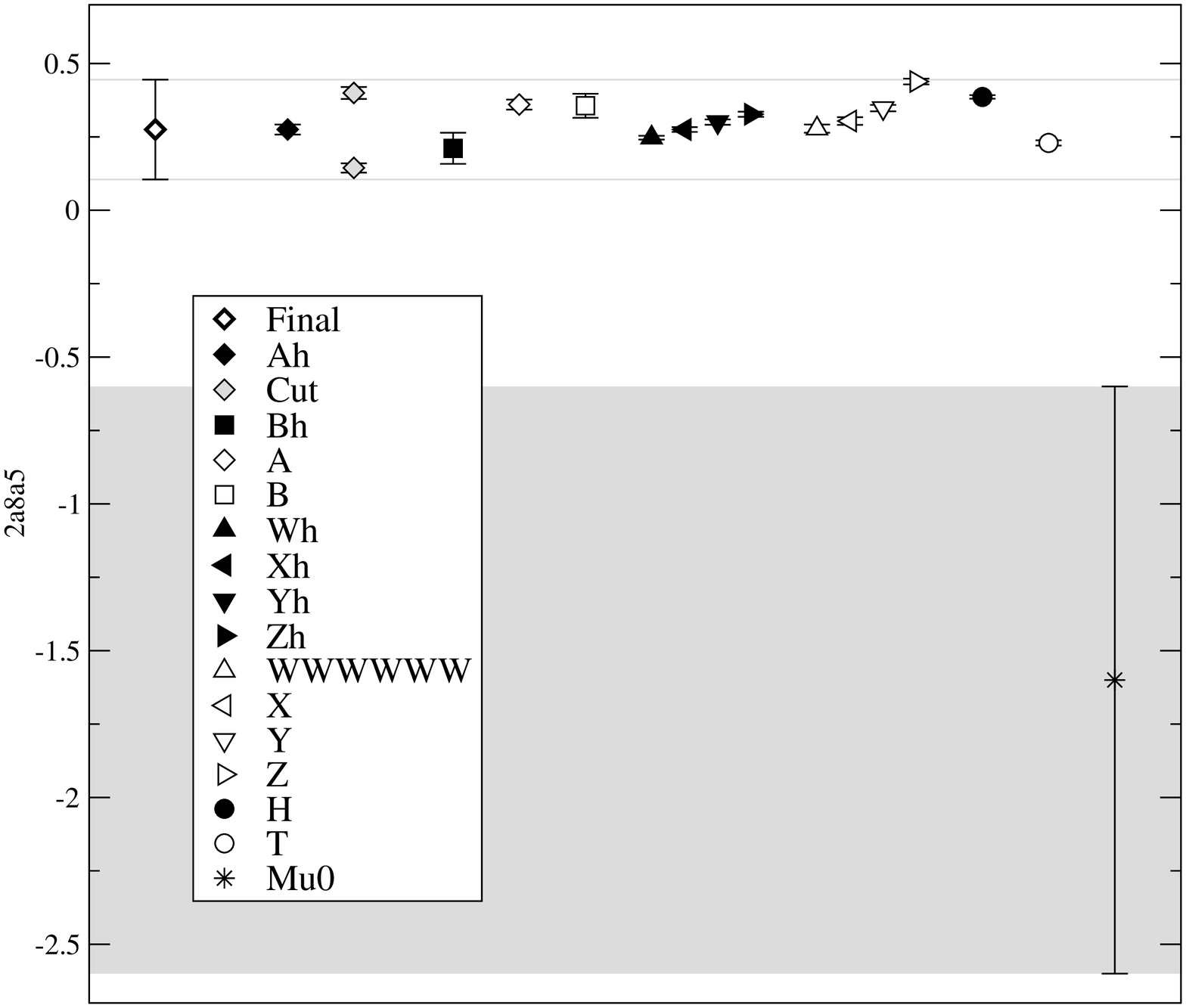}
\mycaption{The compiled results for $2 \alpha_8 - \alpha_5$.  The bold open
diamond corresponds to our final result and our quoted statistical and
systematic error.  All other error bars are statistical.  The gray diamonds
correspond to the result from ensemble \ensemble{A ~ hyp} after shifting the
valence-quark-mass cutoff $\Lambda_{m_V} / m_{Q_K}$ by $\pm 0.015$.  The filled
and open circles correspond to the results of the simultaneous fits of
ensembles \ensemble{W ~ hyp} through \ensemble{Z ~ hyp} and \ensemble{W}
through \ensemble{Z} respectively.  The burst corresponds to the range of
values which allow a massless up quark \eqref{cb:equation}.  All other points
represent the values obtained for $2 \alpha_8 - \alpha_5$ from their
corresponding ensemble.}
\label{ao:figure}
\end{figure}

The investigation of Section \ref{y:section} demonstrated that the choice of
valence-quark-mass cutoff is a significant source of systematic uncertainty.
While this uncertainty is likely related to flavor symmetry breaking, we will
add in quadrature the variation due to changing our valence-quark-mass cutoff
as if it were an independent error source.  It was determined in Section
\ref{y:section} that shifting the valence-quark-mass cutoff $\Lambda_{m_V} /
m_{Q_K}$ by $\pm 0.15$ results in a variation of $2 \alpha_8 - \alpha_5$ equal
to $\pm 0.13$.  We will use this variation as our estimate of the systematic
error due to our somewhat arbitrary choice of valence-quark-mass cutoff.

Presenting all error sources together, our result becomes:
\begin{equation}
2 \alpha_8 - \alpha_5 ~ = ~ 0.275 ~ \pm ~ 0.017 ~ \pm ~ 0.064 ~ \pm ~ 0.085 ~
\pm ~ 0.028 ~ \pm ~ 0.13
\end{equation}
Adding all sources in quadrature gives our final result:
\begin{equation}
2 \alpha_8 - \alpha_5 = 0.28 \pm 0.17
\end{equation}
In terms of the standard GL coefficient normalization, this corresponds to:
\begin{equation}
2 L_8 - L_5 = \bigl( 0.22 \pm 0.14 \bigr) \times 10^{-3}
\label{ch:equation}
\end{equation}
Note that this lies outside the range in which a massless up quark is allowed
\eqref{ca:equation}.

The compiled results for $2 \alpha_8 - \alpha_5$ from all ensembles, as well as
both our final result and the range which allows a massless up quark, can be
found in Figure \ref{ao:figure}.

\subsection{Light-Quark-Mass Ratio}
As discussed in Section \ref{ag:section}, the GL coefficients of our partially
quenched calculation are the same coefficients that appear in the physical
chiral Lagrangian.  As such, we can use our partially quenched results for $2
L_8 - L_5$, along with NLO ChPT as presented in Chapter \ref{n:section}, to
determine the light-quark-mass ratio of the Standard Model.

Using our result \eqref{ch:equation} in \eqref{h:equation} generates a value
for the NLO correction $\Delta_M$:
\begin{equation}
\Delta_M ~ = ~ -0.0919 ~ \pm ~ 0.029 ~ \pm ~ 0.0084
\label{cg:equation}
\end{equation}
where the first uncertainty is due to error in our result
and the second is based on an assumption that the unaccounted for NNLO
corrections are on the order of $\Delta_M^2$.  Error in other inputs to the
calculation of $\Delta_M$ are overwhelmed by the uncertainties given.  Using
Dashen's Theorem to account for the QED contributions to the physical meson
masses, instead of \eqref{w:equation}, does not affect $\Delta_M$ at this level
of precision.

Using our result for $2 L_8 - L_5$ \eqref{ch:equation} in \eqref{i:equation}
allows us to generate a value for the light-quark-mass ratio:
\begin{equation}
\frac{m_u}{m_d} ~ = ~ 0.408 ~ \pm ~ 0.027 ~ \pm ~ 0.008 ~ \pm ~ 0.021
\label{ce:equation}
\end{equation}
where the first uncertainty is due to error in our calculated result, the
second comes from an assumption that the NNLO corrections to $\Delta_M$ are on
the order of $\Delta_M^2$, and the third is due to uncertainty in $\Delta_E$.

Summarizing the repercussions of our calculated value for $2 L_8 - L_5$:
\begin{gather}
\Delta_M = 0.092 \pm 0.030 \\
2 L_8 - L_5 = \bigl( 0.22 \pm 0.14 \bigr) \times 10^{-3} \\
\begin{align}
L_7 = \bigl( -0.23 \pm 0.09 \bigr) \times 10^{-3}
&&
L_8 = \bigl( 0.36 \pm 0.24 \bigr) \times 10^{-3}
\end{align} \\
\frac{m_u}{m_d} = 0.408 \pm 0.035
\label{ci:equation}
\end{gather}
where the error given in \eqref{cg:equation} and \eqref{ce:equation} has been
added in quadrature.  In order to determine $L_7$ and $L_8$, we have used the
value for $L_5$ determined from the physical meson-decay-constant ratio
\eqref{p:equation} and the value for $12 L_7 + 6 L_8 - L_5$ determined from the
physical deviation from the Gell-Mann-Okubo relation \eqref{s:equation}, adding
in quadrature the experimental uncertainty in those values to the error in our
calculation.  Note that we do not use our calculated value for $L_5$ as
presented in Section \ref{ar:section}.

These numbers can be compared to the generally accepted set of values,
\eqref{cl:equation} through \eqref{cm:equation}, which are obtained using
various model-dependent assumptions and data from beyond the light-meson
sector.

The light-quark-mass ratio which results from using Dashen's theorem, instead
of \eqref{w:equation}, is $m_u / m_d = 0.482 \pm 0.026$, where there is now no
uncertainty due to $\Delta_E$.

Our result for the light-quark-mass ratio \eqref{ci:equation} can be compared
with the prediction generated by the generally accepted value for $\Delta_M$
\eqref{cl:equation}:  $m_u / m_d = 0.608 \pm 0.056$.  The literature provides
additional predictions.  Leutwyler \cite{Leutwyler:1996qg} gives a ratio of
$m_u / m_d = 0.553 \pm 0.043$, while Amoros et al.\ \cite{Amoros:2001cp} give a
ratio of $m_u / m_d = 0.46 \pm 0.09$ \eqref{cn:equation}.  Our result is
similar to these values, yet somewhat smaller in each case.  Note that our
error bars are smaller than all quoted cases.  In addition, we are of the
opinion that our error bars are the first that can be well trusted, as all
others are attempts to account for uncontrolled systematic error due to
theoretical assumptions.

The {\it Review of Particle Physics} \cite{PDBook} quotes a very broad range
for the light-quark-mass ratio:  $0.2 < m_u / m_d < 0.7$.  We fall well within
this range.

Our first-principles calculation of the low-energy constants of ChPT
demonstrates that the relevant coefficients are too low to allow for the
scenario in which the up quark is massless and strong NLO terms emulate a
non-zero mass.  As the massless-quark solution is seen here to be unlikely, the
strong $\CP$ problem remains unanswered.

\section{Secondary Results}
The secondary results of our study include values for the GL coefficients
$L_5$, $L_4$, and $L_6$.

\subsection{$L_5$} \label{ar:section}
Our results for the GL coefficient $L_5$ are extremely vulnerable to systematic
error and vary significantly across the ensembles studied.  Estimating the
systematic error using the same method as Section \ref{ah:section} results in:
\begin{equation}
\alpha_5 ~ = ~ 0.240 ~ \pm ~ 0.033 ~ \pm ~ 0.62 ~ \pm ~ 1.8 ~ \pm ~ 0.54 ~ \pm
~ 0.008
\end{equation}
where the listed uncertainties are due to statistical error, reducing the
lattice volume, hypercubic blocking, doubling the lattice spacing, and shifting
the valence-quark-mass cutoff.  While it is not clear with such large variation
that adding in quadrature is justified, doing so results in $\alpha_5 = 0.2 \pm
2.0$.  In terms of the standard GL coefficient normalization, this corresponds
to:
\begin{equation}
L_5 = \bigl( 0.2 \pm 1.6 \bigr) \times 10^{-3}
\label{cj:equation}
\end{equation}

This value falls within the range obtained for $L_5$ from the
meson-decay-constant ratio \eqref{p:equation}.

\subsection{$L_4$ and $L_6$}
Due to the coarse lattice spacing of ensembles \ensemble{W} through
\ensemble{Z}, our results for the GL coefficient combinations $2 \alpha_6 -
\alpha_4$ and $\alpha_4$ are guaranteed to be contaminated by strong systematic
error.  However, with only a single ensemble set to work with, we are unable to
make any qualitative estimates of the error.  Thus we present our values for
these coefficients only as preliminary results, using the values from the
hypercubic-blocked ensemble set and quoting only their statistical error bars:
\begin{align}
L_4 & = \bigl( -1.145 \pm 0.026^\text{(stat)} \bigr) \times 10^{-3}
\label{co:equation} \\
L_6 & = \bigl( -0.680 \pm 0.013^\text{(stat)} \bigr) \times 10^{-3}
\label{ck:equation}
\end{align}

These values both miss their generally accepted ranges, $L_4 = (
-0.3 \pm 0.5 ) \times 10^{-3}$ and $L_6 = ( -0.2 \pm 0.3 ) \times 10^{-3}$,
which are determined using large-$N_c$ considerations \cite{Gasser:1985gg}.

\section{Quenching Effects}
Analyzing the results produced by ensemble \ensemble{Q ~ hyp}, we see that
quenching had a smaller systematic effect on $2 \alpha_8 - \alpha_5$ and
$\alpha_5$ than any other source of systematic error studied.  Clearly, at this
level of precision, quenched and partially quenched ensembles do not generate
distinctly different valence-quark-mass dependencies for the pion mass and
decay constant, despite the differing predictions of quenched and partially
quenched ChPT.  This indicates that, while the $N_f$ dependence of ChPT is
not explicit and thus the GL coefficients are unknown functions of the number
of dynamical quark flavors, that functional dependence is very slight.

\chapter{Summary and Outlook} \label{ab:section}
The low-energy constants of the chiral Lagrangian, the Gasser-Leutwyler
coefficients, are a critical element in the understanding of low-energy QCD.
Yet current theoretical estimates and experimental measurements of the GL
coefficients, even those unaffected by the Kaplan-Manohar ambiguity, have
errors from 10\% to 160\% \cite{Bijnens:1994qh}.  For many of the coefficients,
their current error bars remain as large as they were at the first instance of
their calculation \cite{Gasser:1985gg}.  Given LQCD's capacity to calculate
these coefficients directly, with no uncontrolled approximations, it is clear
that work in this area is warranted.  This may in fact prove to be one of those
rare, yet increasingly common, situations in which lattice techniques have some
chance of providing the greater community with the most trusted predictions
available.

Our study definitively calculates a single combination of the GL coefficients,
determining a value for $2 L_8 - L_5$ which rules out the massless-up-quark
solution to the strong $\CP$ problem.  The culmination of our study is a value
for the light-quark-mass ratio:
\begin{equation}
\frac{m_u}{m_d} = 0.408 \pm 0.035
\end{equation}

This work is far from the closing word in the lattice study of Gasser-Leutwyler
coefficients.  In truth it is just the first step in what will likely be a
long-term and comprehensive study of the coefficients by the lattice community.

From a short-term and more pragmatic perspective, there are several aspects of
our study ripe for improvement.  In addition, new theoretical work is on the
horizon which should dramatically improve lattice calculations of the GL
coefficients.

\section{Improved Systematics}
Clearly, future calculations would do well to improve on the systematics of our
study through superior ensembles, utilizing either more sophisticated actions
or simply more sophisticated computers.

Despite the large uncertainty in the QED contribution, the greatest source of
error in the light-quark-mass ratio \eqref{ce:equation} remains the systematic
error in our calculation of $2 L_8 - L_5$.  Improved systematics thus have the
potential to reduce the error bars by up to a factor of root two.

Additionally, a reduction in systematic error may prove to bring the
fluctuations in $L_5$ under control, leading to a result in which we could have
reasonable confidence.  As $L_5$ is not subject to the KM ambiguity, this would
allow for a comparison between lattice and experimental results.

In the case of the coefficients $L_4$ and $L_6$, our study was not broad enough
to estimate the magnitude of our error.  Producing results with reasonable and
well-estimated systematic error would require the availability of a number of
quality ensemble sets, with each set including ensembles across a range of
dynamical quark masses.  

\subsection{High-Quality Publicly Available Ensembles}
When our study began, high-quality ensembles at $N_f = 3$ were not publicly
available.  We had no choice but to use our relatively limited computer
resources to generate the requisite ensembles.  As a consequence the resulting
ensembles, with their mediocre lattice extent and unimproved action, are far
from cutting edge.

Today however, thanks to the MILC collaboration \cite{b:url:cite} in
conjunction with the Gauge Connection \cite{a:url:cite}, a set of ten $20^3
\times 64$, $N_f = 3$ ensembles are available for public use.  These ensembles
span a range of dynamical quark masses and were generated using an improved
gauge action and the highly improved $a^2$-tad staggered action
\cite{Bernard:2001av}.  The application of our analysis to these ensembles
would not only produce reduced systematic error in the calculation of $2 L_8 -
L_5$, but as the ensembles have matched lattice spacings, would also allow for
an accurate calculation of $L_4$ and $L_6$.

Such a spirit of sharing is extremely valuable to those of us who do not lead
the community in computational resources.  Beyond that, the community itself
profits, as the number of individuals capable of significant and impacting
research is dramatically increased.

\section{Dynamical Hypercubic Blocking}
In our study we reduced the systematic error due to flavor symmetry breaking by
hypercubic blocking after the generation of our ensembles.  In effect we used
hypercubic-blocked valence quarks in conjunction with thin-link dynamical
quarks.  While this technique has merit, clearly a more consistent approach
would be to use hypercubic-blocked dynamical quarks during ensemble creation.
This approach should also further decrease flavor-symmetry-breaking error.

Following this logic, Flemming \cite{Fleming:2002ms} has plans for a direct
continuation of our work via the creation of ensembles using practical
algorithms for dynamical hypercubic-blocked fermions, which are only now
becoming available \cite{Alexandru:2002jr, Alexandru:2002sw,
Hasenfratz:2002pt}.

\section{Staggered Chiral Perturbation Theory} \label{am:section}
A recent and very exciting theoretical development germane to the connection
between LQCD and ChPT is staggered Chiral Perturbation Theory (sChPT)
\cite{Bernard:2001yj, Aubin:2002ss}.

In standard ChPT only a single structure breaks flavor symmetry: the quark mass
matrix.  In the context of staggered fermions, however, the theory contains
additional flavor-breaking structures which arise at finite lattice spacing.
In sChPT these new flavor-breaking elements are accounted for through the
introduction of additional terms to the Lagrangian at each order.  The chiral
Lagrangian becomes an expansion in three parameters instead of two:  meson
momentum, meson mass, and lattice spacing.

A complication arises in sChPT when work is done at $N_f \neq 4$.  At the core
of the staggered formulation are four flavors.  Thus, even when a fractional
power of the fermionic determinant is used to set $N_f \neq 4$, as discussed in
Section \ref{aj:section}, the flavor symmetry breaking due to finite lattice
spacing retains its four-flavor structure.  As a consequence, sChPT can only be
constructed for theories which include some multiple of four flavors.  In order
to apply the results of sChPT to theories with some other number of flavors,
such as $N_f = 3$, the strength of meson-loop graphs must be adjusted by hand.

As discussed in Section \ref{ai:section}, flavor symmetry breaking in staggered
fermions leads to a non-degeneracy of the sixteen light mesons, splitting them
into five levels.  The sChPT expression for the local pion mass conforms to
this scenario, containing additional terms which are the result of meson loops
failing to cancel exactly, as they would in the degenerate continuum case.  As
such, Bernard \cite{Bernard:2001yj} demonstrates that accounting for a majority
of the new finite-lattice-spacing terms does not require the calculation of a
full set of fresh low-energy constants.  Instead, it requires only a
calculation of the four meson-mass splittings.

However, two of these new terms, those which allow mesons to switch flavor
content via insertions into a propagator, have coefficients which can not be
directly measured.  Thus, they must be left as free parameters in a fit.
Preliminary attempts at such fits of the local pion mass \cite{Aubin:2002ss}
have proved unstable, with the two parameters running to unnaturally large and
opposite values.  Introducing priors to the fit could stabilize results.  Also,
as the same two coefficients appear in all sChPT expressions, fitting multiple
quantities simultaneously, such as the pion mass and decay constant, may bind
their values.  It is worth noting that these coefficients can not be determined
once and for all, unlike the GL coefficients, as they are a function of the
action.  Different improved actions have different flavor symmetry breaking,
and thus the coefficients will take on different values.

Because the lattice-spacing dependence of the staggered chiral Lagrangian is
known and explicit up to whatever order we chose, and because in the continuum
limit staggered quarks are equivalent to continuum quarks, a calculation of
the low-energy constants of sChPT corresponds exactly to a calculation of
physical ChPT's GL coefficients.

Preliminary results for sChPT are promising \cite{Aubin:2002ss}.  They
demonstrate that in cases where the local pion mass's dependence on the quark
mass does not follow the form predicted by continuum ChPT, a phenomenon
clearly evident in our data, the dependence matches closely the predictions
of sChPT.  As such, use of sChPT should result in a drastic reduction of the
systematic errors in lattice calculations of the GL coefficients.

In fact the advent of sChPT has an impact beyond simply the accurate
measurement of GL coefficients.  sChPT provides, for any quantities to which it
is relevant, the appropriate simultaneous extrapolation from
numerically-favorable quark masses and finite lattice spacing to physical quark
masses and the continuum limit.  

\section{Quantum Electrodynamic Corrections}
A significant percentage of the uncertainty in the light-quark-mass ratio is
due to uncertainty in the magnitude of the QED contribution to the pion mass.
While not addressed in this study, lattice techniques exist which allow for the
calculation of this contribution \cite{Duncan:1996xy, Duncan:1997sq}.  In a
situation where attempts to significantly reduce the systematic error in $2 L_8
- L_5$ are successful, an accurate lattice calculation of $\Delta_E$ may prove
valuable.

\appendix

\renewcommand{\theequation}{A.\arabic{equation}}
\setcounter{equation}{0}
\renewcommand{\thechapter}{A}

\chapter{Notation}
So as to leave the main text uncluttered, we list several notational
conventions here.

Throughout, we use the pion decay constant normalization $f_\pi \simeq 92.4 \,
\text{MeV}$.  This differs from the other common normalization by a factor of
root two, $\sqrt{2} f_\pi \simeq 130.7 \, \text{MeV}$.

When the standard normalization for the GL coefficients is being used, they are
denoted by $L_i$.  While this is the normalization which generally appears in a
chiral Lagrangian, most NLO expressions for observable quantities are made
cleaner through the use of a second normalization which we denote by
$\alpha_i$.  The normalizations are related by the expression:
\begin{equation}
\alpha_i = 8 ( 4 \pi )^2 L_i
\end{equation}

Traces over color indices are denoted by $\trace$, while traces over other
indices, generally flavor indices, are denoted by $\Trace$.

Integration over all space is denoted by:
\begin{equation}
\int_x \equiv \iiiint d^4x
\end{equation}
while integration over all momenta is denoted by:
\begin{equation}
\int_k \equiv \iiiint \frac{d^4 k}{(2 \pi)^4}
\end{equation}

When denoting dimensionful quark masses, we use a lower-case subscript such as
$m_q$.  When denoting a dimensionless lattice quark mass, we use a
corresponding upper-case subscript such as $m_Q$.  For all quantities other
than quark mass, we use a check over the variable to denote its unitless
counterpart.  For example in the case of the local pion mass, $M_{\pi_5} = a
\U{M}_{\pi_5}$.

The generators of the various $SU(N)$ Lie algebras are represented by
$\lambda^a$ for color transformations and $\tau^a$ for flavor transformations.
They are traceless Hermitian $N \otimes N$ matrices, which are normalized
according to:
\begin{align}
\Tr{ \tau^a \tau^b } = \frac{1}{2} \delta^{ab}
& &
\tr{ \lambda^a \lambda^b } = \frac{1}{2} \delta^{ab}
\end{align}
and satisfy the commutation relations:
\begin{align}
\bigl[ \tau^a , \tau^b \bigr] = i f^{abc} \tau^c
& &
\bigl[ \lambda^a , \lambda^b \bigr] = i g^{abc} \lambda^c
\end{align}
where $f^{abc}$ and $g^{abc}$ are the appropriate structure constants of the
algebras, which are completely antisymmetric and real.  $\lambda^a$ should not
be confused with the $SU(N)$ Gell-Mann matrices, as $\lambda^a = \frac{1}{2}
\lambda_\text{Gell-Mann}^a$.

The Lorentz tensor $\epsilon_{\mu\nu\alpha\beta}$ is defined to be totally
antisymmetric, with:
\begin{equation}
\epsilon_{1234} = 1
\end{equation}

The Euclidean-space Dirac matrices $\gamma^\mu$ are defined to satisfy the
anticommutation relation:
\begin{equation}
\{ \gamma^\mu , \gamma^\nu \} = 2 g^{\mu\nu}_E = 2 \delta^{\mu\nu}
\end{equation}
where $g^{\mu\nu}_E$ is the flat Euclidean-space metric.  Unlike the
Minkowski-space Dirac matrices, the Euclidean-space matrices are Hermitian:
\begin{equation}
\gamma^\mu = \gamma^{\mu\dagger}
\end{equation}
The Dirac matrices can be used to construct the generators of a spinor
representation of the Lorentz group.  Given a Lorentz transformation:
\begin{align}
x^\mu & \pad{\symarrow} {x'}^\mu = \Lambda^\mu_\nu x^\nu \\
& \qquad \Lambda^\mu_\nu = e^{ -\frac{i}{2} \omega_{\alpha\beta}
(\mathcal{J}^{\alpha\beta})^\mu_\nu} \in O(4)_\text{Lorentz}
\end{align}
where $\mathcal{J}^{\alpha\beta}$ are the generators of Lorentz rotations on
Lorentz 4-vectors and $\omega_{\alpha\beta}$ is an antisymmetric tensor which
parameterizes the transformation, a spinor transforms as:
\begin{align}
\psi & \pad{\symarrow} \psi' = S(\Lambda) \psi \\
\bar{\psi} & \pad{\symarrow} \bar{\psi}' = \bar{\psi} S^{-1}(\Lambda) \\
& \qquad S(\Lambda) = e^{ -\frac{1}{8} \omega_{\alpha\beta} [
\gamma^\alpha , \gamma^\beta ]} \in O(4)_\text{spin}
\end{align}
The Dirac matrices themselves transform correctly under the Lorentz group both
as Lorentz 4-vectors and as spin space matrices:
\begin{eqnarray}
S(\Lambda) \gamma^\mu S^{-1}(\Lambda) = \Lambda^\mu_\nu \gamma^\nu
\end{eqnarray}
Using the Dirac matrices, the matrix $\gamma^5$ is defined as:
\begin{equation}
\gamma^5 = \gamma^1 \gamma^2 \gamma^3 \gamma^4
\end{equation}
which anticommutes with the other four Dirac matrices:
\begin{align}
\{ \gamma^5 , \gamma^\mu \} = 0
& &
\gamma^5 \gamma^5 = \Eins
\end{align}

We define the discretized directional derivative $\discder_\mu$ as:
\begin{align}
\discder_\mu \varphi(x) & \equiv \frac{ \varphi ( x + a \muh ) - \varphi
( x - a \muh ) }{2a} \notag \\
& = \frac{1}{2a} \Bigl( e^{a \partial_\mu} - e^{-a \partial_\mu} \Bigr)
\varphi(x) \notag \\
& = \frac{1}{a} \Bigl( \sinh a \partial_\mu \Bigr) \varphi(x) \notag \\[1mm]
& =  \partial_\mu \varphi(x) + O(a^2)
\label{cs:equation}
\end{align}
while we defined the second discretized directional derivative $\discder^2_\mu$
as:
\begin{align}
\discder^2_\mu \varphi(x) & \equiv \frac{ \varphi ( x + a \muh ) - 2 \varphi(x)
+ \varphi ( x - a \muh ) }{a^2} \notag \\
& = \frac{1}{a^2} \Bigl( e^{a \partial_\mu} - 2 + e^{-a \partial_\mu} \Bigr)
\varphi(x) \notag \\
& = \frac{1}{a^2} \Bigl( 1 + a \partial_\mu + \half a^2 \partial^2_\mu - 2 + 1
- a \partial_\mu + \half a^2 \partial^2_\mu \Bigr) \varphi(x) + O(a^4) \notag
\\[1mm]
& = \partial^2_\mu \varphi(x) + O(a^4)
\label{ab:equation}
\end{align}

\renewcommand{\theequation}{B.\arabic{equation}}
\setcounter{equation}{0}
\renewcommand{\thechapter}{B}

\chapter{Staggered Identities}

\section{Bilinear Sum} \label{k:section}
For the sources of our bilinear correlators, we use a linear combination of all
bilinears with a given distance binary four-vector $\mathcal{D}_{\mathcal{S},
\mathcal{F}} = \mathcal{D}$.  Such a combination has a simple form when
expressed in terms of $\chi$ and $\chib$:
\begin{align}
\sum_\mathcal{R} \mathcal{J}_{\mathcal{R} + \mathcal{D}, \mathcal{R}; h} & =
\sum_\mathcal{R} \Qb_h \Bilinear{\gamma_\mathcal{D}
\gamma_\mathcal{R}}{\xi_\mathcal{R}} Q_h \notag \\
& = \sum_\mathcal{R} \Tr{ \Qb_h \gamma_\mathcal{D} \gamma_\mathcal{R} Q_h
\gamma^\dagger_\mathcal{R} } \notag \\
& = 4 \Tr{ \Qb_h \gamma_\mathcal{D}} \Tr{ Q_h } \vphantom{\sum_A}
\notag \\
& = \sum_A \sum_B \Tr{ \Gamma^\dagger_A \gamma_\mathcal{D}} \Tr{ \Gamma_B }
\chib_{h + A} \chi_{h + B} \notag \\
& = 16 \sum_A \sum_B \delta_{A, \mathcal{D}} \delta_{B, 0} \chib_{h + A}
\chi_{h + B} \notag \\
& = 16 \chib_{h + \mathcal{D}} \chi_{h}
\end{align}
where we have used the identity
\begin{equation}
\sum_C \bigl( \Gamma^\dagger_C \bigr)_{b \beta} \bigl( \Gamma_C \bigr)_{\alpha
a} = 4 \delta_{b,a} \delta_{\beta,\alpha}
\end{equation}
or equivalently
\begin{equation}
\sum_C \Gamma^\dagger_A \Gamma_C \Gamma_B \Gamma^\dagger_C = 4 \Gamma^\dagger_A
\Trace \Gamma_B
\end{equation}
The linear combination leaves only a single contraction, that between $\chib$
at the corner of the hypercube offset from the lowest corner by $\mathcal{D}$
and $\chi$ at the lowest corner.  Adding the gauge links required by an
interacting theory, we have:
\begin{equation}
\sum_\mathcal{R} \mathcal{J}_{\mathcal{R}, \mathcal{R} + \mathcal{D}; h} = 16
\chib_{h + \mathcal{D}} \mathcal{U}_{\mathcal{D}, 0; h} \chi_h
\label{cu:equation}
\end{equation}
where $\mathcal{U}_{A, B; h}$ is defined after \eqref{ak:equation}.  In the
case of local bilinears, $\mathcal{D} = 0$, the contraction is contained
completely by the lowest corner of the hypercube:
\begin{equation}
\sum_\mathcal{R} \mathcal{J}_{\mathcal{R}, \mathcal{R}; h} = 16 \chib_h
\chi_{h}
\label{am:equation}
\end{equation}

\section{Transpose of the Staggered Interaction Matrix}
The transpose of the staggered fermion interaction matrix $M^S \fconfig{U}^T$
arises as the antiquark propagator in our calculation of bilinear correlators.
The adjoint matrix can be expressed in terms of $M^S \fconfig{U}$ in a manner
similar to that used for the naive interaction matrix \eqref{aq:equation}:
\begin{align}
M^S \fconfig{U}^\dagger_{n,m} & = \frac{1}{2} \sum_\mu \eta_{\mu;m} \Bigl[
U^{}_{\mu;m} \delta^{}_{m; n - \muh} - U^\dagger_{\mu; m - \muh} \delta^{}_{m,
n + \muh} \Bigr] + m_Q \delta_{m,n} \notag \\
& = - \frac{1}{2} \sum_\mu \Bigl[ \eta^{}_{\mu; n - \muh} U^{\dagger}_{\mu; m -
\muh} \delta^{}_{m; n + \muh} - \eta^{}_{\mu; n + \muh} U^{}_{\mu;m}
\delta^{}_{m; n - \muh} \Bigr] + m_Q \delta_{n,m} \notag \\
& = - \frac{1}{2} \sum_\mu \eta_{\mu;n} \Bigl[ U^{\dagger}_{\mu;n}
\delta^{}_{n, m - \muh} - U^{}_{\mu; n - \muh} \delta^{}_{n; m + \muh} \Bigr] +
m_Q \delta_{n,m} \notag \\
& = \epsilon_n \epsilon_m \biggl\{ \frac{1}{2} \sum_\mu \eta_{\mu;n} \Bigl[
U^\dagger_{\mu;n} \delta^{}_{n, m - \muh} - U^{}_{\mu; n - \muh} \delta^{}_{n;
m + \muh} \Bigr] + m_Q \delta_{n,m} \biggr\} \notag \\
& = \epsilon_n M^S_{n,m} \fconfig{U} \epsilon_m
\label{an:equation}
\end{align}
This mirrors closely the result for naive fermions, with $\epsilon_n$ filling
the roll of $\gamma_5$.

\section{Non-Local Bilinear Correlators} \label{j:section}
In order to study the non-local staggered mesons, the correlators between
non-local bilinears must be calculated.  We now express non-local bilinear
correlators explicitly in terms of quantities which are straightforward to
calculate using lattice techniques.  That is, the inverse staggered fermion
interaction matrix applied to various fermion field vectors.  A similar
discussion limited to local bilinears is found in Section \ref{f:section}.

Our general bilinear correlator, using a wall sink to overlap only with zero
momentum states, is:
\begin{equation}
\Bigl\bra \sum_{\substack{\vec{g} \\ g_4 = t}} \mathcal{J}_{\mathcal{S},
\mathcal{F}; g} \mathcal{J}_{\mathcal{S}, \mathcal{F}; 0} \Bigr\ket
\end{equation}
We replace the bilinear in our source with a linear combination of all
bilinears with distance vector $\mathcal{D} = \mathcal{D}_{\mathcal{S},
\mathcal{F}} = \mathcal{S} + \mathcal{F}$:
\begin{equation}
\frac{1}{16} \Bigl\bra \sum_{\substack{\vec{g} \\ g_4 = t}}
\mathcal{J}_{\mathcal{S}, \mathcal{F}; g} \sum_\mathcal{R}
\mathcal{J}_{\mathcal{R + \mathcal{D}}, \mathcal{R}; 0} \Bigr\ket
\end{equation}
We replace our single-bilinear source with a wall of bilinears at the
appropriate time slice:
\begin{equation}
C_{\mathcal{S}, \mathcal{F}; t} = \frac{1}{16} \Bigl\bra
\sum_{\substack{\vec{g} \\ g_4 = t}} \mathcal{J}_{\mathcal{S}, \mathcal{F}; g}
\sum_{\substack{\vec{h} \\ h_4 = 0}} \sum_\mathcal{R} \mathcal{J}_{\mathcal{R}
+ \mathcal{D}, \mathcal{R}; h} \Bigr\ket
\end{equation}

Using \eqref{ak:equation}, \eqref{ag:equation}, and \eqref{cu:equation},
we express the correlator in terms of $\chi$ and $\chib$:
\begin{align}
C_{\mathcal{S}, \mathcal{F}; t} & = \Bigl\bra \sum_{\substack{\vec{g} \\ g_4 =
t}} \sum_B (-)^{\varphi_{\mathcal{S}, \mathcal{F}; B}} \notag \\
& \hphantom{= \Bigl\bra} \quad \times \sum_{\substack{\vec{h} \\ h_4 = 0}}
\sum_{a,b,c,d} \chib_\dsub{g + B}{a} \mathcal{U}_\dsub{B, B + \mathcal{D};
g}{a,b} \chi_\dsub{g + B + \mathcal{D}}{b} \chib_\dsub{h + \mathcal{D}}{d}
\mathcal{U}_\dsub{\mathcal{D}, 0; h}{d,c} \chi_\dsub{h}{c} \Bigr\ket
\end{align}
Integrating over the Grassmann variables using Wick contractions, we express
the correlator in terms of the inverse of the interaction matrix:
% all this gymnastics just to get wick.sty working :(
\newsavebox{\layoutdsubga}
\savebox{\layoutdsubga}{$\dsub{g + B}{a}$}
\newsavebox{\layoutdsubgb}
\savebox{\layoutdsubgb}{$\dsub{g + B + \mathcal{D}}{b}$}
\newsavebox{\layoutdsubhc}
\savebox{\layoutdsubhc}{$\dsub{h + \mathcal{D}}{d}$}
\newsavebox{\layoutdsububa}
\savebox{\layoutdsububa}{$\dsub{B, B + \mathcal{D}; g}{a,b}$}
\newsavebox{\layoutdsubuzc}
\savebox{\layoutdsubuzc}{$\dsub{\mathcal{D}, 0; h}{d,c}$}
\newsavebox{\layoutU}
\savebox{\layoutU}{$\mathcal{U}$}
\newsavebox{\layoutwicknl}
\savebox{\layoutwicknl}{$\overwick{2}{ <+\chi_{\usebox{\layoutdsubgb}}
>+{\usebox{\layoutchib}}_{\usebox{\layoutdsubhc}}}$}
\begin{align}
C_{\mathcal{S}, \mathcal{F}; t} & = \Bigl\bra \sum_{\substack{\vec{g} \\ g_4 =
t}} \sum_B (-)^{\varphi_{\mathcal{S}, \mathcal{F}; B}} \notag \\
& \hphantom{= \Bigl\bra} \quad \times \sum_{\substack{\vec{h} \\ h_4 = 0}}
\sum_{a,b,c,d} \overwick{1}{ <+{\usebox{\layoutchib}}_{\usebox{\layoutdsubga}}
\usebox{\layoutU}_{\usebox{\layoutdsububa}} \usebox{\layoutwicknl}
\usebox{\layoutU}_{\usebox{\layoutdsubuzc}} >+\chi_{\usebox{\layoutdsubh}}}
\Bigr\ket \notag \\
& = \Bigl\bra \sum_{\substack{\vec{g} \\ g_4 = t}} \sum_B
(-)^{\varphi_{\mathcal{S}, \mathcal{F}; B}} \notag \\
& \hphantom{= \Bigl\bra} \quad \times \sum_{\substack{\vec{h} \\ h_4 = 0}}
\sum_{a,b,c,d} \mathcal{U}_\dsub{B, B + \mathcal{D}; g}{a,b}
\mathcal{U}_\dsub{\mathcal{D}, 0; h}{d,c} M^S \fconfig{U}^{-1}_\dsub{g + B,
h}{a,c} ( M^S \fconfig{U}^T )^{-1}_\dsub{g + B + \mathcal{D}, h +
\mathcal{D}}{b,d} \Bigr\ket \notag \\
& = \Bigl\bra \sum_{\substack{\vec{g} \\ g_4 = t}} \sum_B
(-)^{\varphi_{\mathcal{S} + 5, \mathcal{F} + 5; B}} \notag \\
& \hphantom{= \Bigl\bra} \quad \times \sum_{\substack{\vec{h} \\ h_4 = 0}}
\sum_{a,b,c,d} \mathcal{U}_\dsub{B, B + \mathcal{D}; g}{a,b}
\mathcal{U}_\dsub{\mathcal{D}, 0; h}{d,c} M^S \fconfig{U}^{-1}_\dsub{g + B,
h}{a,c} M^S \fconfig{U}^{-1*}_\dsub{g + B + \mathcal{D}, h + \mathcal{D}}{b,d}
\Bigr\ket
\end{align}

In order to reduce the number of required applications of the inverse
interaction matrix, we calculate the $2 N_c$ field vectors $X^{(c)}$ and
$Y^{(c)}$:
\begin{align}
X^{(c)}_\dsub{n}{a} & = \sum_{\substack{\vec{h} \\ h_4 = 0}}
M^S \fconfig{U}^{-1}_\dsub{n,h}{a,c} \\
Y^{(c)}_\dsub{m}{b} & = \sum_d \sum_{\substack{\vec{h} \\ h_4 = 0}}
M^S \fconfig{U}^{-1}_\dsub{m,h + \mathcal{D}}{b,d} \mathcal{U}^*_\dsub{\mathcal{D}, 0;
h}{d,c}
\end{align}
To calculate $X^{(c)}$ we construct the field vector $W^{(c)}$, which equals
one only at color $c$ in the lowest corner of each hypercube on the time slice
$n_4 = 0$ and zero elsewhere.  The result of applying the inverse interaction
matrix to $W^{(c)}$ is $X^{(c)}$:
\begin{equation}
X^{(c)}_\dsub{n}{a} = \bigl( M^S \fconfig{U}^{-1} W^{(c)} \bigr)_\dsub{n}{a}
\end{equation}
To calculate $Y^{(c)}$ we first apply a swap operator $S_\mathcal{D}$ to
$W^{(c)}$ and then the inverse interaction matrix.  The resulting field vector
is $Y^{(c)}$:
\begin{equation}
Y^{(c)}_\dsub{m}{b} = \bigl( M^S \fconfig{U}^{-1} S_\mathcal{D} W^{(c)}
\bigr)_\dsub{m}{b}
\end{equation}
The swap operator $S_\mathcal{D}$ is similar to the bilinear operator
$\bilinear{\mathcal{S}}{\mathcal{F}}$ defined by \eqref{ah:equation}, but does
not apply phase factors to the field vector.  It swaps corners separated by the
offset vector $\mathcal{D}$ within each hypercube, and applies the appropriate
color matrix for that movement:
\begin{equation}
\bigl( S_\mathcal{D} \bigr)_{n,m} = \delta_{h + A + \mathcal{D}, m}
\mathcal{U}^*_{A, A + \mathcal{D}; h}
\end{equation}
where $h$ denotes the hypercube containing $n$, and $A$ denotes its position
within that hypercube:
\begin{align}
h = 2 \Bigl\lfloor \frac{n}{2} \Bigr\rfloor
&&
A = n - h
\end{align}

The unique random phase associated with each lattice site due to local gauge
freedom washes out the position-off-diagonal terms in the product of $X^{(c)}$
and $Y^{(c)}$:
\begin{align}
Y^{(c)*}_\dsub{m}{b} X^{(c)}_\dsub{n}{a} & = \sum_d \sum_{\substack{\vec{h} \\
h_4 = 0}} M^S \fconfig{U}^{-1*}_\dsub{m,h + \mathcal{D}}{b,d}
\mathcal{U}_\dsub{\mathcal{D}, 0; h}{d,c} \sum_{\substack{\vec{f} \\ f_4 = 0}}
M^S \fconfig{U}^{-1}_\dsub{n,f}{a,c} \notag \\
& = \sum_d \sum_{\substack{\vec{h} \\ h_4 = 0}} M^S \fconfig{U}^{-1*}_\dsub{m,h
+ \mathcal{D}}{b,d} \mathcal{U}_\dsub{\mathcal{D}, 0; h}{d,c} M^S
\fconfig{U}^{-1}_\dsub{n,h}{a,c}
\end{align}
Thus, once $X^{(c)}$ and $Y^{(c)}$ have been calculated, we can construct our
correlator:
\begin{align}
C_{\mathcal{S}, \mathcal{F}; t} & = \Bigl\bra \sum_c \sum_{\substack{\vec{g} \\
g_4 = t}} \sum_B (-)^{\varphi_{\mathcal{S} + 5, \mathcal{F} + 5;
B}} \sum_{a,b} \mathcal{U}_\dsub{B, B + \mathcal{D}; g}{a,b} Y^{(c)*}_\dsub{g +
B + \mathcal{D}}{b} X^{(c)}_\dsub{g + B}{a} \Bigr\ket \notag \\
& = \Bigl\bra \sum_c \sum_{\substack{\vec{g} \\ g_4 = t}} \sum_B
(-)^{\varphi_{\mathcal{S} + 5, \mathcal{F} + 5; B}} \sum_{a,b}
X^{(c)}_\dsub{g + B}{a} \mathcal{U}_\dsub{B, B + \mathcal{D}; g}{a,b}
Y^{(c)*}_\dsub{g + B + \mathcal{D}}{b} \Bigr\ket
\end{align}
or more concisely
\begin{align}
C_{\mathcal{S}, \mathcal{F}; t} & = \Bigl\bra \sum_c \sum_{\substack{\vec{g} \\
g_4 = t}} \sum_B (-)^{\varphi_{\mathcal{S} + 5, \mathcal{F} + 5;
B}} \sum_{a,b} X^{(c)}_\dsub{g + B}{a} \bigl( S^*_\mathcal{D} \bigr)_\dsub{g +
B, g + B + \mathcal{D}}{a,b} Y^{(c)*}_\dsub{g + B + \mathcal{D}}{b} \Bigr\ket
\notag \\
& = \Bigl\bra \sum_c \sum_{\substack{\vec{g} \\ g_4 = t}} \sum_B
(-)^{\varphi_{\mathcal{S} + 5, \mathcal{F} + 5; B}} \sum_a
X^{(c)}_\dsub{g + B}{a} \bigl( S_\mathcal{D} Y^{(c)} \bigr)^*_\dsub{g + B}{a}
\Bigr\ket
\end{align}
The correlator is a contraction of $X^{(c)}$ and $S_\mathcal{D} Y^{(c)}$ summed
only over the time slices $t$ and $t + 1$.

Note that the field vectors $X^{(c)}$ and $S_\mathcal{D} Y^{(c)}$ are
independent of $t$ and depend only on the distance vector of the bilinear,
$\mathcal{D} = \mathcal{S} + \mathcal{F}$.  Thus, once they are known, we can
calculate all bilinear correlators of distance vector $\mathcal{D}$ at every
time separation with no additional inversions.

\renewcommand{\theequation}{C.\arabic{equation}}
\setcounter{equation}{0}
\renewcommand{\thechapter}{C}

\chapter{$SU(3)$ Projection} \label{l:section}
The application of hypercubic blocking, as discussed in Section
\ref{h:section}, requires the projection of an arbitrary $3 \times 3$ matrix
onto the group $SU(3)$.  We describe here the algorithm used for this
projection.  For clarity we will use capital letters to denote $3 \times 3$
matrices and lower-case letters to denote $2 \times 2$ matrices.

Given the $3 \times 3$ complex matrix $M$, we wish to find the nearest $SU(3)$
matrix $G$:
\begin{equation}
\proj_{SU(3)} \bigl[ M \bigr] \equiv G
\end{equation}
We define the nearest group element to be the one which maximizes:
\begin{equation}
\tr{ G M^\dagger }
\end{equation}
In order to maximize this trace, we must choose a $G$ such that $G M^\dagger$
is as nearly proportional to identity as possible.  For $SU(2)$ this problem
can be solved in closed form.  While this is not true for $SU(3)$, we can break
the problem down into an iterative procedure of repeatedly applying the
closed-form solution on $SU(2)$ subgroups.  If we use a set of subgroups which
span the full $SU(3)$ group, the process will converge on the correct $G$.

We begin with a guess for $G$:
\begin{equation}
G_1 = \identity
\end{equation}
This then gives us an initial residual matrix $R$:
\begin{equation}
R_1 = G_1 M^\dagger = M^\dagger
\end{equation}
The residual matrix is the matrix we wish to make proportional to identity.

We now enter an iterative process in which each iteration pushes $R$
closer to identity and leads us to a more refined value for $G$.  At the
start of each iteration we choose an $SU(2)$ subgroup to work within.  In
practice we choose from among three $SU(2)$ subgroups, using each in turn.
We extract the appropriate $2 \times 2$ matrix $r$ from $R$:
\begin{gather}
r = r^{(a)} \\
\begin{aligned}
r^{(1)} = \begin{bmatrix} R_{n;11} & R_{n;12} \\ R_{n;21} & R_{n;22}
\end{bmatrix}
&&
r^{(2)} = \begin{bmatrix} R_{n;11} & R_{n;13} \\ R_{n;31} & R_{n;33}
\end{bmatrix}
&&
r^{(3)} = \begin{bmatrix} R_{n;22} & R_{n;23} \\ R_{n;32} & R_{n;33}
\end{bmatrix}
\end{aligned}
\end{gather}
where $a$ identifies the subgroup chosen.

In the context of the $SU(2)$ subgroup, there exists a closed-form expression
for the group element $u$ nearest the matrix $r$.  We calculate directly the
unnormalized coefficients $\tilde{\alpha}_\mu$ of the matrix $u$:
\begin{align}
\tilde{\alpha}_4 = \Real \biggl\{ \frac{1}{2} \Tr{ r } \biggr\}
&&
\tilde{\alpha}_i = \Real \biggl\{ -\frac{i}{2} \Tr{ r \sigma_i } \biggr\}
\end{align}
where $\sigma_i$ are the Pauli matrices:
\begin{align}
\sigma_1 \equiv \begin{bmatrix} 0 & 1 \\ 1 & 0 \end{bmatrix}
&&
\sigma_2 \equiv \begin{bmatrix} 0 & -i \\ i & 0 \end{bmatrix}
&&
\sigma_3 \equiv \begin{bmatrix} 1 & 0 \\ 0 & -1 \end{bmatrix}
\end{align}
After normalizing the coefficients:
\begin{equation}
\alpha_\mu = \frac{\tilde{\alpha}_\mu}{\lvert \tilde{\alpha} \rvert}
\end{equation}
$u$ is constructed:
\begin{equation}
u = \alpha_4 \identity + i \sum_i \alpha_i \sigma_i
\end{equation}
Clearly, if $u$ is the closest $SU(2)$ group element to $r$, applying the
inverse of $u$ to $r$ will result in a matrix as near to proportional to
identity as is possible.

We now return to the full $3 \times 3$ matrices in order to complete the
iteration:
\begin{equation}
U = \left\{
\begin{aligned}
\begin{bmatrix}
u_{11} & u_{12} & 0 \\ u_{21} & u_{22} & 0 \\ 0 & 0 & 1
\end{bmatrix} \qquad a = 1 \\[3mm]
\begin{bmatrix}
u_{11} & 0 & u_{12} \\ 0 & 1 & 0 \\ u_{21} & 0 & u_{22}
\end{bmatrix} \qquad a = 2 \\[3mm]
\begin{bmatrix}
1 & 0 & 0 \\ 0 & u_{11} & u_{12} \\ 0 & u_{21} & u_{22}
\end{bmatrix} \qquad a = 3
\end{aligned}
\right.
\end{equation}
We move $R$ closer to identity, and refine our value for $G$, by applying to $G$
the inverse of $U$.  As $U$ is unitary, its inverse is simply its adjoint:
\begin{align}
G_{n + 1} & = U^\dagger G_n \\
R_{n + 1} & = U^\dagger R_n = G_{n + 1} M^\dagger
\end{align}
At this point we have completed one iteration.  The process is now repeated
using our refined guess for $G$.

When the trace of $R$ stops improving, we have reached our final value for $G$.
We end the iterative process when:
\begin{align}
\bigl\lvert \Trace R_{n + 1} - \Trace R_n \bigr\rvert < \epsilon
\end{align}
where $\epsilon$ is our error tolerance.  In practice we only perform this
test after using all three $SU(2)$ subgroups in turn, as it is possible that
the trace may be flat with respect to one subgroup, while improvement is still
possible along the other directions.

\renewcommand{\theequation}{D.\arabic{equation}}
\setcounter{equation}{0}
\renewcommand{\thechapter}{D}

\chapter{Figures}
\begin{figure}
\centering
\psfrag{C0V}[b][B]{$C_{5, 5; 0} / V$}
\psfrag{C15V}[b][B]{$C_{5, 5; 15} / V$}
\psfrag{traj}[Bl]{trajectory}
\psfrag{Nb}[Bl]{$N_B$}
\psfrag{Nt}[l][l]{$N_T$}
\includegraphics[width=\textwidth,clip=]{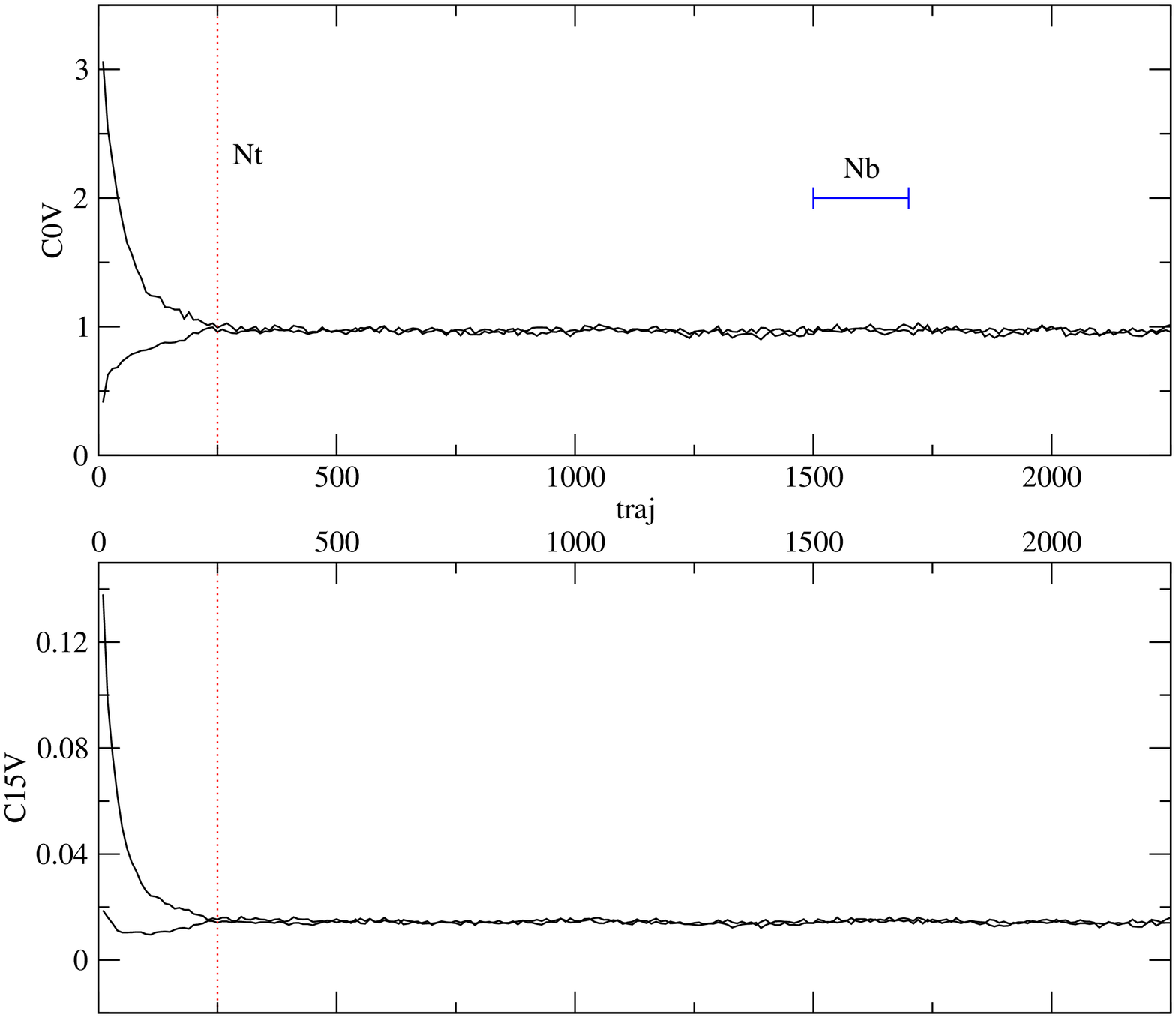}
\mycaption{Markov chain of ensemble \ensemble{A} with thermalization point
$N_T$ and block length $N_B$ shown.  The correlator $C_{5, 5; t}$ is calculated
using $m_V = 0.01$.  Of the ensemble's two Markov chains, the chain which
begins above the equilibrium value has a disordered initial condition.  The
second has an ordered initial condition.}
\label{b:figure}
\end{figure}

\begin{figure}
\centering
\psfrag{C0V}[b][B]{$C_{5, 5; 0} / V$}
\psfrag{C15V}[b][B]{$C_{5, 5; 15} / V$}
\psfrag{traj}[Bl]{trajectory}
\psfrag{Nb}[Bl]{$N_B$}
\psfrag{Nt}[l][l]{$N_T$}
\includegraphics[width=\textwidth,clip=]{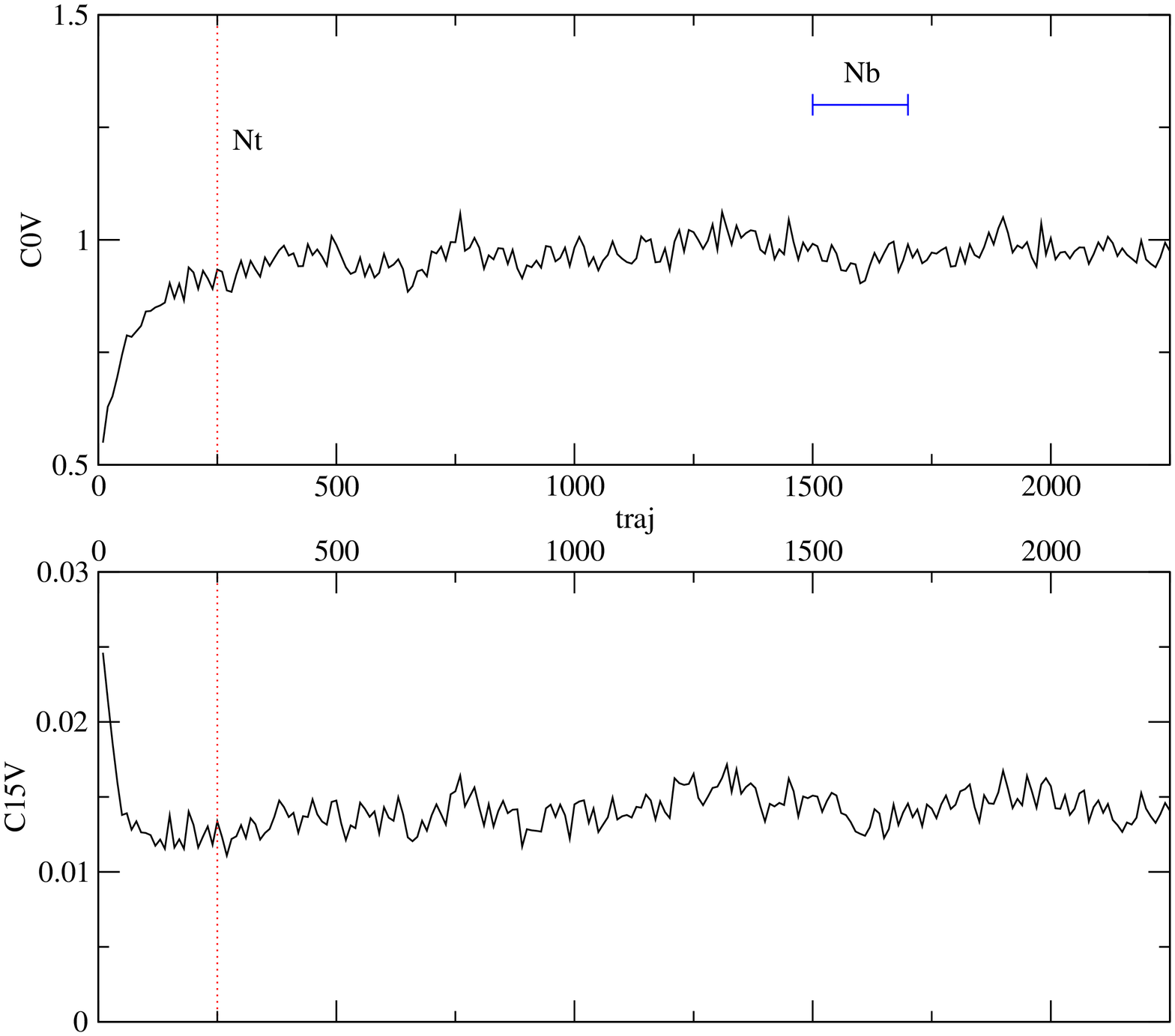}
\mycaption{Markov chain of ensemble \ensemble{B} with thermalization point
$N_T$ and block length $N_B$ shown.  The correlator $C_{5, 5; t}$ is calculated
using $m_V = 0.01$.  The Markov chain has an ordered initial condition.}
\end{figure}

\begin{figure}
\centering
\psfrag{C0V}[b][B]{$C_{5, 5; 0} / V$}
\psfrag{C15V}[b][B]{$C_{5, 5; 15} / V$}
\psfrag{traj}[Bl]{trajectory}
\psfrag{Nb}[Bl]{$N_B$}
\psfrag{Nt}[l][l]{$N_T$}
\includegraphics[width=\textwidth,clip=]{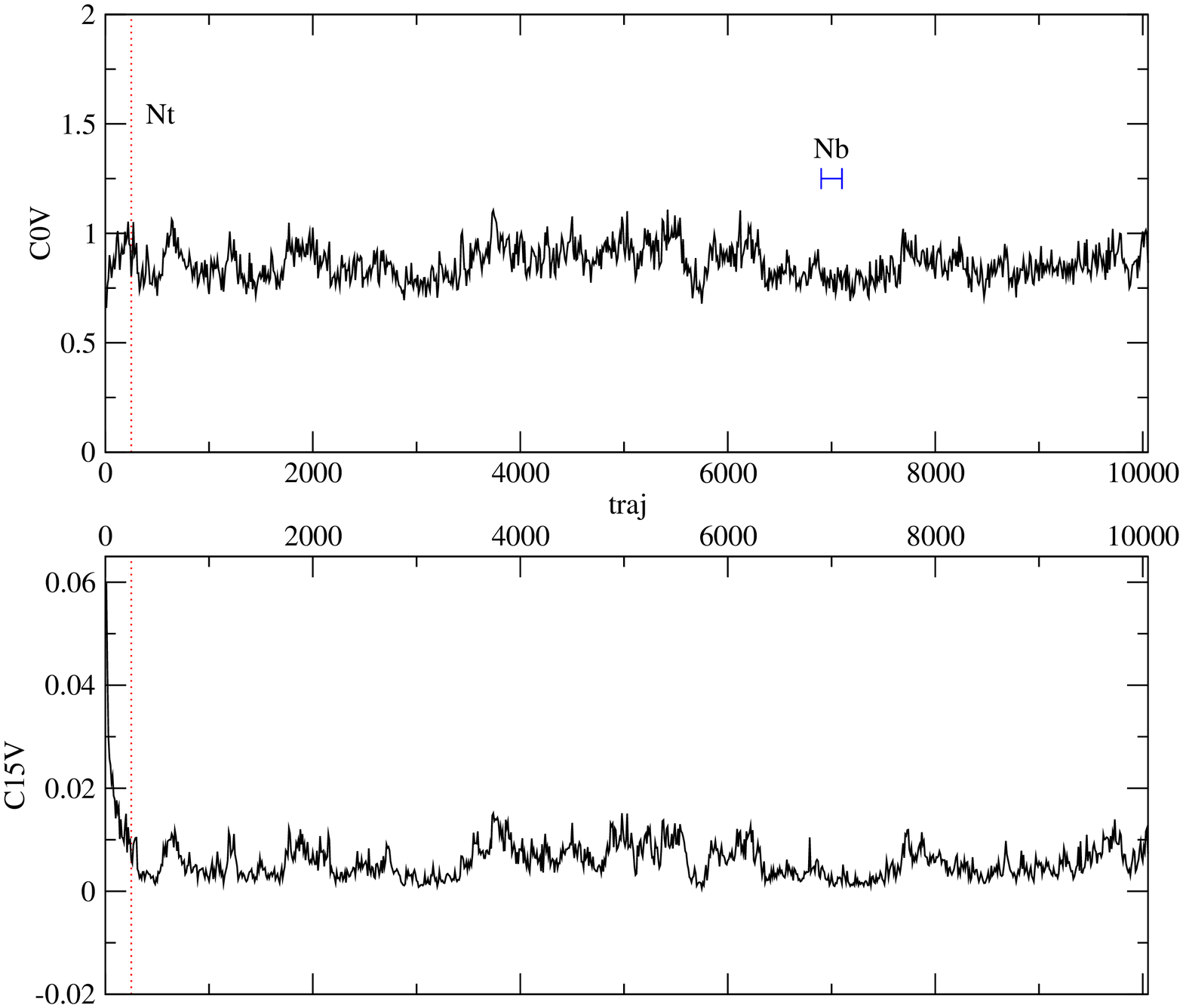}
\mycaption{Markov chain of ensemble \ensemble{C} with thermalization point 
$N_T$ and block length $N_B$ shown.  The correlator $C_{5, 5; t}$ is calculated
using $m_V = 0.01$.  The Markov chain has an ordered initial condition.}
\end{figure}

\begin{figure}
\centering
\psfrag{C0V}[b][B]{$C_{5, 5; 0} / V$}
\psfrag{C15V}[b][B]{$C_{5, 5; 15} / V$}
\psfrag{traj}[Bl]{trajectory}
\psfrag{Nb}[Bl]{$N_B$}
\psfrag{Nt}[l][l]{$N_T$}
\includegraphics[width=\textwidth,clip=]{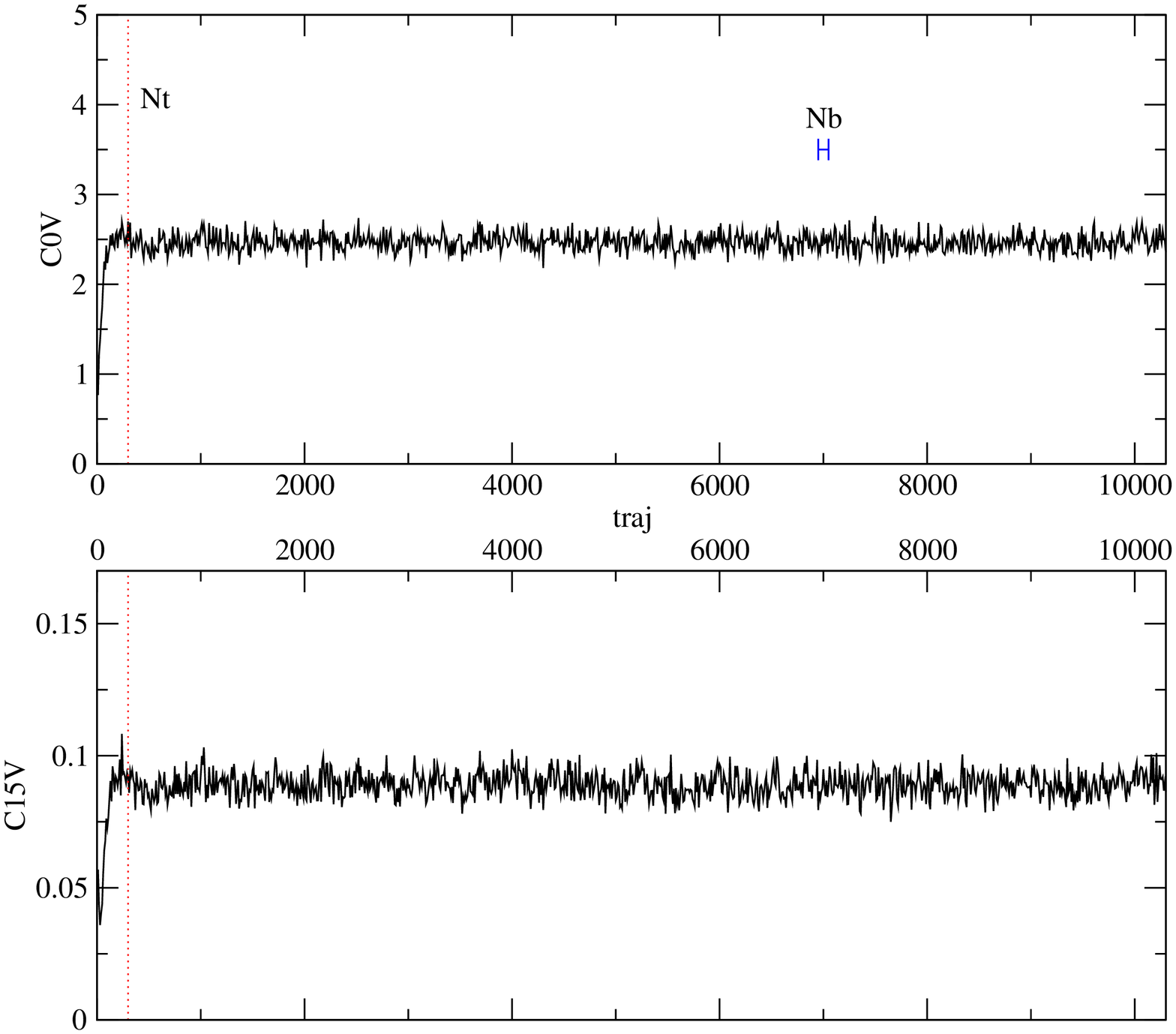}
\mycaption{Markov chain of ensemble \ensemble{W} with thermalization point 
$N_T$ and block length $N_B$ shown.  The correlator $C_{5, 5; t}$ is calculated
using $m_V = 0.01$.  The Markov chain has an ordered initial condition.}
\end{figure}

\begin{figure}
\centering
\psfrag{C0V}[b][B]{$C_{5, 5; 0} / V$}
\psfrag{C15V}[b][B]{$C_{5, 5; 15} / V$}
\psfrag{traj}[Bl]{trajectory}
\psfrag{Nb}[Bl]{$N_B$}
\psfrag{Nt}[l][l]{$N_T$}
\includegraphics[width=\textwidth,clip=]{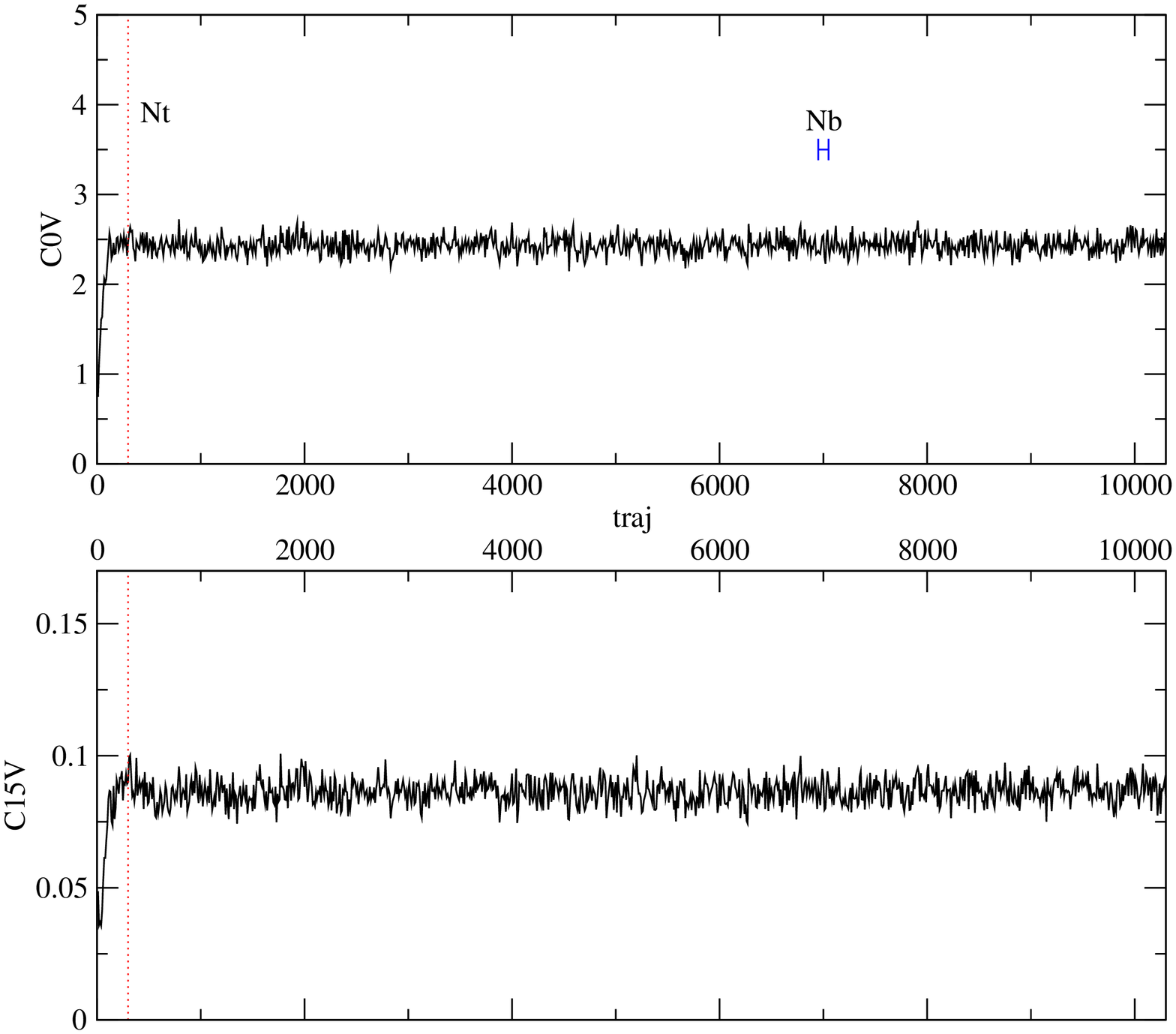}
\mycaption{Markov chain of ensemble \ensemble{X} with thermalization point 
$N_T$ and block length $N_B$ shown.  The correlator $C_{5, 5; t}$ is calculated
using $m_V = 0.01$.  The Markov chain has an ordered initial condition.}
\end{figure}

\begin{figure}
\centering
\psfrag{C0V}[b][B]{$C_{5, 5; 0} / V$}
\psfrag{C15V}[b][B]{$C_{5, 5; 15} / V$}
\psfrag{traj}[Bl]{trajectory}
\psfrag{Nb}[Bl]{$N_B$}
\psfrag{Nt}[l][l]{$N_T$}
\includegraphics[width=\textwidth,clip=]{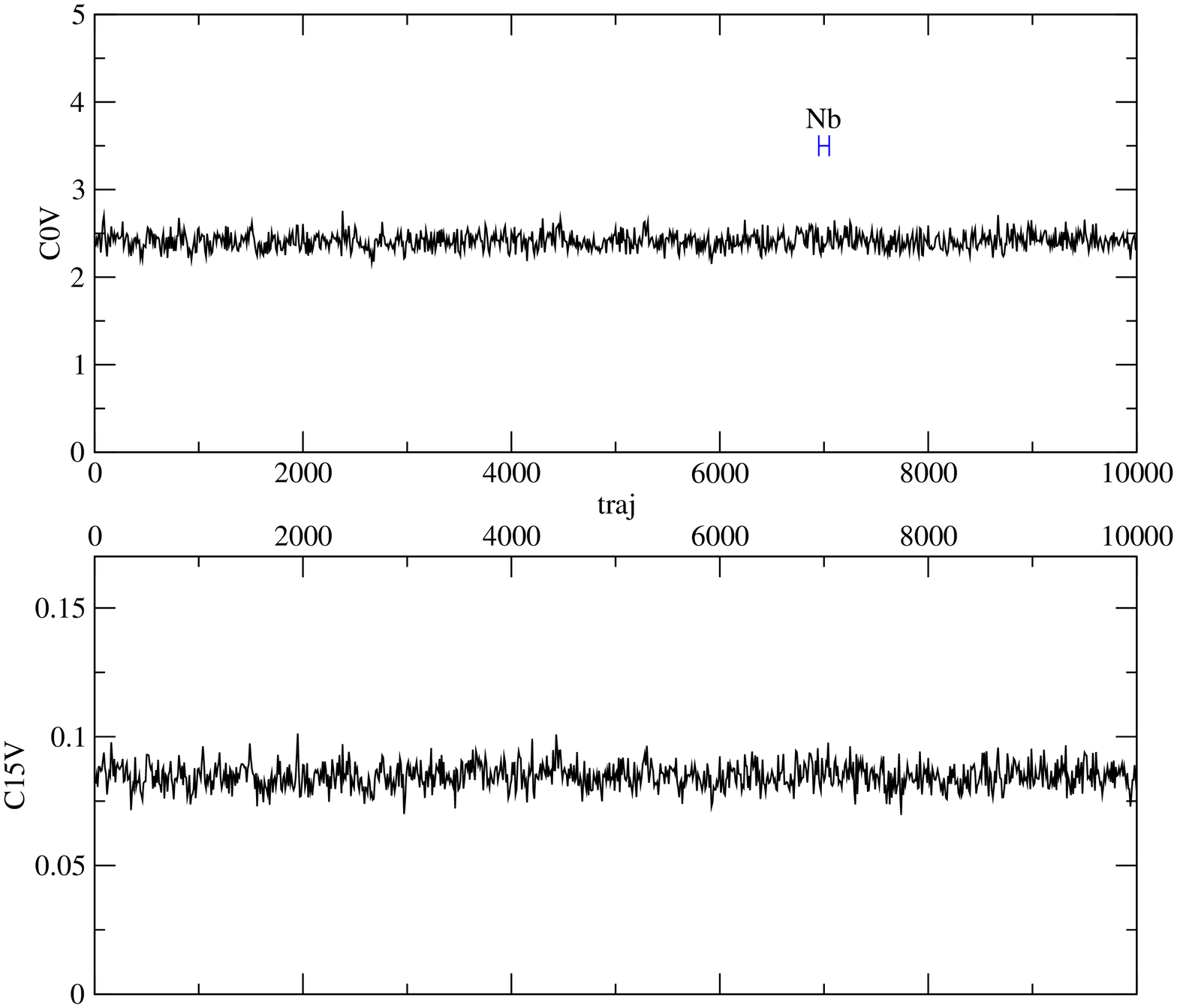}
\mycaption{Markov chain of ensemble \ensemble{Y} with thermalization point 
$N_T$ and block length $N_B$ shown.  The correlator $C_{5, 5; t}$ is calculated
using $m_V = 0.01$.  The Markov chain has an ordered initial condition.}
\end{figure}

\begin{figure}
\centering
\psfrag{C0V}[b][B]{$C_{5, 5; 0} / V$}
\psfrag{C15V}[b][B]{$C_{5, 5; 15} / V$}
\psfrag{traj}[Bl]{trajectory}
\psfrag{Nb}[Bl]{$N_B$}
\psfrag{Nt}[l][l]{$N_T$}
\includegraphics[width=\textwidth,clip=]{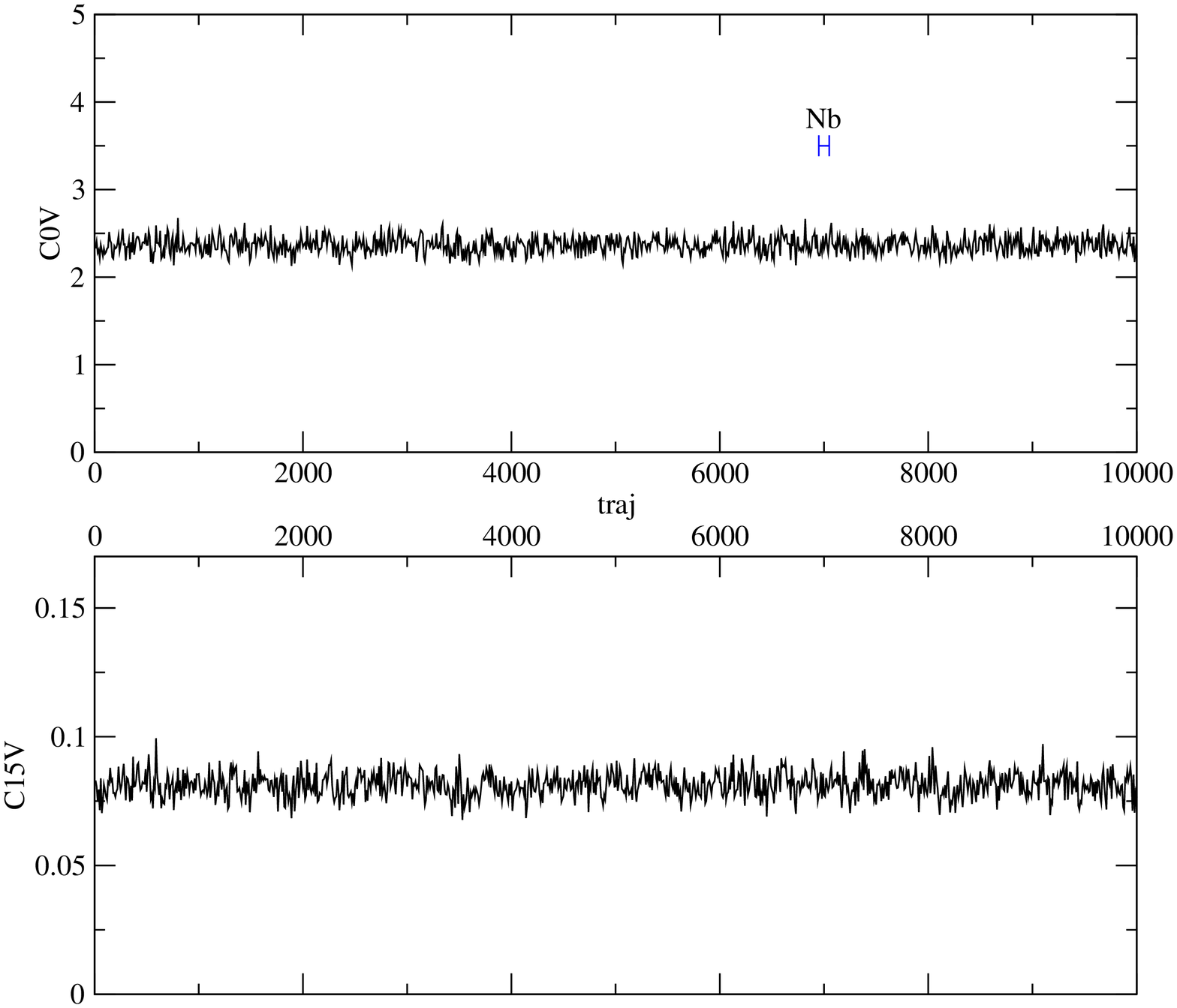}
\mycaption{Markov chain of ensemble \ensemble{Z} with thermalization point 
$N_T$ and block length $N_B$ shown.  The correlator $C_{5, 5; t}$ is calculated
using $m_V = 0.01$.  The Markov chain has an ordered initial condition.}
\label{c:figure}
\end{figure}

\clearpage

\begin{figure}
\centering
\psfrag{t}[t][t]{$t$}
\psfrag{Vs1}[b][B]{$a V_\text{eff} ( s = 1 )$}
\psfrag{Vs4}[b][B]{$a V_\text{eff} ( s = 4 )$}
\psfrag{A}[B][B]{\ensemble{A}}
\psfrag{Ah}[B][B]{\ensemble{A ~ hyp}}
\includegraphics[width=\textwidth,clip=]{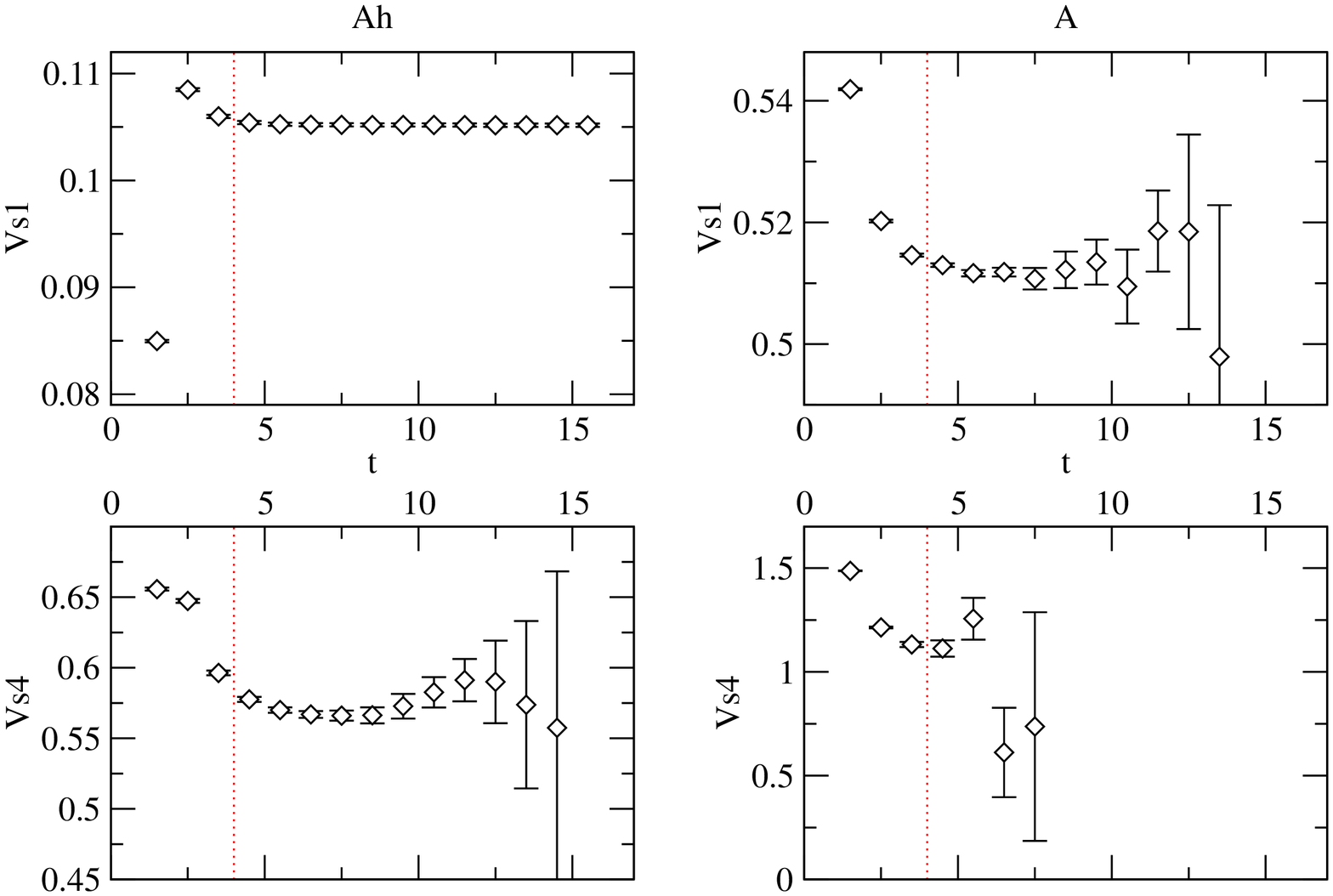}
\mycaption{Effective potential for ensembles \ensemble{A ~ hyp} and
\ensemble{A}.  The minimum time separation chosen is $t_\text{min} = 4$.}
\label{d:figure}
\end{figure}

\begin{figure}
\centering
\psfrag{t}[t][t]{$t$}
\psfrag{Vs1}[b][B]{$a V_\text{eff} ( s = 1 )$}
\psfrag{Vs4}[b][B]{$a V_\text{eff} ( s = 4 )$}
\psfrag{B}[B][B]{\ensemble{B}}
\psfrag{Bh}[B][B]{\ensemble{B ~ hyp}}
\includegraphics[width=\textwidth,clip=]{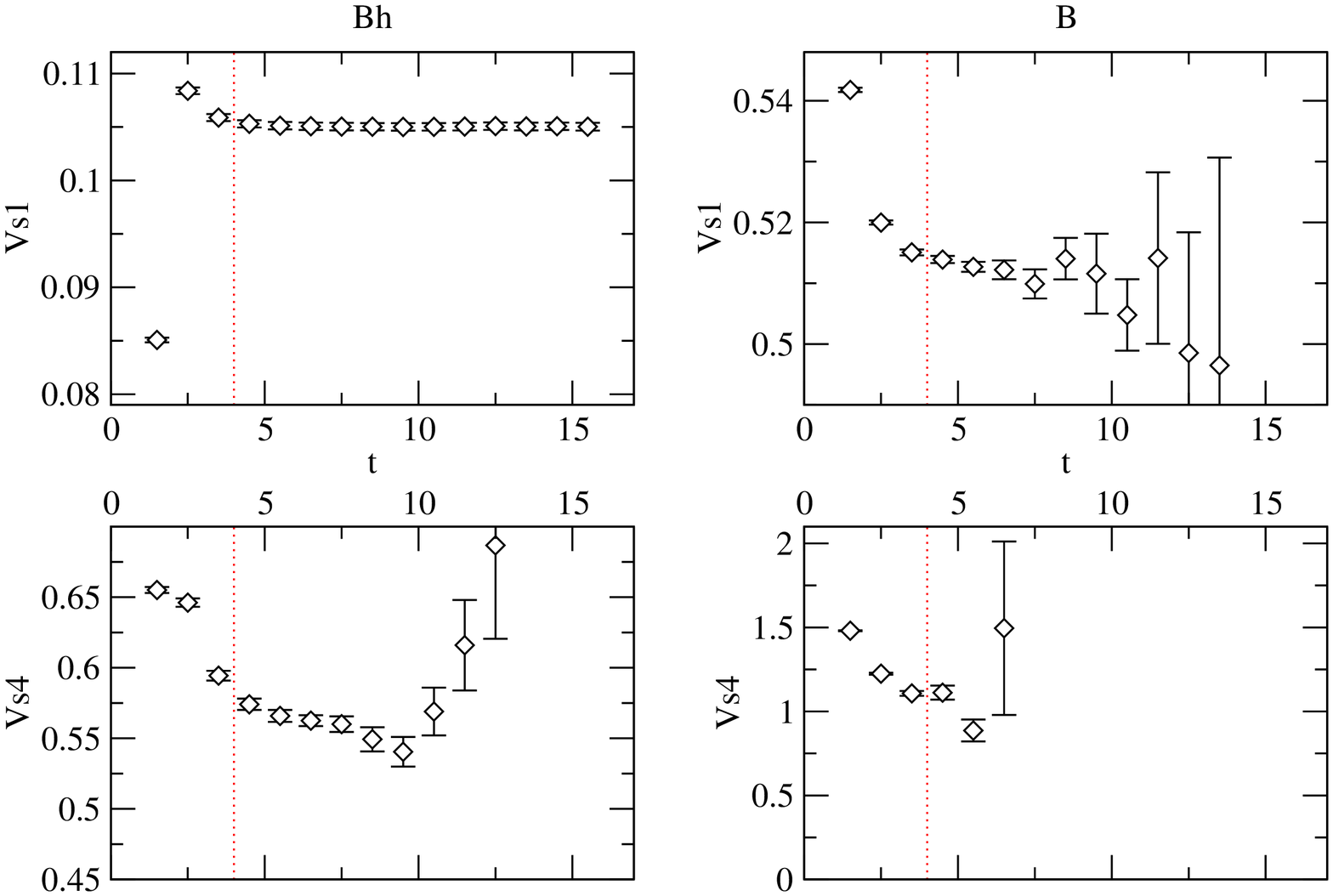}
\mycaption{Effective potential for ensembles \ensemble{B ~ hyp} and
\ensemble{B}.  The minimum time separation chosen is $t_\text{min} = 4$.}
\end{figure}

\begin{figure}
\centering
\psfrag{t}[t][t]{$t$}
\psfrag{Vs1}[b][B]{$a V_\text{eff} ( s = 1 )$}
\psfrag{Vs4}[b][B]{$a V_\text{eff} ( s = 4 )$}
\psfrag{C}[B][B]{\ensemble{C}}
\psfrag{Ch}[B][B]{\ensemble{C ~ hyp}}
\includegraphics[width=\textwidth,clip=]{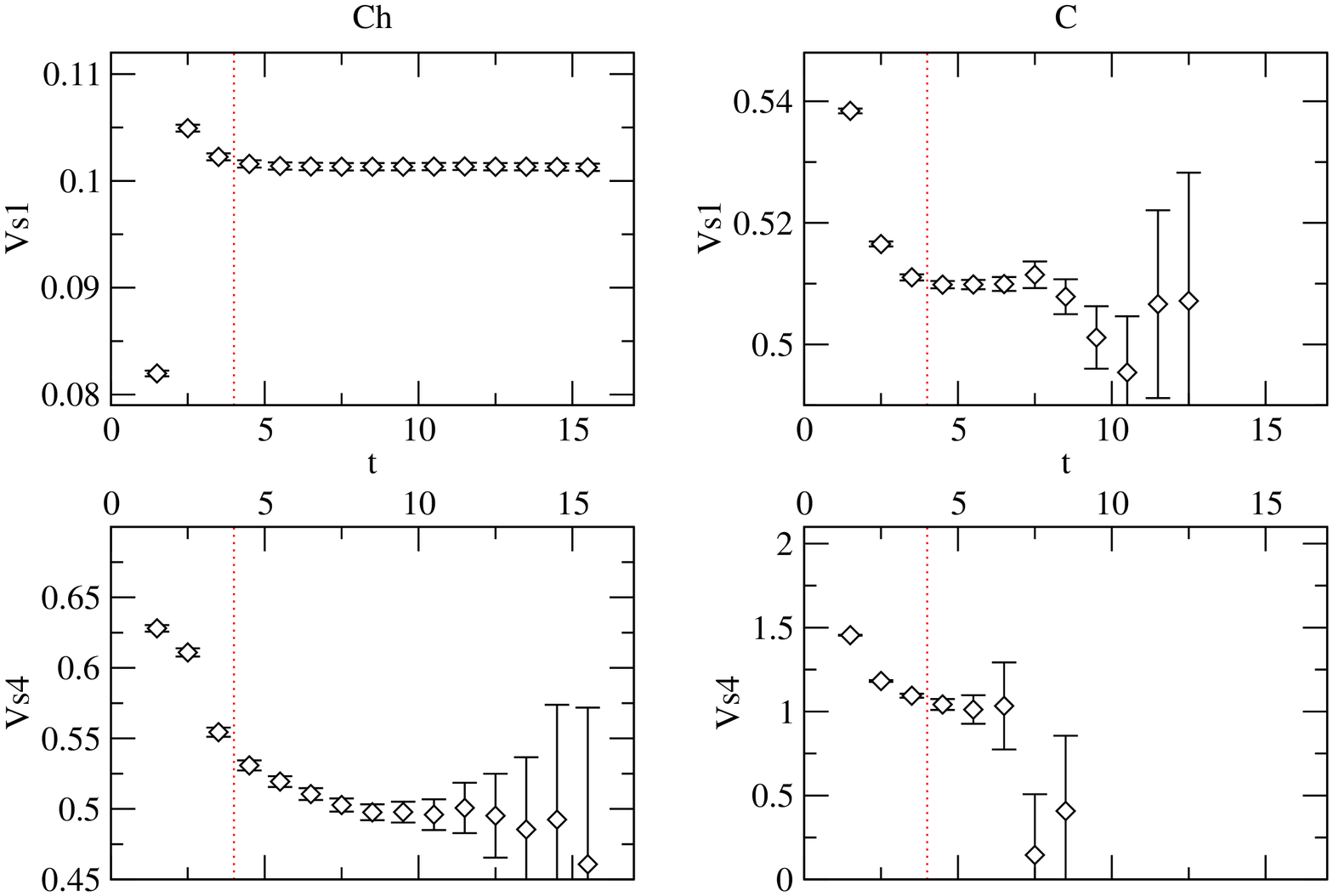}
\mycaption{Effective potential for ensembles \ensemble{C ~ hyp} and
\ensemble{C}.  The minimum time separation chosen is $t_\text{min} = 4$.}
\end{figure}
 
\begin{figure}
\centering
\psfrag{t}[t][t]{$t$}
\psfrag{Vs1}[b][B]{$a V_\text{eff} ( s = 1 )$}
\psfrag{Vs2}[b][B]{$a V_\text{eff} ( s = 2 )$}
\psfrag{W}[B][B]{\ensemble{W}}
\psfrag{Wh}[B][B]{\ensemble{W ~ hyp}}
\includegraphics[width=\textwidth,clip=]{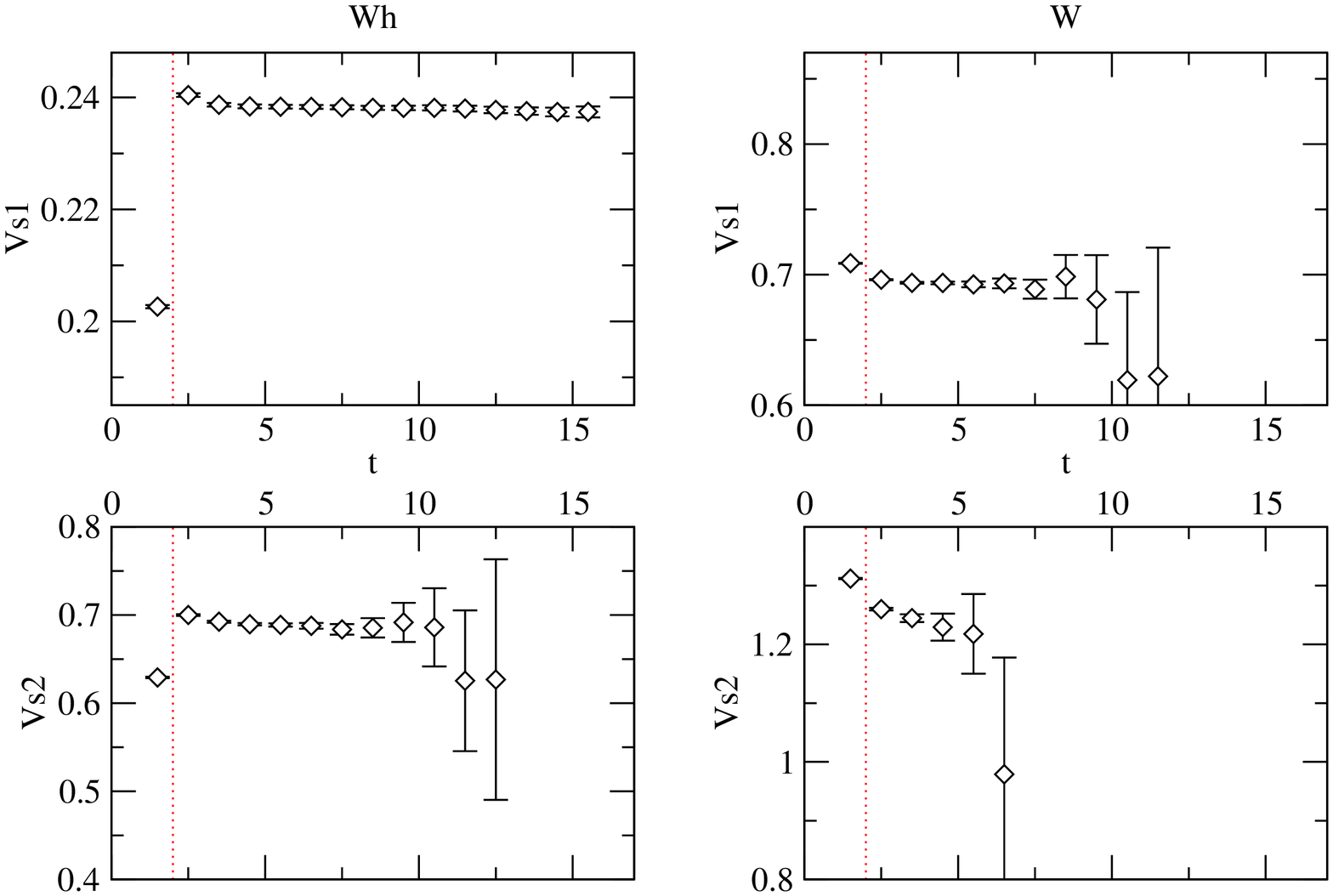}
\mycaption{Effective potential for ensembles \ensemble{W ~ hyp} and
\ensemble{W}.  The minimum time separation chosen is $t_\text{min} = 2$.}
\end{figure}

\begin{figure}
\centering
\psfrag{t}[t][t]{$t$}
\psfrag{Vs1}[b][B]{$a V_\text{eff} ( s = 1 )$}
\psfrag{Vs2}[b][B]{$a V_\text{eff} ( s = 2 )$}
\psfrag{X}[B][B]{\ensemble{X}}
\psfrag{Xh}[B][B]{\ensemble{X ~ hyp}}
\includegraphics[width=\textwidth,clip=]{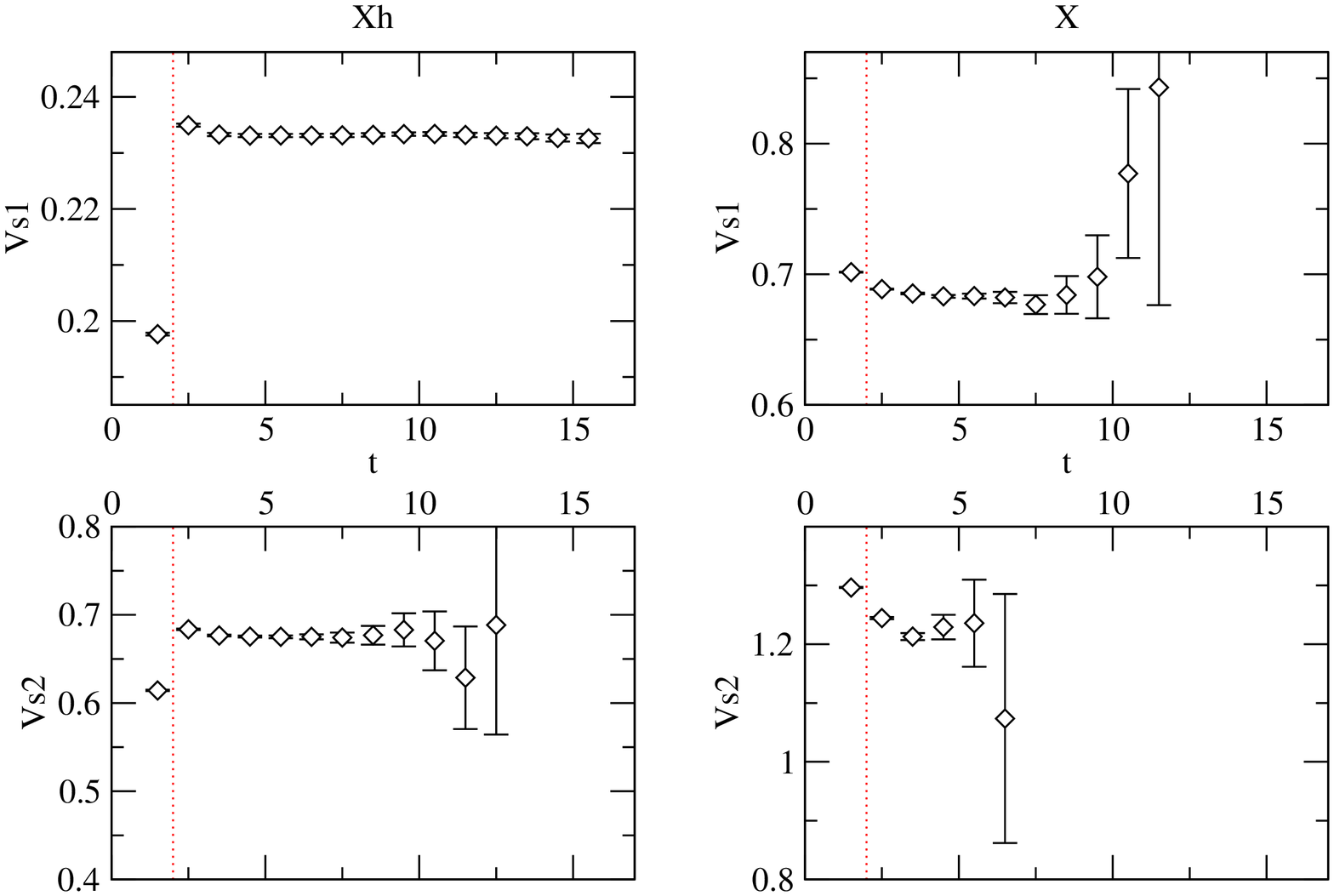}
\mycaption{Effective potential for ensembles \ensemble{X ~ hyp} and
\ensemble{X}.  The minimum time separation chosen is $t_\text{min} = 2$.}
\end{figure}

\begin{figure}
\centering
\psfrag{t}[t][t]{$t$}
\psfrag{Vs1}[b][B]{$a V_\text{eff} ( s = 1 )$}
\psfrag{Vs2}[b][B]{$a V_\text{eff} ( s = 2 )$}
\psfrag{Y}[B][B]{\ensemble{Y}}
\psfrag{Yh}[B][B]{\ensemble{Y ~ hyp}}
\includegraphics[width=\textwidth,clip=]{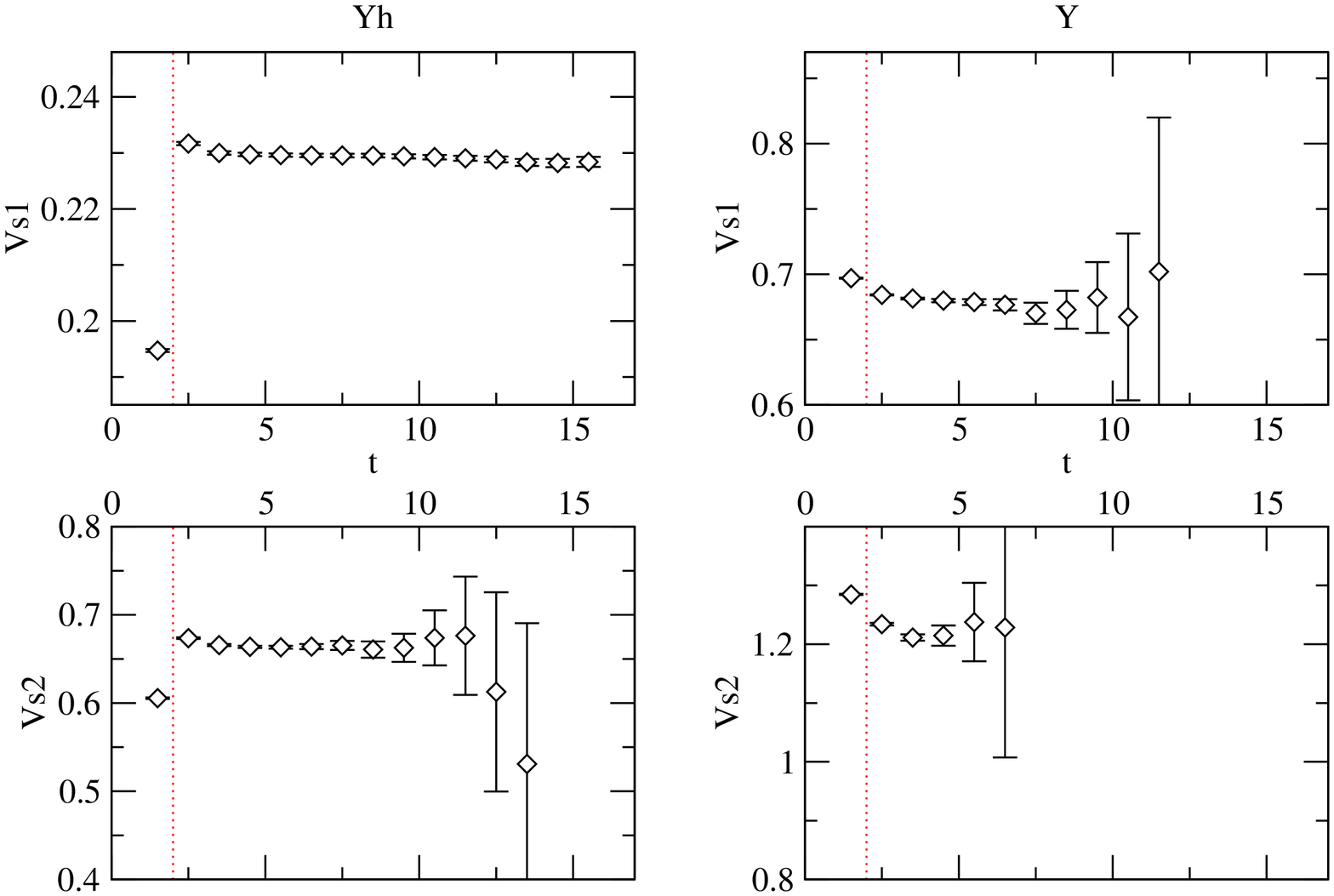}
\mycaption{Effective potential for ensembles \ensemble{Y ~ hyp} and
\ensemble{Y}.  The minimum time separation chosen is $t_\text{min} = 2$.}
\end{figure}

\begin{figure}
\centering
\psfrag{t}[t][t]{$t$}
\psfrag{Vs1}[b][B]{$a V_\text{eff} ( s = 1 )$}
\psfrag{Vs2}[b][B]{$a V_\text{eff} ( s = 2 )$}
\psfrag{Z}[B][B]{\ensemble{Z}}
\psfrag{Zh}[B][B]{\ensemble{Z ~ hyp}}
\includegraphics[width=\textwidth,clip=]{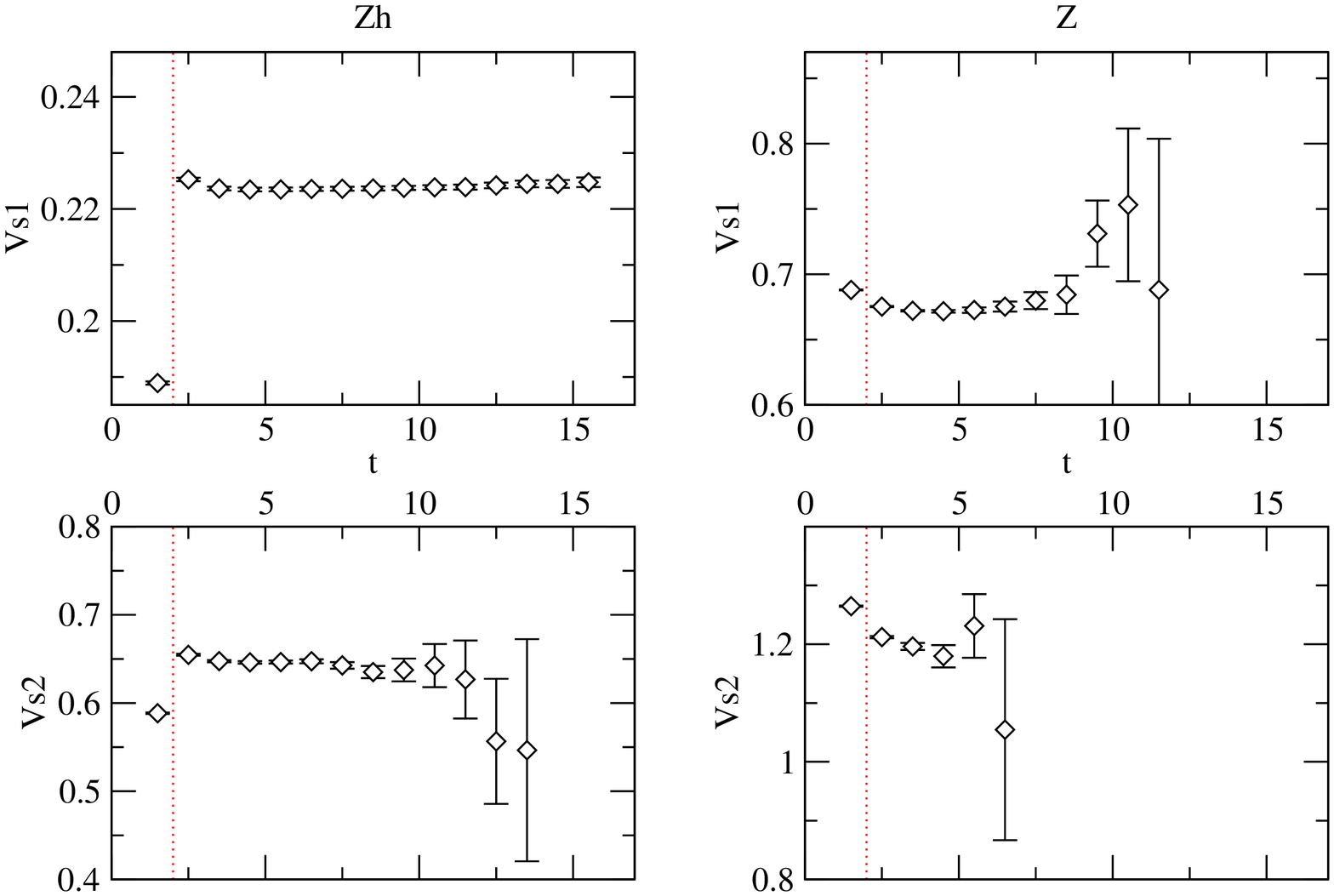}
\mycaption{Effective potential for ensembles \ensemble{Z ~ hyp} and
\ensemble{Z}.  The minimum time separation chosen is $t_\text{min} = 2$.}
\end{figure}

\begin{figure}
\centering
\psfrag{t}[t][t]{$t$}
\psfrag{Vs1}[b][B]{$a V_\text{eff} ( s = 1 )$}
\psfrag{Vs4}[b][B]{$a V_\text{eff} ( s = 4 )$}
\psfrag{Q}[B][B]{\ensemble{Q}}
\psfrag{Qh}[B][B]{\ensemble{Q ~ hyp}}
\includegraphics[width=\textwidth,clip=]{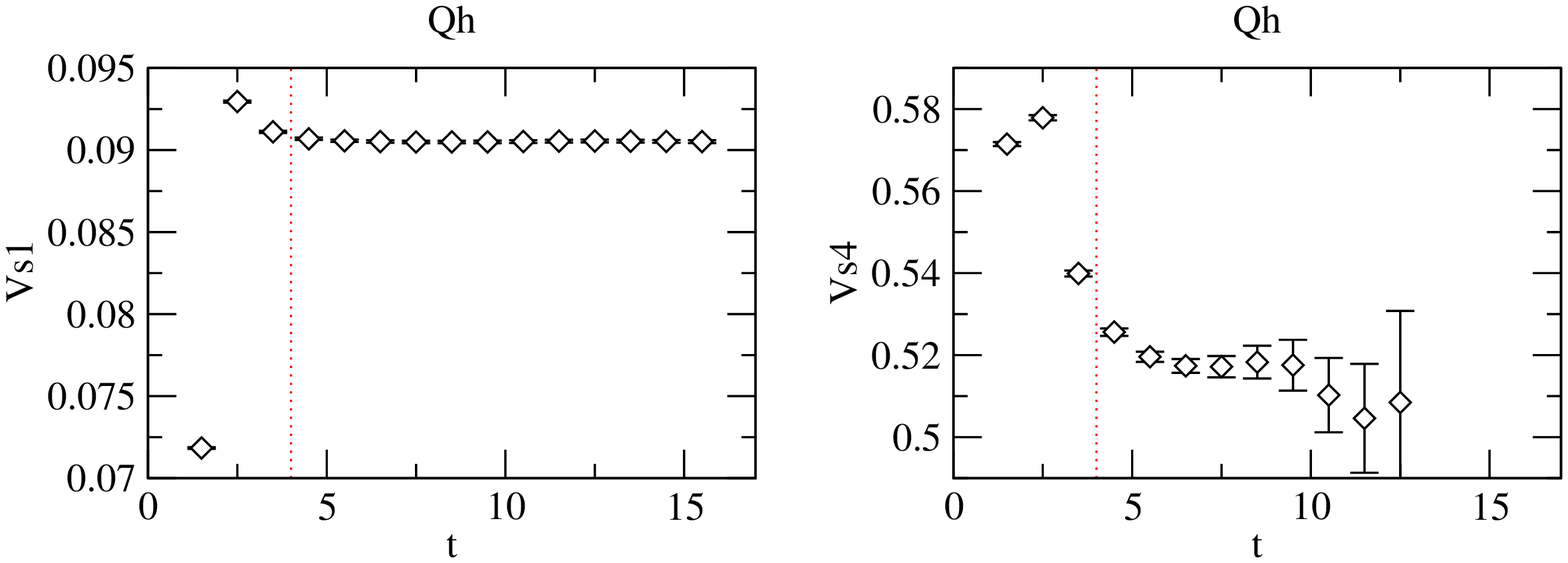}
\mycaption{Effective potential for ensemble \ensemble{Q ~ hyp}.  The minimum time
separation chosen is $t_\text{min} = 4$.}
\label{e:figure}
\end{figure}

\clearpage

\begin{figure}
\centering
\psfrag{aV}[b][B]{\Large $a V_\text{corr}$}
\psfrag{s}[t][t]{\Large $s$}
\psfrag{Ah}[B][B]{\Large \ensemble{A ~ hyp}}
\psfrag{r0a}[r][r]{\Large $r_0 / a$}
\includegraphics[width=\textwidth,clip=]{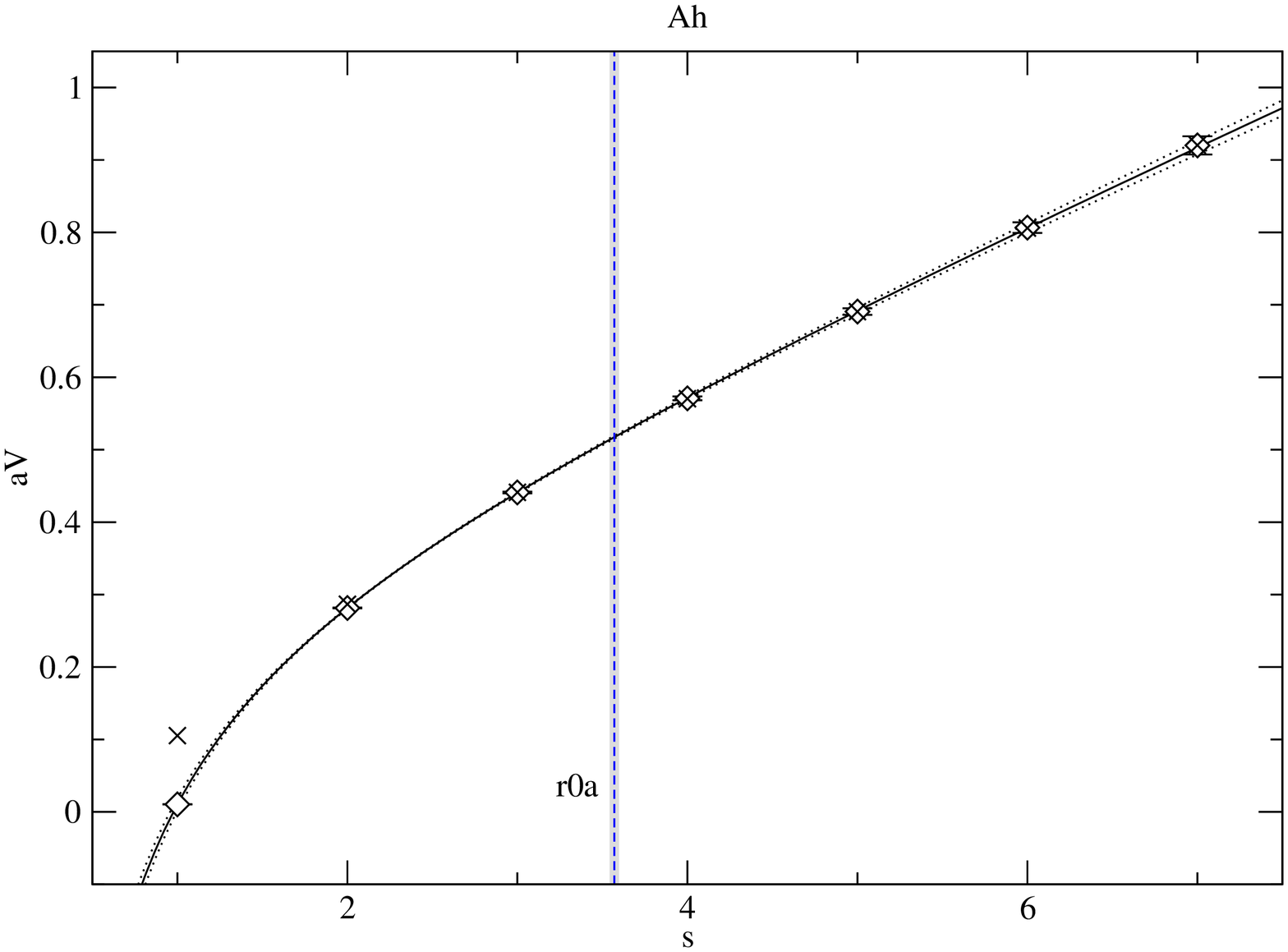}
\mycaption{Static quark potential for ensemble \ensemble{A ~ hyp}.  The
$\times$'s correspond to the uncorrected static quark potential, while the diamonds
correspond to the corrected potential.  The result is $r_0 / a = 3.570(27)$.}
\label{f:figure}
\end{figure}

\begin{figure}
\centering
\psfrag{aV}[b][B]{\Large $a V_\text{corr}$}
\psfrag{s}[t][t]{\Large $s$}
\psfrag{Bh}[B][B]{\Large \ensemble{B ~ hyp}}
\psfrag{r0a}[r][r]{\Large $r_0 / a$}
\includegraphics[width=\textwidth,clip=]{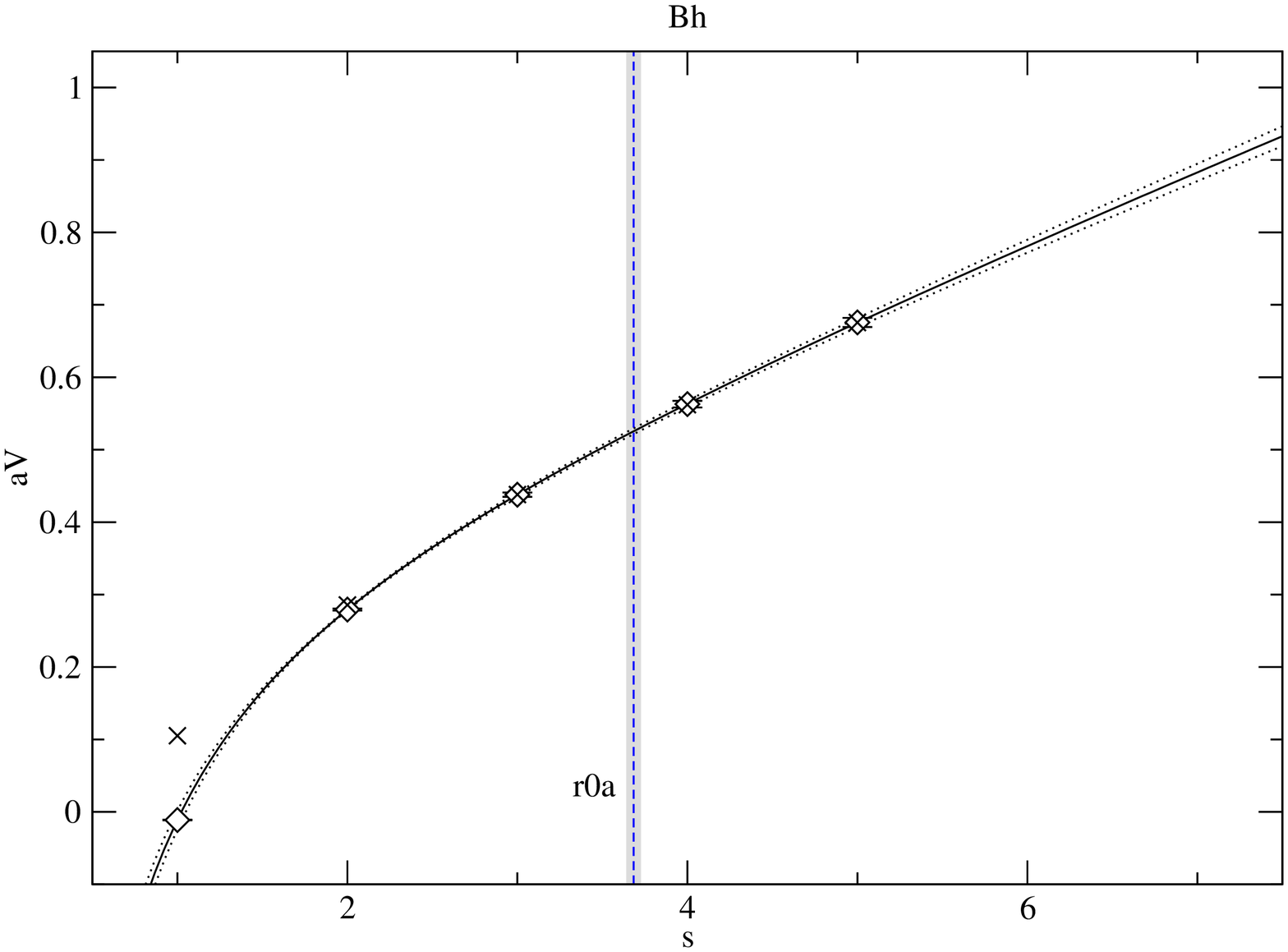}
\mycaption{Static quark potential for ensemble \ensemble{B ~ hyp}.  The
$\times$'s correspond to the uncorrected static quark potential, while the diamonds
correspond to the corrected potential.  The result is $r_0 / a = 3.684(44)$.}
\end{figure}

\begin{figure}
\centering
\psfrag{aV}[b][B]{\Large $a V_\text{corr}$}
\psfrag{s}[t][t]{\Large $s$}
\psfrag{Ch}[B][B]{\Large \ensemble{C ~ hyp}}
\psfrag{r0a}[r][r]{\Large $r_0 / a$}
\includegraphics[width=\textwidth,clip=]{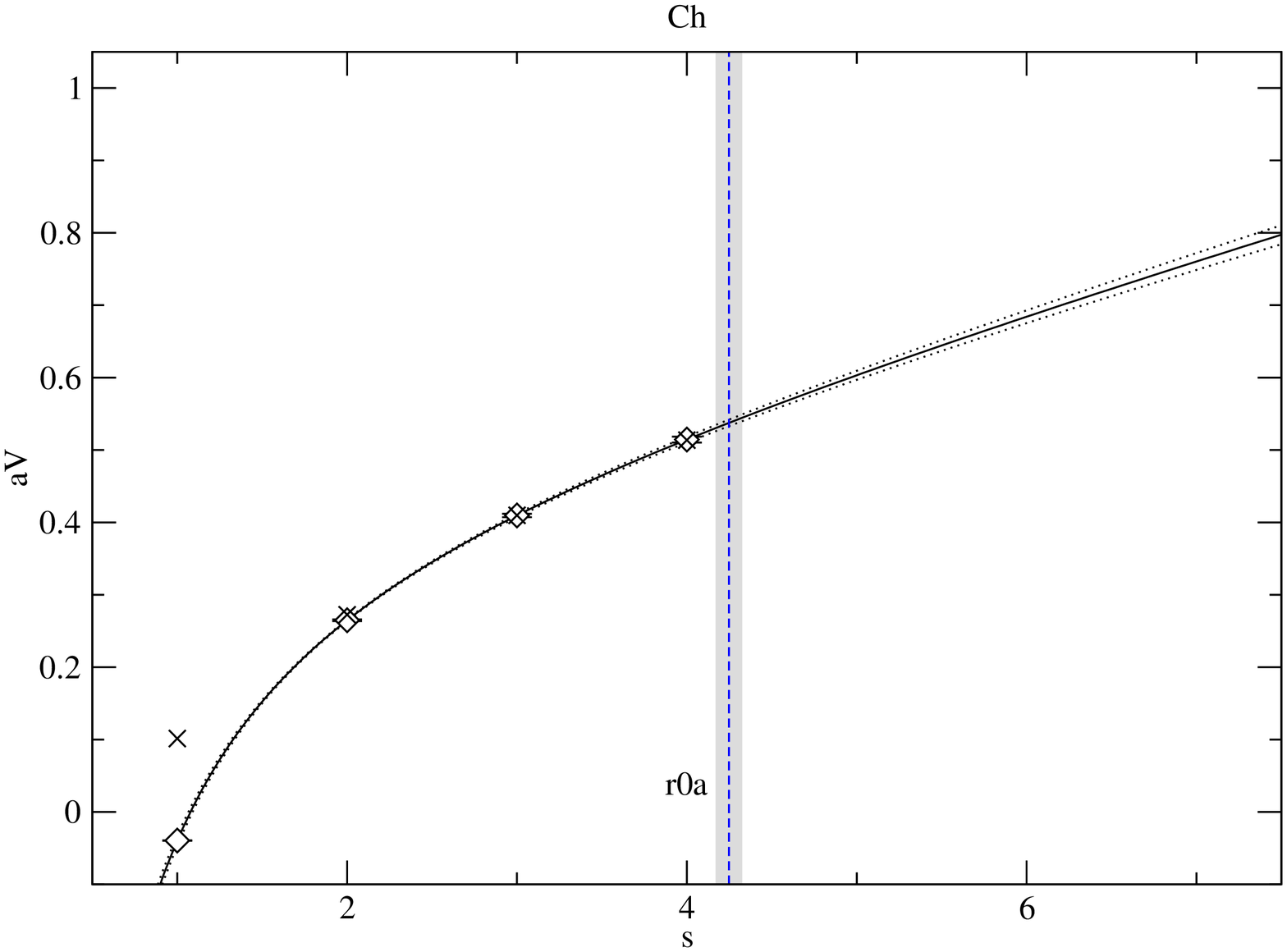}
\mycaption{Static quark potential for ensemble \ensemble{C ~ hyp}.  The
$\times$'s correspond to the uncorrected static quark potential, while the diamonds
correspond to the corrected potential.  The result is $r_0 / a = 4.248(78)$.}
\end{figure}

\begin{figure}
\centering
\psfrag{aV}[b][B]{\Large $a V_\text{corr}$}
\psfrag{s}[t][t]{\Large $s$}
\psfrag{A}[B][B]{\Large \ensemble{A}}
\psfrag{r0a}[r][r]{\Large $r_0 / a$}
\includegraphics[width=\textwidth,clip=]{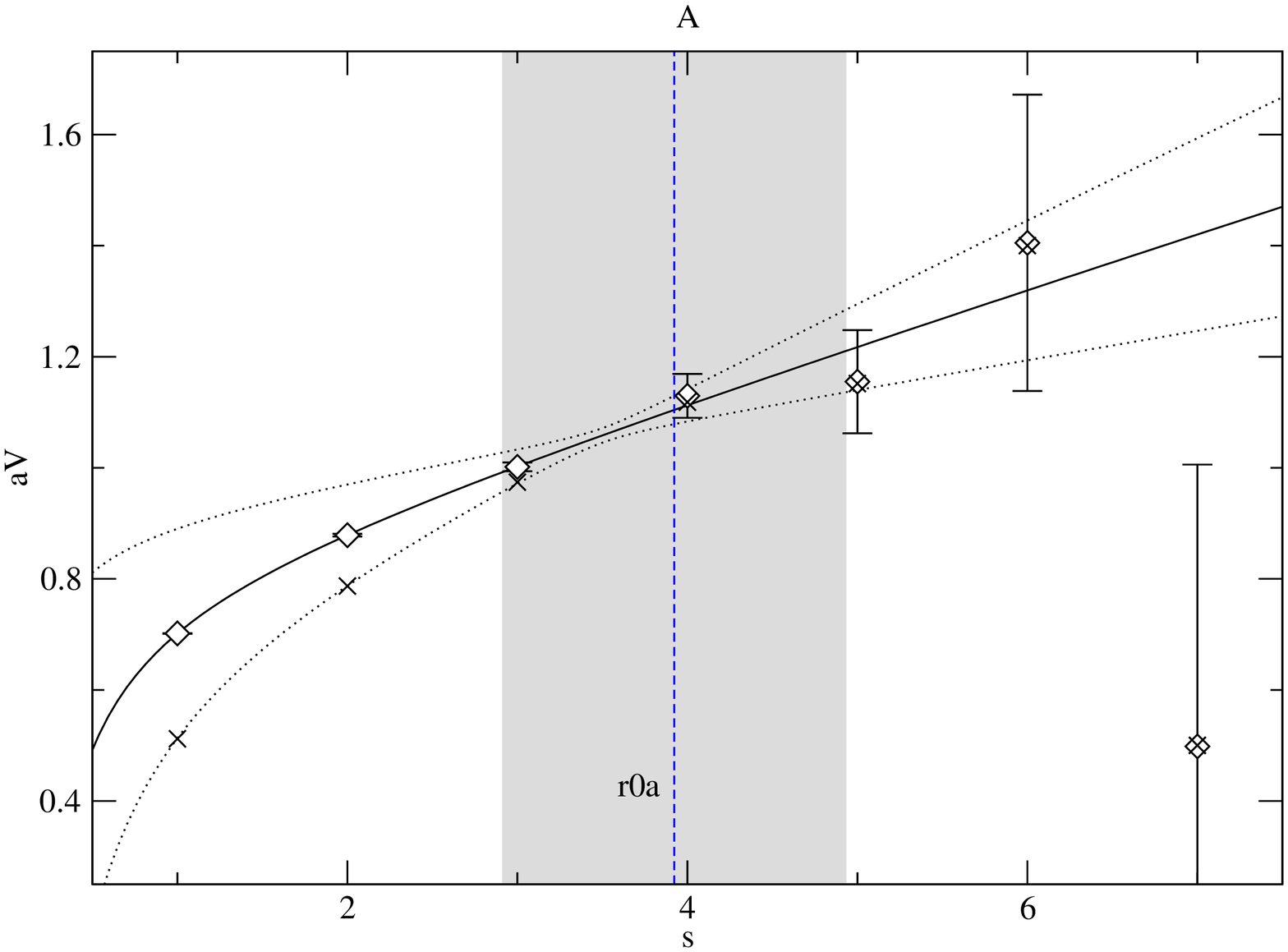}
\mycaption{Static quark potential for ensemble \ensemble{A}.  The $\times$'s
correspond to the uncorrected static quark potential, while the diamonds
correspond to the corrected potential.  The result is $r_0 / a = 3.9(10)$.}
\end{figure}

\begin{figure}
\centering
\psfrag{aV}[b][B]{\Large $a V_\text{corr}$}
\psfrag{s}[t][t]{\Large $s$}
\psfrag{B}[B][B]{\Large \ensemble{B}}
\psfrag{r0a}[r][r]{\Large $r_0 / a$}
\includegraphics[width=\textwidth,clip=]{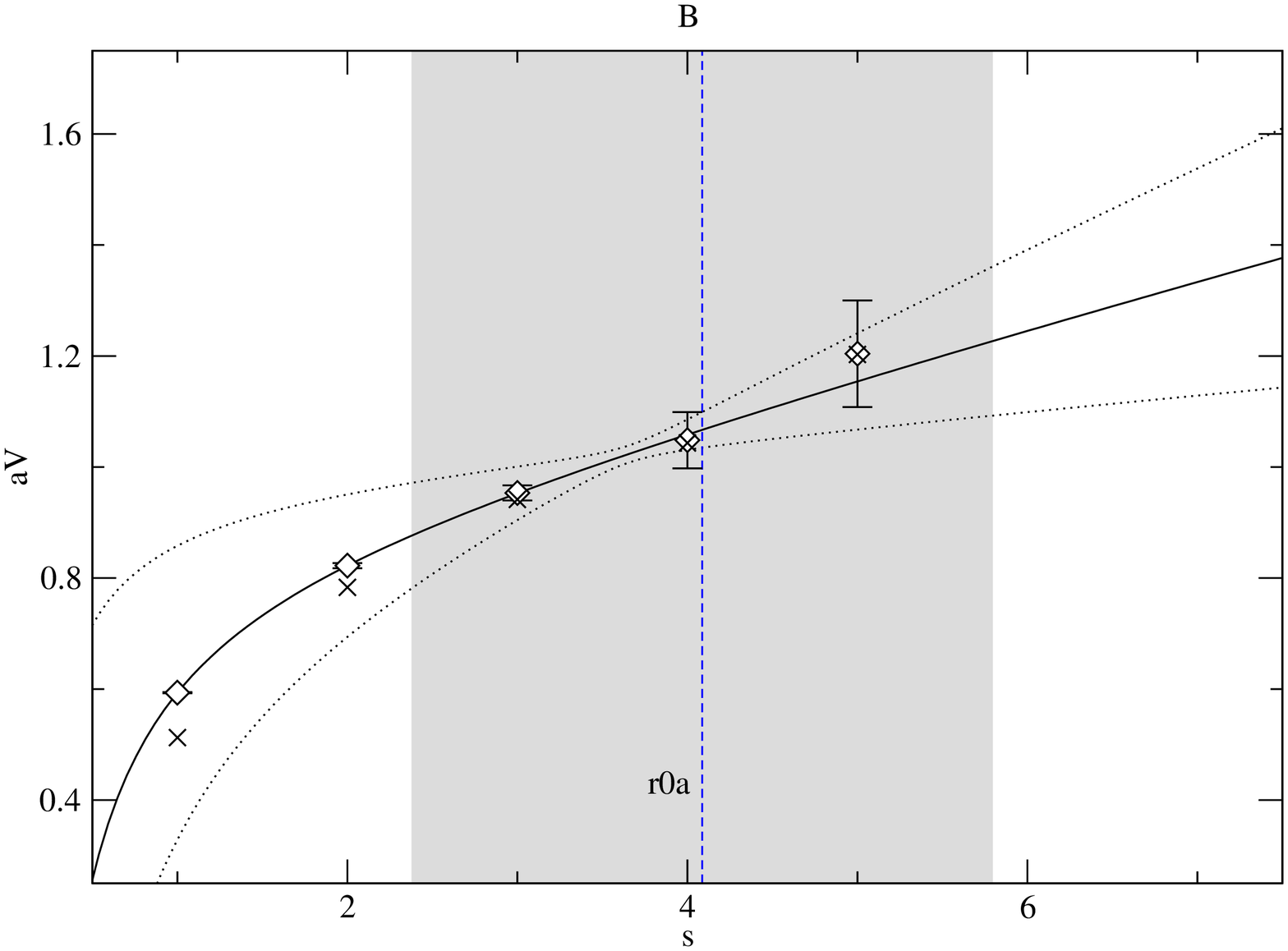}
\mycaption{Static quark potential for ensemble \ensemble{B}.  The $\times$'s
correspond to the uncorrected static quark potential, while the diamonds
correspond to the corrected potential.  The result is $r_0 / a = 4.1(17)$.}
\end{figure}

\begin{figure}
\centering
\psfrag{aV}[b][B]{\Large $a V_\text{corr}$}
\psfrag{s}[t][t]{\Large $s$}
\psfrag{C}[B][B]{\Large \ensemble{C}}
\psfrag{r0a}[r][r]{\Large $r_0 / a$}
\includegraphics[width=\textwidth,clip=]{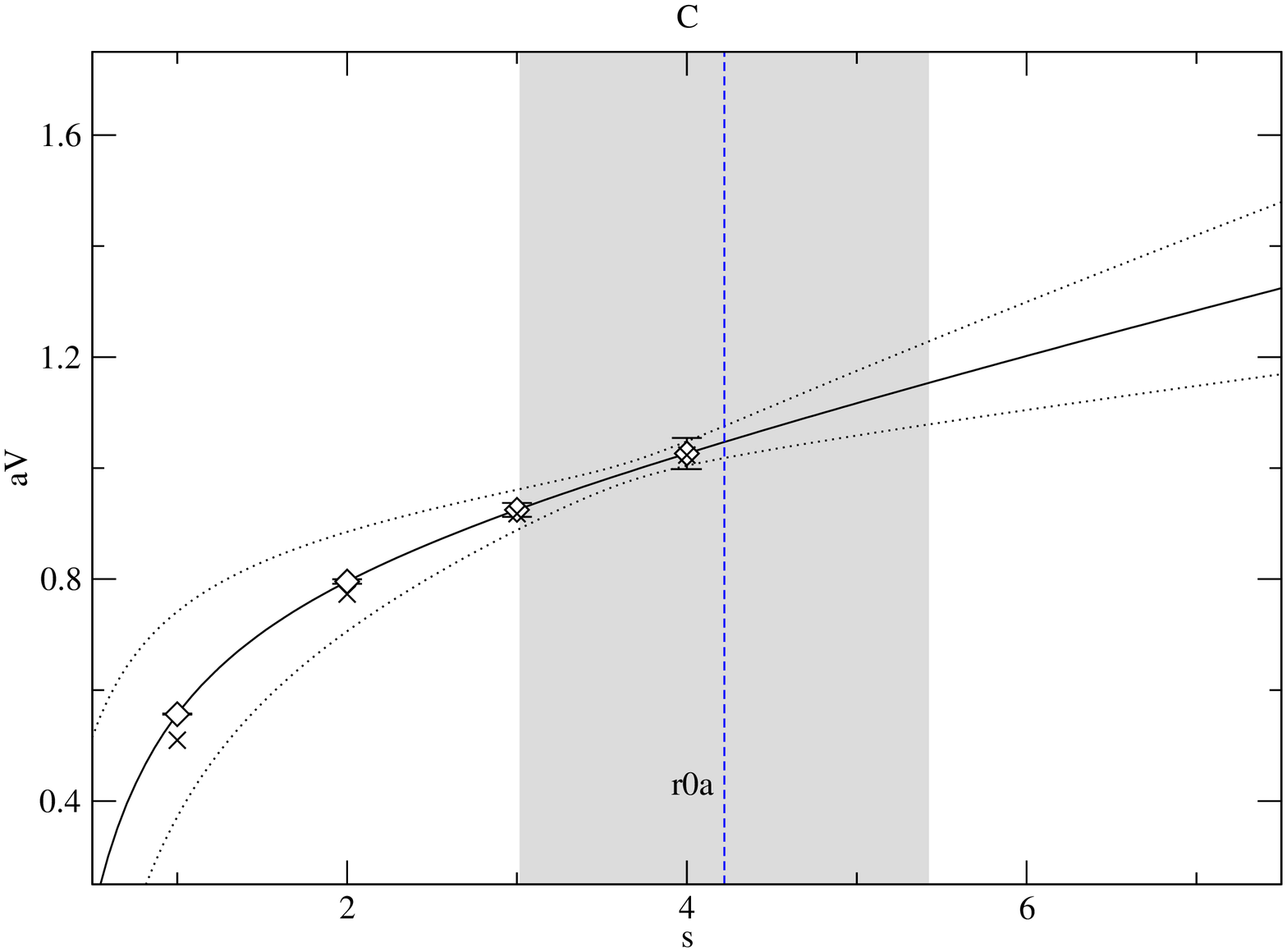}
\mycaption{Static quark potential for ensemble \ensemble{C}.  The $\times$'s
correspond to the uncorrected static quark potential, while the diamonds
correspond to the corrected potential.  The result is $r_0 / a = 4.2(12)$.}
\end{figure}

\begin{figure}
\centering
\psfrag{aV}[b][B]{\Large $a V_\text{corr}$}
\psfrag{s}[t][t]{\Large $s$}
\psfrag{Wh}[B][B]{\Large \ensemble{W ~ hyp}}
\psfrag{r0a}[l][l]{\Large $r_0 / a$}
\includegraphics[width=\textwidth,clip=]{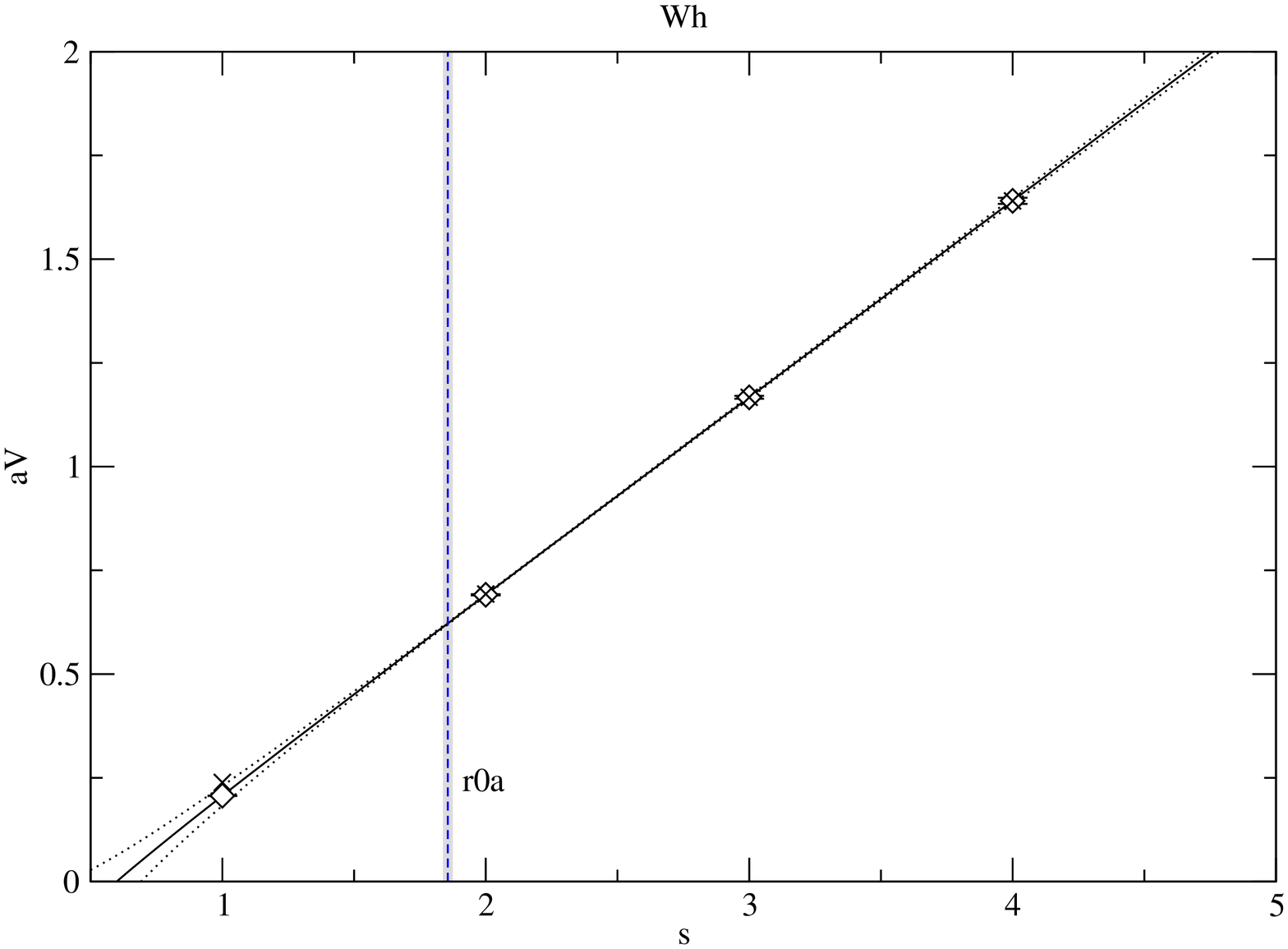}
\mycaption{Static quark potential for ensemble \ensemble{W ~ hyp}.  The
$\times$'s correspond to the uncorrected static quark potential, while the diamonds
correspond to the corrected potential.  The result is $r_0 / a = 1.856(19)$.}
\end{figure}

\begin{figure}
\centering
\psfrag{aV}[b][B]{\Large $a V_\text{corr}$}
\psfrag{s}[t][t]{\Large $s$}
\psfrag{Xh}[B][B]{\Large \ensemble{X ~ hyp}}
\psfrag{r0a}[l][l]{\Large $r_0 / a$}
\includegraphics[width=\textwidth,clip=]{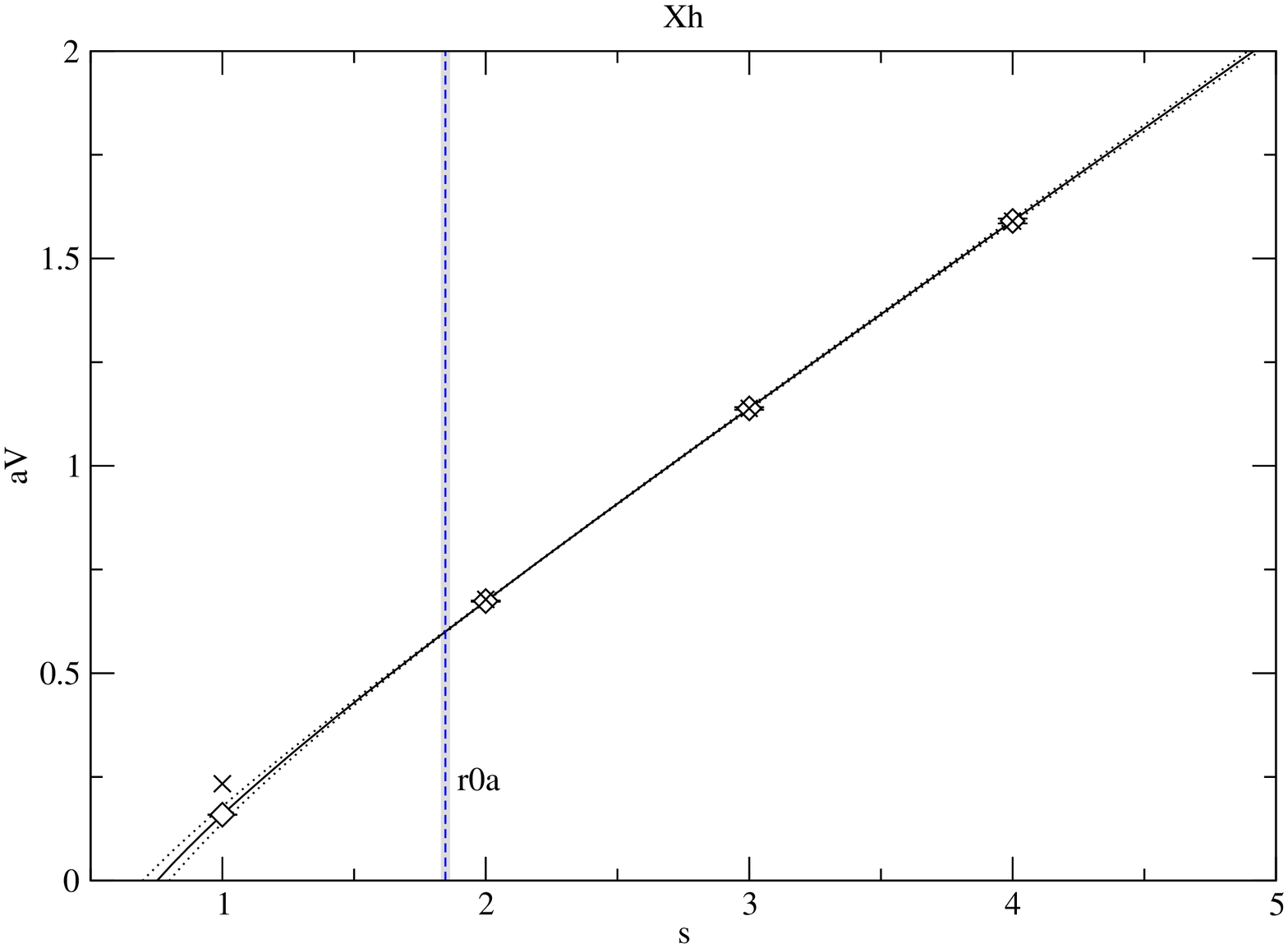}
\mycaption{Static quark potential for ensemble \ensemble{X ~ hyp}.  The
$\times$'s correspond to the uncorrected static quark potential, while the diamonds
correspond to the corrected potential.  The result is $r_0 / a = 1.847(17)$.}
\end{figure}

\begin{figure}
\centering
\psfrag{aV}[b][B]{\Large $a V_\text{corr}$}
\psfrag{s}[t][t]{\Large $s$}
\psfrag{Yh}[B][B]{\Large \ensemble{Y ~ hyp}}
\psfrag{r0a}[l][l]{\Large $r_0 / a$}
\includegraphics[width=\textwidth,clip=]{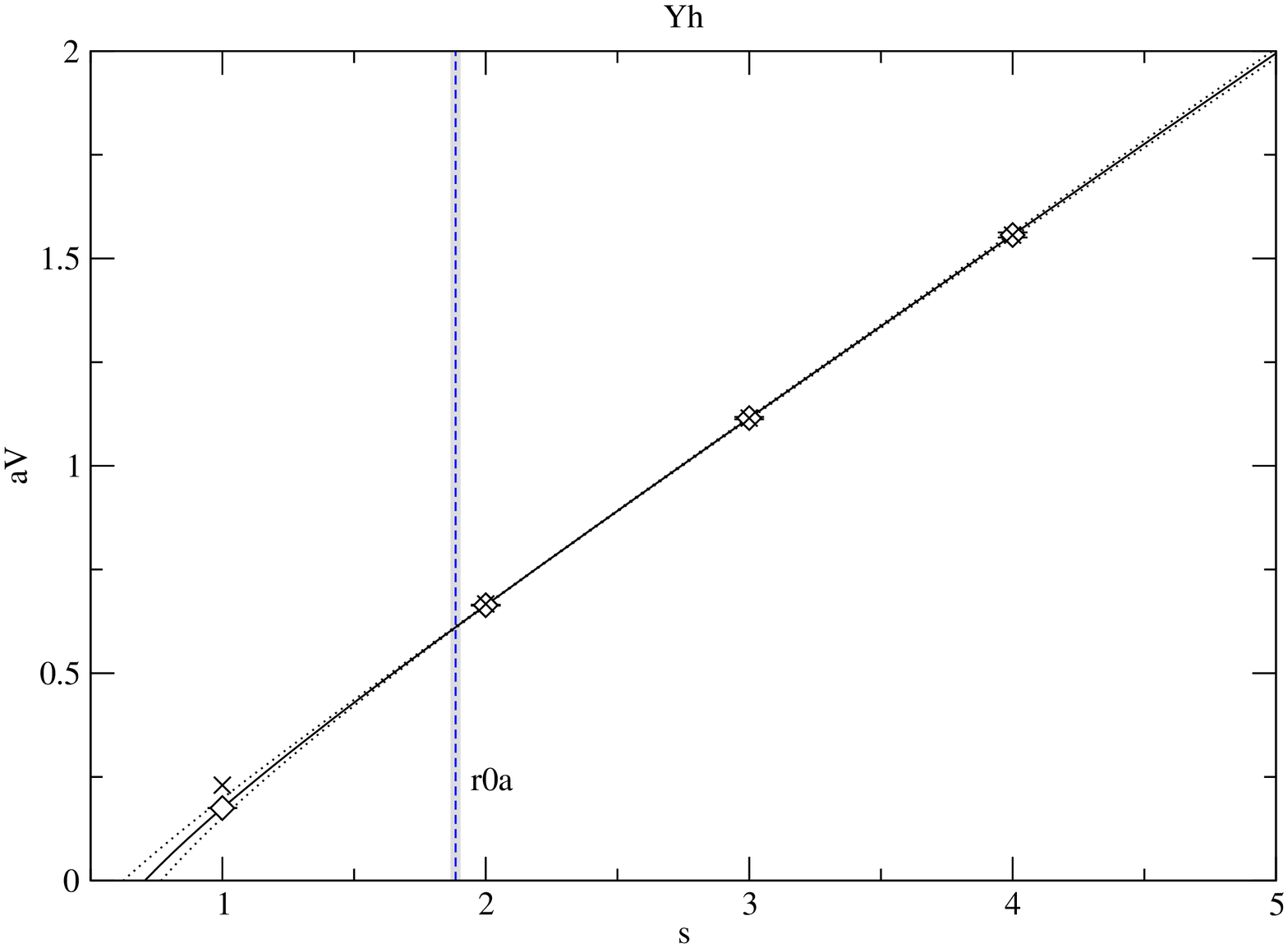}
\mycaption{Static quark potential for ensemble \ensemble{Y ~ hyp}.  The
$\times$'s correspond to the uncorrected static quark potential, while the diamonds
correspond to the corrected potential.  The result is $r_0 / a = 1.885(19)$.}
\end{figure}

\begin{figure}
\centering
\psfrag{aV}[b][B]{\Large $a V_\text{corr}$}
\psfrag{s}[t][t]{\Large $s$}
\psfrag{Zh}[B][B]{\Large \ensemble{Z ~ hyp}}
\psfrag{r0a}[l][l]{\Large $r_0 / a$}
\includegraphics[width=\textwidth,clip=]{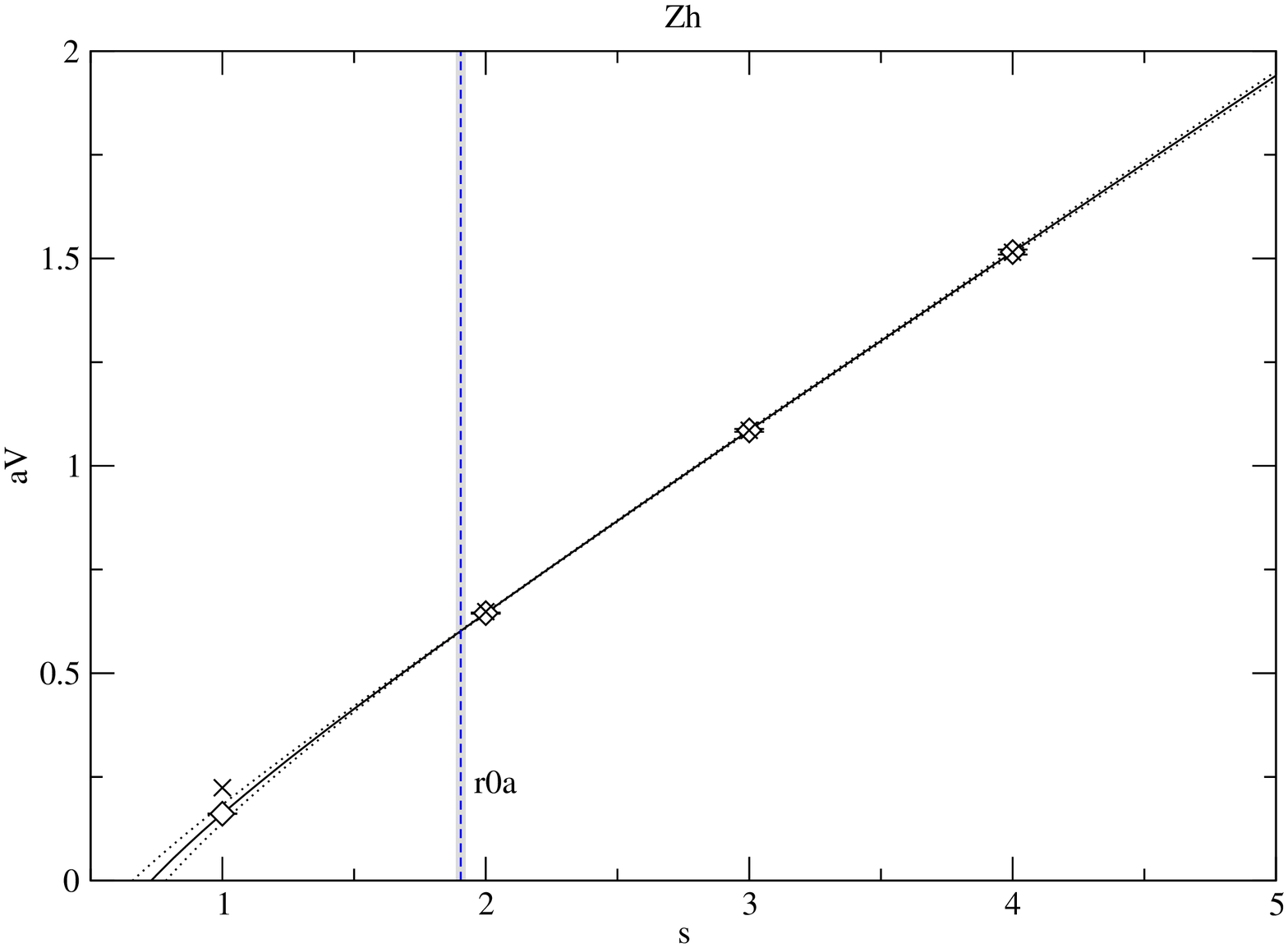}
\mycaption{Static quark potential for ensemble \ensemble{Z ~ hyp}.  The
$\times$'s correspond to the uncorrected static quark potential, while the diamonds
correspond to the corrected potential.  The result is $r_0 / a = 1.905(19)$.}
\end{figure}

\begin{figure}
\centering
\psfrag{aV}[b][B]{\Large $a V_\text{corr}$}
\psfrag{s}[t][t]{\Large $s$}
\psfrag{W}[B][B]{\Large \ensemble{W}}
\psfrag{r0a}[l][l]{\Large $r_0 / a$}
\includegraphics[width=\textwidth,clip=]{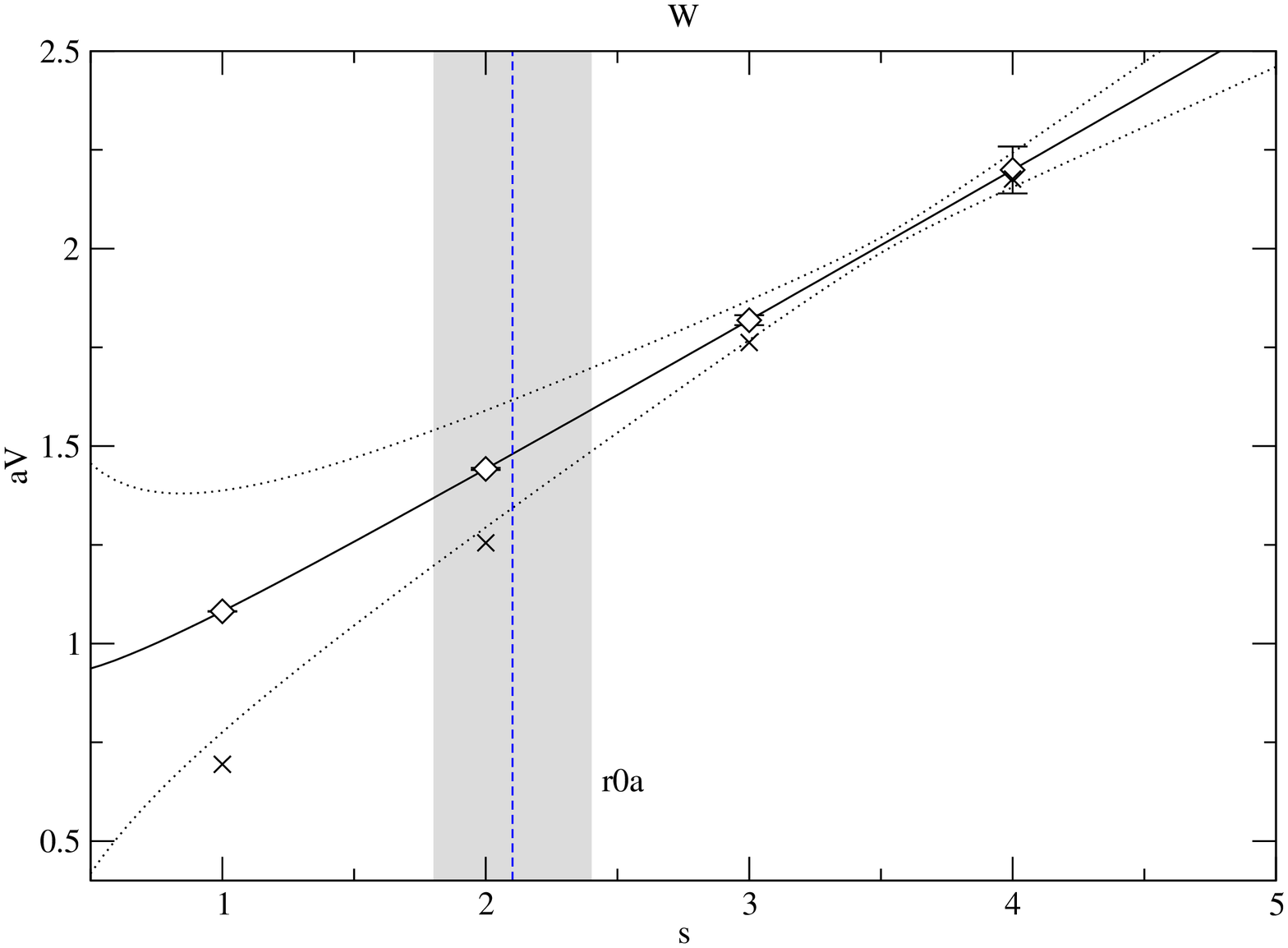}
\mycaption{Static quark potential for ensemble \ensemble{W}.  The $\times$'s
correspond to the uncorrected static quark potential, while the diamonds
correspond to the corrected potential.  The result is $r_0 / a = 2.10(30)$.}
\label{h:figure}
\end{figure}

\begin{figure}
\centering
\psfrag{aV}[b][B]{\Large $a V_\text{corr}$}
\psfrag{s}[t][t]{\Large $s$}
\psfrag{X}[B][B]{\Large \ensemble{X}}
\psfrag{r0a}[l][l]{\Large $r_0 / a$}
\includegraphics[width=\textwidth,clip=]{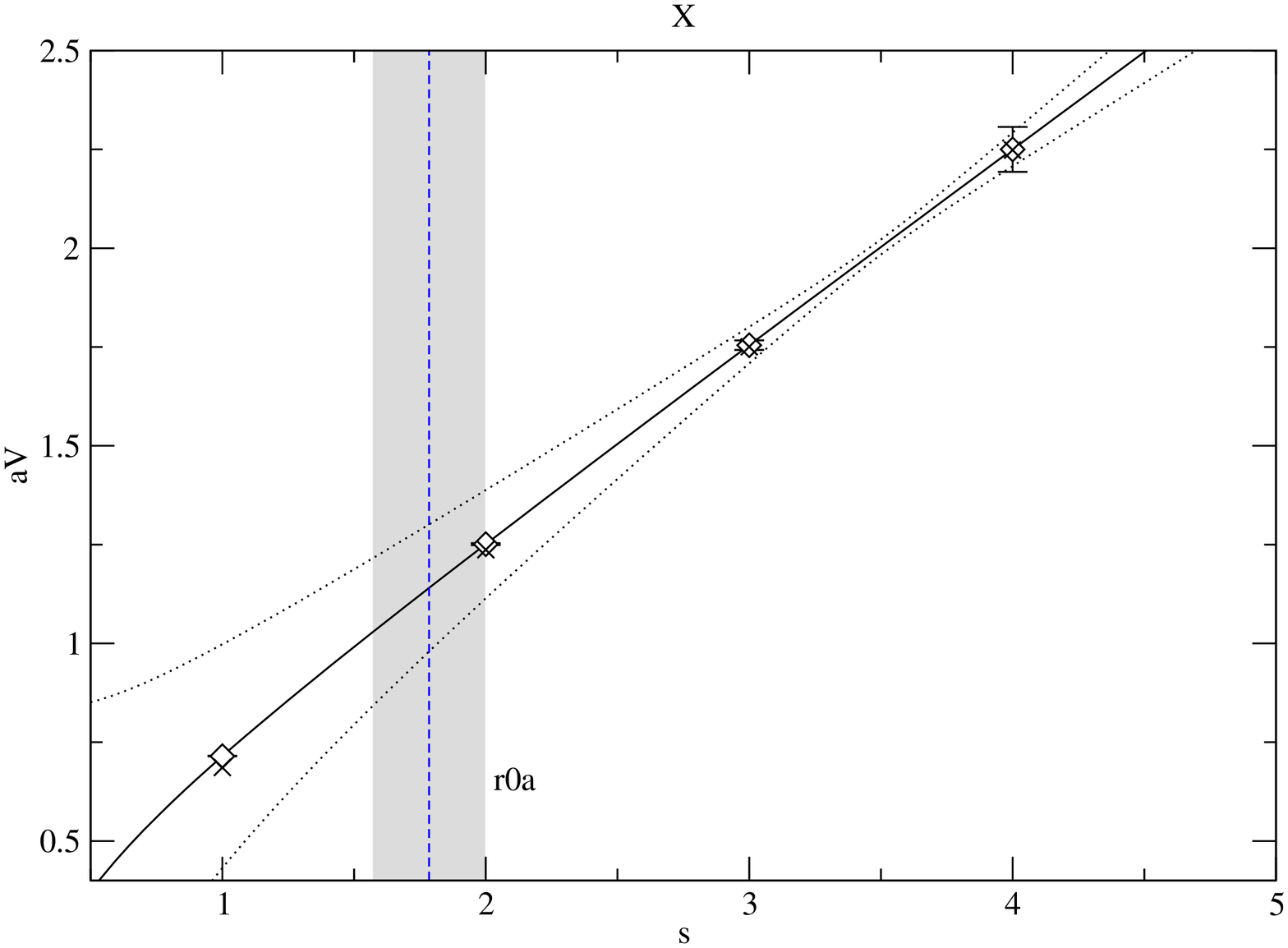}
\mycaption{Static quark potential for ensemble \ensemble{X}.  The $\times$'s
correspond to the uncorrected static quark potential, while the diamonds
correspond to the corrected potential.  The result is $r_0 / a = 1.78(21)$.}
\end{figure}

\begin{figure}
\centering
\psfrag{aV}[b][B]{\Large $a V_\text{corr}$}
\psfrag{s}[t][t]{\Large $s$}
\psfrag{Y}[B][B]{\Large \ensemble{Y}}
\psfrag{r0a}[l][l]{\Large $r_0 / a$}
\includegraphics[width=\textwidth,clip=]{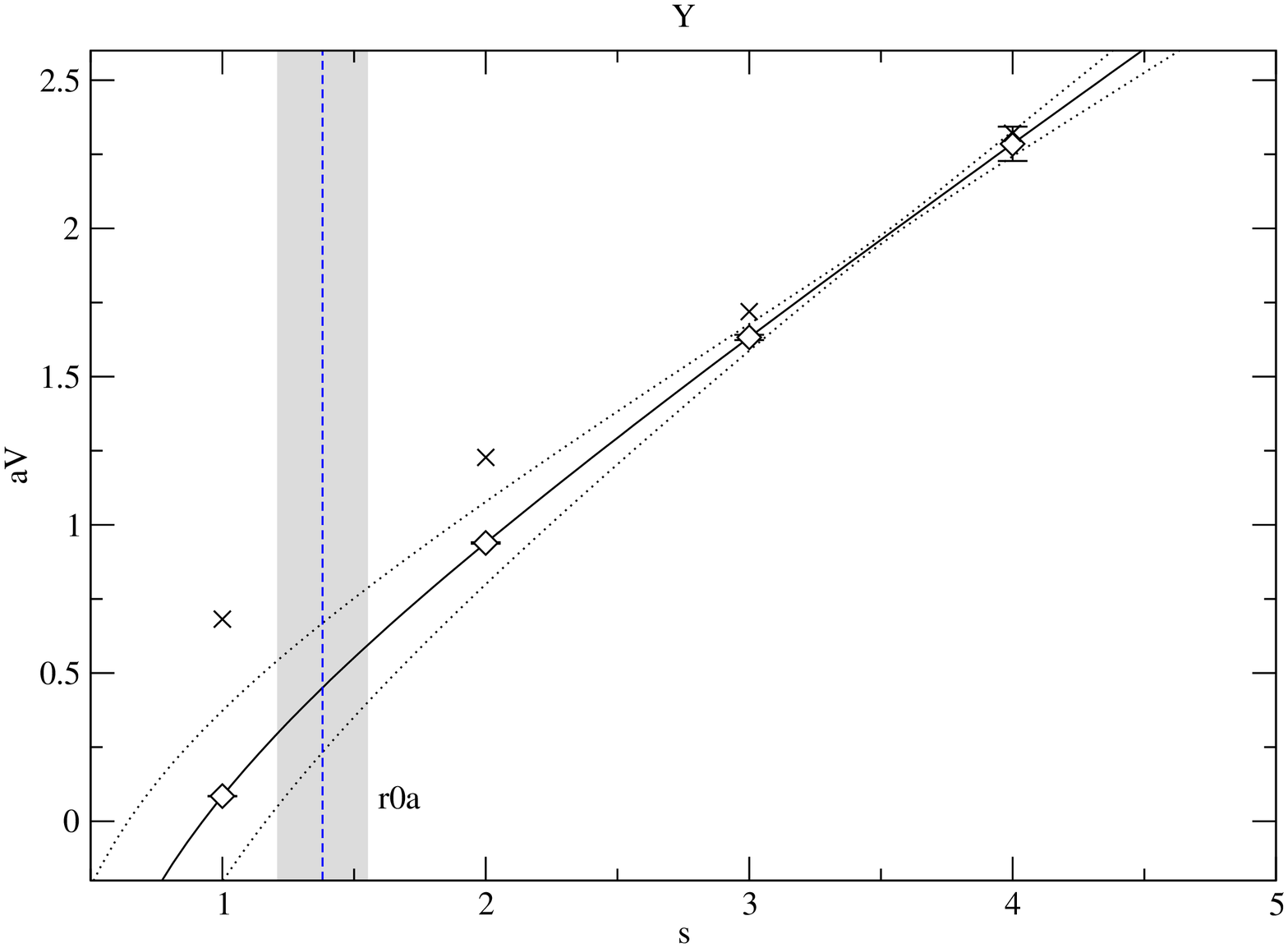}
\mycaption{Static quark potential for ensemble \ensemble{Y}.  The $\times$'s
correspond to the uncorrected static quark potential, while the diamonds
correspond to the corrected potential.  The result is $r_0 / a = 1.38(17)$.}
\end{figure}

\begin{figure}
\centering
\psfrag{aV}[b][B]{\Large $a V_\text{corr}$}
\psfrag{s}[t][t]{\Large $s$}
\psfrag{Z}[B][B]{\Large \ensemble{Z}}
\psfrag{r0a}[l][l]{\Large $r_0 / a$}
\includegraphics[width=\textwidth,clip=]{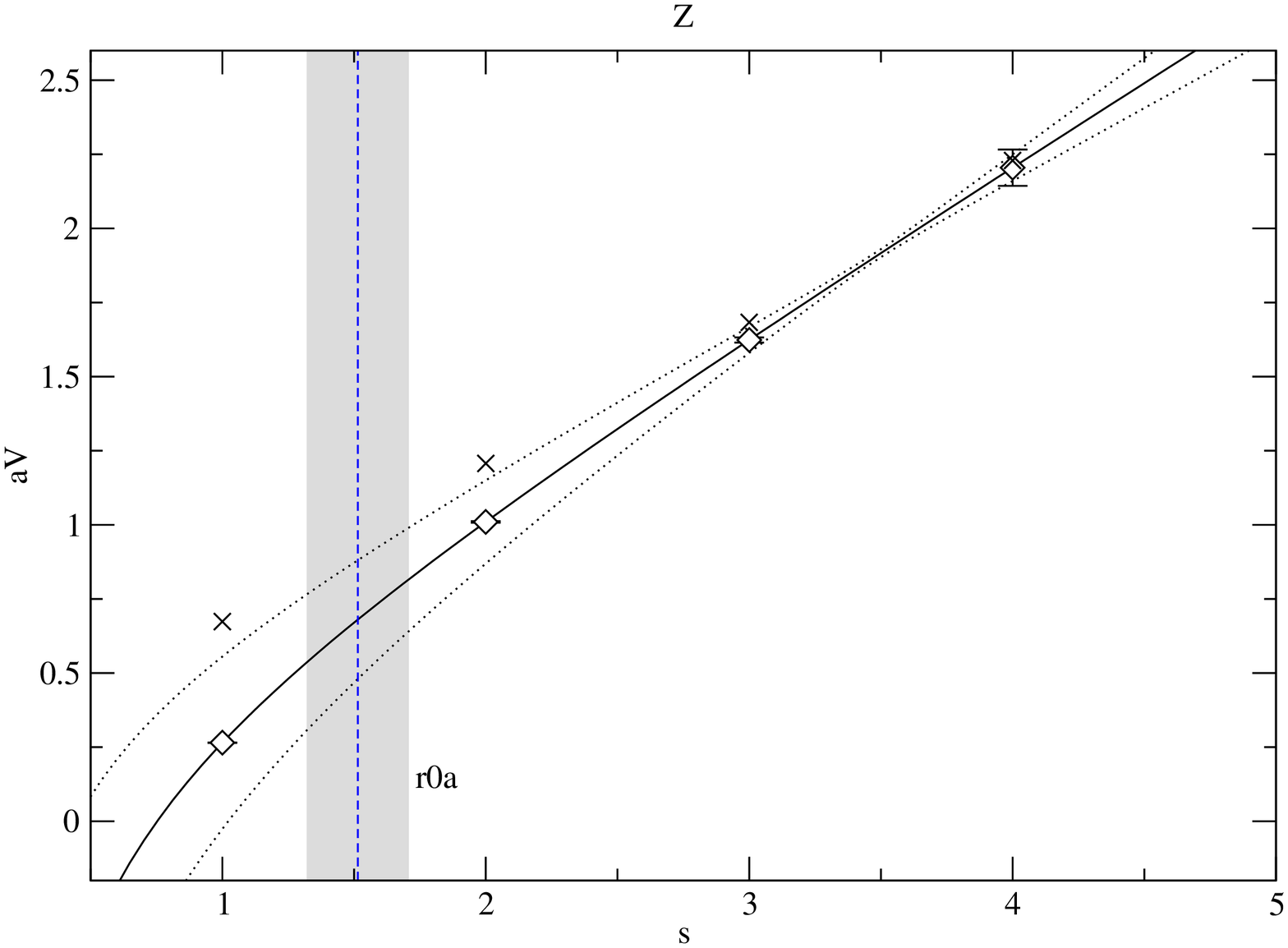}
\mycaption{Static quark potential for ensemble \ensemble{Z}.  The $\times$'s
correspond to the uncorrected static quark potential, while the diamonds
correspond to the corrected potential.  The result is $r_0 / a = 1.51(19)$.}
\label{i:figure}
\end{figure}

\begin{figure}
\centering
\psfrag{aV}[b][B]{\Large $a V_\text{corr}$}
\psfrag{s}[t][t]{\Large $s$}
\psfrag{Qh}[B][B]{\Large \ensemble{Q ~ hyp}}
\psfrag{r0a}[r][r]{\Large $r_0 / a$}
\includegraphics[width=\textwidth,clip=]{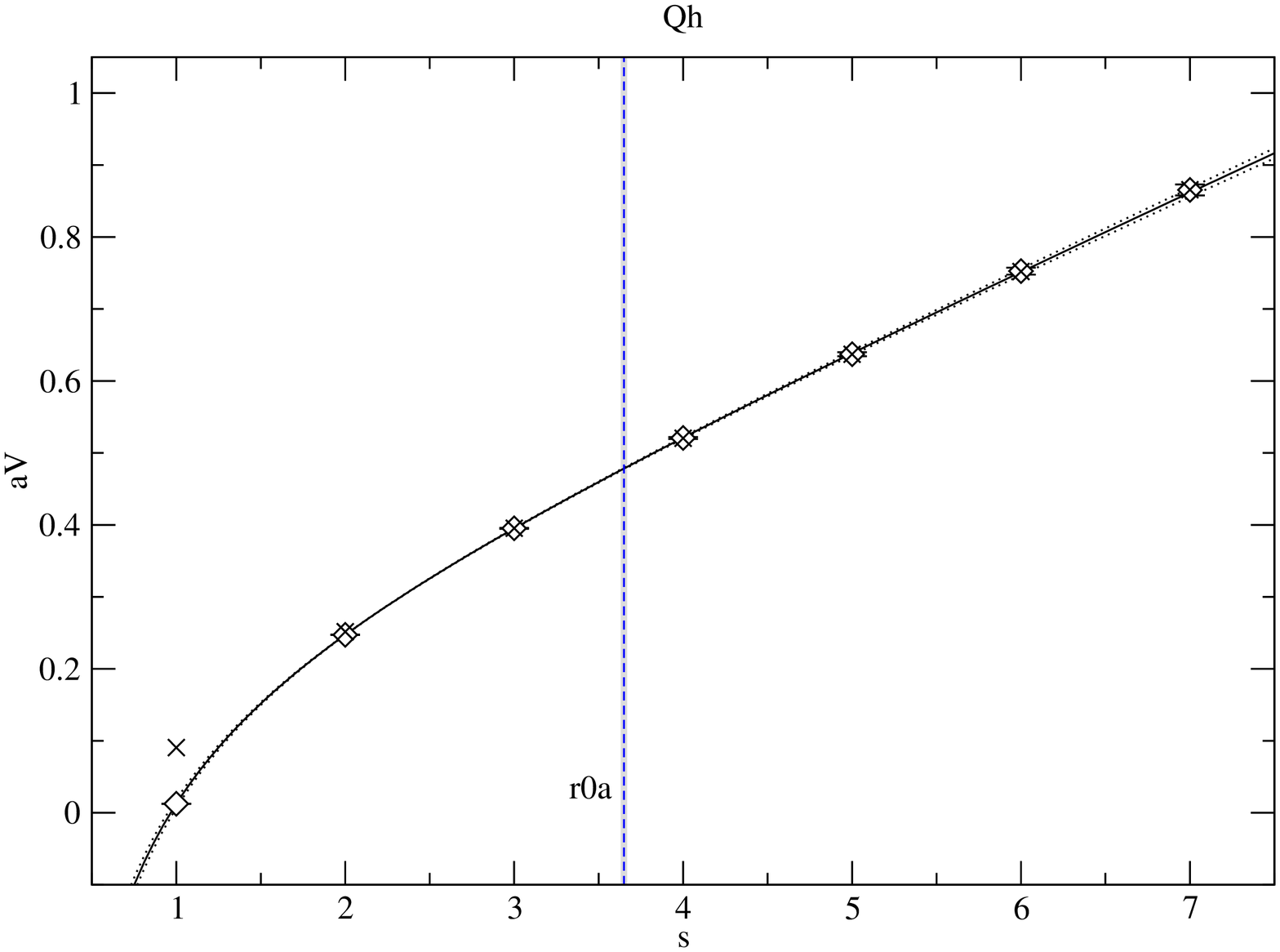}
\mycaption{Static quark potential for ensemble \ensemble{Q ~ hyp}.  The
$\times$'s correspond to the uncorrected static quark potential, while the diamonds
correspond to the corrected potential.  The result is $r_0 / a = 3.650(19)$.}
\label{g:figure}
\end{figure}

\clearpage

\begin{figure}
\centering
\psfrag{tmin}[t][t]{$t_\text{min}$}
\psfrag{r0a}[b][B]{$r_0 / a$}
\psfrag{A}[B][B]{\ensemble{A}}
\psfrag{Ah}[B][B]{\ensemble{A ~ hyp}}
\psfrag{B}[B][B]{\ensemble{B}}
\psfrag{Bh}[B][B]{\ensemble{B ~ hyp}}
\psfrag{C}[B][B]{\ensemble{C}}
\psfrag{Ch}[B][B]{\ensemble{C ~ hyp}}
\includegraphics[width=\textwidth,clip=]{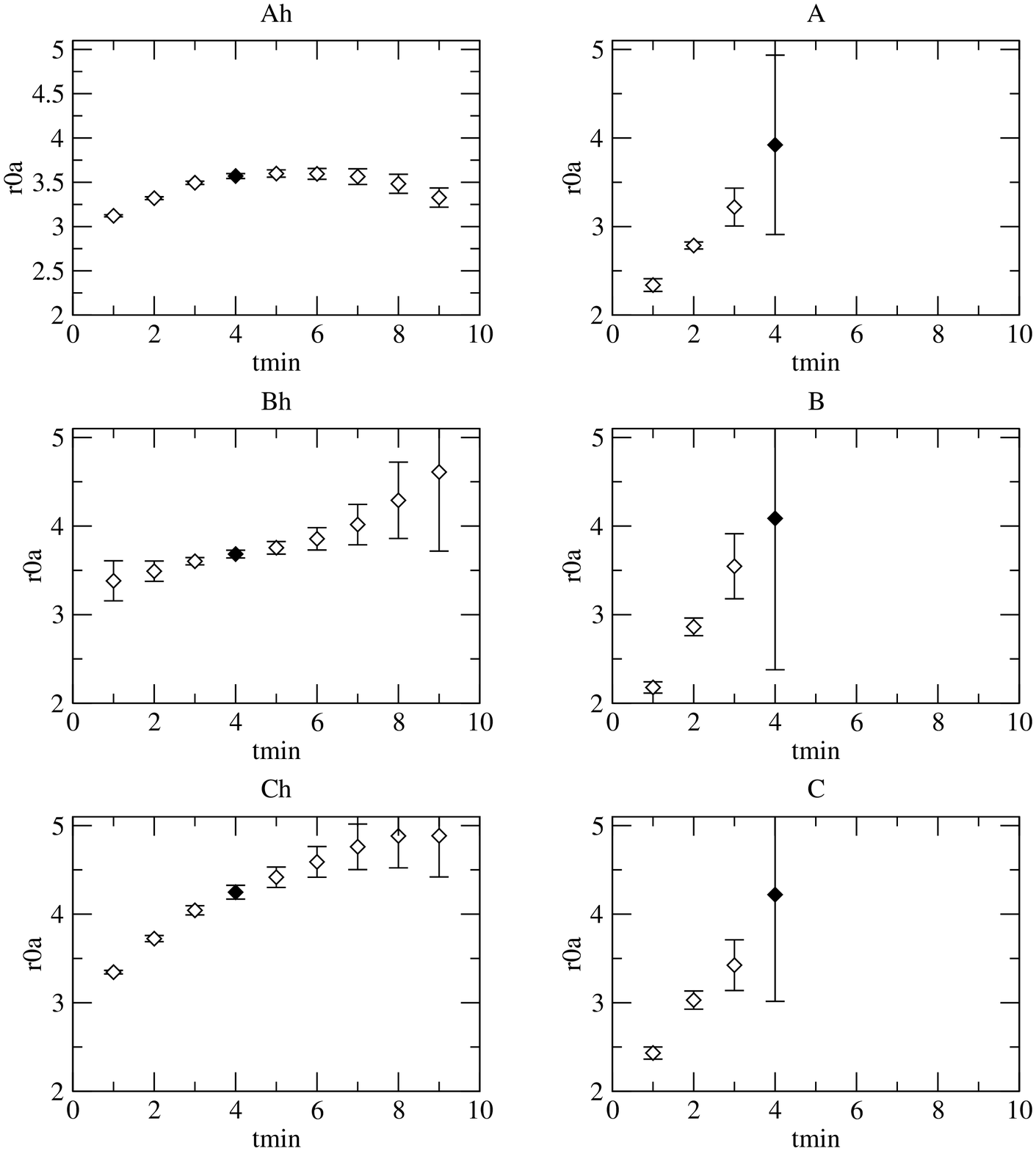}
\mycaption{Dependence of the Sommer scale on $t_\text{min}$ for ensembles
\ensemble{A ~ hyp}, \ensemble{A}, \ensemble{B ~ hyp}, \ensemble{B}, \ensemble{C
~ hyp}, and \ensemble{C}.  The filled diamond corresponds to $t_\text{min}$
used in the final fit.}
\label{j:figure}
\end{figure}

\begin{figure}
\centering
\psfrag{tmin}[t][t]{$t_\text{min}$}
\psfrag{r0a}[b][B]{$r_0 / a$}
\psfrag{W}[B][B]{\ensemble{W}}
\psfrag{Wh}[B][B]{\ensemble{W ~ hyp}}
\psfrag{X}[B][B]{\ensemble{X}}
\psfrag{Xh}[B][B]{\ensemble{X ~ hyp}}
\includegraphics[width=\textwidth,clip=]{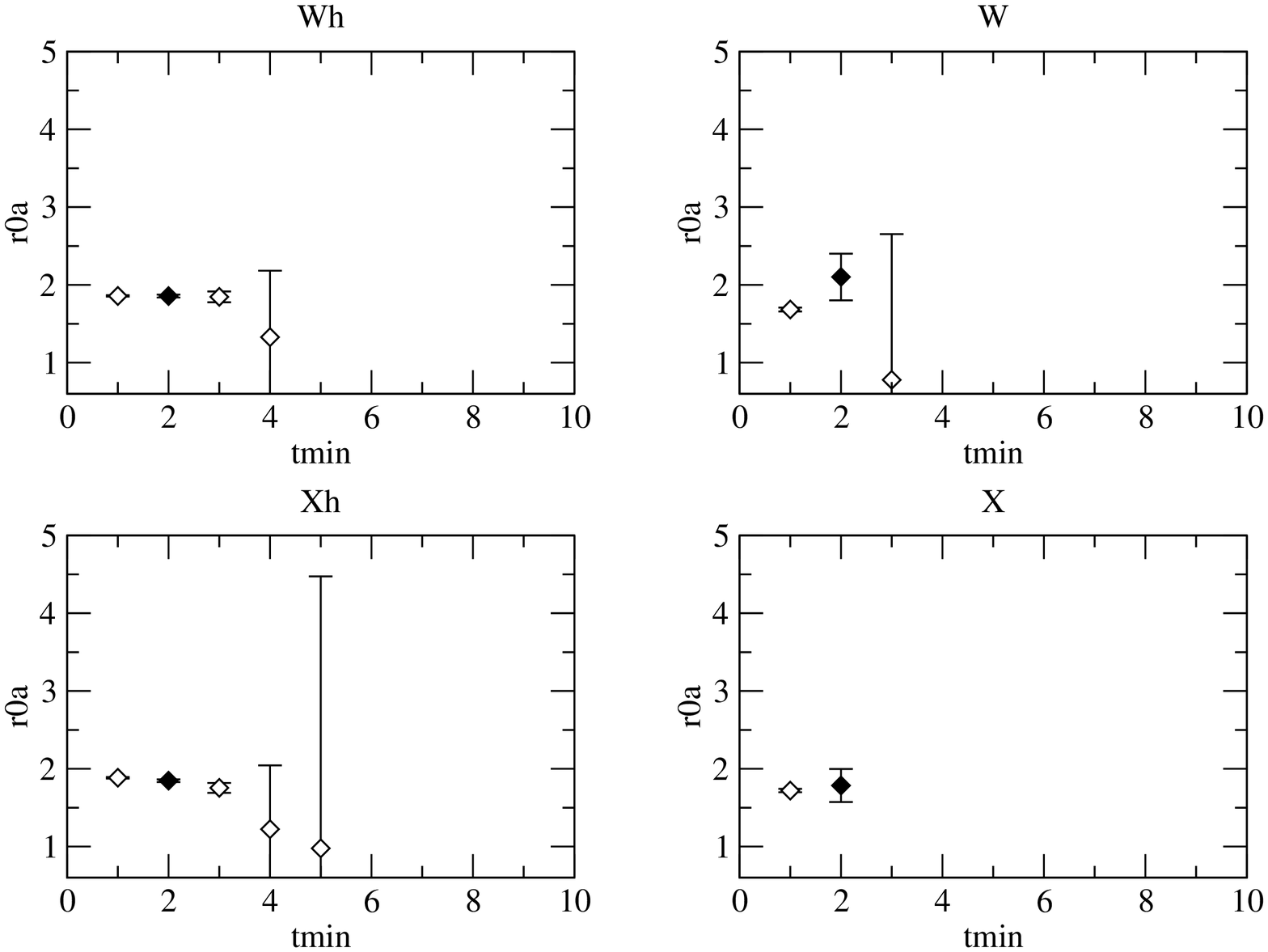}
\mycaption{Dependence of the Sommer scale on $t_\text{min}$ for ensembles
\ensemble{W ~ hyp}, \ensemble{W}, \ensemble{X ~ hyp}, and \ensemble{X}.  The
filled diamond corresponds to $t_\text{min}$ used in the final fit.}
\end{figure}
 
\begin{figure}
\centering
\psfrag{tmin}[t][t]{$t_\text{min}$}
\psfrag{r0a}[b][B]{$r_0 / a$}
\psfrag{Y}[B][B]{\ensemble{Y}}
\psfrag{Yh}[B][B]{\ensemble{Y ~ hyp}}
\psfrag{Z}[B][B]{\ensemble{Z}}
\psfrag{Zh}[B][B]{\ensemble{Z ~ hyp}}
\includegraphics[width=\textwidth,clip=]{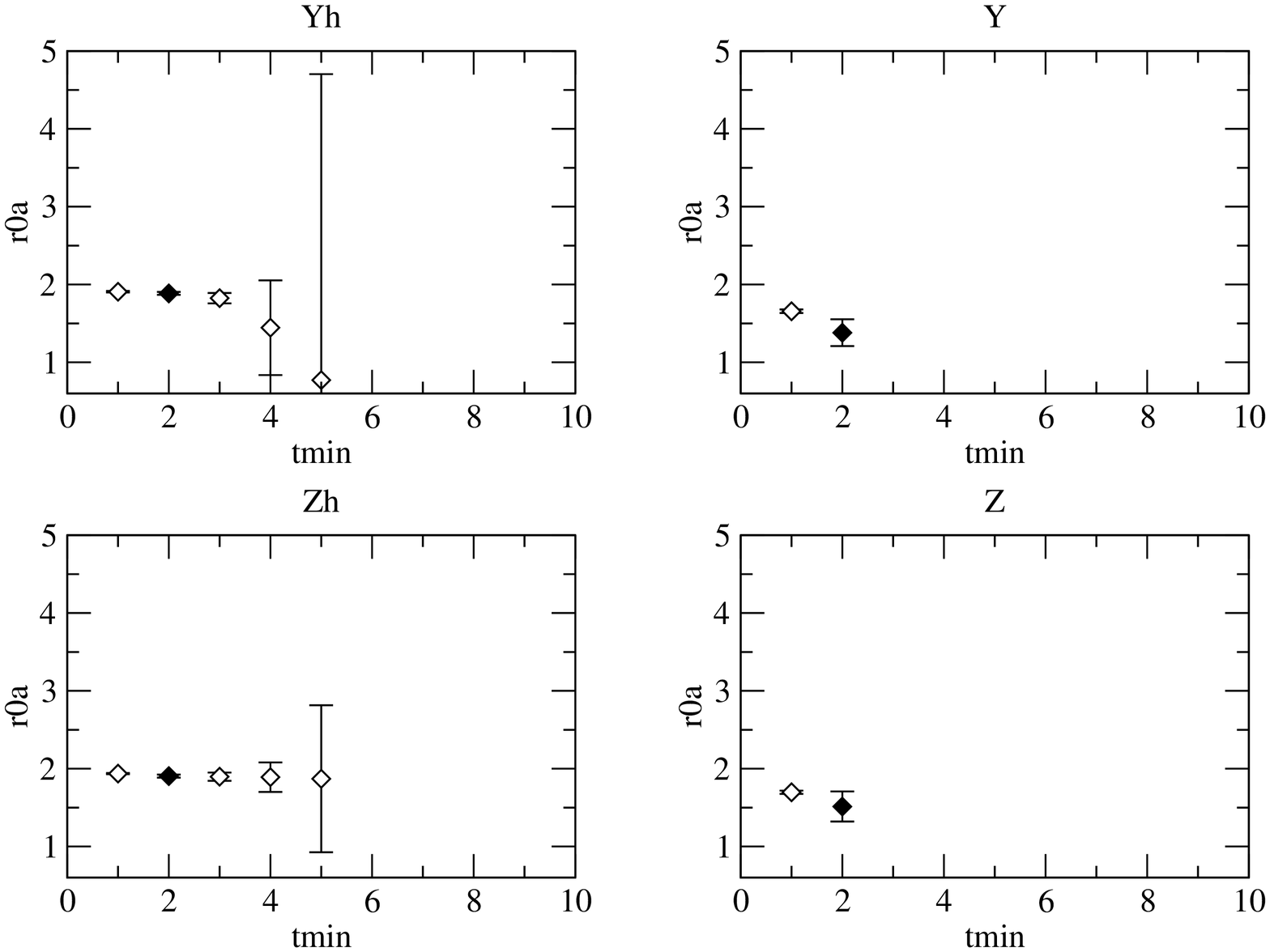}
\mycaption{Dependence of the Sommer scale on $t_\text{min}$ for ensembles
\ensemble{Y ~ hyp}, \ensemble{Y}, \ensemble{Z ~ hyp}, and \ensemble{Z}.  The
filled diamond corresponds to $t_\text{min}$ used in the final fit.}
\end{figure}

\begin{figure}
\centering
\psfrag{tmin}[t][t]{$t_\text{min}$}
\psfrag{r0a}[b][B]{$r_0 / a$}
\psfrag{Qh}[B][B]{\ensemble{Q ~ hyp}}
\includegraphics[width=0.5\textwidth,clip=]{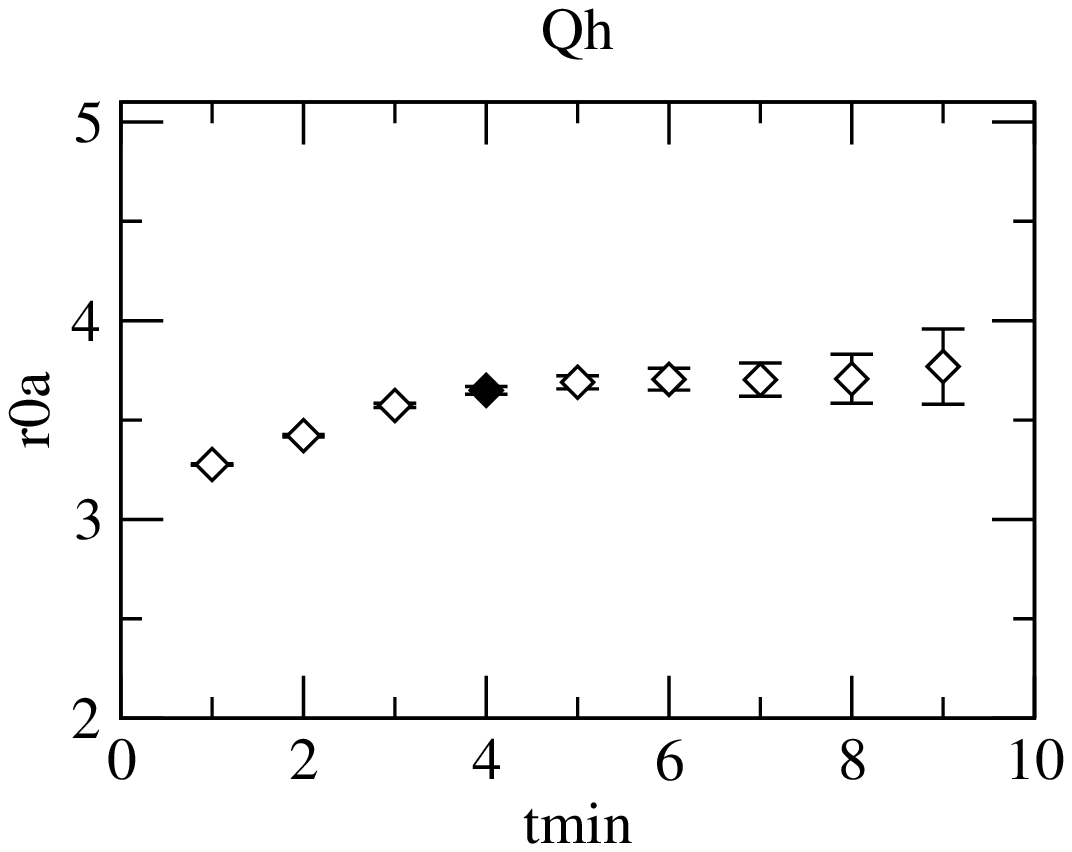}
\mycaption{Dependence of the Sommer scale on $t_\text{min}$ for ensemble
\ensemble{Q ~ hyp}.  The filled diamond corresponds to $t_\text{min}$ used in
the final fit.}
\label{k:figure}
\end{figure}

\clearpage

\begin{figure}
\centering
\psfrag{tmax}[t][t]{$t_\text{max}$}
\psfrag{r0a}[b][B]{$r_0 / a$}
\psfrag{A}[B][B]{\ensemble{A}}
\psfrag{Ah}[B][B]{\ensemble{A ~ hyp}}
\psfrag{B}[B][B]{\ensemble{B}}
\psfrag{Bh}[B][B]{\ensemble{B ~ hyp}}
\psfrag{C}[B][B]{\ensemble{C}}
\psfrag{Ch}[B][B]{\ensemble{C ~ hyp}}
\includegraphics[width=\textwidth,clip=]{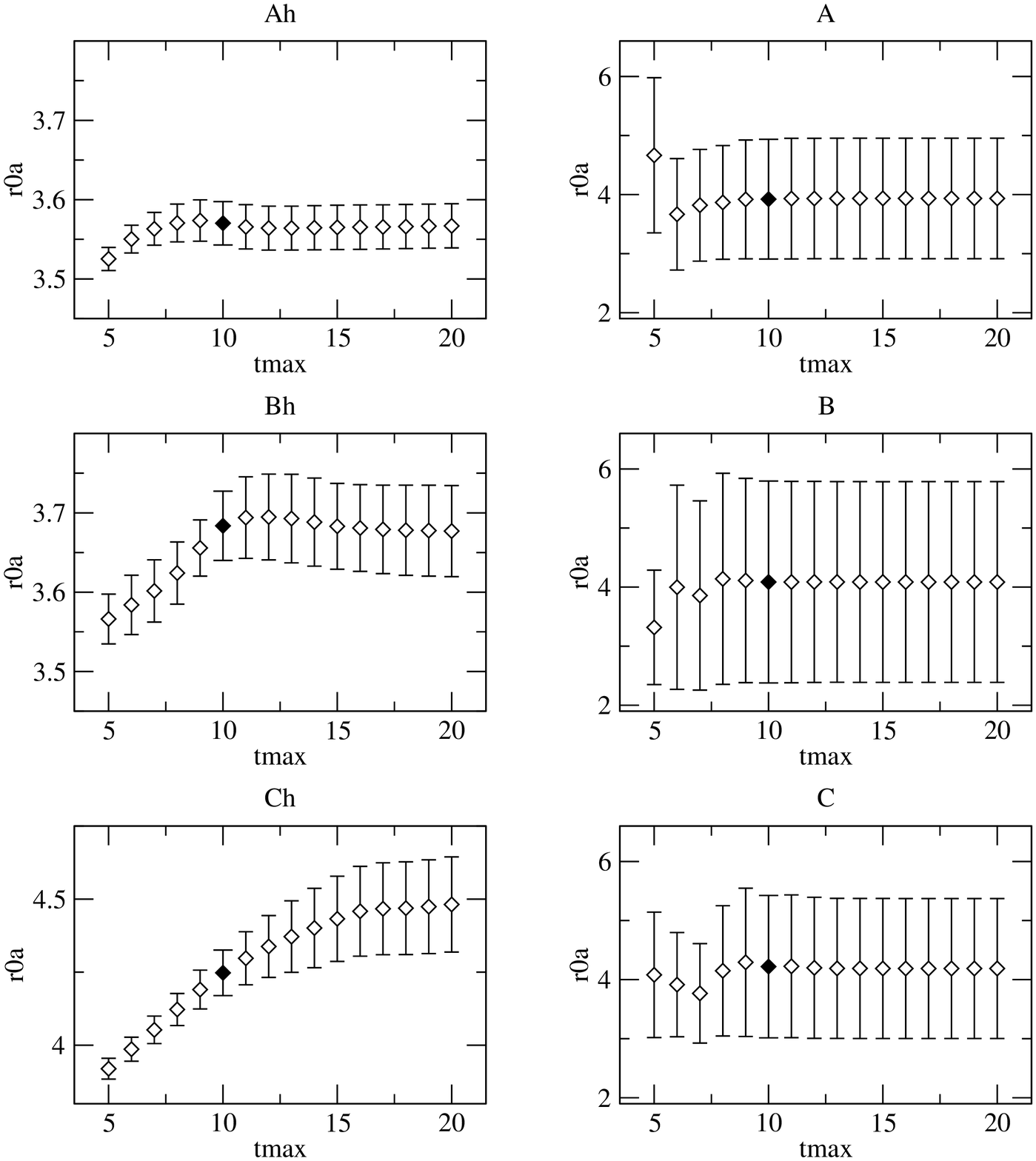}
\mycaption{Dependence of the Sommer scale on $t_\text{max}$ for ensembles
\ensemble{A ~ hyp}, \ensemble{A}, \ensemble{B ~ hyp}, \ensemble{B}, \ensemble{C
~ hyp}, and \ensemble{C}.  The filled diamond corresponds to $t_\text{max}$
used in the final fit.}
\label{l:figure}
\end{figure}

\begin{figure}
\centering
\psfrag{tmax}[t][t]{$t_\text{max}$}
\psfrag{r0a}[b][B]{$r_0 / a$}
\psfrag{W}[B][B]{\ensemble{W}}
\psfrag{Wh}[B][B]{\ensemble{W ~ hyp}}
\psfrag{X}[B][B]{\ensemble{X}}
\psfrag{Xh}[B][B]{\ensemble{X ~ hyp}}
\includegraphics[width=\textwidth,clip=]{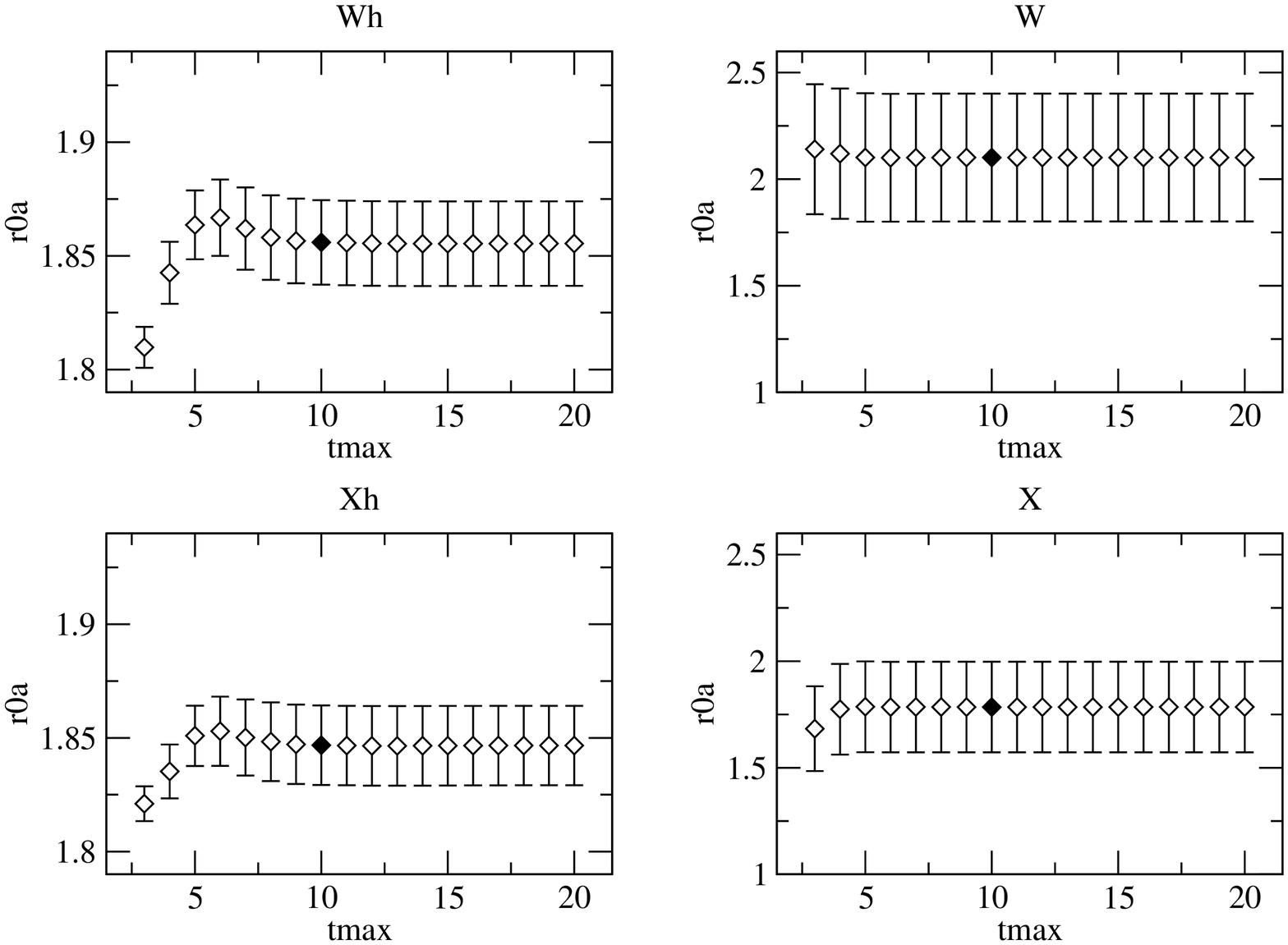}
\mycaption{Dependence of the Sommer scale on $t_\text{max}$ for ensembles
\ensemble{W ~ hyp}, \ensemble{W}, \ensemble{X ~ hyp}, and \ensemble{X}.  The
filled diamond corresponds to $t_\text{max}$ used in the final fit.}
\end{figure}

\begin{figure}
\centering
\psfrag{tmax}[t][t]{$t_\text{max}$}
\psfrag{r0a}[b][B]{$r_0 / a$}
\psfrag{Y}[B][B]{\ensemble{Y}}
\psfrag{Yh}[B][B]{\ensemble{Y ~ hyp}}
\psfrag{Z}[B][B]{\ensemble{Z}}
\psfrag{Zh}[B][B]{\ensemble{Z ~ hyp}}
\includegraphics[width=\textwidth,clip=]{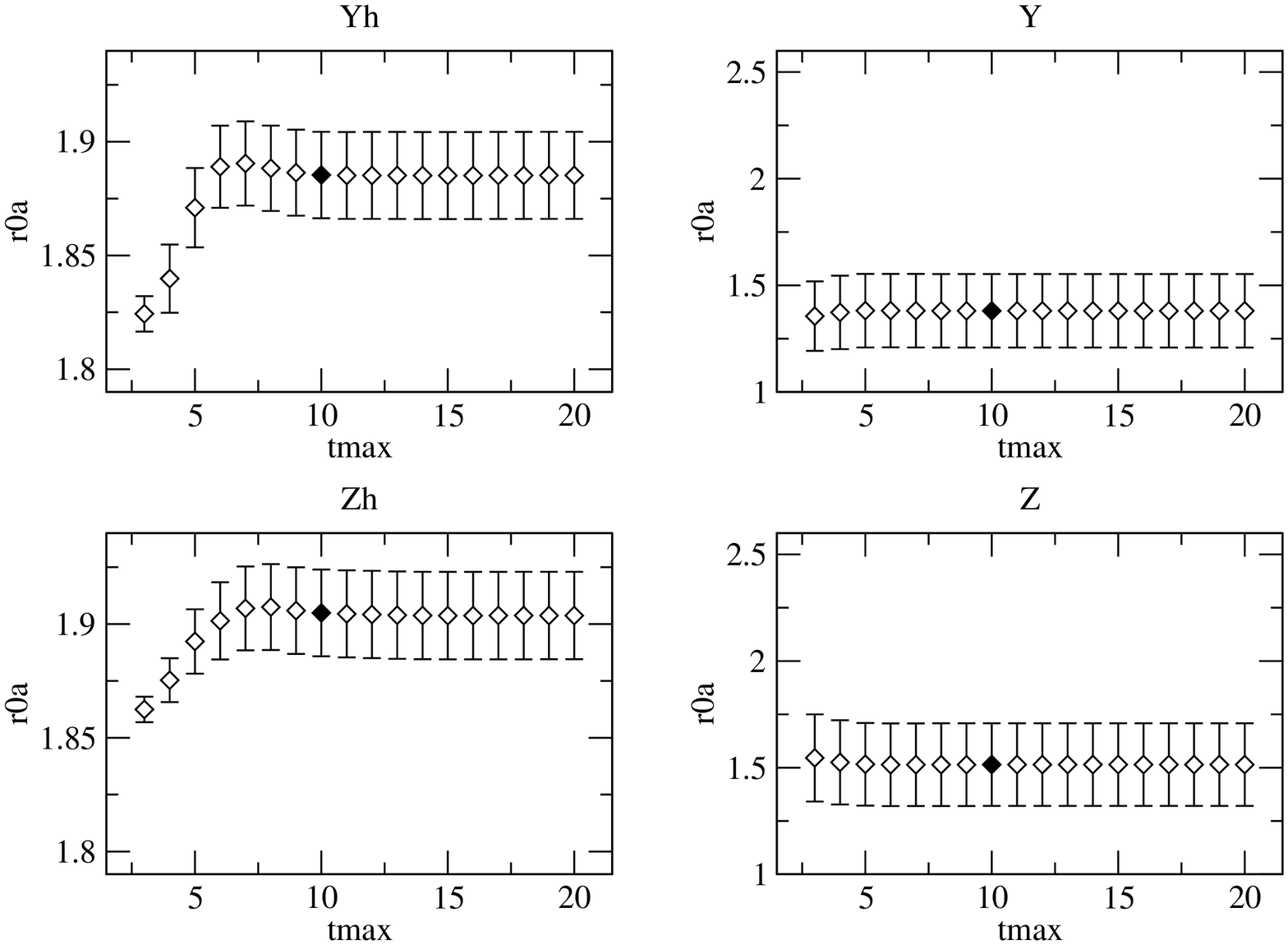}
\mycaption{Dependence of the Sommer scale on $t_\text{max}$ for ensembles
\ensemble{Y ~ hyp}, \ensemble{Y}, \ensemble{Z ~ hyp}, and \ensemble{Z}.  The
filled diamond corresponds to $t_\text{max}$ used in the final fit.}
\end{figure}

\begin{figure}
\centering
\psfrag{tmax}[t][t]{$t_\text{max}$}
\psfrag{r0a}[b][B]{$r_0 / a$}
\psfrag{Qh}[B][B]{\ensemble{Q ~ hyp}}
\includegraphics[width=0.5\textwidth,clip=]{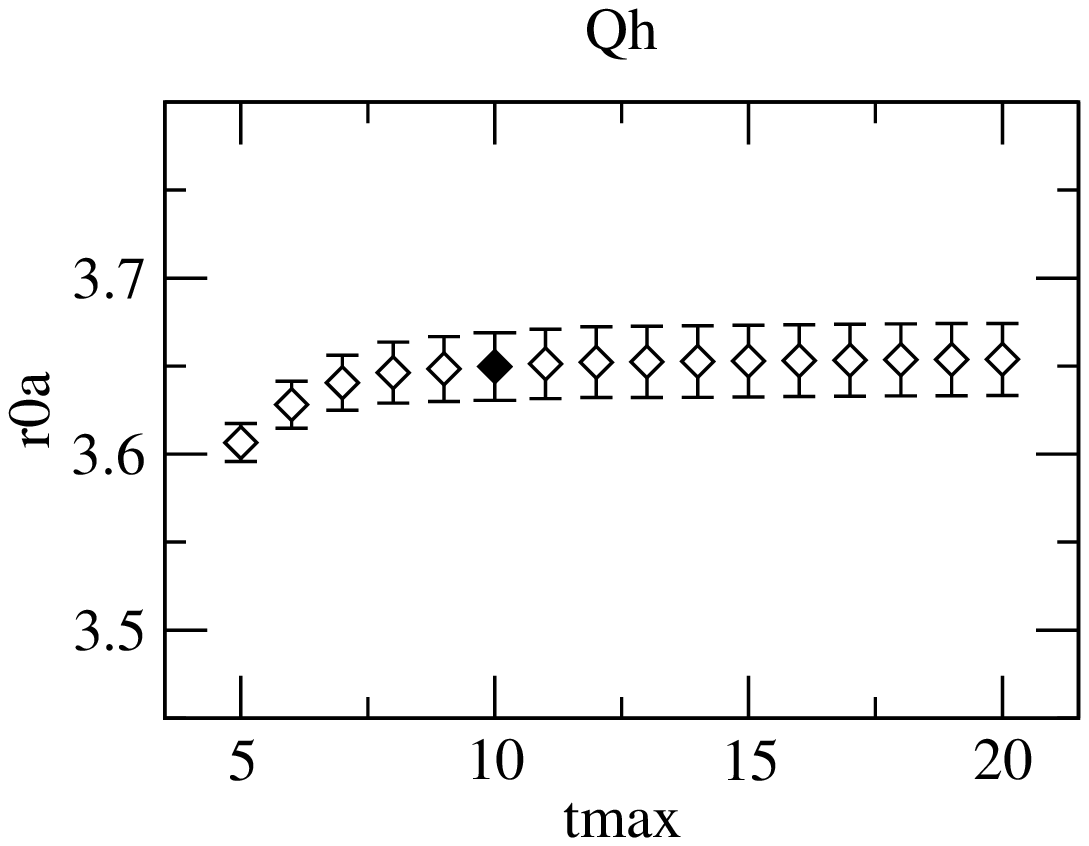}
\mycaption{Dependence of the Sommer scale on $t_\text{max}$ for ensemble
\ensemble{Q ~ hyp}.  The filled diamond corresponds to $t_\text{max}$ used in
the final fit.}
\label{m:figure}
\end{figure}

\clearpage

\begin{figure}
\centering
\psfrag{smax}[t][t]{$s_\text{max}$}
\psfrag{r0a}[b][B]{$r_0 / a$}
\psfrag{A}[B][B]{\ensemble{A}}
\psfrag{Ah}[B][B]{\ensemble{A ~ hyp}}
\psfrag{B}[B][B]{\ensemble{B}}
\psfrag{Bh}[B][B]{\ensemble{B ~ hyp}}
\psfrag{C}[B][B]{\ensemble{C}}
\psfrag{Ch}[B][B]{\ensemble{C ~ hyp}}
\includegraphics[width=\textwidth,clip=]{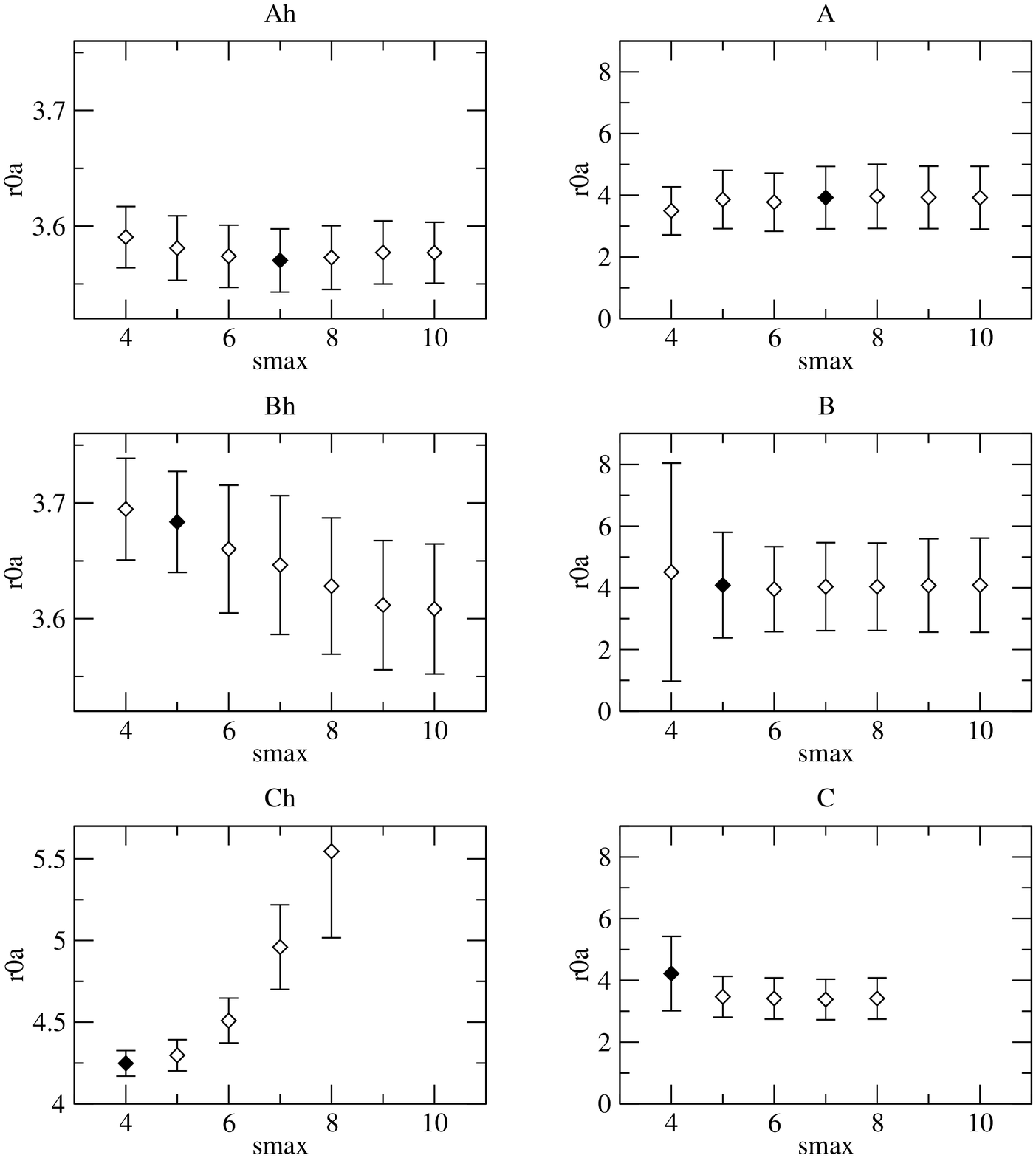}
\mycaption{Dependence of the Sommer scale on $s_\text{min}$ for ensembles
\ensemble{A ~ hyp}, \ensemble{A}, \ensemble{B ~ hyp}, \ensemble{B}, \ensemble{C
~ hyp}, and \ensemble{C}.  The filled diamond corresponds to $s_\text{max}$
used in the final fit.}
\label{n:figure}
\end{figure}

\begin{figure}
\centering
\psfrag{smax}[t][t]{$s_\text{max}$}
\psfrag{r0a}[b][B]{$r_0 / a$}
\psfrag{W}[B][B]{\ensemble{W}}
\psfrag{Wh}[B][B]{\ensemble{W ~ hyp}}
\psfrag{X}[B][B]{\ensemble{X}}
\psfrag{Xh}[B][B]{\ensemble{X ~ hyp}}
\includegraphics[width=\textwidth,clip=]{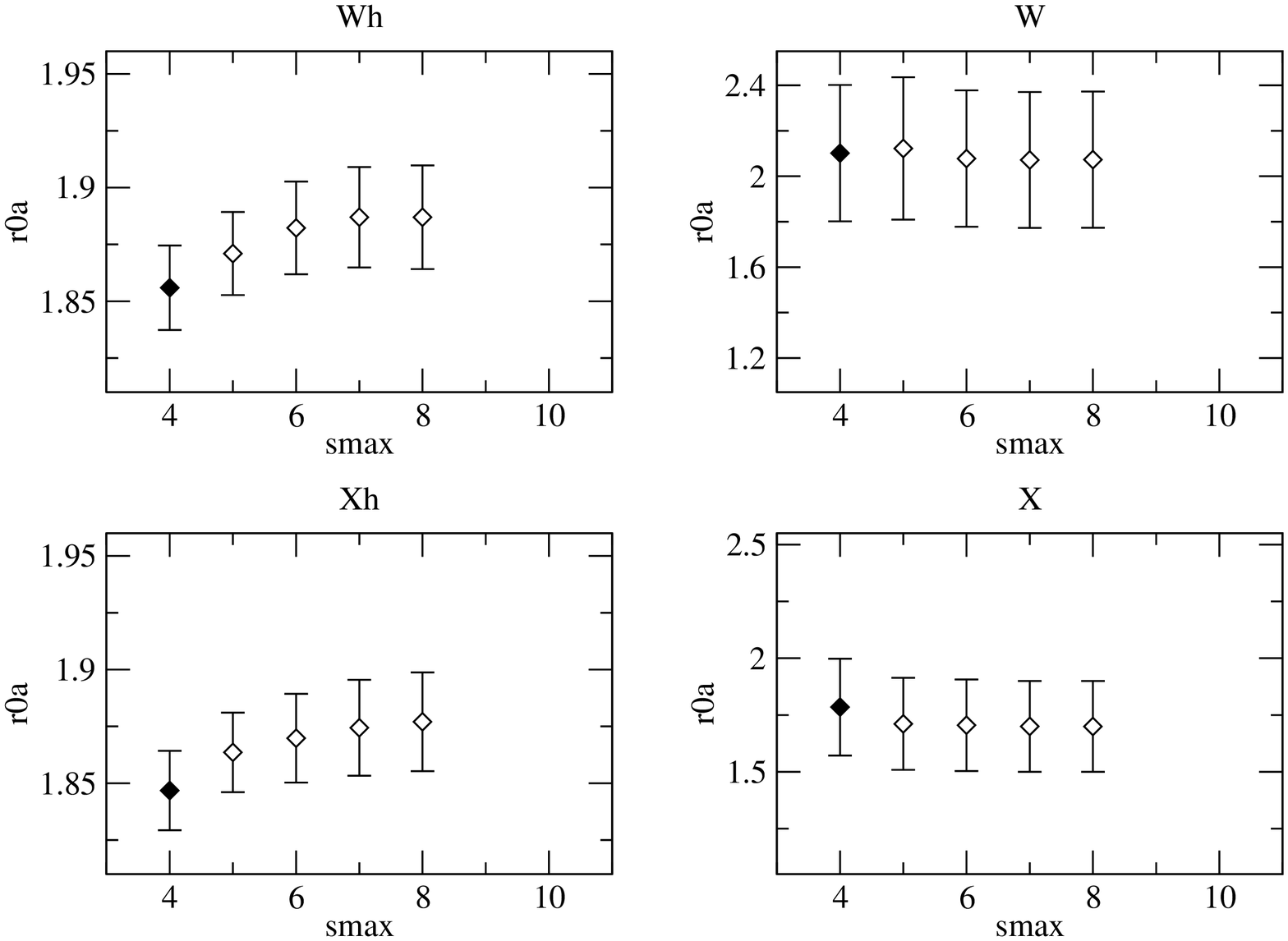}
\mycaption{Dependence of the Sommer scale on $s_\text{min}$ for ensembles
\ensemble{W ~ hyp}, \ensemble{W}, \ensemble{X ~ hyp}, and \ensemble{X}.  The
filled diamond corresponds to $s_\text{max}$ used in the final fit.}
\end{figure}

\begin{figure}
\centering
\psfrag{smax}[t][t]{$s_\text{max}$}
\psfrag{r0a}[b][B]{$r_0 / a$}
\psfrag{Y}[B][B]{\ensemble{Y}}
\psfrag{Yh}[B][B]{\ensemble{Y ~ hyp}}
\psfrag{Z}[B][B]{\ensemble{Z}}
\psfrag{Zh}[B][B]{\ensemble{Z ~ hyp}}
\includegraphics[width=\textwidth,clip=]{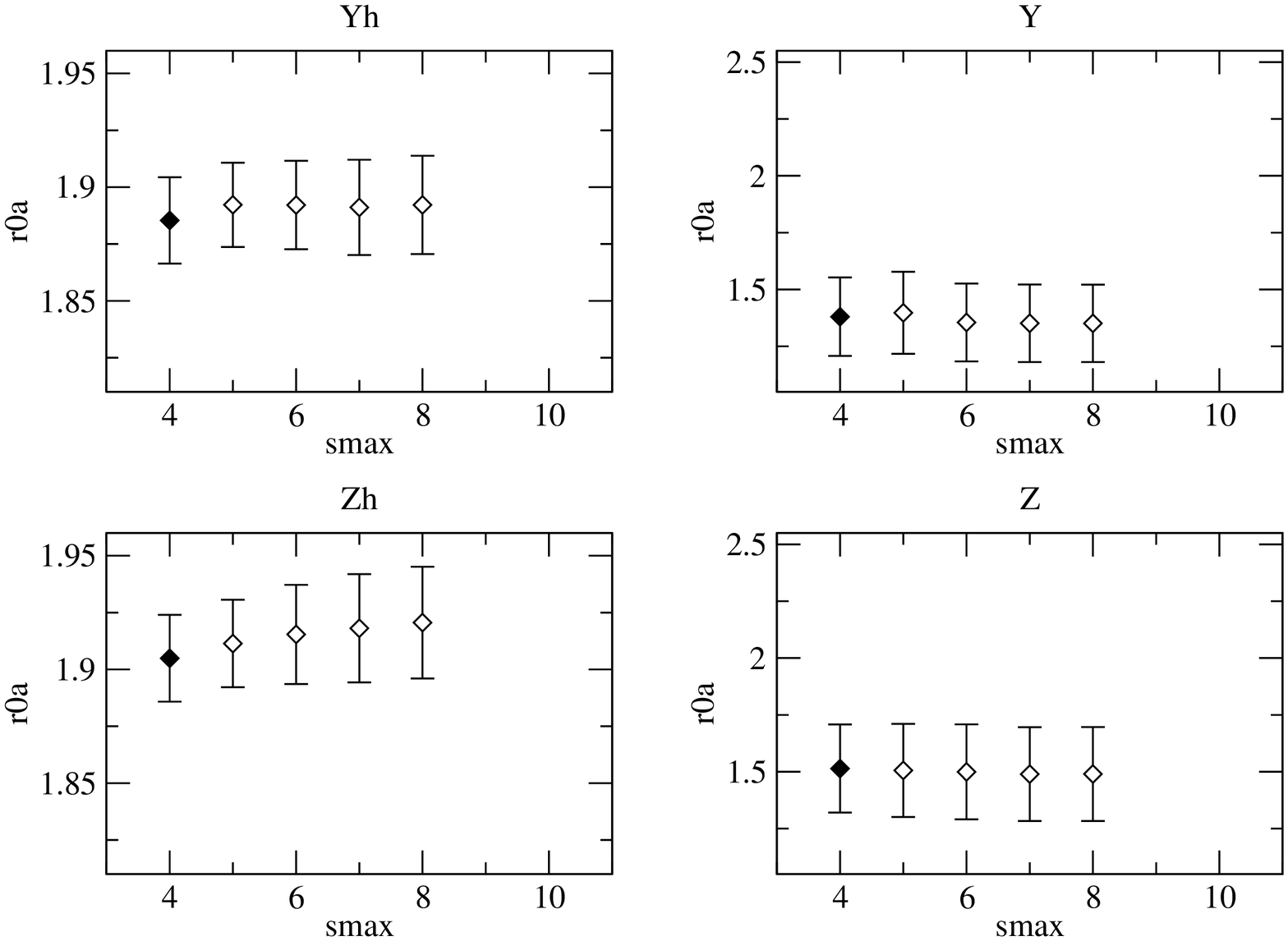}
\mycaption{Dependence of the Sommer scale on $s_\text{min}$ for ensembles
\ensemble{Y ~ hyp}, \ensemble{Y}, \ensemble{Z ~ hyp}, and \ensemble{Z}.  The
filled diamond corresponds to $s_\text{max}$ used in the final fit.}
\end{figure}

\begin{figure}
\centering
\psfrag{smax}[t][t]{$s_\text{max}$}
\psfrag{r0a}[b][B]{$r_0 / a$}
\psfrag{Qh}[B][B]{\ensemble{Q ~ hyp}}
\includegraphics[width=0.5\textwidth,clip=]{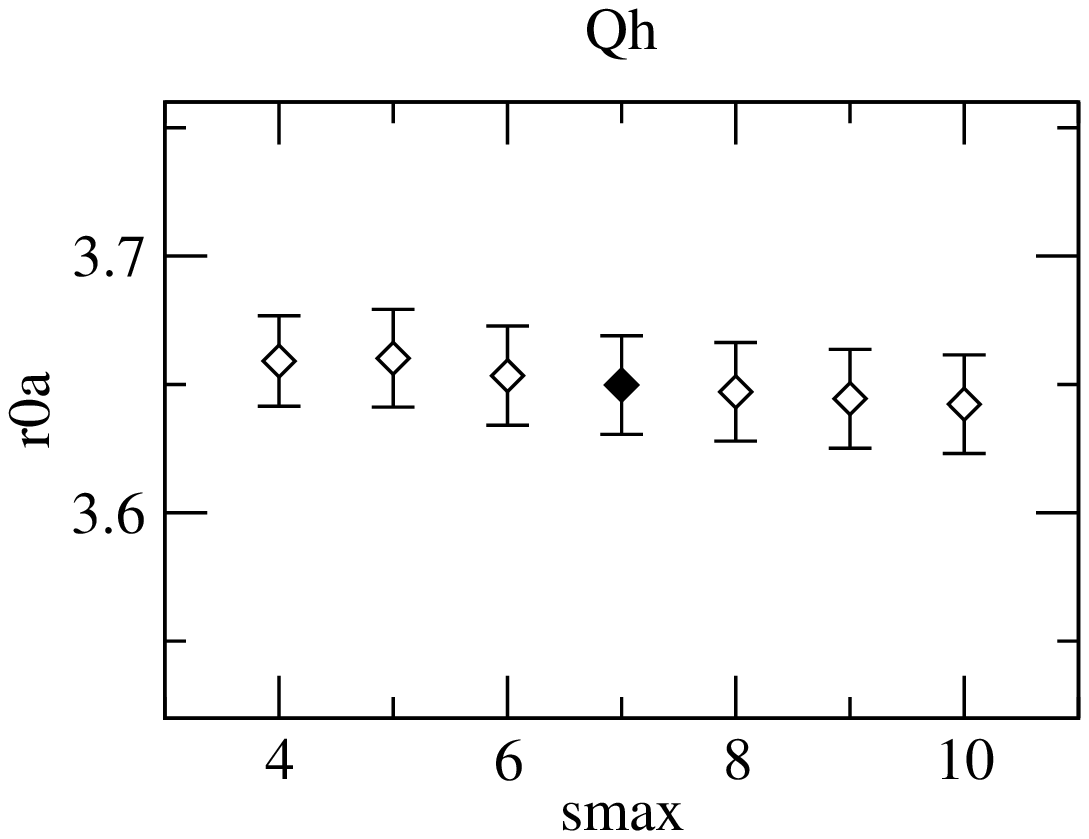}
\mycaption{Dependence of the Sommer scale on $s_\text{min}$ for ensemble
\ensemble{Q ~ hyp}.  The filled diamond corresponds to $s_\text{max}$ used in
the final fit.}
\label{o:figure}
\end{figure}

\clearpage

\begin{figure}
\centering
\psfrag{t}[t][t]{$t$}
\psfrag{Meff}[b][B]{$a M_\text{eff}$}
\psfrag{A}[B][B]{\ensemble{A}}
\psfrag{Ah}[B][B]{\ensemble{A ~ hyp}}
\psfrag{B}[B][B]{\ensemble{B}}
\psfrag{Bh}[B][B]{\ensemble{B ~ hyp}}
\psfrag{C}[B][B]{\ensemble{C}}
\psfrag{Ch}[B][B]{\ensemble{C ~ hyp}}
\includegraphics[width=\textwidth,clip=]{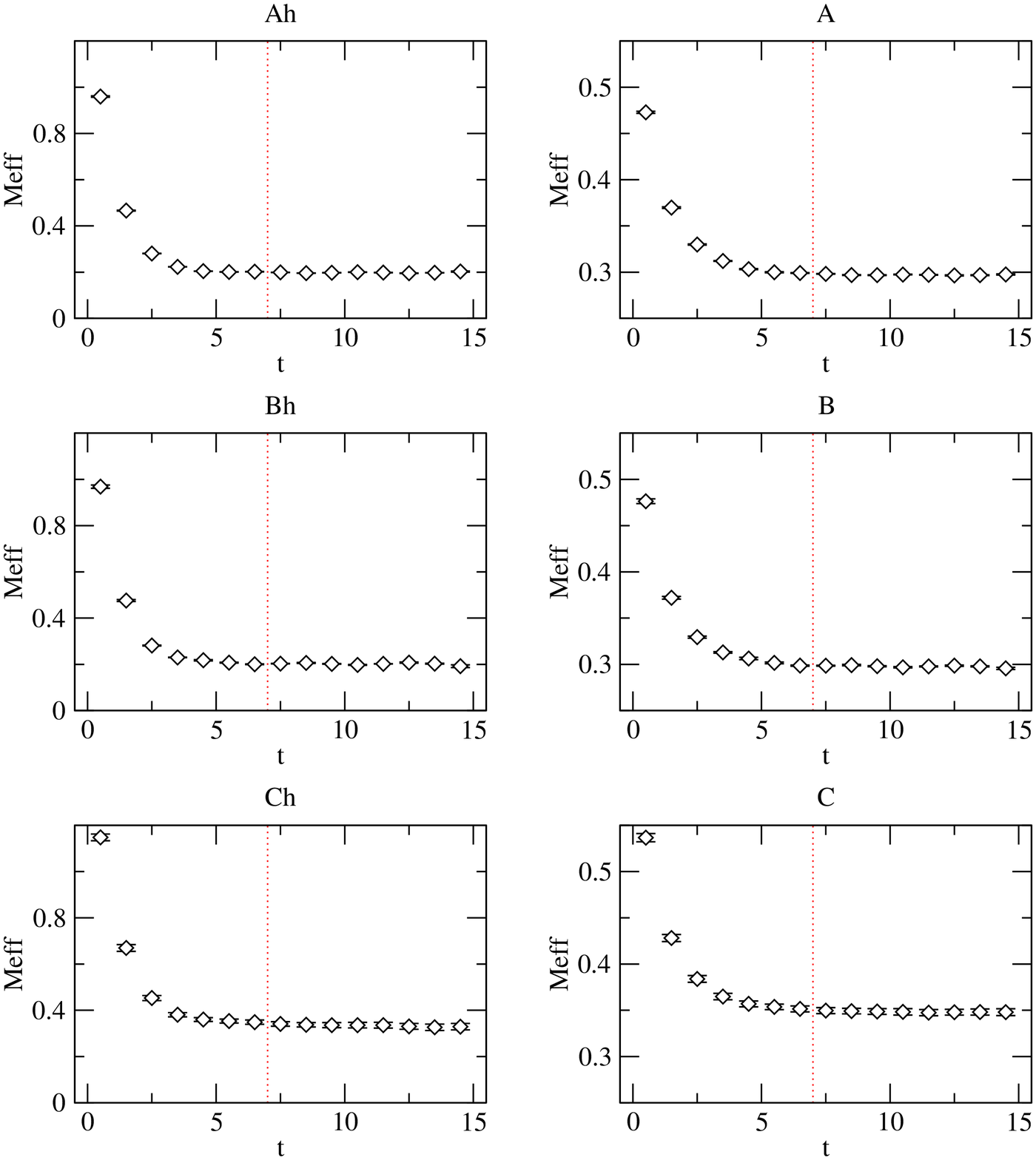}
\mycaption{Effective pion mass for ensembles \ensemble{A ~ hyp}, \ensemble{A},
\ensemble{B ~ hyp}, \ensemble{B}, \ensemble{C ~ hyp}, and \ensemble{C} at $m_V
= 0.01$.  The minimum time separation chosen is $t_\text{min} = 7$.}
\label{r:figure}
\end{figure}

\begin{figure}
\centering
\psfrag{t}[t][t]{$t$}
\psfrag{Meff}[b][B]{$a M_\text{eff}$}
\psfrag{W}[B][B]{\ensemble{W}}
\psfrag{Wh}[B][B]{\ensemble{W ~ hyp}}
\psfrag{X}[B][B]{\ensemble{X}}
\psfrag{Xh}[B][B]{\ensemble{X ~ hyp}}
\includegraphics[width=\textwidth,clip=]{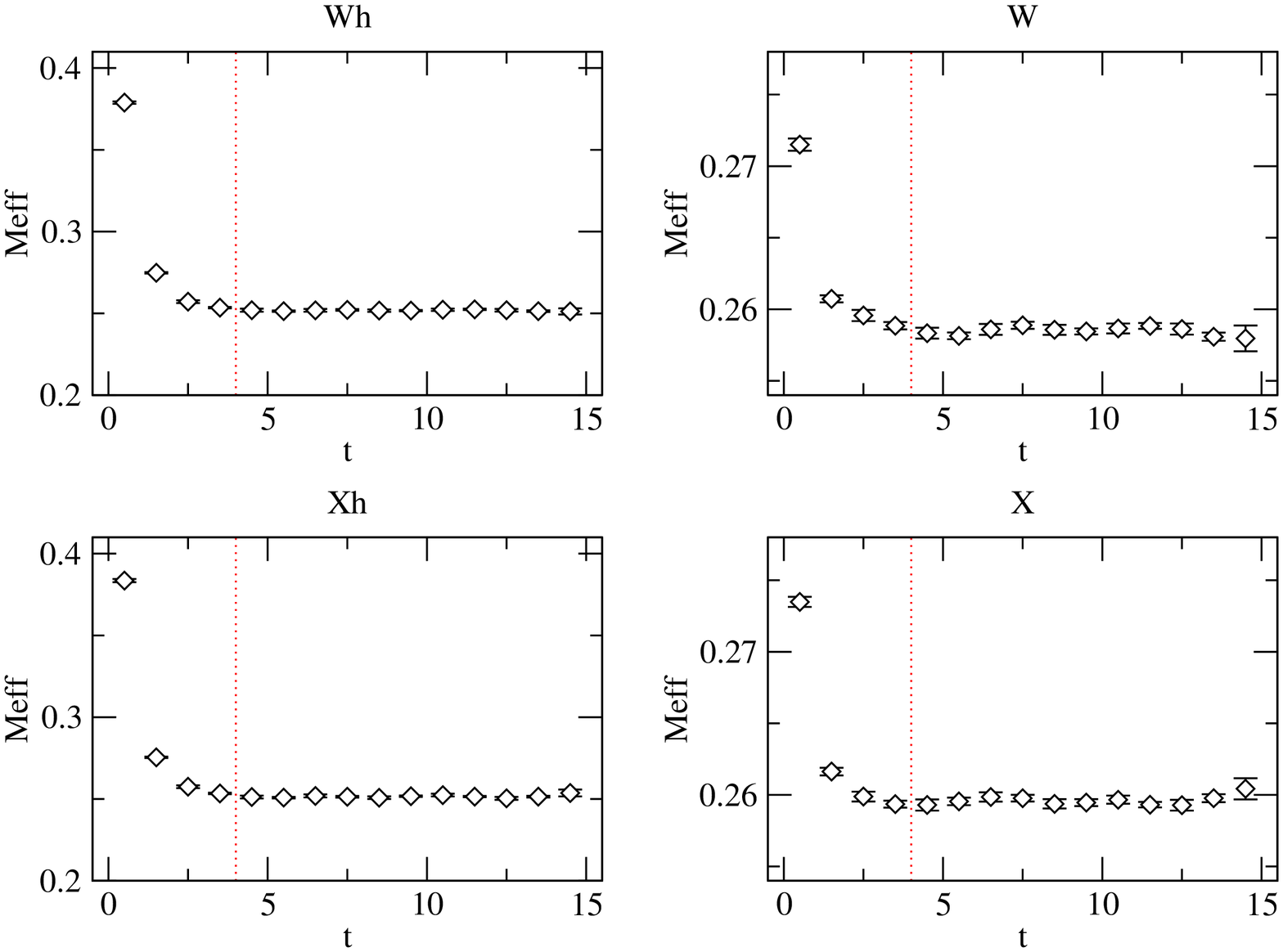}
\mycaption{Effective pion mass for ensembles \ensemble{W ~ hyp}, \ensemble{W},
\ensemble{Z ~ hyp}, and \ensemble{Z} at $m_V = 0.01$.  The minimum time
separation chosen is $t_\text{min} = 4$.}
\end{figure}

\begin{figure}
\centering
\psfrag{t}[t][t]{$t$}
\psfrag{Meff}[b][B]{$a M_\text{eff}$}
\psfrag{Y}[B][B]{\ensemble{Y}}
\psfrag{Yh}[B][B]{\ensemble{Y ~ hyp}}
\psfrag{Z}[B][B]{\ensemble{Z}}
\psfrag{Zh}[B][B]{\ensemble{Z ~ hyp}}
\includegraphics[width=\textwidth,clip=]{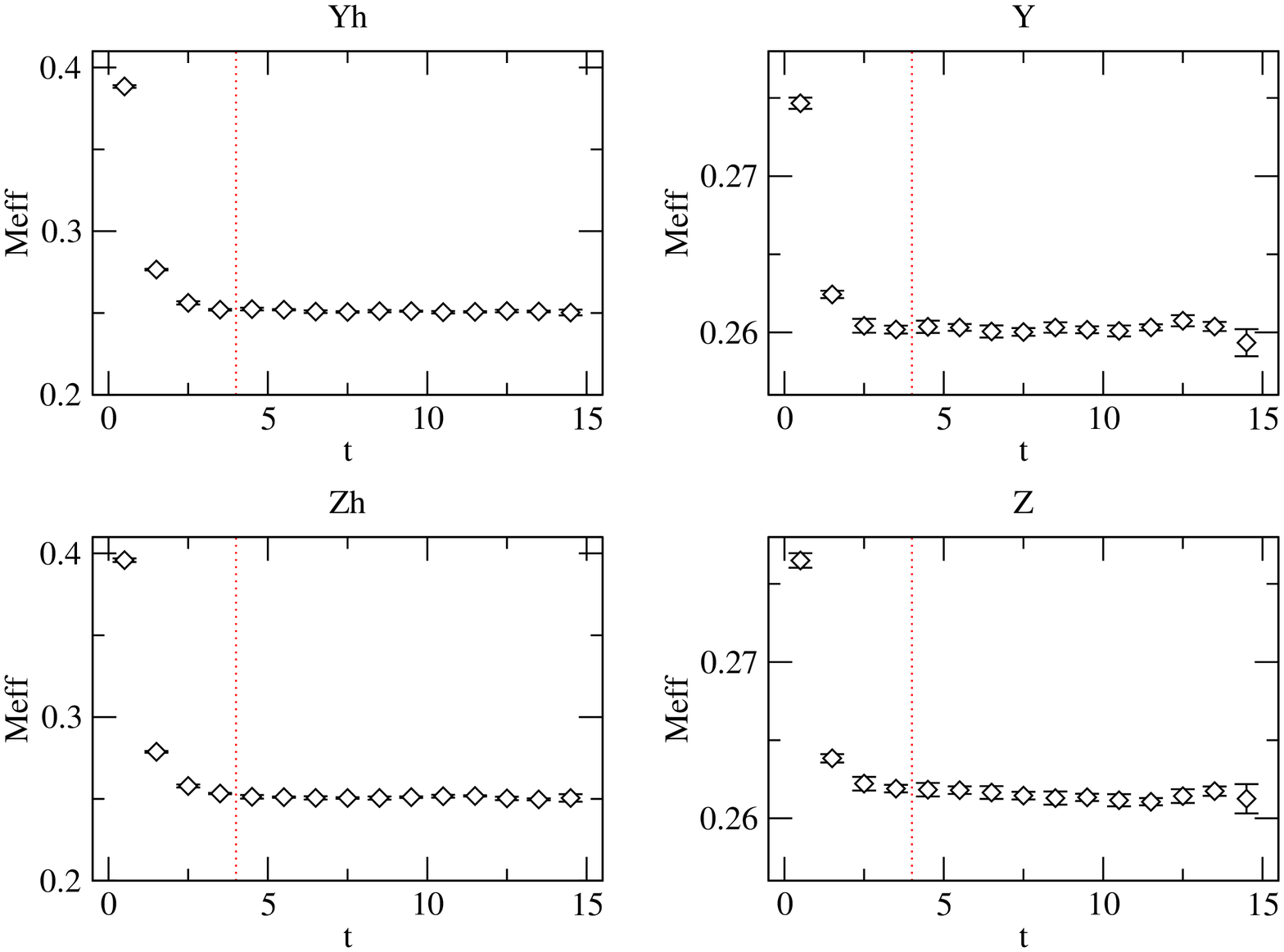}
\mycaption{Effective pion mass for ensembles \ensemble{Y ~ hyp}, \ensemble{Y},
\ensemble{Z ~ hyp}, and \ensemble{Z} at $m_V = 0.01$.  The minimum time
separation chosen is $t_\text{min} = 4$.}
\end{figure}

\begin{figure}
\centering
\psfrag{t}[t][t]{$t$}
\psfrag{Meff}[b][B]{$a M_\text{eff}$}
\psfrag{Q}[B][B]{\ensemble{Q}}
\psfrag{Qh}[B][B]{\ensemble{Q ~ hyp}}
\includegraphics[width=0.5\textwidth,clip=]{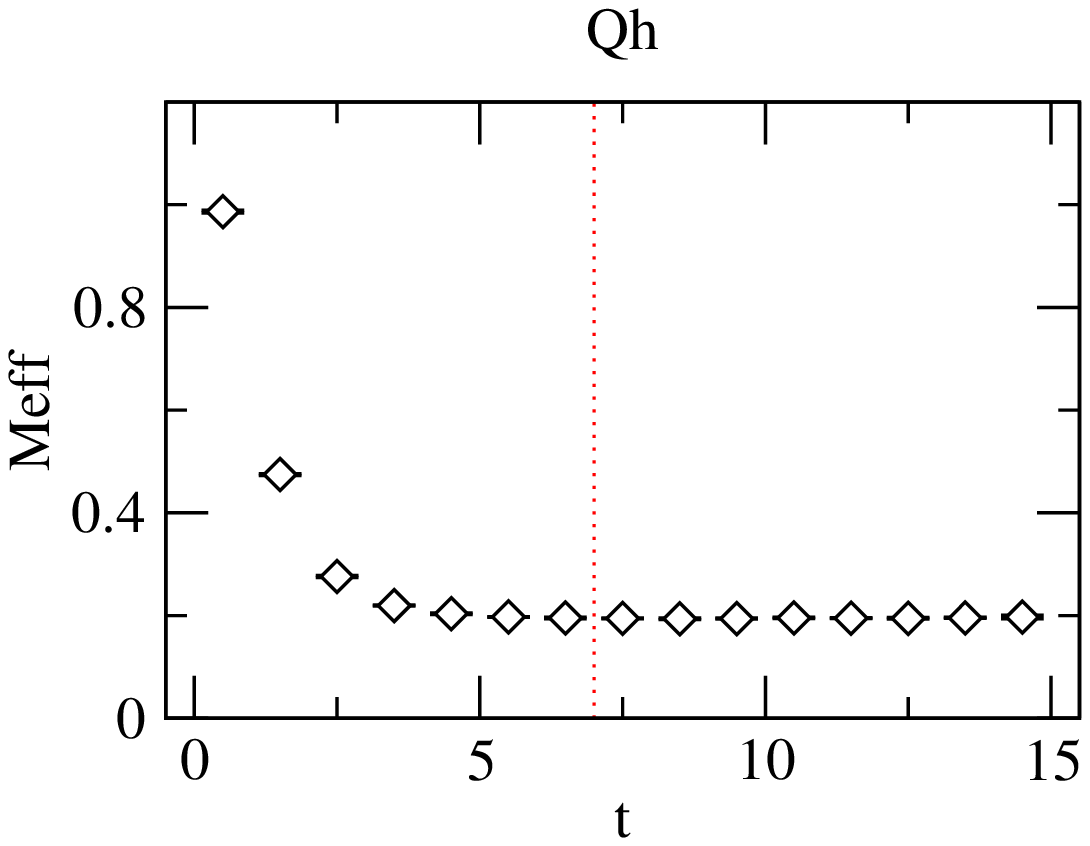}
\mycaption{Effective pion mass for ensemble \ensemble{Q ~ hyp} at $m_V = 0.01$.
The minimum time separation chosen is $t_\text{min} = 7$.}
\label{s:figure}
\end{figure}

\clearpage

\begin{figure}
\centering
\psfrag{mV}[t][t]{$m_V$}
\psfrag{Mpi2}[b][B]{$M^2_{\pi_5} \, (\text{GeV}^2)$}
\psfrag{A}[B][B]{\ensemble{A}}
\psfrag{Ah}[B][B]{\ensemble{A ~ hyp}}
\psfrag{B}[B][B]{\ensemble{B}}
\psfrag{Bh}[B][B]{\ensemble{B ~ hyp}}
\psfrag{C}[B][B]{\ensemble{C}}
\psfrag{Ch}[B][B]{\ensemble{C ~ hyp}}
\includegraphics[width=\textwidth,clip=]{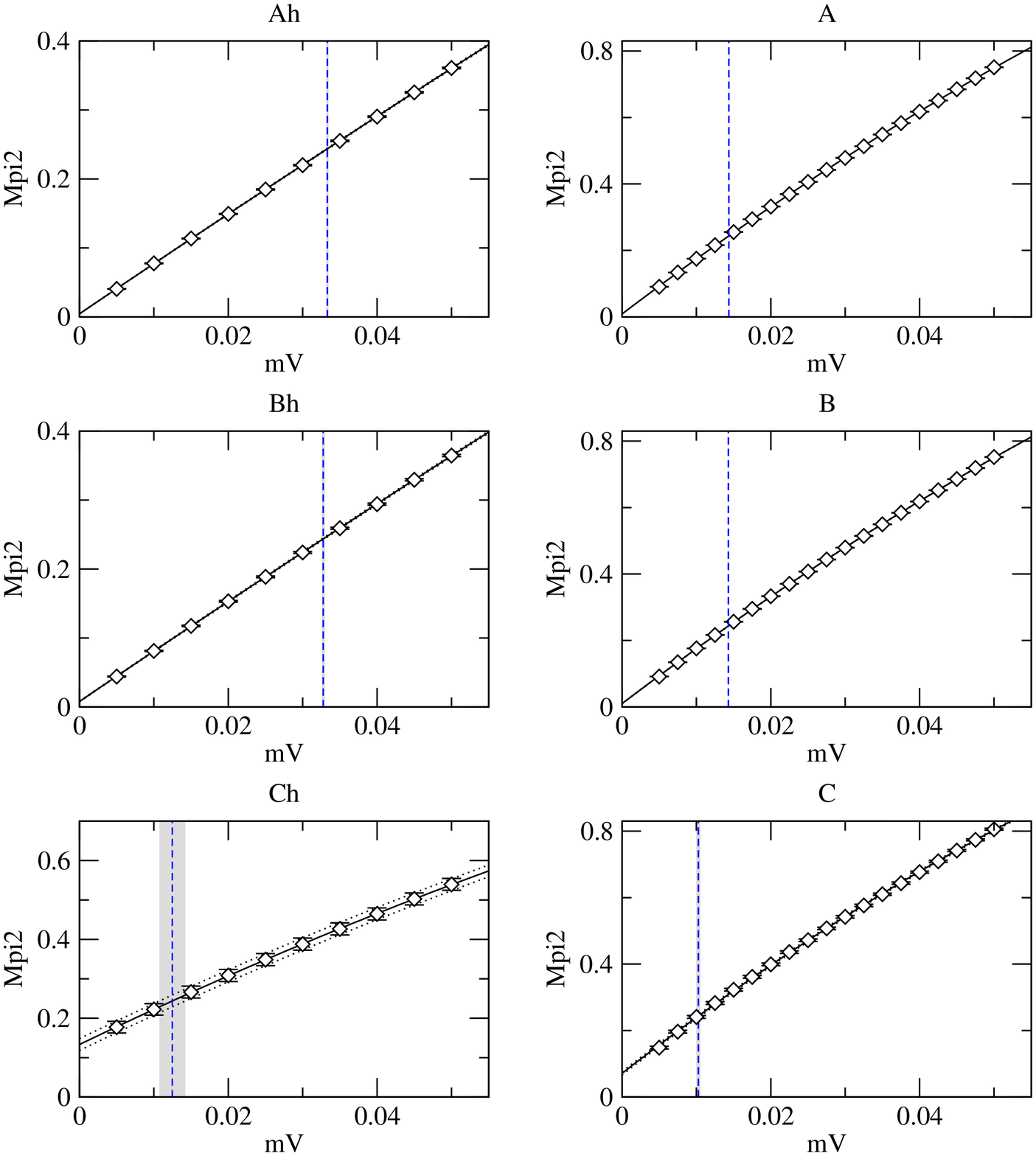}
\mycaption{Quadratic pion-mass fit and kaon quark-mass threshold for ensembles
\ensemble{A ~ hyp}, \ensemble{A}, \ensemble{B ~ hyp}, \ensemble{B}, \ensemble{C
~ hyp}, and \ensemble{C}.  The dotted vertical lines correspond to the
determined values of $m_{Q_K}$.}
\label{p:figure}
\end{figure}

\begin{figure}
\centering
\psfrag{mV}[t][t]{$m_V$}
\psfrag{Mpi2}[b][B]{$M^2_{\pi_5} \, (\text{GeV}^2)$}
\psfrag{W}[B][B]{\ensemble{W}}
\psfrag{Wh}[B][B]{\ensemble{W ~ hyp}}
\psfrag{X}[B][B]{\ensemble{X}}
\psfrag{Xh}[B][B]{\ensemble{X ~ hyp}}
\includegraphics[width=\textwidth,clip=]{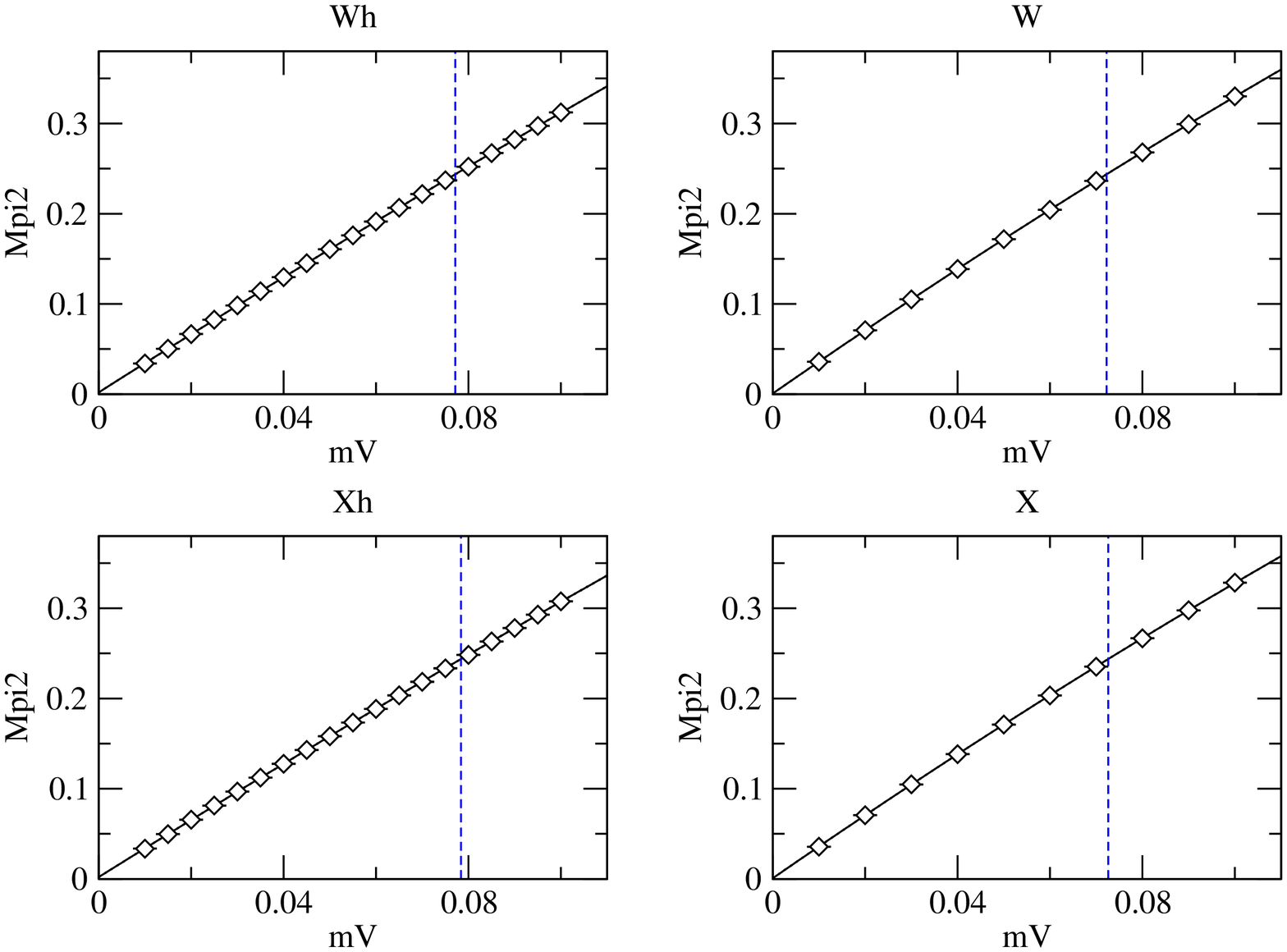}
\mycaption{Quadratic pion-mass fit and kaon quark-mass threshold for ensembles
\ensemble{W ~ hyp}, \ensemble{W}, \ensemble{X ~ hyp}, and \ensemble{X}.  The
dotted vertical lines correspond to the determined values of $m_{Q_K}$.}
\end{figure}

\begin{figure}
\centering
\psfrag{mV}[t][t]{$m_V$}
\psfrag{Mpi2}[b][B]{$M^2_{\pi_5} \, (\text{GeV}^2)$}
\psfrag{Y}[B][B]{\ensemble{Y}}
\psfrag{Yh}[B][B]{\ensemble{Y ~ hyp}}
\psfrag{Z}[B][B]{\ensemble{Z}}
\psfrag{Zh}[B][B]{\ensemble{Z ~ hyp}}
\includegraphics[width=\textwidth,clip=]{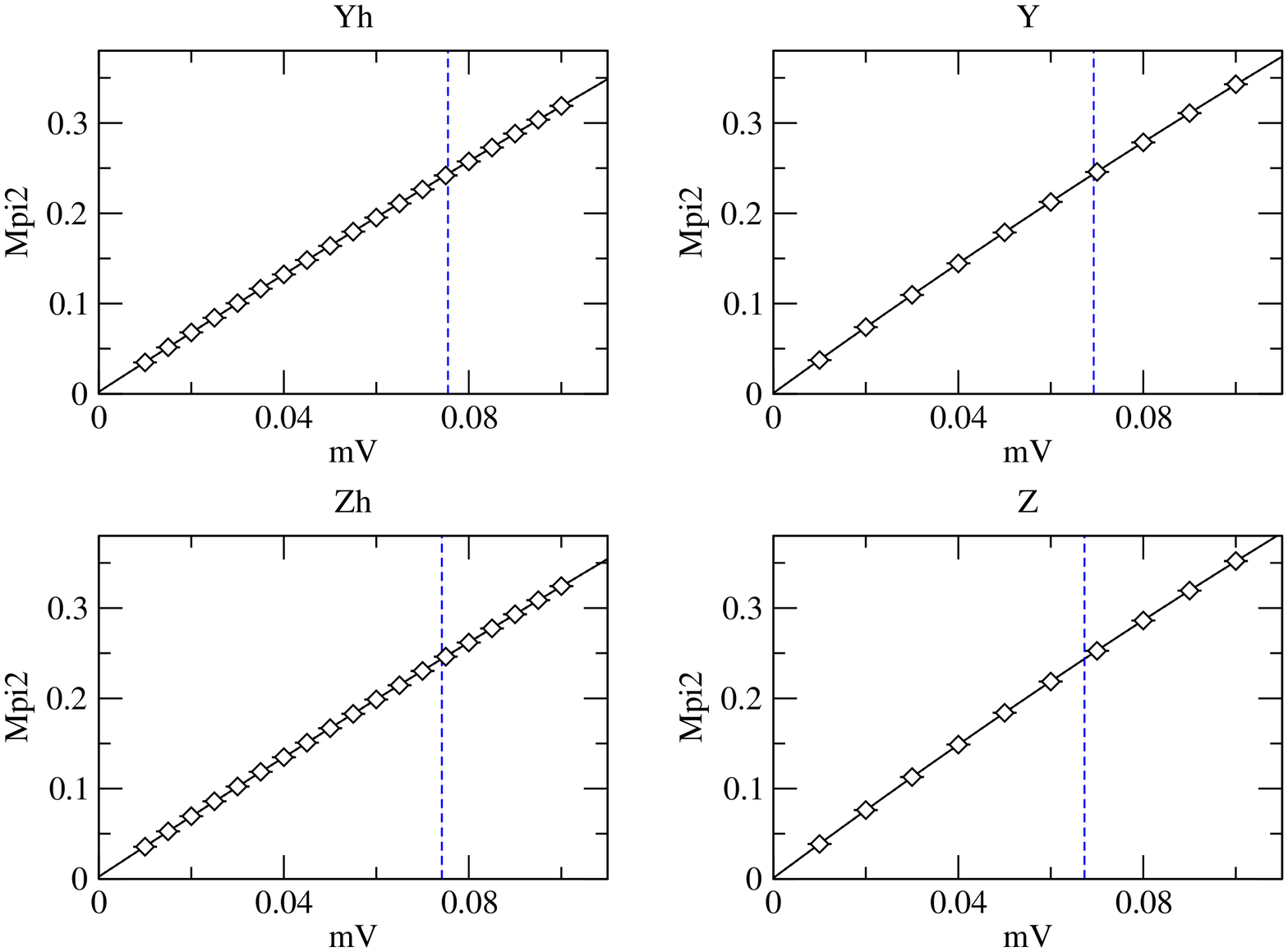}
\mycaption{Quadratic pion-mass fit and kaon quark-mass threshold for ensembles
\ensemble{Y ~ hyp}, \ensemble{Y}, \ensemble{Z ~ hyp}, and \ensemble{Z}.  The
dotted vertical lines correspond to the determined values of $m_{Q_K}$.}
\end{figure}

\begin{figure}
\centering
\psfrag{mV}[t][t]{$m_V$}
\psfrag{Mpi2}[b][B]{$M^2_{\pi_5} \, (\text{GeV}^2)$}
\psfrag{Q}[B][B]{\ensemble{Q}}
\psfrag{Qh}[B][B]{\ensemble{Q ~ hyp}}
\includegraphics[width=0.5\textwidth,clip=]{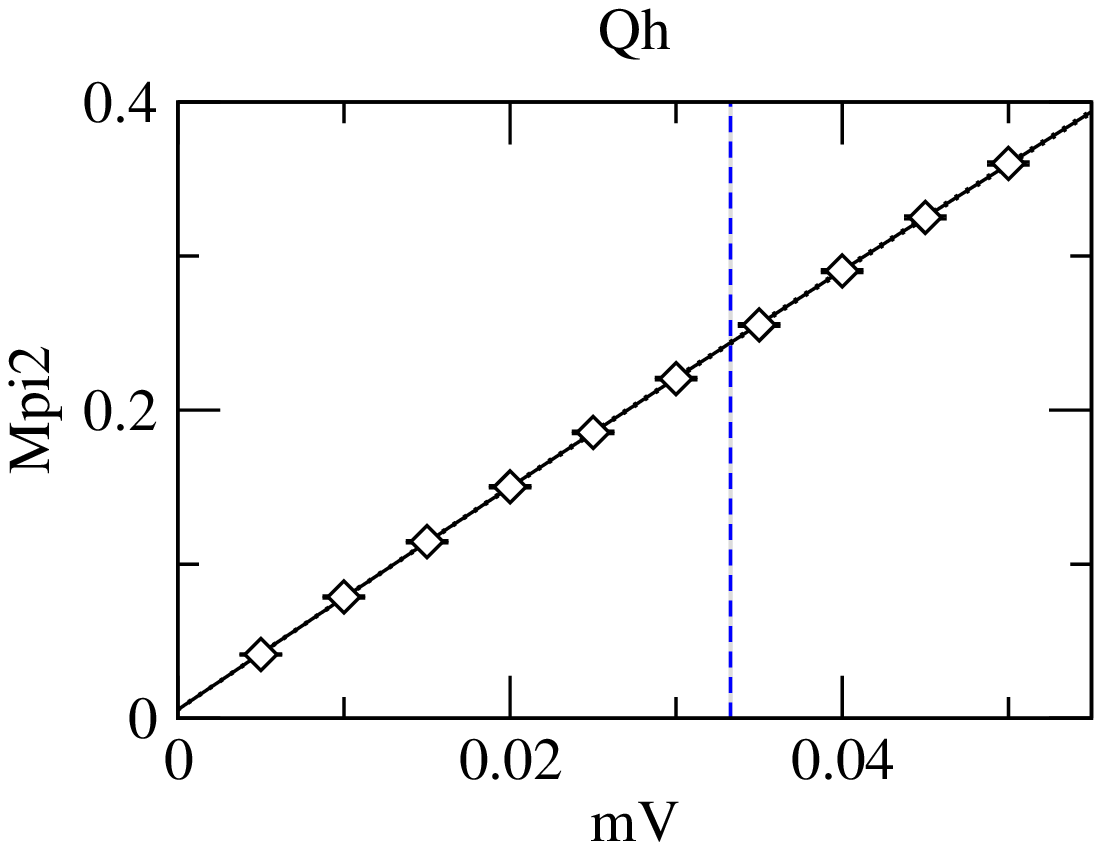}
\mycaption{Quadratic pion-mass fit and kaon quark-mass threshold for ensemble
\ensemble{Q ~ hyp}.  The dotted vertical line corresponds to the determined
value of $m_{Q_K}$.}
\label{q:figure}
\end{figure}

\clearpage

\begin{figure}
\centering
\psfrag{X2dof}[b][B]{\Large $\chi^2 / \text{dof}$}
\psfrag{LmV}[t][t]{\Large $\Lambda_{m_V}$}
\psfrag{Ah}[B][B]{\Large \ensemble{A ~ hyp}}
\includegraphics[width=\textwidth,clip=]{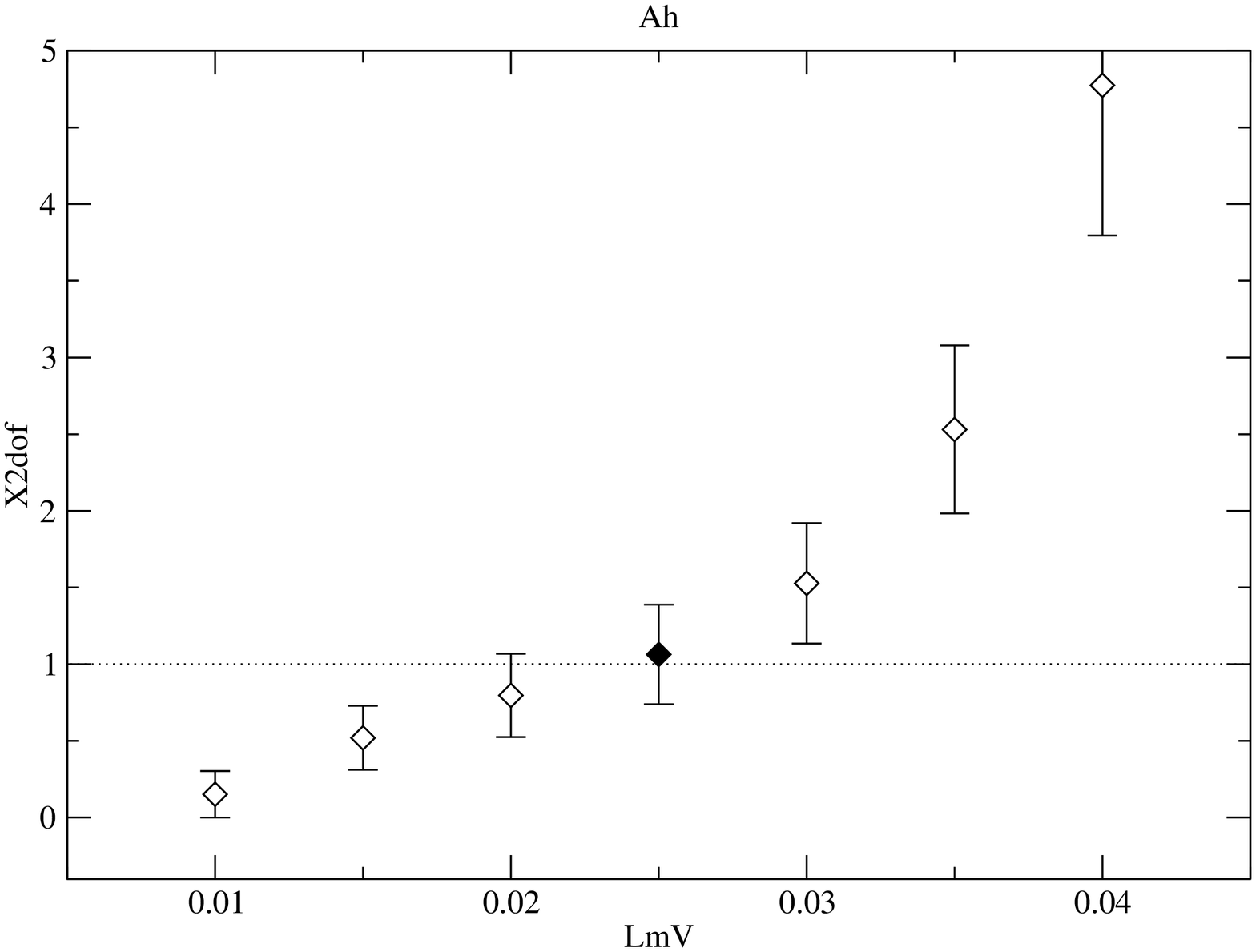}
\mycaption{$\chi^2$ per degree of freedom over a range of valence-quark-mass
cutoffs for ensemble \ensemble{A ~ hyp}.  The filled diamond corresponds to
the cutoff used in the final fit.}
\label{t:figure}
\end{figure}

\clearpage

\begin{figure}
\centering
\psfrag{mVmK}[t][t]{\Large $m_V / m_{Q_K}$}
\psfrag{Mpi2}[b][B]{\Large $M^2_{\pi_5} \, (\text{GeV}^2)$}
\psfrag{Ah}[B][B]{\Large \ensemble{A ~ hyp}}
\includegraphics[width=\textwidth,clip=]{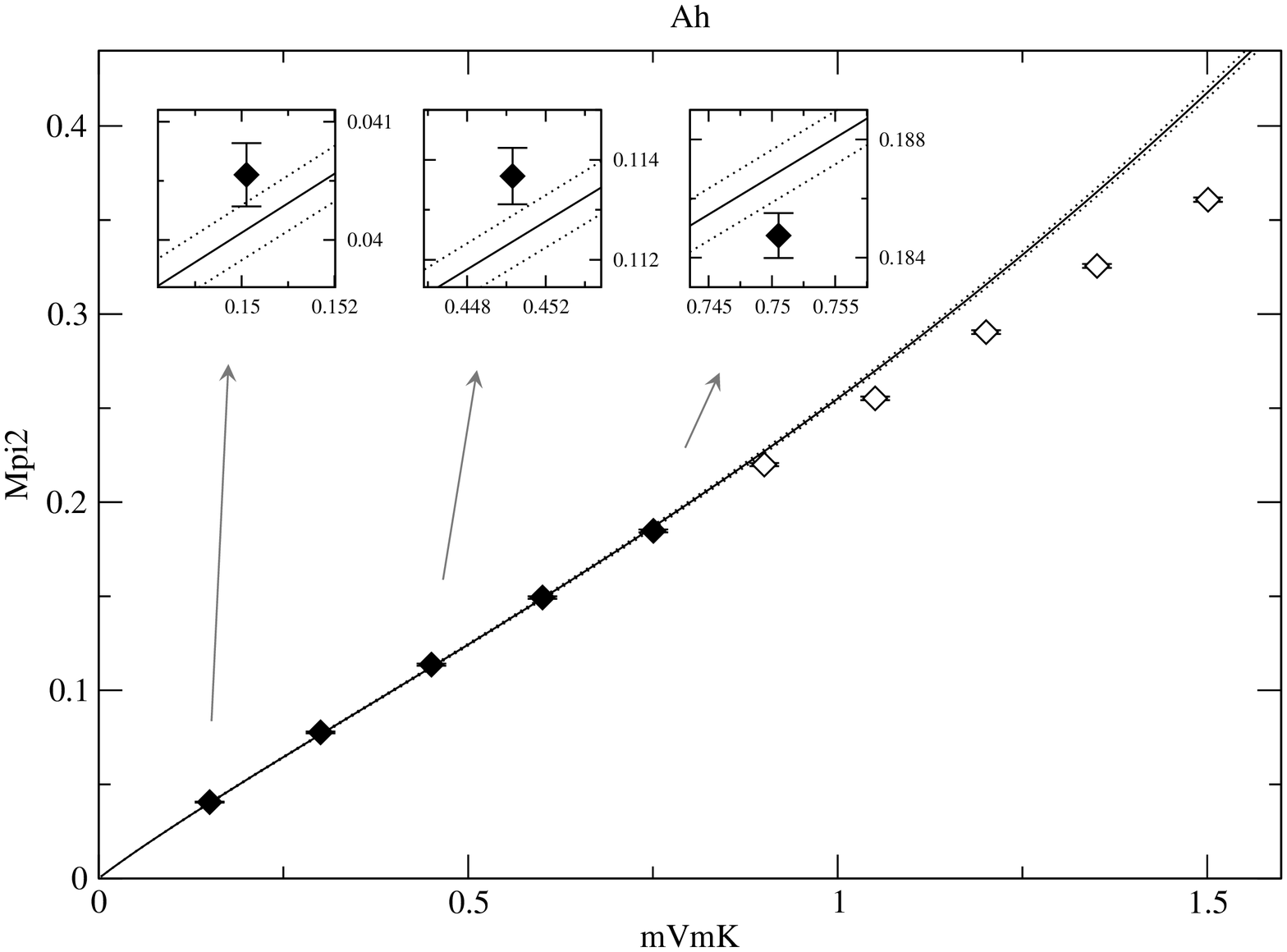}
\mycaption{Pion mass squared versus valence quark mass from the pqChPT
fit of ensemble \ensemble{A ~ hyp}.  Filled diamonds
correspond to valence-quark-mass values within the fit range, while open
diamonds correspond to those values beyond it.}
\label{u:figure}
\end{figure}

\begin{figure}
\centering
\psfrag{mVmK}[t][t]{\Large $m_V / m_{Q_K}$}
\psfrag{fpi}[b][B]{\Large $f_{\pi_5} \, (\text{MeV})$}
\psfrag{Ah}[B][B]{\Large \ensemble{A ~ hyp}}
\includegraphics[width=\textwidth,clip=]{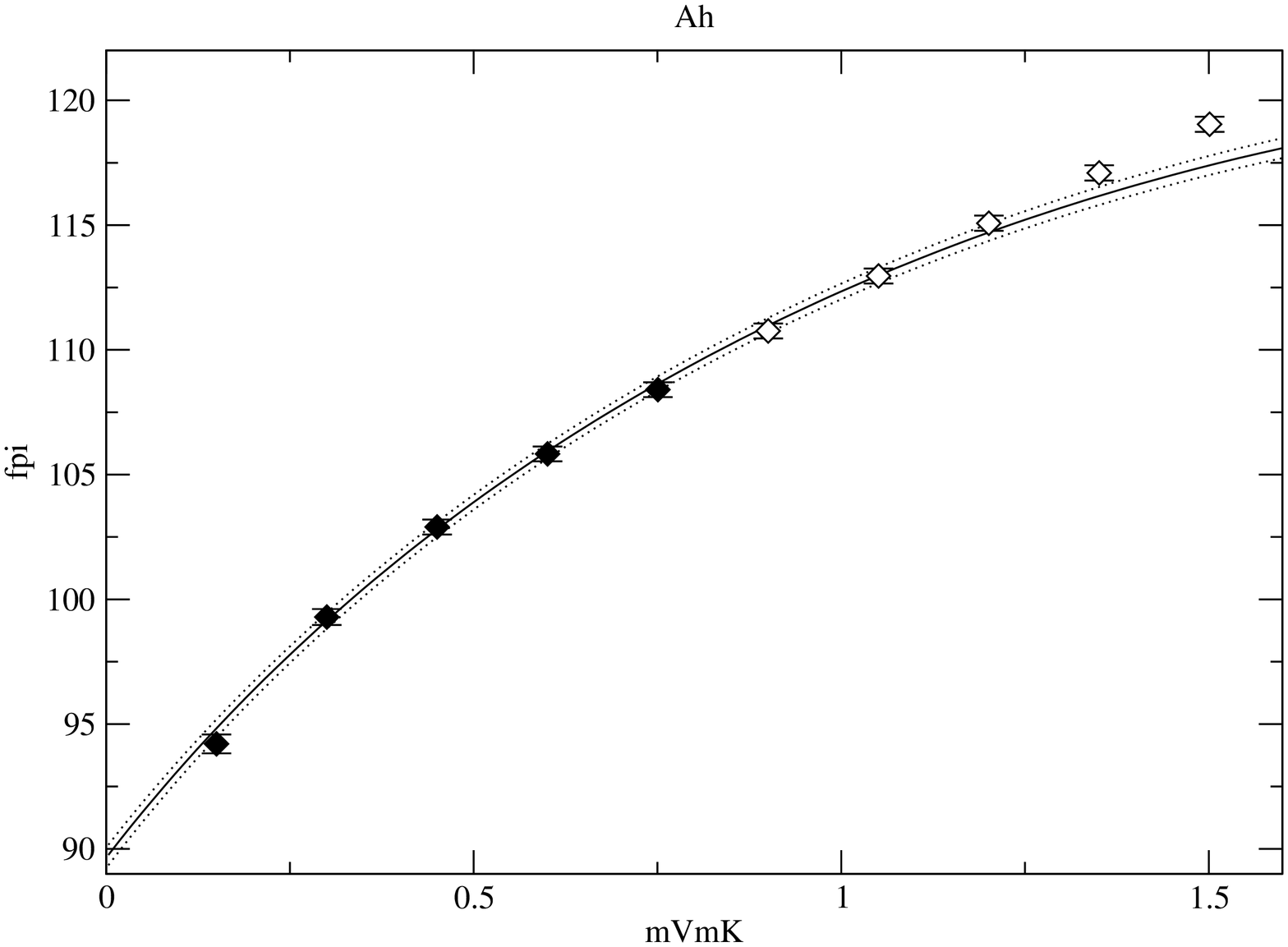}
\mycaption{Pion decay constant versus valence quark mass from the pqChPT
fit of ensemble \ensemble{A ~ hyp}.  Filled diamonds
correspond to valence-quark-mass values within the fit range, while open
diamonds correspond to those values beyond it.}
\label{v:figure}
\end{figure}

\begin{figure}
\centering
\psfrag{mVmK}[t][t]{\Large $m_V / m_{Q_K}$}
\psfrag{RM}[b][B]{\Large $R_M$}
\psfrag{Ah}[B][B]{\Large \ensemble{A ~ hyp}}
\includegraphics[width=\textwidth,clip=]{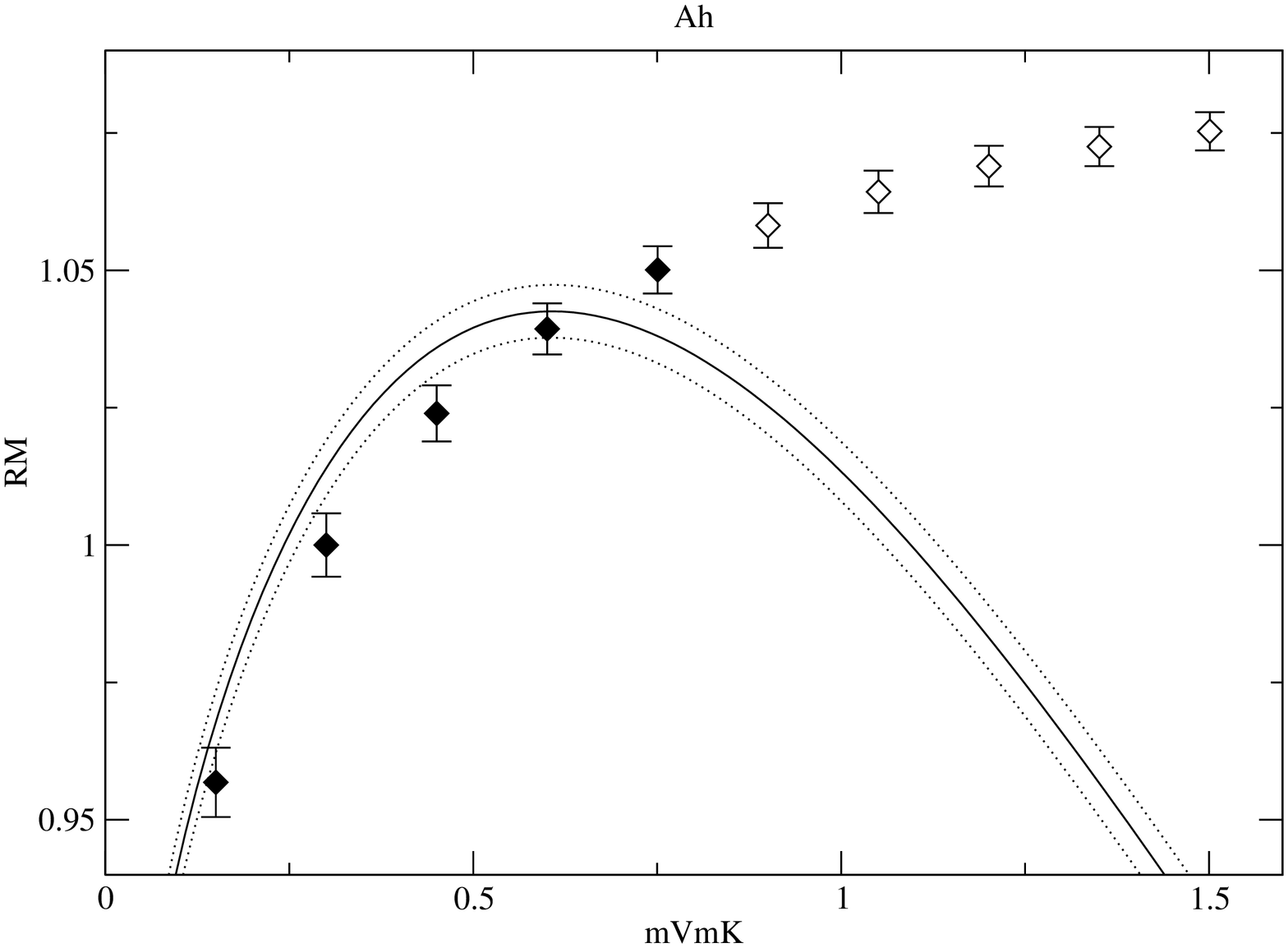}
\mycaption{$R_M$ versus valence quark mass from the pqChPT fit of the
correlator of ensemble \ensemble{A ~ hyp}.  Filled diamonds
correspond to valence-quark-mass values within the fit range, while open
diamonds correspond to those values beyond it.}
\label{w:figure}
\end{figure}

\clearpage

\begin{figure}
\centering
\psfrag{mVmK}[t][t]{$m_V / m_{Q_K}$}
\psfrag{RM}[b][B]{$R_M$}
\psfrag{Mpi2}[b][B]{$M^2_{\pi_5} \, (\text{GeV}^2)$}
\psfrag{fpi}[b][B]{$f_{\pi_5} \, (\text{MeV})$}
\psfrag{Ah}[B][B]{\Large \ensemble{A ~ hyp}}
\includegraphics[width=\textwidth,clip=]{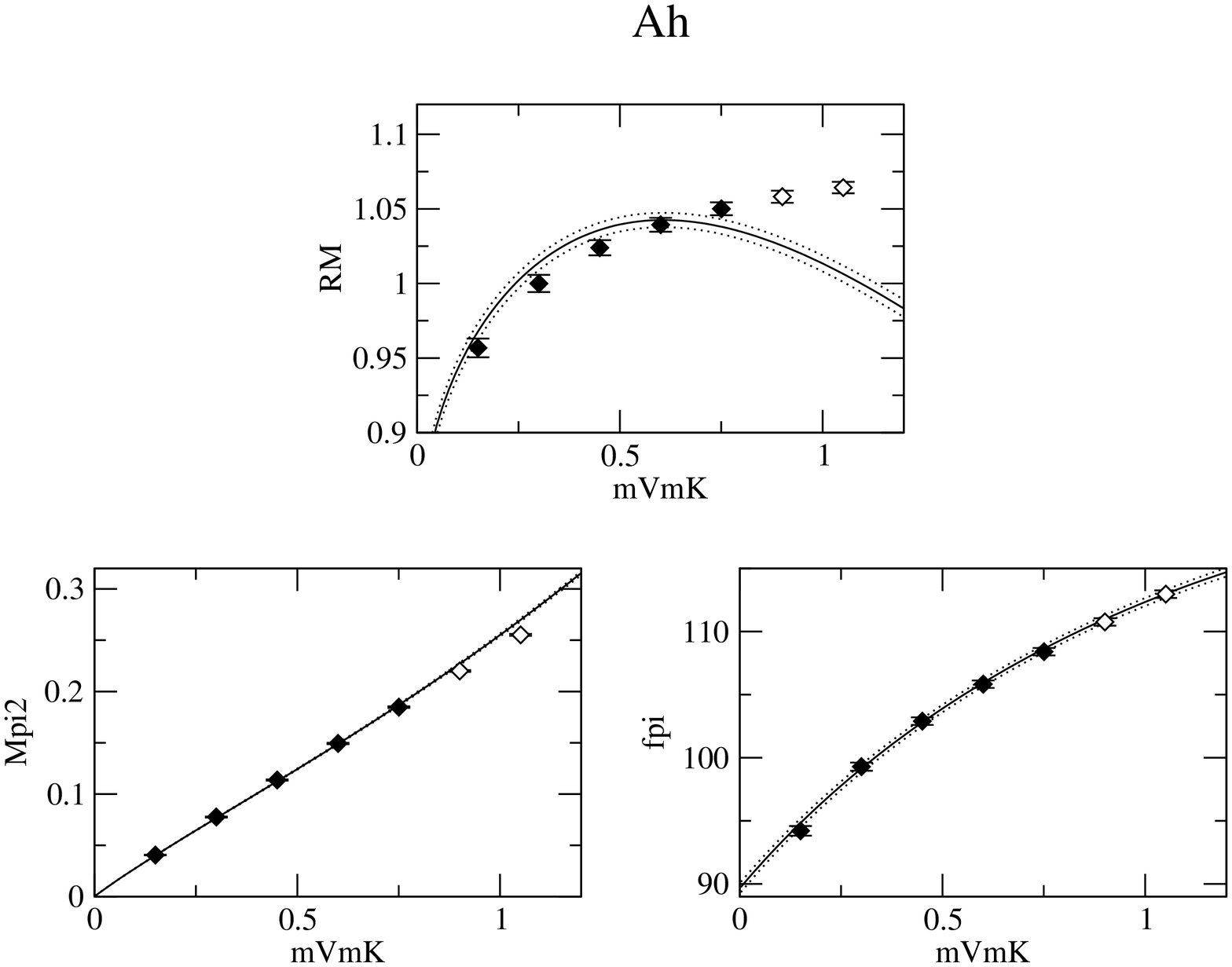}
\mycaption{Results from the pqChPT fit of ensemble
\ensemble{A ~ hyp}.  Filled diamonds correspond to valence-quark-mass values
within the fit range, while open diamonds correspond to those values beyond
it.}
\label{y:figure}
\end{figure}

\begin{figure}
\centering
\psfrag{mVmK}[t][t]{$m_V / m_{Q_K}$}
\psfrag{RM}[b][B]{$R_M$}
\psfrag{Mpi2}[b][B]{$M^2_{\pi_5} \, (\text{GeV}^2)$}
\psfrag{fpi}[b][B]{$f_{\pi_5} \, (\text{MeV})$}
\psfrag{Bh}[B][B]{\Large \ensemble{B ~ hyp}}
\includegraphics[width=\textwidth,clip=]{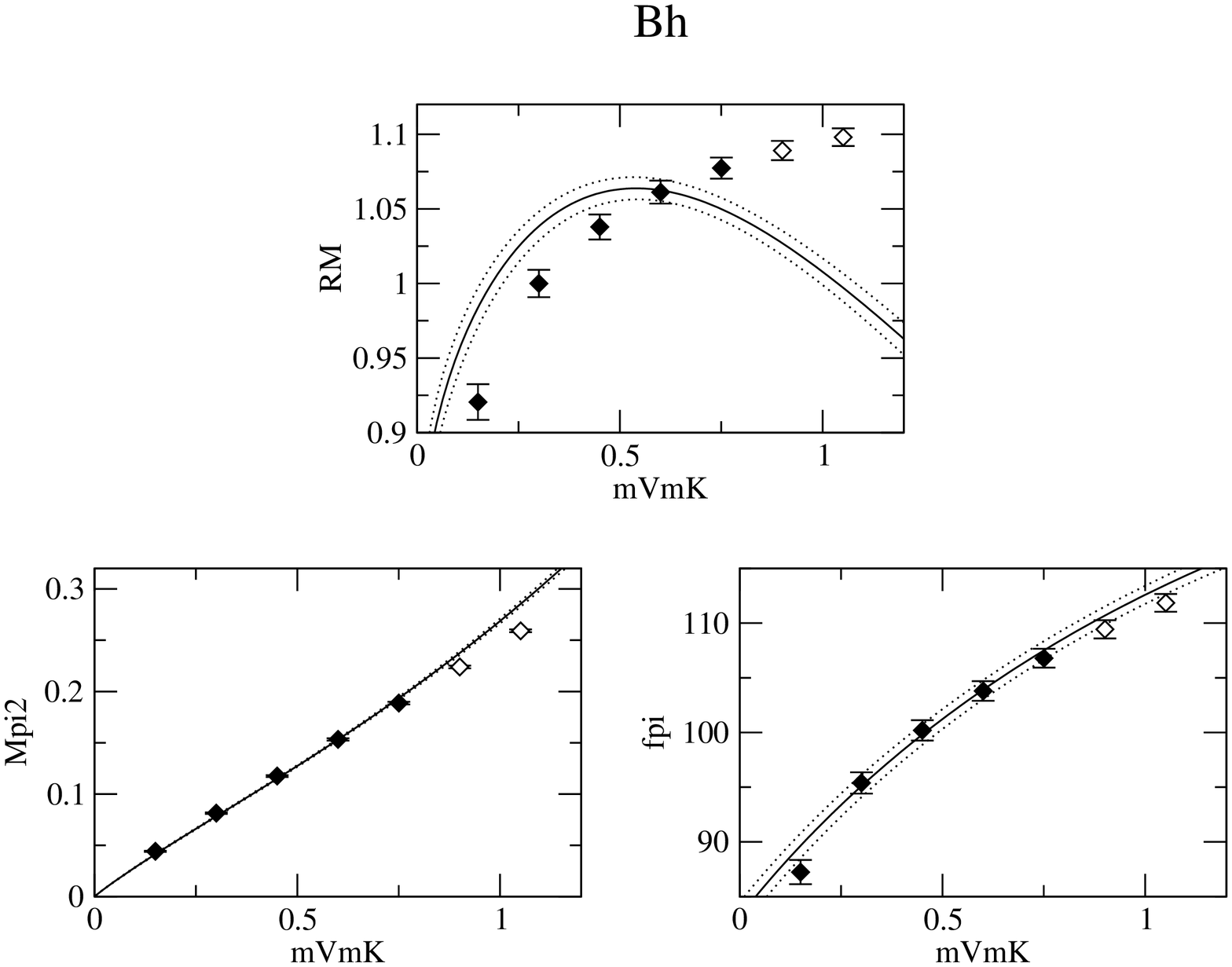}
\mycaption{Results from the pqChPT fit of ensemble
\ensemble{B ~ hyp}.  Filled diamonds correspond to valence-quark-mass values
within the fit range, while open diamonds correspond to those values beyond
it.}
\end{figure}

\begin{figure}
\centering
\psfrag{mVmK}[t][t]{$m_V / m_{Q_K}$}
\psfrag{RM}[b][B]{$R_M$}
\psfrag{Mpi2}[b][B]{$M^2_{\pi_5} \, (\text{GeV}^2)$}
\psfrag{fpi}[b][B]{$f_{\pi_5} \, (\text{MeV})$}
\psfrag{A}[B][B]{\Large \ensemble{A}}
\includegraphics[width=\textwidth,clip=]{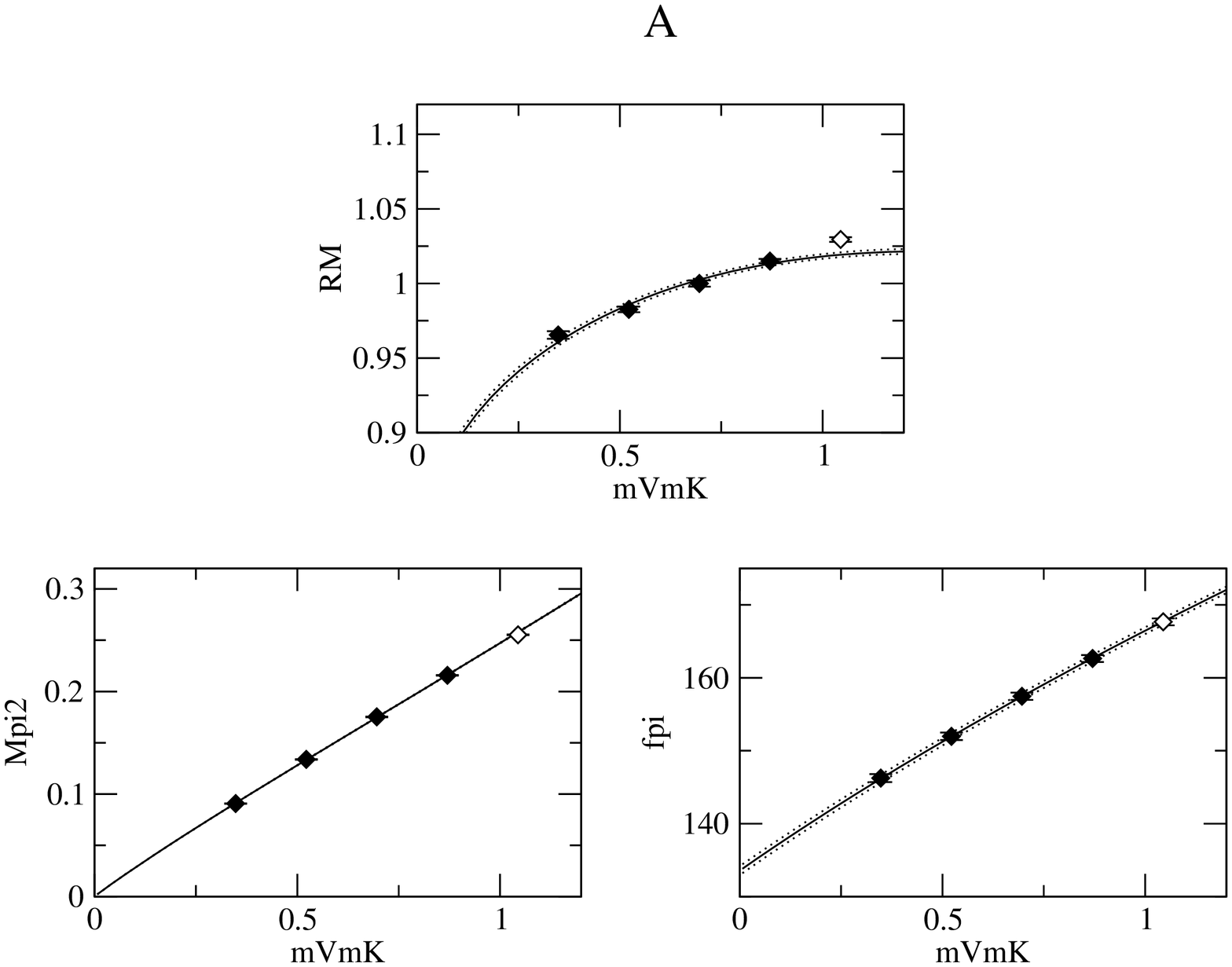}
\mycaption{Results from the pqChPT fit of ensemble
\ensemble{A}.  Filled diamonds correspond to valence-quark-mass values within
the fit range, while open diamonds correspond to those values beyond it.} 
\end{figure}

\begin{figure}
\centering
\psfrag{mVmK}[t][t]{$m_V / m_{Q_K}$}
\psfrag{RM}[b][B]{$R_M$}
\psfrag{Mpi2}[b][B]{$M^2_{\pi_5} \, (\text{GeV}^2)$}
\psfrag{fpi}[b][B]{$f_{\pi_5} \, (\text{MeV})$}
\psfrag{B}[B][B]{\Large \ensemble{B}}
\includegraphics[width=\textwidth,clip=]{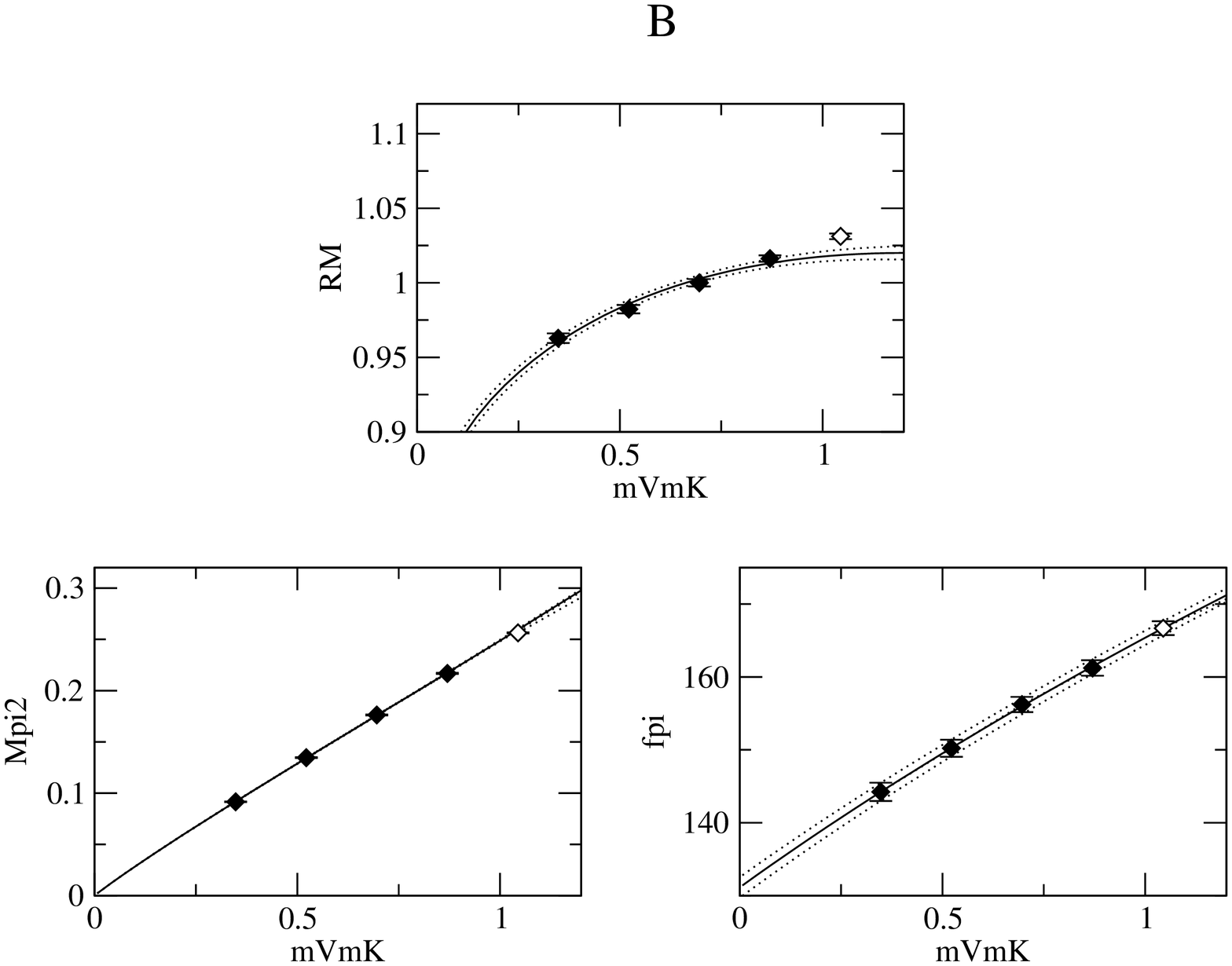}
\mycaption{Results from the pqChPT fit of ensemble
\ensemble{B}.  Filled diamonds correspond to valence-quark-mass values within
the fit range, while open diamonds correspond to those values beyond it.}
\end{figure}

\begin{figure}
\centering
\psfrag{mVmK}[t][t]{$m_V / m_{Q_K}$}
\psfrag{RM}[b][B]{$R_M$}
\psfrag{Mpi2}[b][B]{$M^2_{\pi_5} \, (\text{GeV}^2)$}
\psfrag{fpi}[b][B]{$f_{\pi_5} \, (\text{MeV})$}
\psfrag{Wh}[B][B]{\Large \ensemble{W ~ hyp}}
\includegraphics[width=\textwidth,clip=]{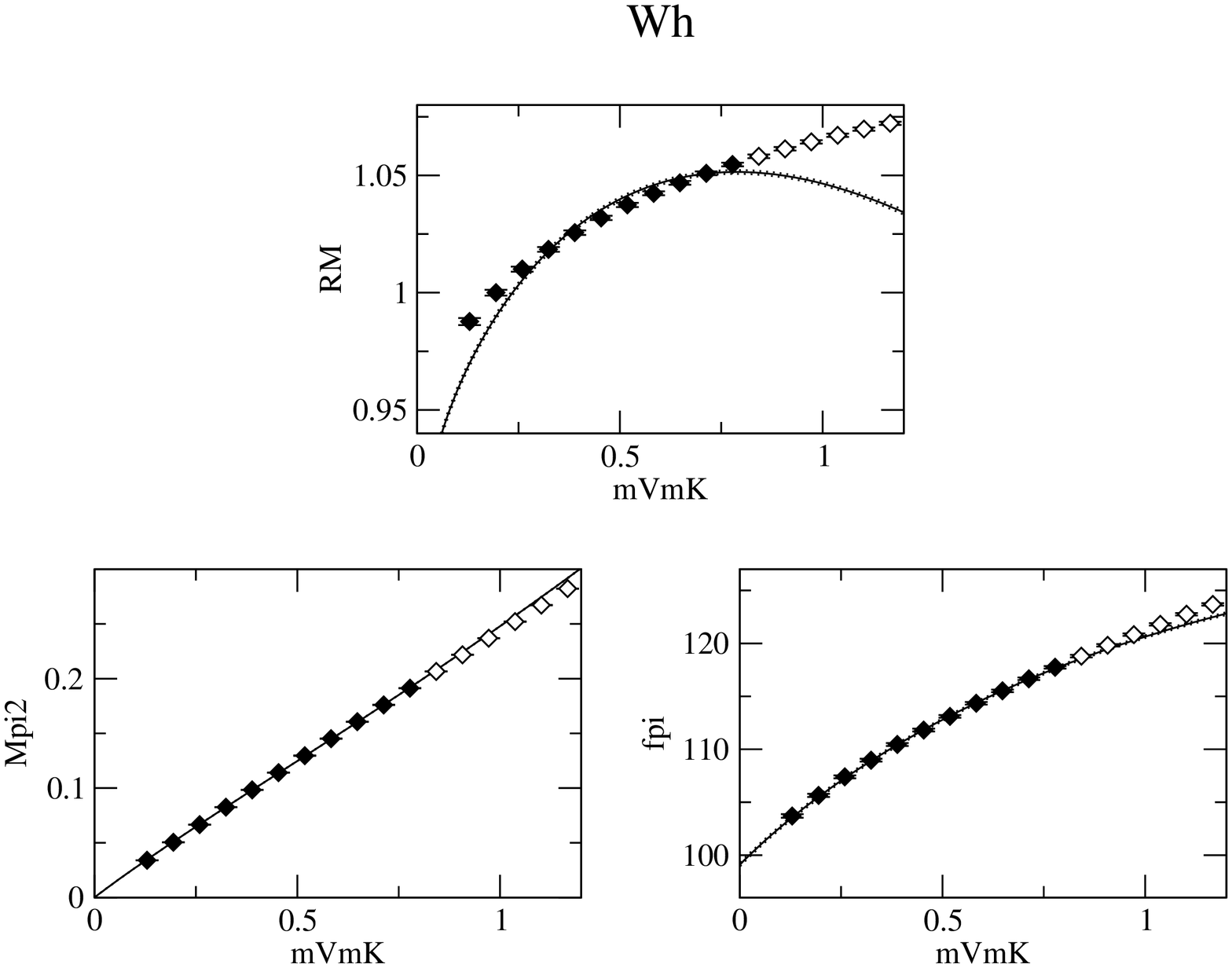}
\mycaption{Results from the pqChPT fit of ensemble
\ensemble{W ~ hyp}.  Filled diamonds correspond to valence-quark-mass values
within the fit range, while open diamonds correspond to those values beyond
it.} 
\end{figure}

\begin{figure}
\centering
\psfrag{mVmK}[t][t]{$m_V / m_{Q_K}$}
\psfrag{RM}[b][B]{$R_M$}
\psfrag{Mpi2}[b][B]{$M^2_{\pi_5} \, (\text{GeV}^2)$}
\psfrag{fpi}[b][B]{$f_{\pi_5} \, (\text{MeV})$}
\psfrag{Xh}[B][B]{\Large \ensemble{X ~ hyp}}
\includegraphics[width=\textwidth,clip=]{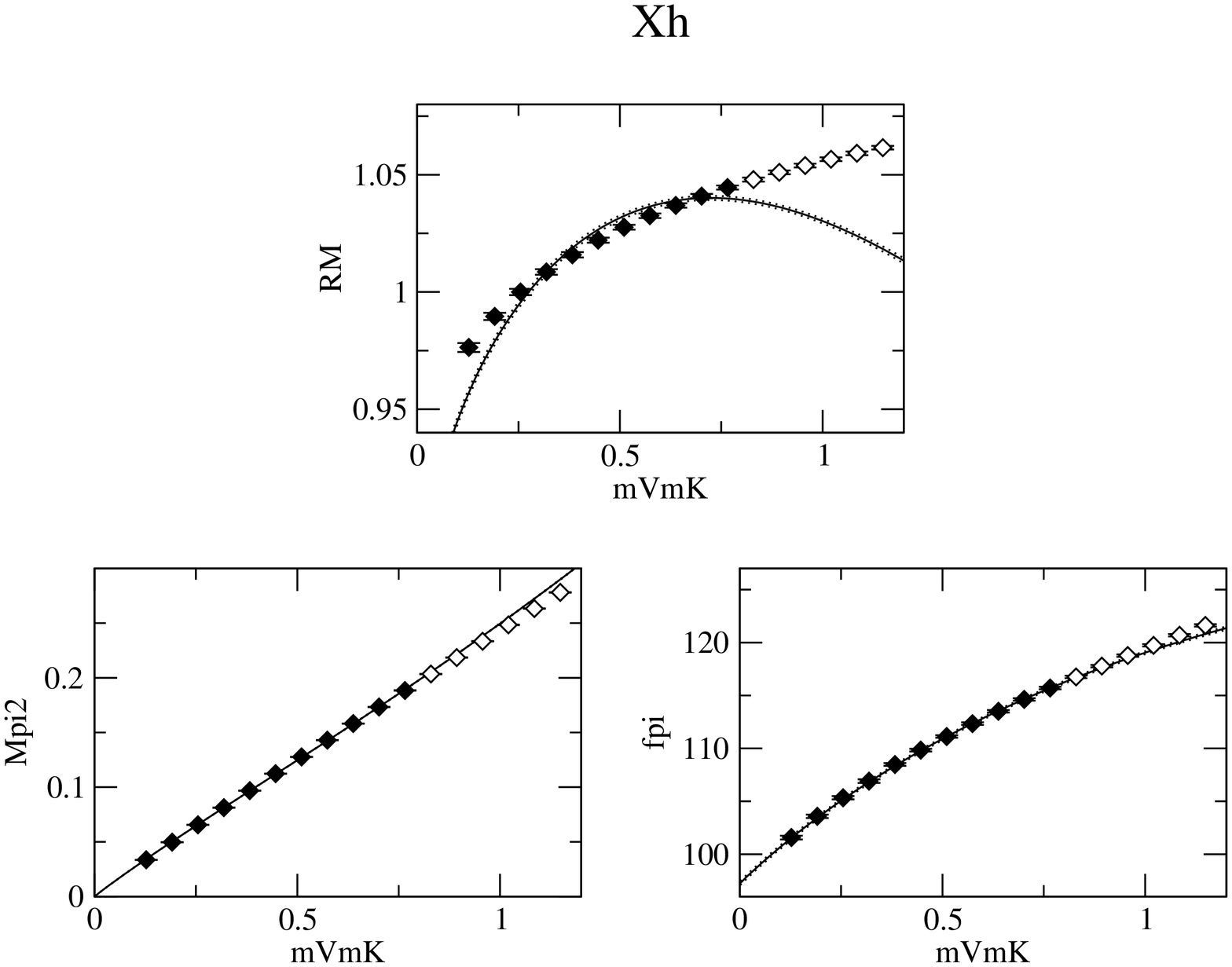}
\mycaption{Results from the pqChPT fit of ensemble
\ensemble{X ~ hyp}.  Filled diamonds correspond to valence-quark-mass values
within the fit range, while open diamonds correspond to those values beyond
it.} 
\end{figure}

\begin{figure}
\centering
\psfrag{mVmK}[t][t]{$m_V / m_{Q_K}$}
\psfrag{RM}[b][B]{$R_M$}
\psfrag{Mpi2}[b][B]{$M^2_{\pi_5} \, (\text{GeV}^2)$}
\psfrag{fpi}[b][B]{$f_{\pi_5} \, (\text{MeV})$}
\psfrag{Yh}[B][B]{\Large \ensemble{Y ~ hyp}}
\includegraphics[width=\textwidth,clip=]{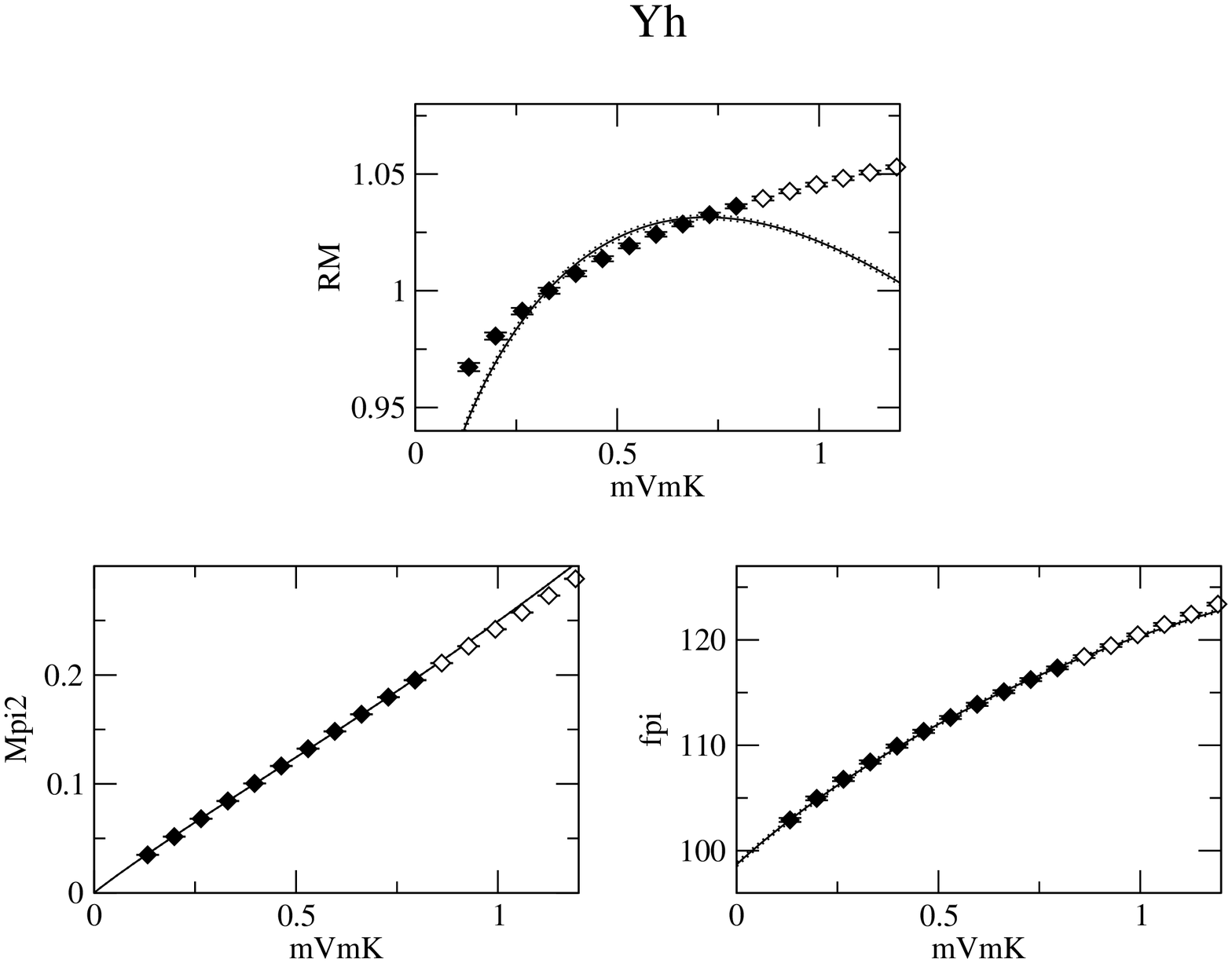}
\mycaption{Results from the pqChPT fit of ensemble
\ensemble{Y ~ hyp}.  Filled diamonds correspond to valence-quark-mass values
within the fit range, while open diamonds correspond to those values beyond
it.} 
\end{figure}

\begin{figure}
\centering
\psfrag{mVmK}[t][t]{$m_V / m_{Q_K}$}
\psfrag{RM}[b][B]{$R_M$}
\psfrag{Mpi2}[b][B]{$M^2_{\pi_5} \, (\text{GeV}^2)$}
\psfrag{fpi}[b][B]{$f_{\pi_5} \, (\text{MeV})$}
\psfrag{Zh}[B][B]{\Large \ensemble{Z ~ hyp}}
\includegraphics[width=\textwidth,clip=]{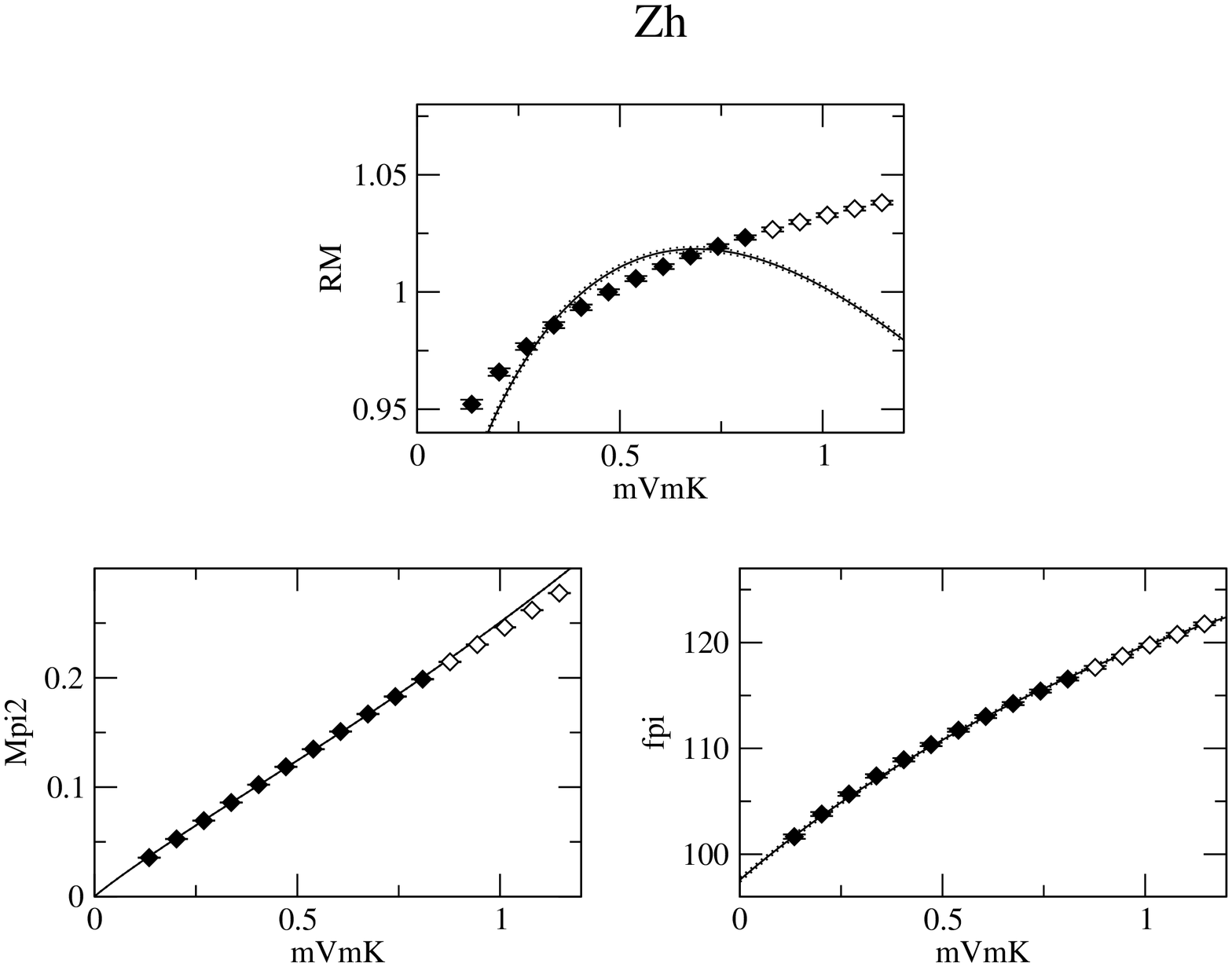}
\mycaption{Results from the pqChPT fit of ensemble
\ensemble{Z ~ hyp}.  Filled diamonds correspond to valence-quark-mass values
within the fit range, while open diamonds correspond to those values beyond
it.} 
\end{figure}

\begin{figure}
\centering
\psfrag{mVmK}[t][t]{$m_V / m_{Q_K}$}
\psfrag{RM}[b][B]{$R_M$}
\psfrag{Mpi2}[b][B]{$M^2_{\pi_5} \, (\text{GeV}^2)$}
\psfrag{fpi}[b][B]{$f_{\pi_5} \, (\text{MeV})$}
\psfrag{W}[B][B]{\Large \ensemble{W}}
\includegraphics[width=\textwidth,clip=]{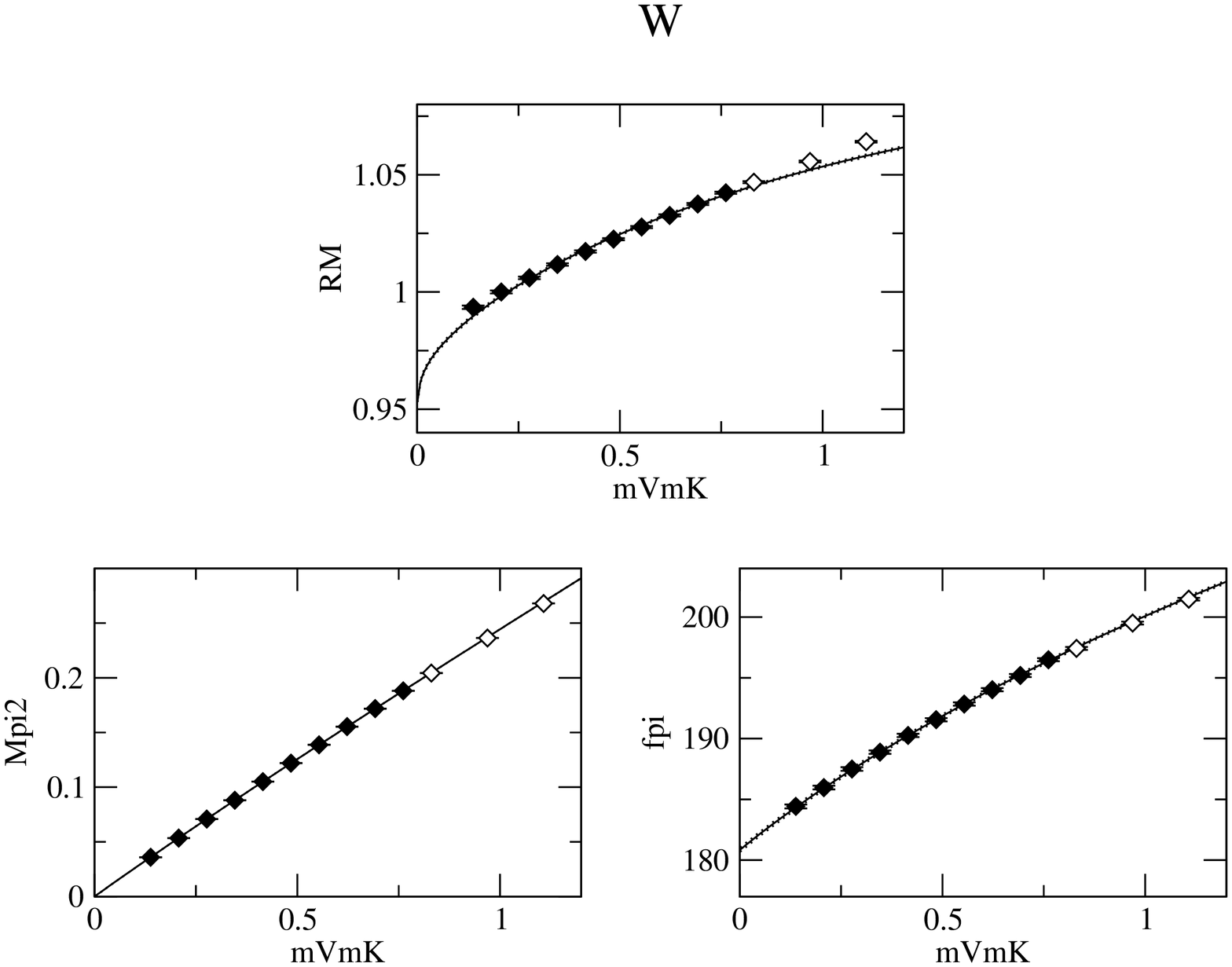}
\mycaption{Results from the pqChPT fit of ensemble
\ensemble{W}.  Filled diamonds correspond to valence-quark-mass values within
the fit range, while open diamonds correspond to those values beyond it.} 
\end{figure}

\begin{figure}
\centering
\psfrag{mVmK}[t][t]{$m_V / m_{Q_K}$}
\psfrag{RM}[b][B]{$R_M$}
\psfrag{Mpi2}[b][B]{$M^2_{\pi_5} \, (\text{GeV}^2)$}
\psfrag{fpi}[b][B]{$f_{\pi_5} \, (\text{MeV})$}
\psfrag{X}[B][B]{\Large \ensemble{X}}
\includegraphics[width=\textwidth,clip=]{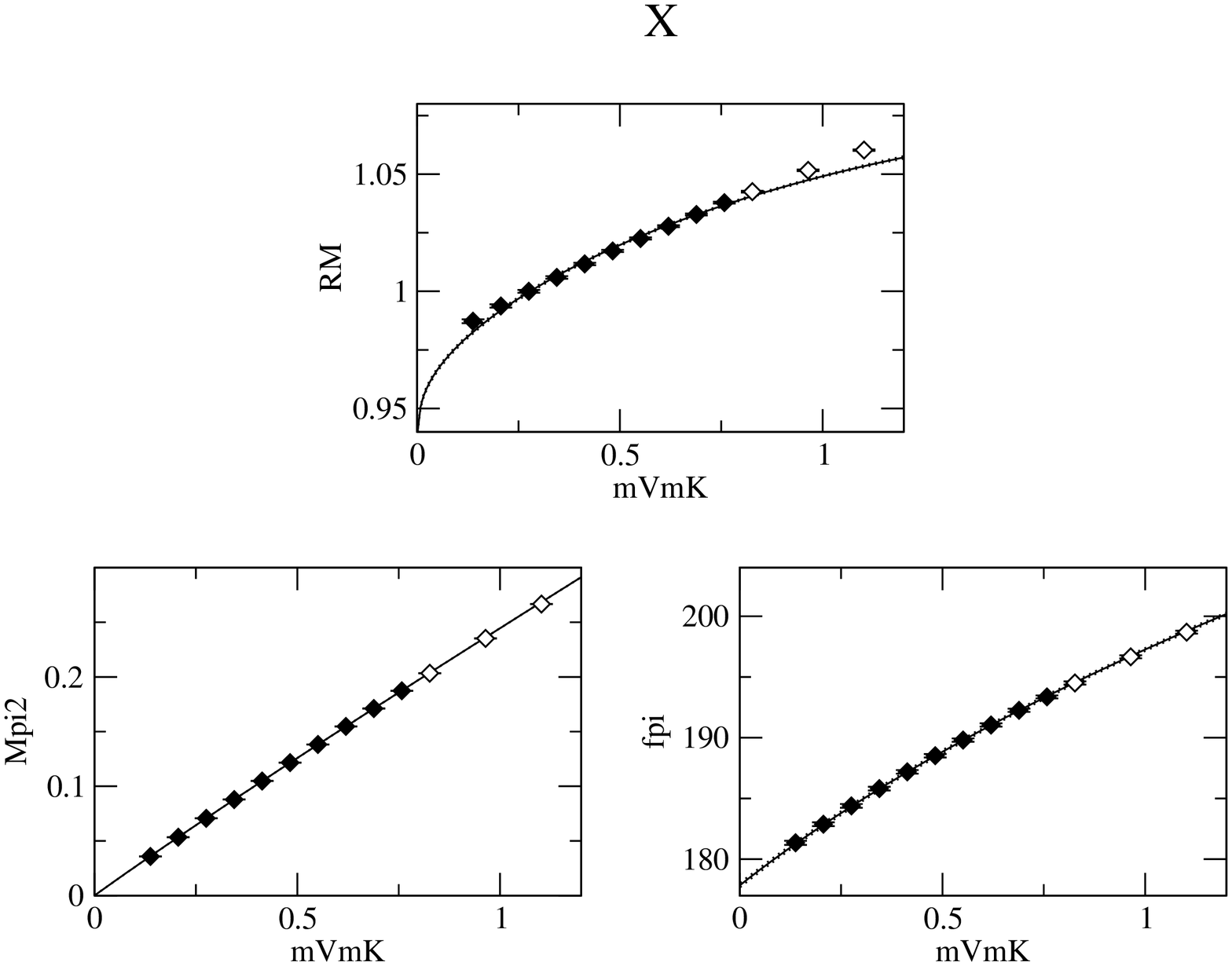}
\mycaption{Results from the pqChPT fit of ensemble
\ensemble{X}.  Filled diamonds correspond to valence-quark-mass values within
the fit range, while open diamonds correspond to those values beyond it.} 
\end{figure}

\begin{figure}
\centering
\psfrag{mVmK}[t][t]{$m_V / m_{Q_K}$}
\psfrag{RM}[b][B]{$R_M$}
\psfrag{Mpi2}[b][B]{$M^2_{\pi_5} \, (\text{GeV}^2)$}
\psfrag{fpi}[b][B]{$f_{\pi_5} \, (\text{MeV})$}
\psfrag{Y}[B][B]{\Large \ensemble{Y}}
\includegraphics[width=\textwidth,clip=]{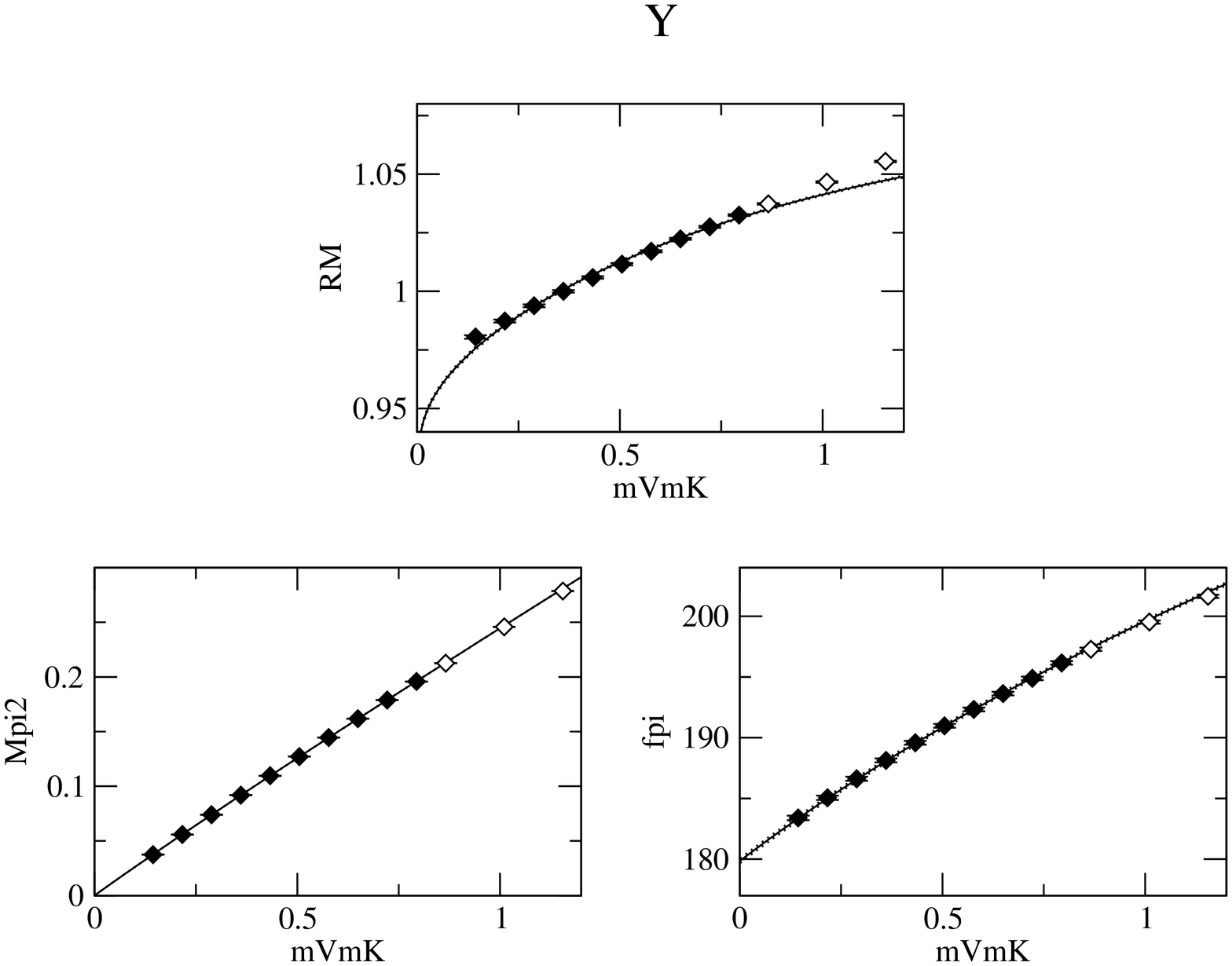}
\mycaption{Results from the pqChPT fit of ensemble
\ensemble{Y}.  Filled diamonds correspond to valence-quark-mass values within
the fit range, while open diamonds correspond to those values beyond it.} 
\end{figure}

\begin{figure}
\centering
\psfrag{mVmK}[t][t]{$m_V / m_{Q_K}$}
\psfrag{RM}[b][B]{$R_M$}
\psfrag{Mpi2}[b][B]{$M^2_{\pi_5} \, (\text{GeV}^2)$}
\psfrag{fpi}[b][B]{$f_{\pi_5} \, (\text{MeV})$}
\psfrag{Z}[B][B]{\Large \ensemble{Z}}
\includegraphics[width=\textwidth,clip=]{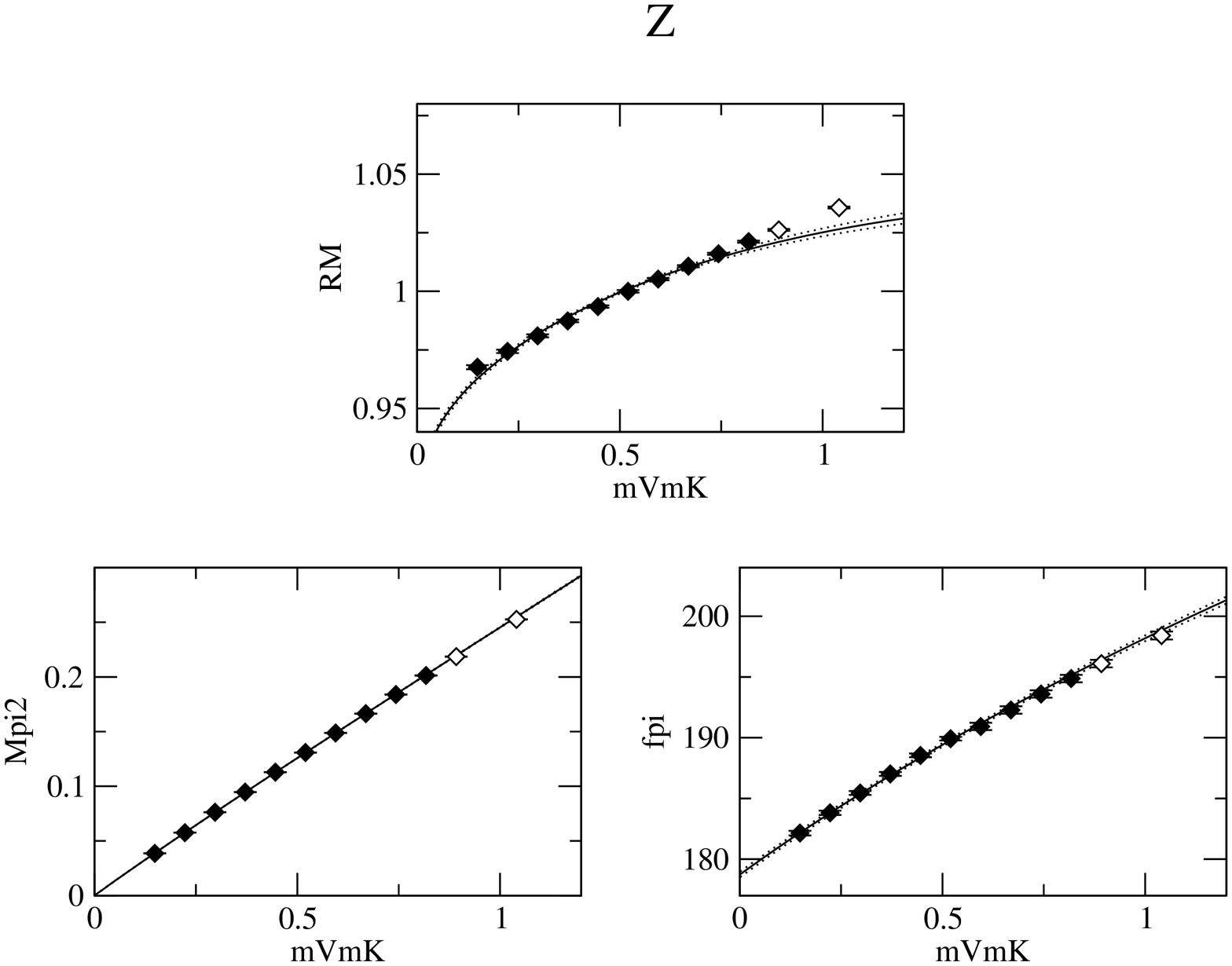}
\mycaption{Results from the pqChPT fit of ensemble
\ensemble{Z}.  Filled diamonds correspond to valence-quark-mass values within
the fit range, while open diamonds correspond to those values beyond it.} 
\end{figure}

\begin{figure}
\centering
\psfrag{mVmK}[t][t]{$m_V / m_{Q_K}$}
\psfrag{RM}[b][B]{$R_M$}
\psfrag{Mpi2}[b][B]{$M^2_{\pi_5} \, (\text{GeV}^2)$}
\psfrag{fpi}[b][B]{$f_{\pi_5} \, (\text{MeV})$}
\psfrag{Qh}[B][B]{\Large \ensemble{Q ~ hyp}}
\includegraphics[width=\textwidth,clip=]{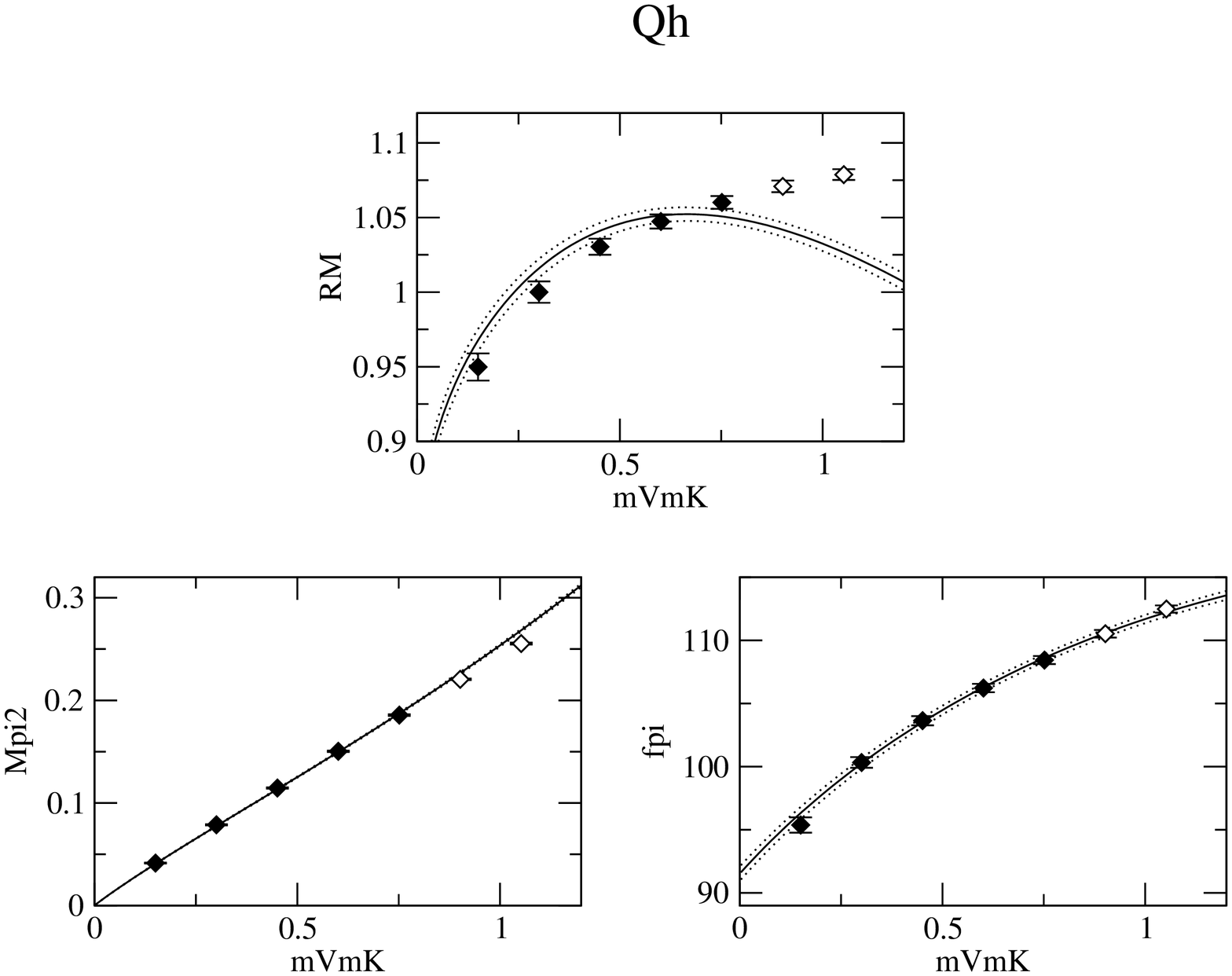}
\mycaption{Results from the pqChPT fit of ensemble
\ensemble{Q ~ hyp}.  Filled diamonds correspond to valence-quark-mass values
within the fit range, while open diamonds correspond to those values beyond
it.} 
\label{z:figure}
\end{figure}

\clearpage

\begin{figure}
\centering
\psfrag{mVmK}[t][t]{\Large $m_V / m_{Q_K}$}
\psfrag{RM}[b][B]{\Large $R_M$}
\psfrag{Ah}[B][B]{\Large \ensemble{A ~ hyp}}
\includegraphics[width=\textwidth,clip=]{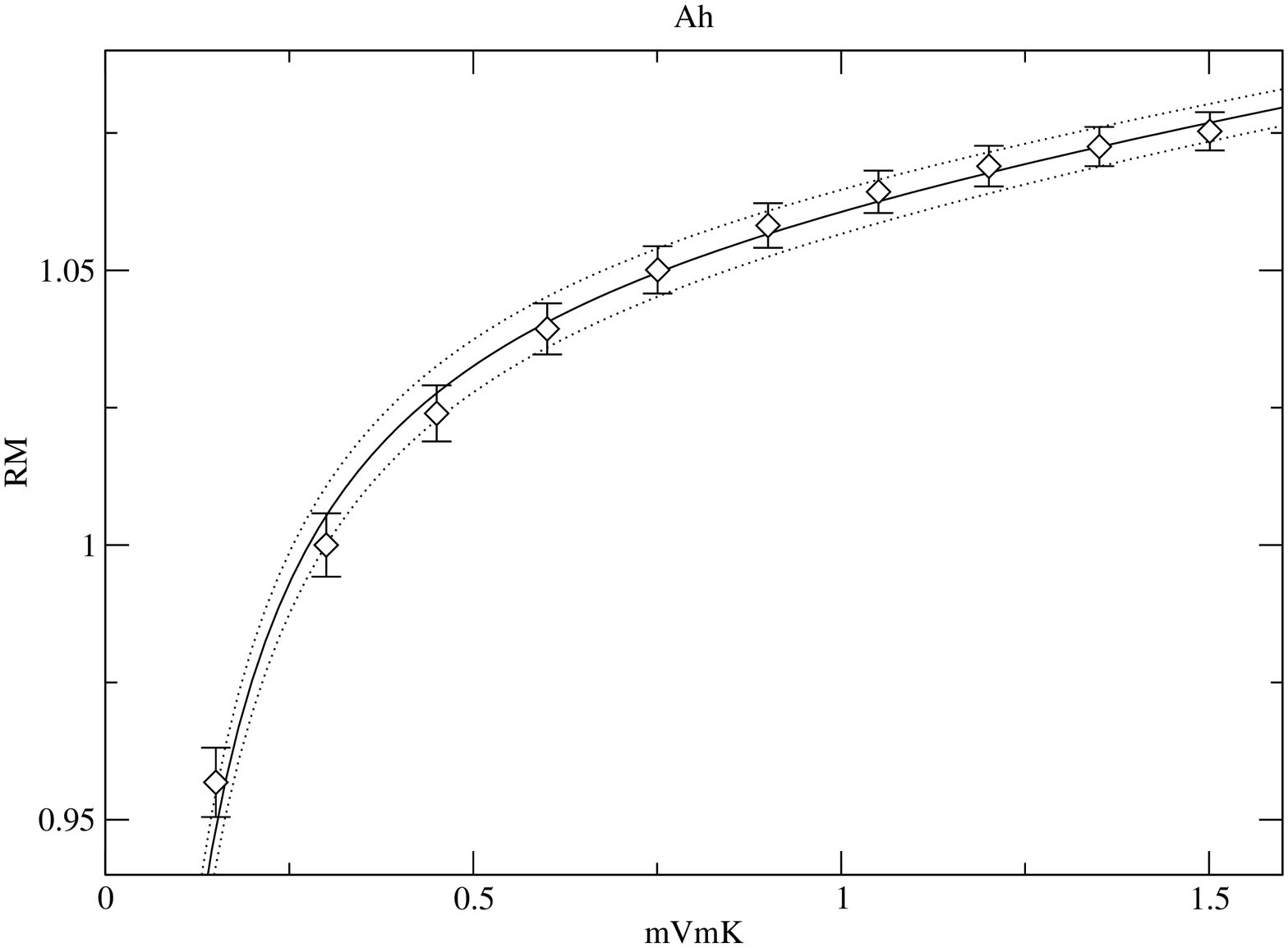}
\mycaption{$R_M$ versus valence quark mass from the quadratic fit of the
correlator of ensemble \ensemble{A ~ hyp}.}
\label{ab:figure}
\end{figure}

\begin{figure}
\centering
\psfrag{mVmK}[t][t]{\Large $m_V / m_{Q_K}$}
\psfrag{RM}[b][B]{\Large $R_M$}
\psfrag{Ah}[B][B]{\Large \ensemble{A ~ hyp}}
\includegraphics[width=\textwidth,clip=]{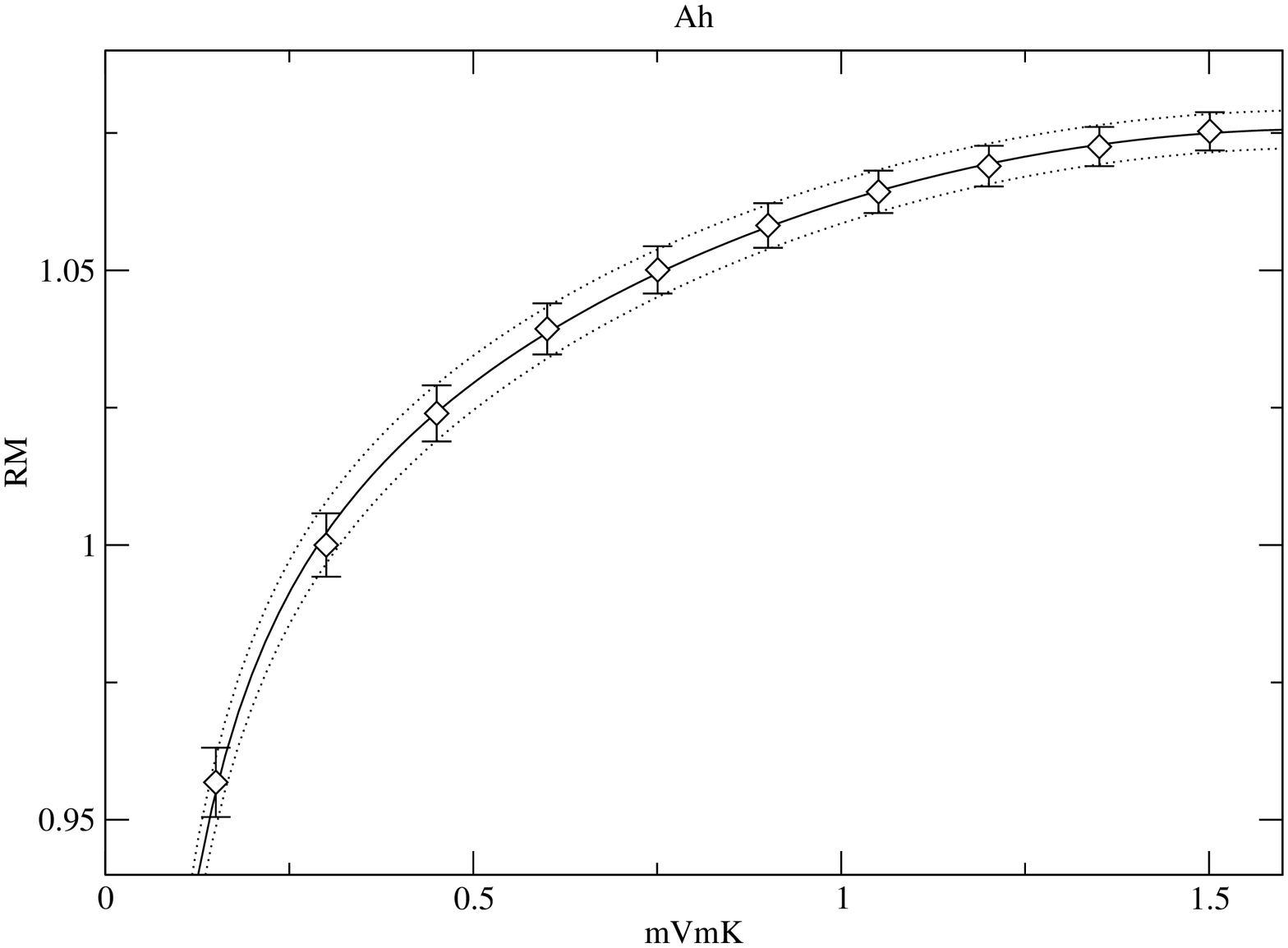}
\mycaption{$R_M$ versus valence quark mass from the cubic fit of the correlator
of ensemble \ensemble{A ~ hyp}.}
\label{ac:figure}
\end{figure}

\clearpage

\begin{figure}
\centering
\psfrag{mVmK}[t][t]{$m_V / m_{Q_K}$}
\psfrag{RM}[b][B]{$R_M$}
\psfrag{Mpi2}[b][B]{$M^2_{\pi_5} \, (\text{GeV}^2)$}
\psfrag{fpi}[b][B]{$f_{\pi_5} \, (\text{MeV})$}
\psfrag{Ch}[B][B]{\Large \ensemble{C ~ hyp}}
\includegraphics[width=\textwidth,clip=]{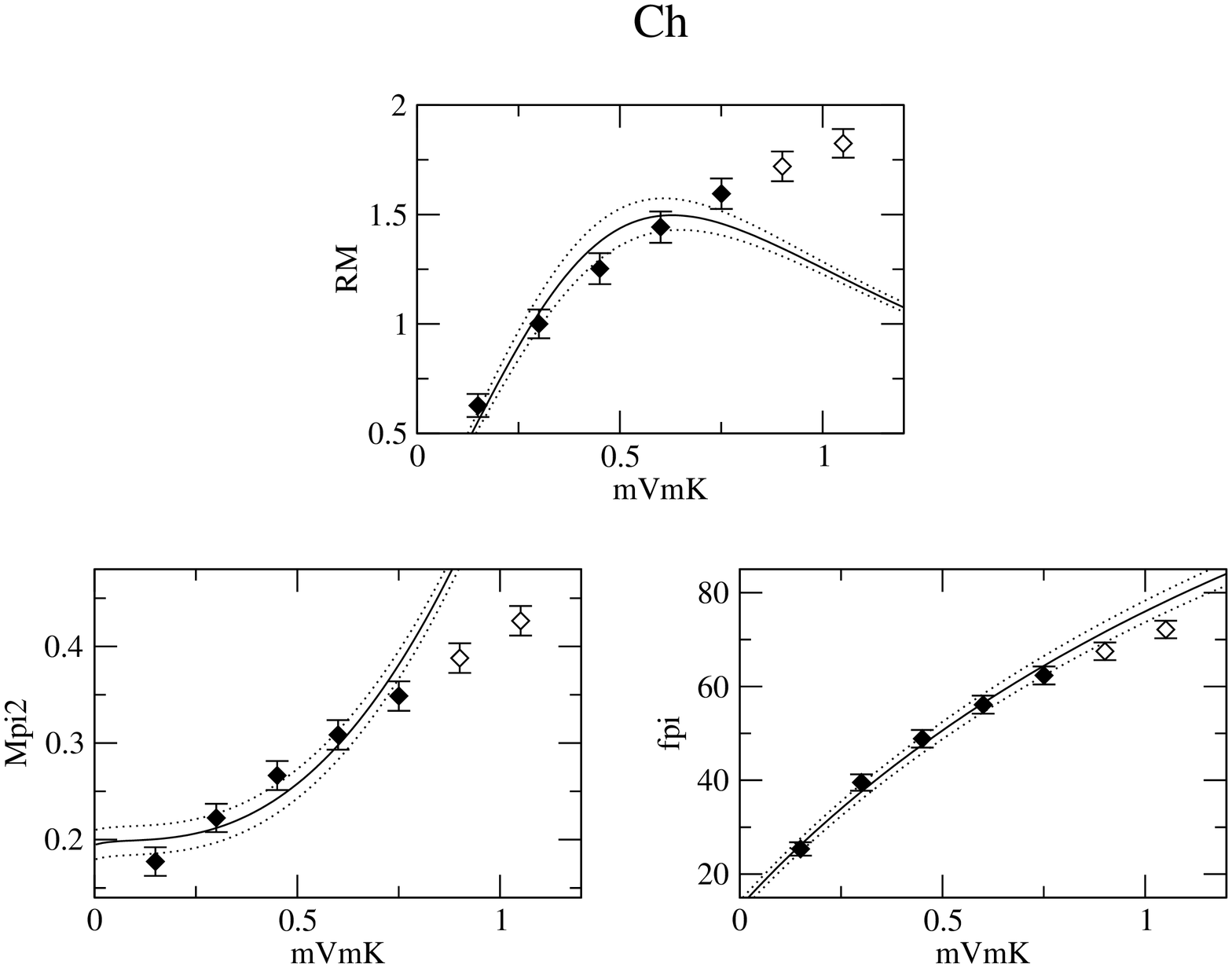}
\mycaption{Results from the pqChPT fit of ensemble
\ensemble{C ~ hyp}.  Filled diamonds correspond to valence-quark-mass values
within the fit range, while open diamonds correspond to those values beyond
it.} 
\label{ad:figure}
\end{figure}

\begin{figure}
\centering
\psfrag{mVmK}[t][t]{$m_V / m_{Q_K}$}
\psfrag{RM}[b][B]{$R_M$}
\psfrag{Mpi2}[b][B]{$M^2_{\pi_5} \, (\text{GeV}^2)$}
\psfrag{fpi}[b][B]{$f_{\pi_5} \, (\text{MeV})$}
\psfrag{C}[B][B]{\Large \ensemble{C}}
\includegraphics[width=\textwidth,clip=]{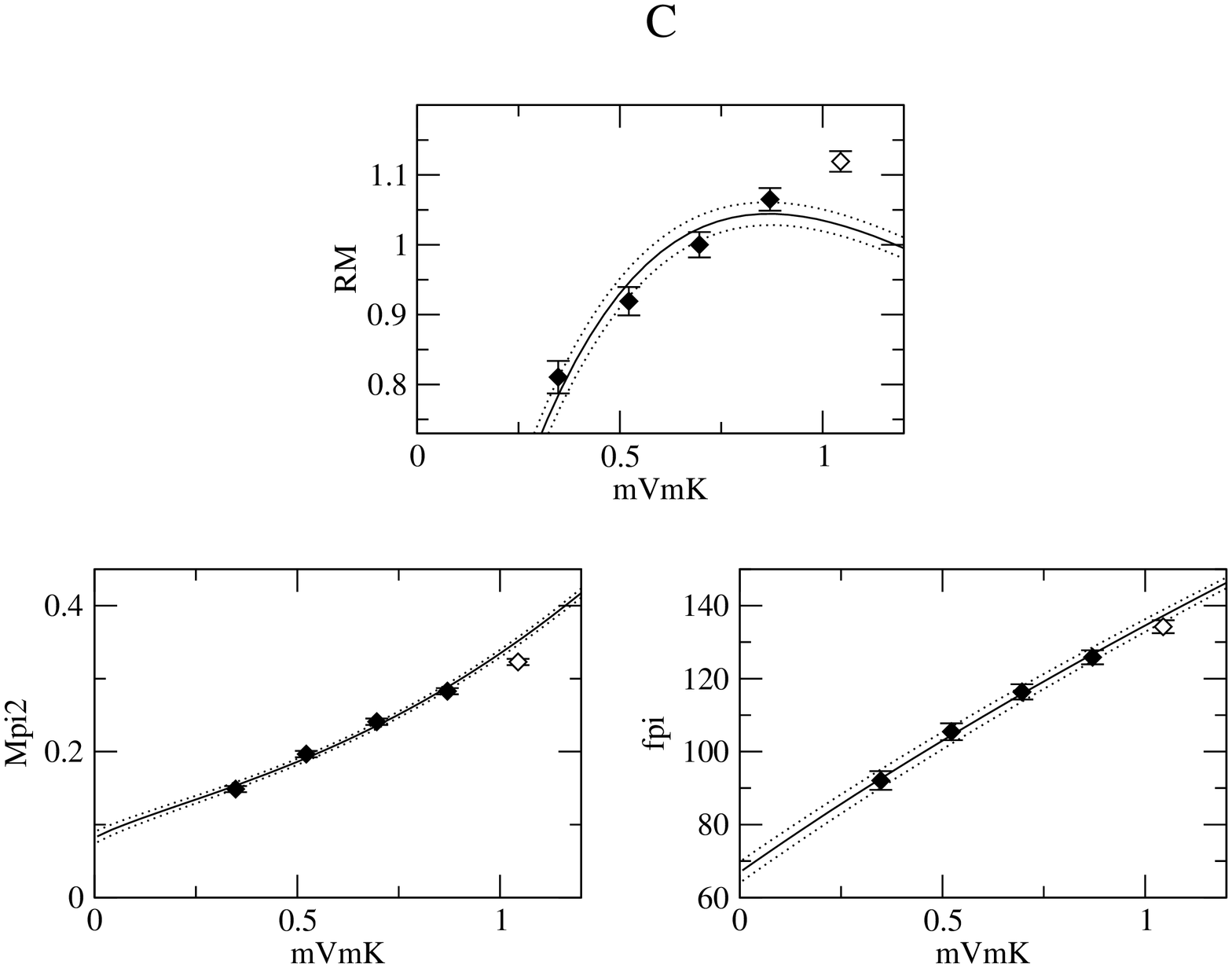}
\mycaption{Results from the pqChPT fit of ensemble
\ensemble{C}.  Filled diamonds correspond to valence-quark-mass values within
the fit range, while open diamonds correspond to those values beyond it.} 
\label{ae:figure}
\end{figure}

\clearpage

\begin{figure}
\centering
\psfrag{LvmK}[B][B]{$\Lambda_{m_V} / m_{Q_K}$}
\psfrag{z}[b][B]{$\uz$}
\psfrag{f}[b][B]{$\uf$}
\psfrag{2a8a5}[b][B]{$2 \alpha_8 - \alpha_5$}
\psfrag{a5}[b][B]{$\alpha_5$}
\psfrag{Ah}[B][B]{\Large \ensemble{A ~ hyp}}
\includegraphics[width=\textwidth,clip=]{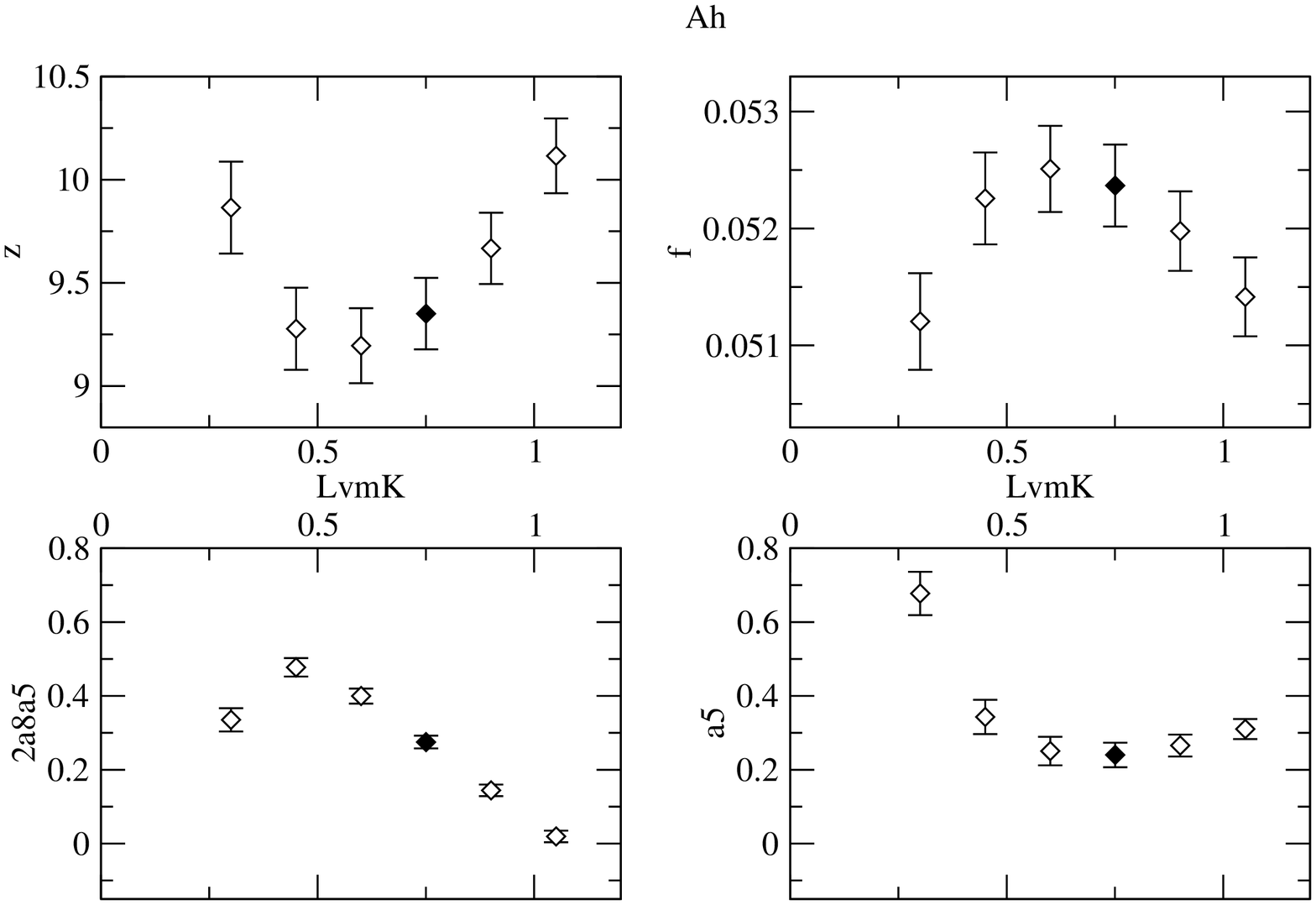}
\mycaption{Dependence of the results of the pqChPT fit of
ensemble \ensemble{A ~ hyp} on the valence-quark-mass cutoff.  The filled
diamond corresponds to the cutoff used in the final fit.}
\label{af:figure}
\end{figure}

\begin{figure}
\centering
\psfrag{Lv}[t][t]{$\Lambda_{m_V} / m_{Q_K}$}
\psfrag{2a8a5}[b][B]{$2 \alpha_8 - \alpha_5$}
\psfrag{A}[B][B]{\ensemble{A}}
\psfrag{Ah}[B][B]{\ensemble{A ~ hyp}}
\psfrag{B}[B][B]{\ensemble{B}}
\psfrag{Bh}[B][B]{\ensemble{B ~ hyp}}
\psfrag{C}[B][B]{\ensemble{C}}
\psfrag{Ch}[B][B]{\ensemble{C ~ hyp}}
\includegraphics[width=\textwidth,clip=]{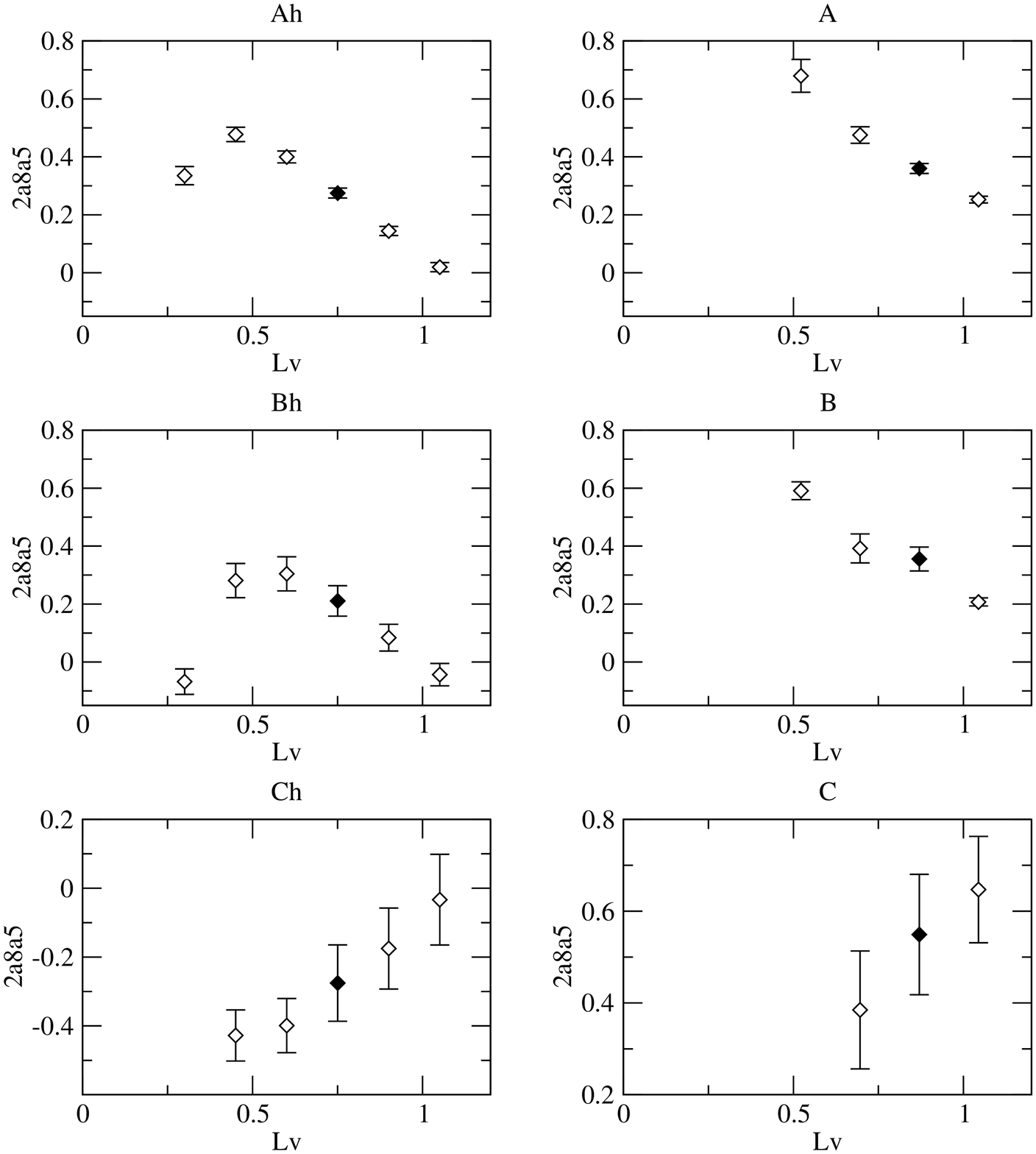}
\mycaption{Dependence of the results of the pqChPT fit of
ensembles \ensemble{A ~ hyp}, \ensemble{A}, \ensemble{B ~ hyp}, \ensemble{B},
\ensemble{C ~ hyp}, and \ensemble{C} on the valence-quark-mass cutoff.  The
filled diamond corresponds to the cutoff used in the final fit.}
\label{ah:figure}
\end{figure}

\begin{figure}
\centering
\psfrag{Lv}[t][t]{$\Lambda_{m_V} / m_{Q_K}$}
\psfrag{2a8a5}[b][B]{$2 \alpha_8 - \alpha_5$}
\psfrag{W}[B][B]{\ensemble{W}}
\psfrag{Wh}[B][B]{\ensemble{W ~ hyp}}
\psfrag{X}[B][B]{\ensemble{X}}
\psfrag{Xh}[B][B]{\ensemble{X ~ hyp}}
\includegraphics[width=\textwidth,clip=]{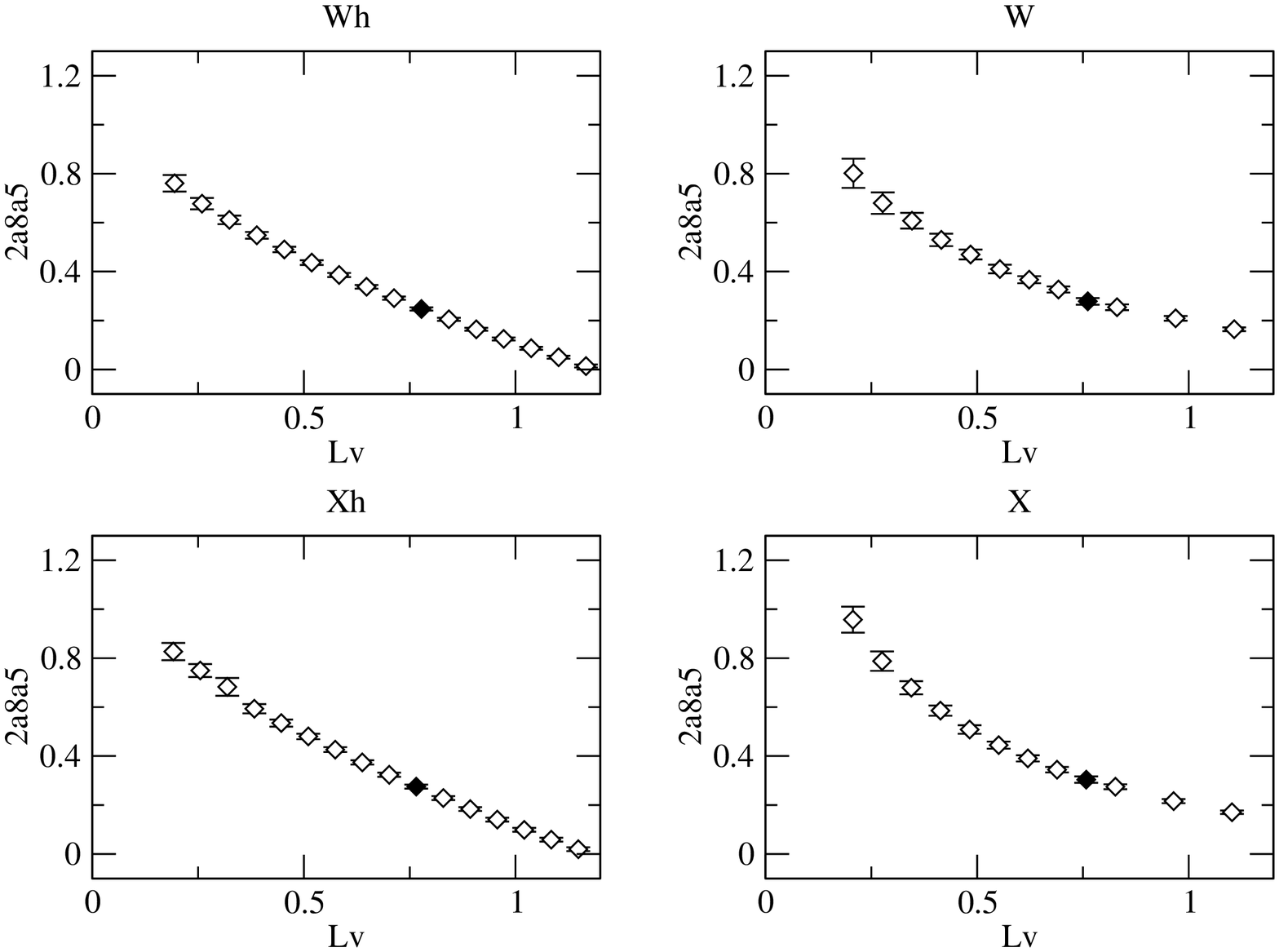}
\mycaption{Dependence of the results of the pqChPT fit of
ensembles \ensemble{W ~ hyp}, \ensemble{Z}, \ensemble{W ~ hyp}, and
\ensemble{W} on the valence-quark-mass cutoff.  The filled diamond corresponds
to the cutoff used in the final fit.}
\end{figure}

\begin{figure}
\centering
\psfrag{Lv}[t][t]{$\Lambda_{m_V} / m_{Q_K}$}
\psfrag{2a8a5}[b][B]{$2 \alpha_8 - \alpha_5$}
\psfrag{Y}[B][B]{\ensemble{Y}}
\psfrag{Yh}[B][B]{\ensemble{Y ~ hyp}}
\psfrag{Z}[B][B]{\ensemble{Z}}
\psfrag{Zh}[B][B]{\ensemble{Z ~ hyp}}
\includegraphics[width=\textwidth,clip=]{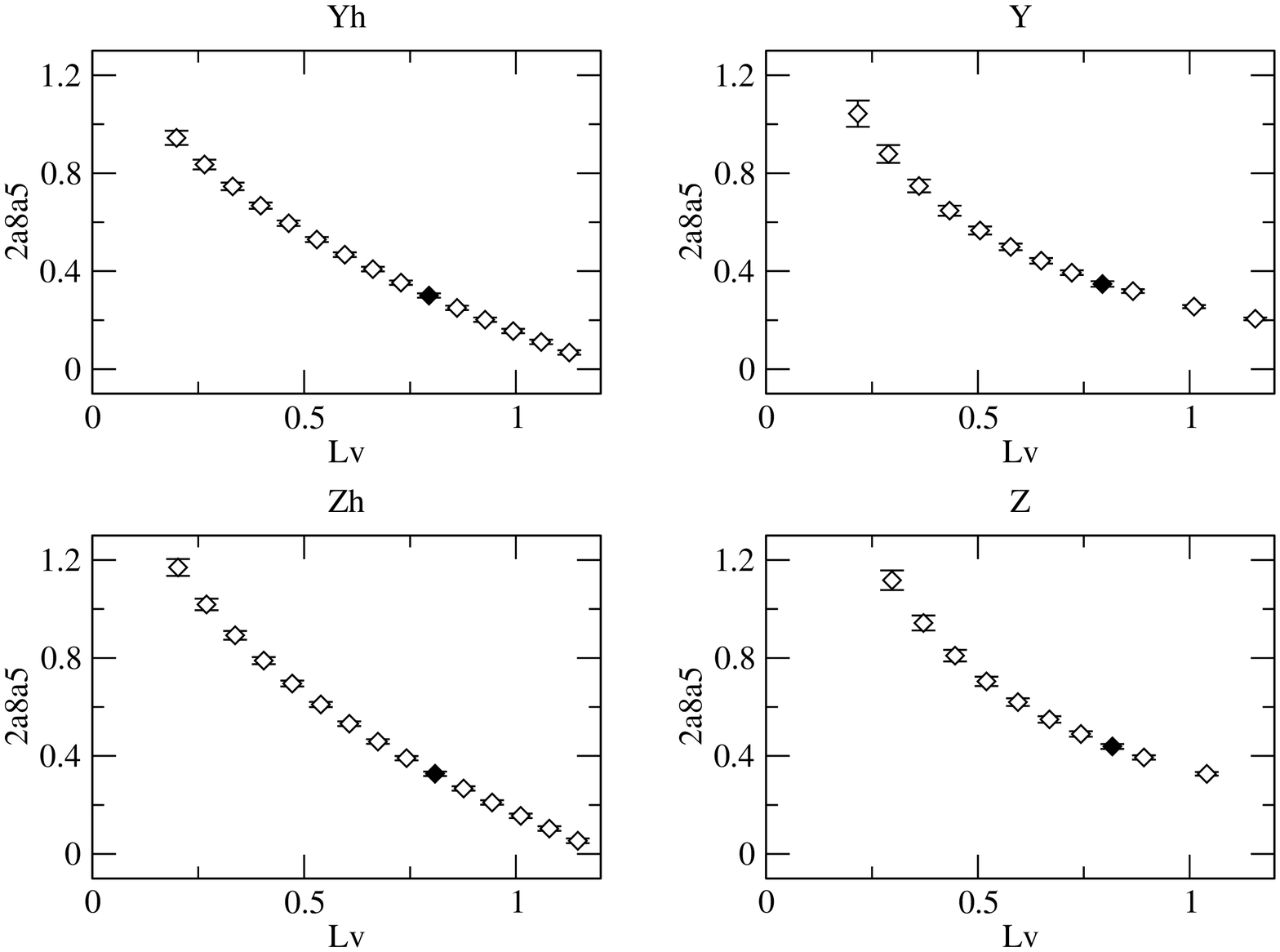}
\mycaption{Dependence of the results of the pqChPT fit of
ensembles \ensemble{Y ~ hyp}, \ensemble{Y}, \ensemble{Z ~ hyp}, and
\ensemble{Z} on the valence-quark-mass cutoff.  The filled diamond corresponds
to the cutoff used in the final fit.}
\end{figure}

\begin{figure}
\centering
\psfrag{Lv}[t][t]{$\Lambda_{m_V} / m_{Q_K}$}
\psfrag{2a8a5}[b][B]{$2 \alpha_8 - \alpha_5$}
\psfrag{Qh}[B][B]{\ensemble{Q ~ hyp}}
\includegraphics[width=0.46\textwidth,clip=]{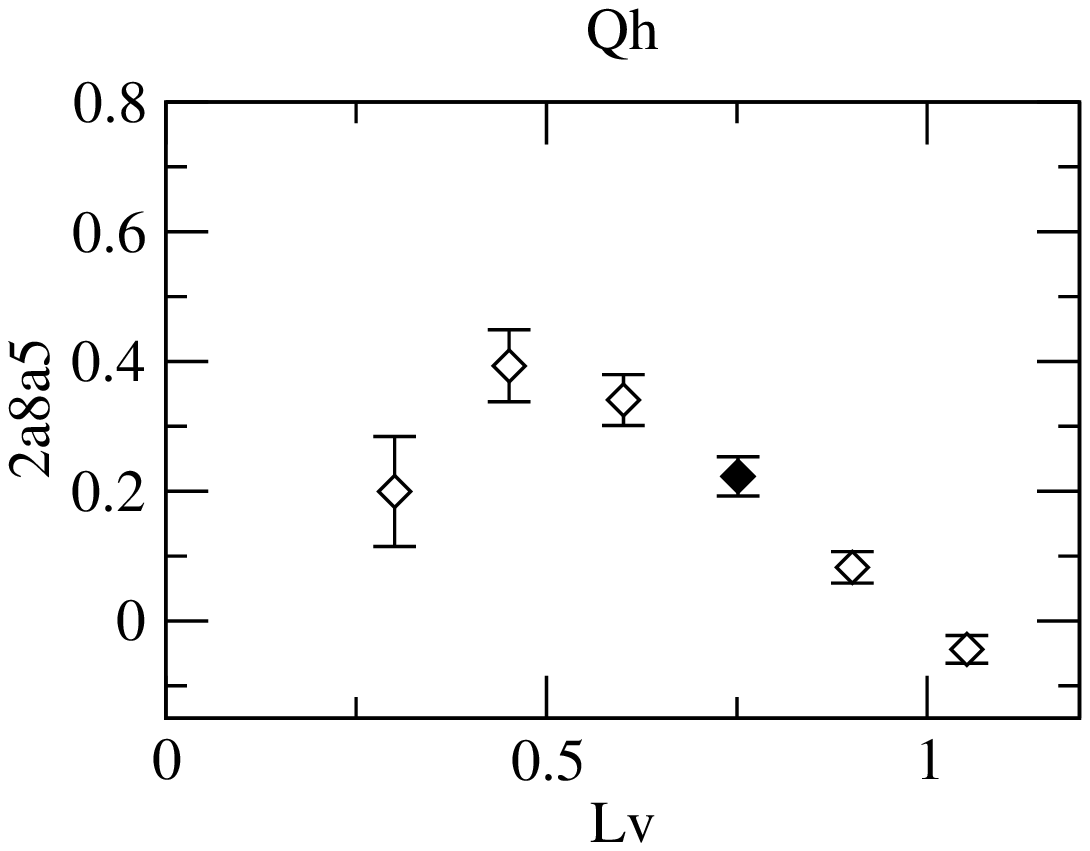}
\mycaption{Dependence of the results of the pqChPT fit of ensemble \ensemble{Q
~ hyp} on the valence-quark-mass cutoff.  The filled diamond corresponds to the
cutoff used in the final fit.}
\label{ag:figure}
\end{figure}

\clearpage

\begin{figure}
\centering
\psfrag{mVmbK}[t][t]{\Large $m_V / \avg{m}_{Q_K}$}
\psfrag{RM}[b][B]{\Large $R_M$}
\psfrag{WhXhYhZh}[b][B]{
\begin{tabular}{cc}
\Large \ensemble{W ~ hyp} - \ensemble{Z ~ hyp} \\
\large (simultaneous)
\end{tabular}}
\psfrag{WWWWW}[l][l]{\large $m_S = 0.015$}
\psfrag{Xh}[l][l]{\large $m_S = 0.02$}
\psfrag{Yh}[l][l]{\large $m_S = 0.025$}
\psfrag{Zh}[l][l]{\large $m_S = 0.035$}
\includegraphics[width=\textwidth,clip=]{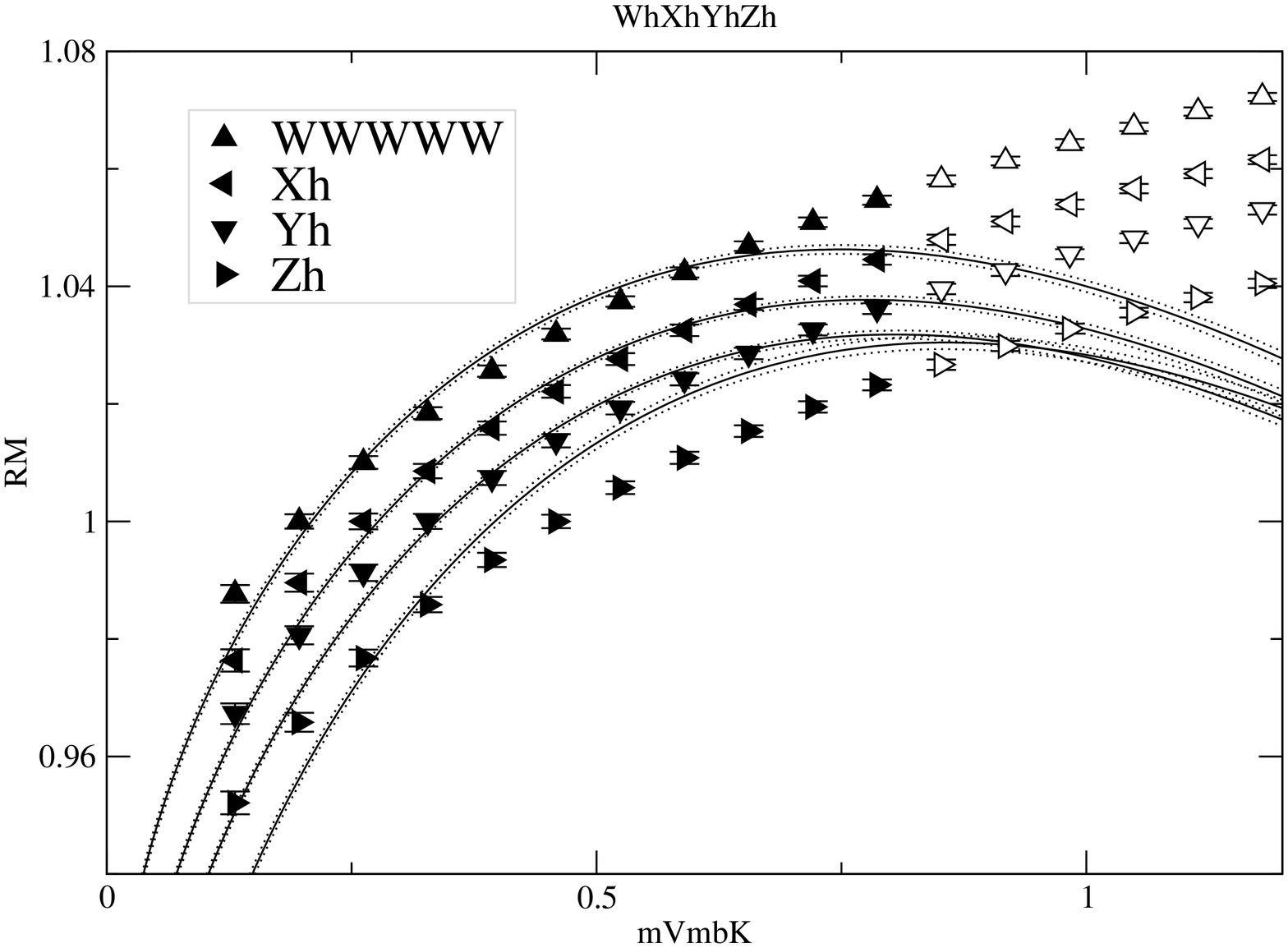}
\mycaption{$R_M$ versus valence quark mass from the simultaneous pqChPT fit of
the correlators of ensembles \ensemble{W ~ hyp} through \ensemble{Z ~ hyp}.
Filled diamonds correspond to valence-quark-mass values within the fit range,
while open diamonds correspond to those values beyond it.  An average of the
ensembles' values for $m_{Q_K}$ is used.}
\label{ai:figure}
\end{figure}

\begin{figure}
\centering
\psfrag{mVmbK}[t][t]{\Large $m_V / \avg{m}_{Q_K}$}
\psfrag{fpi}[b][B]{\Large $f_{\pi_5} \, (\text{MeV})$}
\psfrag{WhXhYhZh}[b][B]{
\begin{tabular}{cc}
\Large \ensemble{W ~ hyp} - \ensemble{Z ~ hyp} \\
\large (simultaneous)
\end{tabular}}
\psfrag{WWWWW}[l][l]{\large $m_S = 0.015$}
\psfrag{Xh}[l][l]{\large $m_S = 0.02$}
\psfrag{Yh}[l][l]{\large $m_S = 0.025$}
\psfrag{Zh}[l][l]{\large $m_S = 0.035$}
\includegraphics[width=\textwidth,clip=]{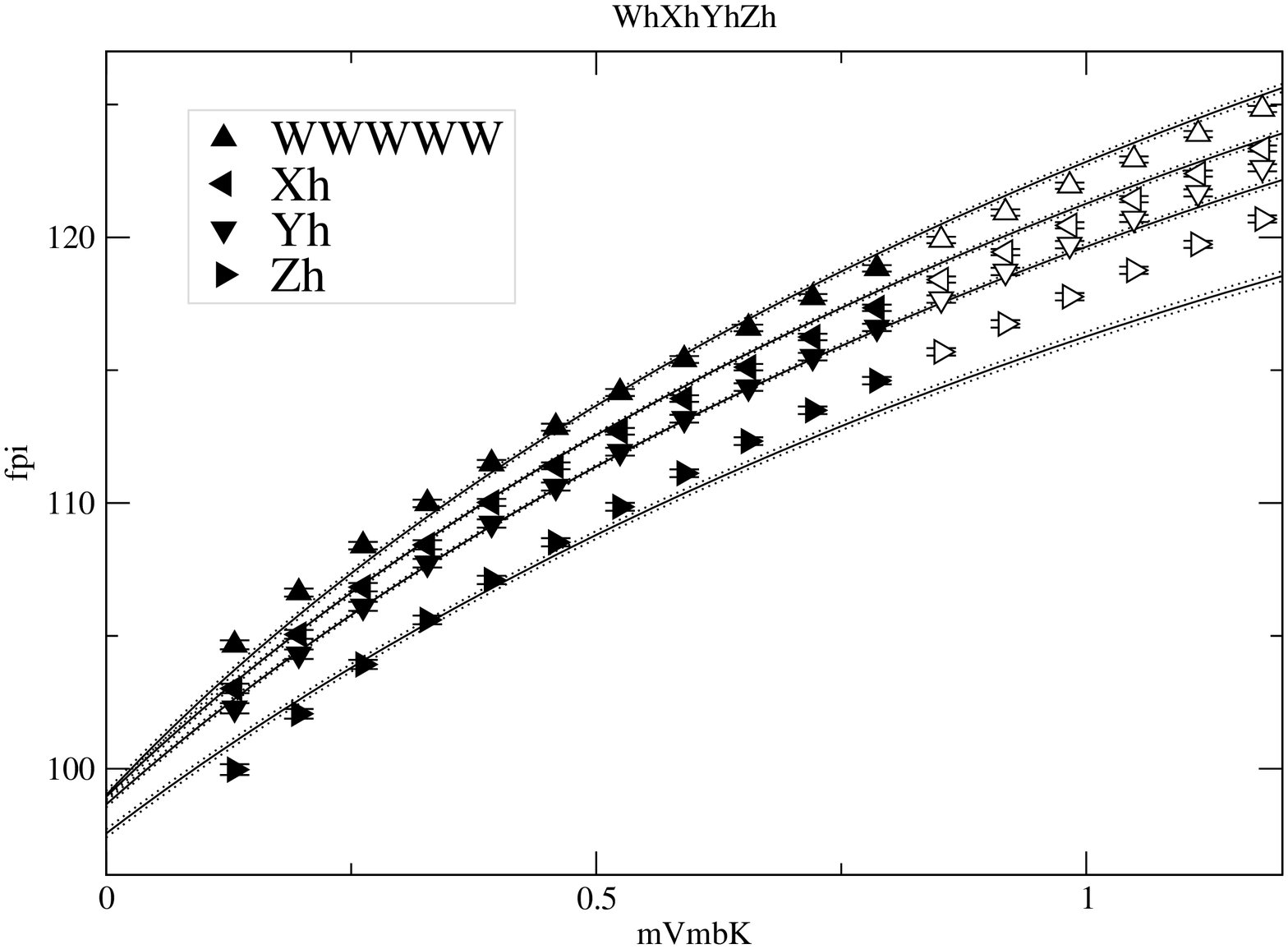}
\mycaption{Pion decay constant versus valence quark mass from the simultaneous
pqChPT fit of ensembles \ensemble{W ~ hyp} through
\ensemble{Z ~ hyp}.  Filled diamonds correspond to valence-quark-mass values
within the fit range, while open diamonds correspond to those values beyond it.
An average of the ensembles' values for $m_{Q_K}$ and lattice spacing are
used.}
\end{figure}

\begin{figure}
\centering
\psfrag{mVmbK}[t][t]{\Large $m_V / \avg{m}_{Q_K}$}
\psfrag{RM}[b][B]{\Large $R_M$}
\psfrag{WXYZ}[b][B]{
\begin{tabular}{cc}
\Large \ensemble{W} - \ensemble{Z} \\
\large (simultaneous)
\end{tabular}}
\psfrag{WWWWW}[l][l]{\large $m_S = 0.015$}
\psfrag{X}[l][l]{\large $m_S = 0.02$}
\psfrag{Y}[l][l]{\large $m_S = 0.025$}
\psfrag{Z}[l][l]{\large $m_S = 0.035$}
\includegraphics[width=\textwidth,clip=]{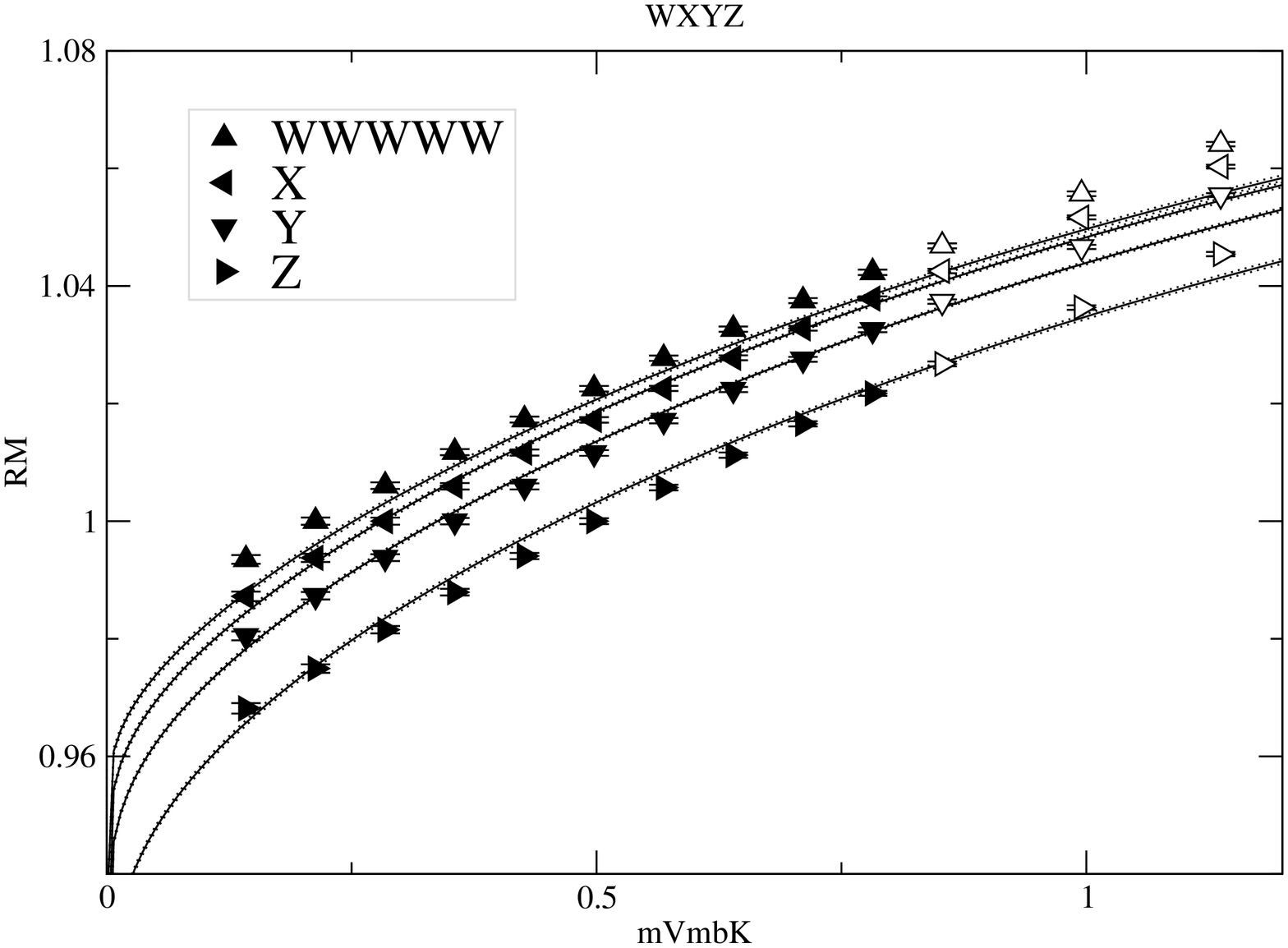}
\mycaption{Pion decay constant versus valence quark mass from the simultaneous
pqChPT fit of ensembles \ensemble{W} through \ensemble{Z}.
Filled diamonds correspond to valence-quark-mass values within the fit range,
while open diamonds correspond to those values beyond it.  An average of the
ensembles' values for $m_{Q_K}$ and lattice spacing are used.}
\end{figure}

\begin{figure}
\centering
\psfrag{mVmbK}[t][t]{\Large $m_V / \avg{m}_{Q_K}$}
\psfrag{fpi}[b][B]{\Large $f_{\pi_5} \, (\text{MeV})$}
\psfrag{WXYZ}[b][B]{
\begin{tabular}{cc}
\Large \ensemble{W} - \ensemble{Z} \\
\large (simultaneous)
\end{tabular}}
\psfrag{WWWWW}[l][l]{\large $m_S = 0.015$}
\psfrag{X}[l][l]{\large $m_S = 0.02$}
\psfrag{Y}[l][l]{\large $m_S = 0.025$}
\psfrag{Z}[l][l]{\large $m_S = 0.035$}
\includegraphics[width=\textwidth,clip=]{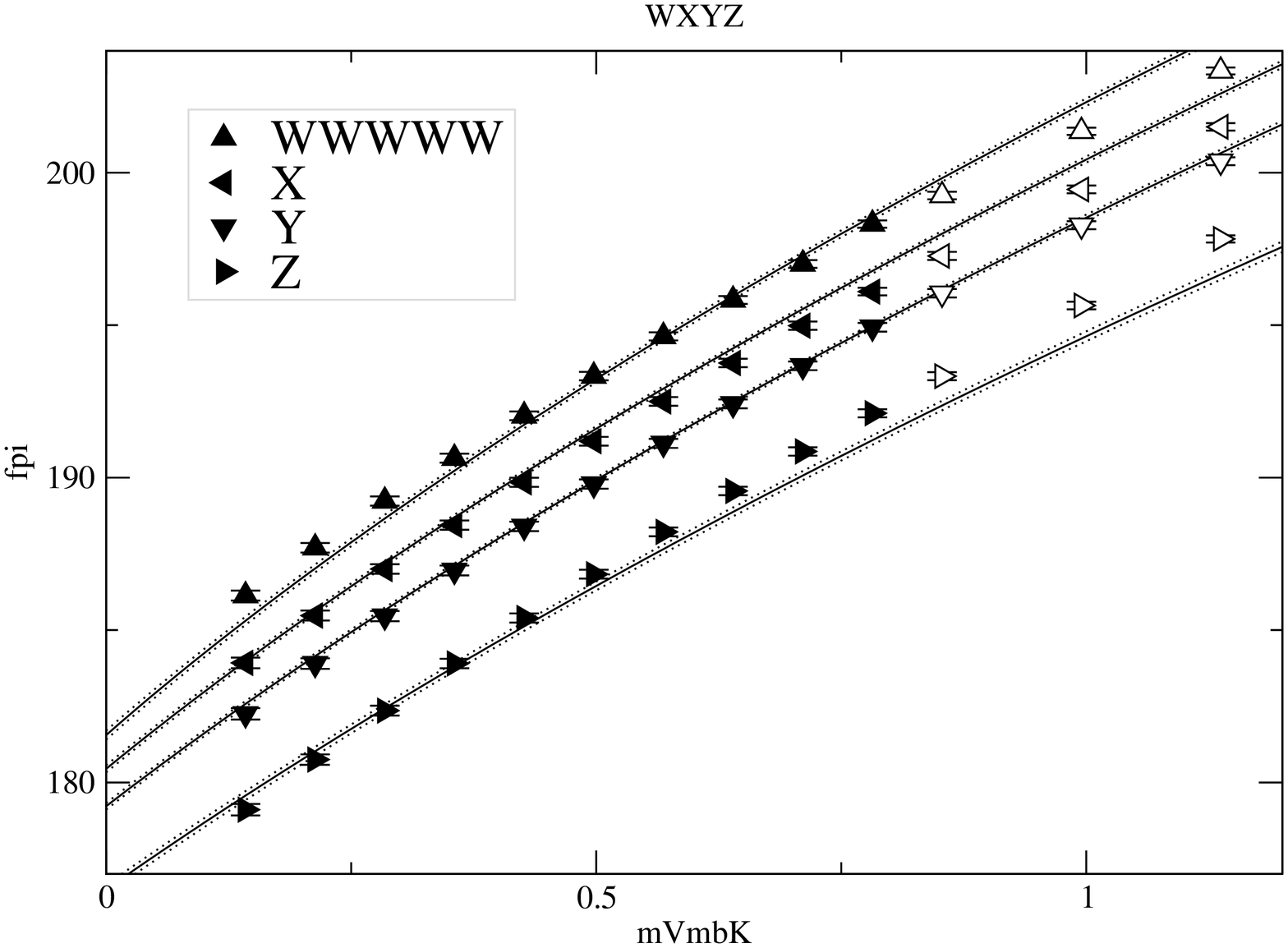}
\mycaption{Pion decay constant versus valence quark mass from the simultaneous
pqChPT fit of ensembles \ensemble{W} through \ensemble{Z}.
Filled diamonds correspond to valence-quark-mass values within the fit range,
while open diamonds correspond to those values beyond it.  An average of the
ensembles' values for $m_{Q_K}$ and lattice spacing are used.}
\label{aj:figure}
\end{figure}

\clearpage

\begin{figure}
\centering
\psfrag{mVmbK}[t][t]{\Large $m_V / \avg{m}_{Q_K}$}
\psfrag{RM}[b][B]{\Large $R_M$}
\psfrag{WhXhYhZh}[b][B]{
\begin{tabular}{cc}
\Large \ensemble{W ~ hyp} - \ensemble{Z ~ hyp} \\
\large (independent)
\end{tabular}}
\psfrag{WWWWW}[l][l]{\large $m_S = 0.015$}
\psfrag{Xh}[l][l]{\large $m_S = 0.02$}
\psfrag{Yh}[l][l]{\large $m_S = 0.025$}
\psfrag{Zh}[l][l]{\large $m_S = 0.035$}
\includegraphics[width=\textwidth,clip=]{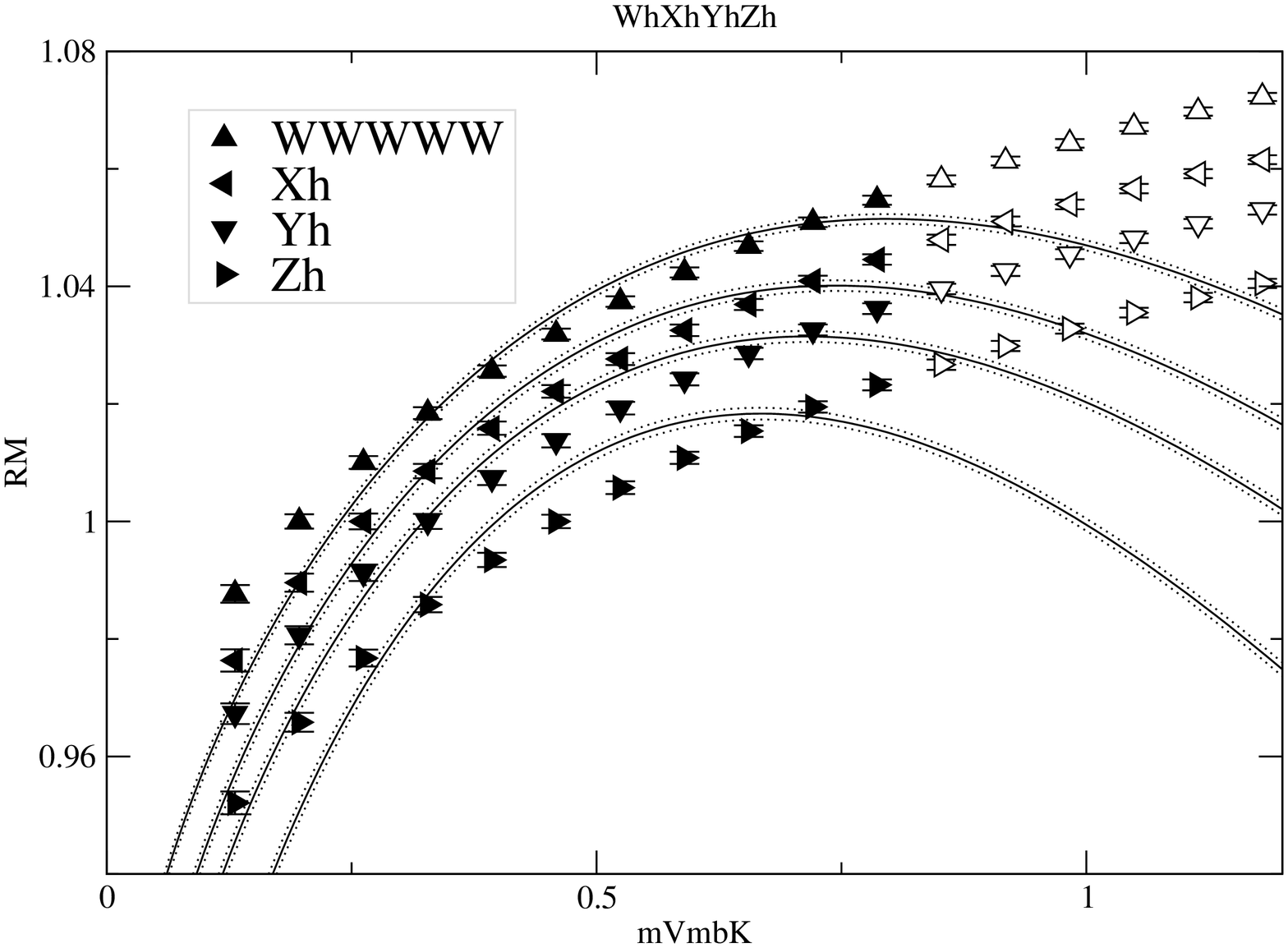}
\mycaption{$R_M$ versus valence quark mass from the four independent pqChPT
fits of ensembles \ensemble{W ~ hyp} through \ensemble{Z ~
hyp}.  Filled diamonds correspond to valence-quark-mass values within the fit
range, while open diamonds correspond to those values beyond it.  An average of
the ensembles' values for $m_{Q_K}$ is used.}
\label{ap:figure}
\end{figure}

\begin{figure}
\centering
\psfrag{mVmbK}[t][t]{\Large $m_V / \avg{m}_{Q_K}$}
\psfrag{fpi}[b][B]{\Large $f_{\pi_5} \, (\text{MeV})$}
\psfrag{WhXhYhZh}[b][B]{
\begin{tabular}{cc}
\Large \ensemble{W ~ hyp} - \ensemble{Z ~ hyp} \\
\large (independent)
\end{tabular}}
\psfrag{WWWWW}[l][l]{\large $m_S = 0.015$}
\psfrag{Xh}[l][l]{\large $m_S = 0.02$}
\psfrag{Yh}[l][l]{\large $m_S = 0.025$}
\psfrag{Zh}[l][l]{\large $m_S = 0.035$}
\includegraphics[width=\textwidth,clip=]{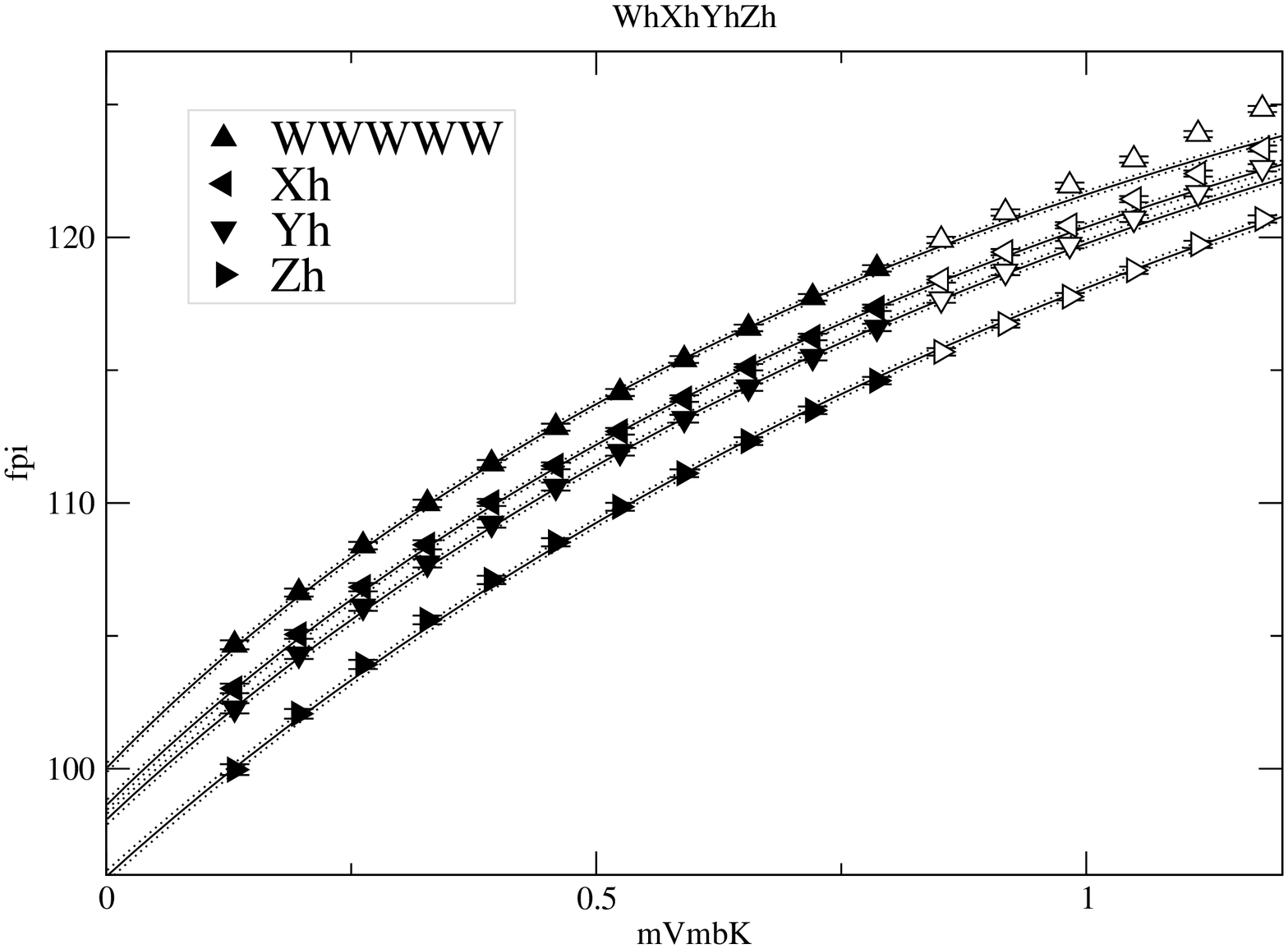}
\mycaption{Pion decay constant versus valence quark mass from the four
independent pqChPT fits of ensembles \ensemble{W ~ hyp}
through \ensemble{Z ~ hyp}.  Filled diamonds correspond to valence-quark-mass
values within the fit range, while open diamonds correspond to those values
beyond it.  An average of the ensembles' values for $m_{Q_K}$ and lattice
spacing are used.}
\end{figure}

\begin{figure}
\centering
\psfrag{mVmbK}[t][t]{\Large $m_V / \avg{m}_{Q_K}$}
\psfrag{RM}[b][B]{\Large $R_M$}
\psfrag{WXYZ}[b][B]{
\begin{tabular}{cc}
\Large \ensemble{W} - \ensemble{Z} \\
\large (independent)
\end{tabular}}
\psfrag{WWWWW}[l][l]{\large $m_S = 0.015$}
\psfrag{X}[l][l]{\large $m_S = 0.02$}
\psfrag{Y}[l][l]{\large $m_S = 0.025$}
\psfrag{Z}[l][l]{\large $m_S = 0.035$}
\includegraphics[width=\textwidth,clip=]{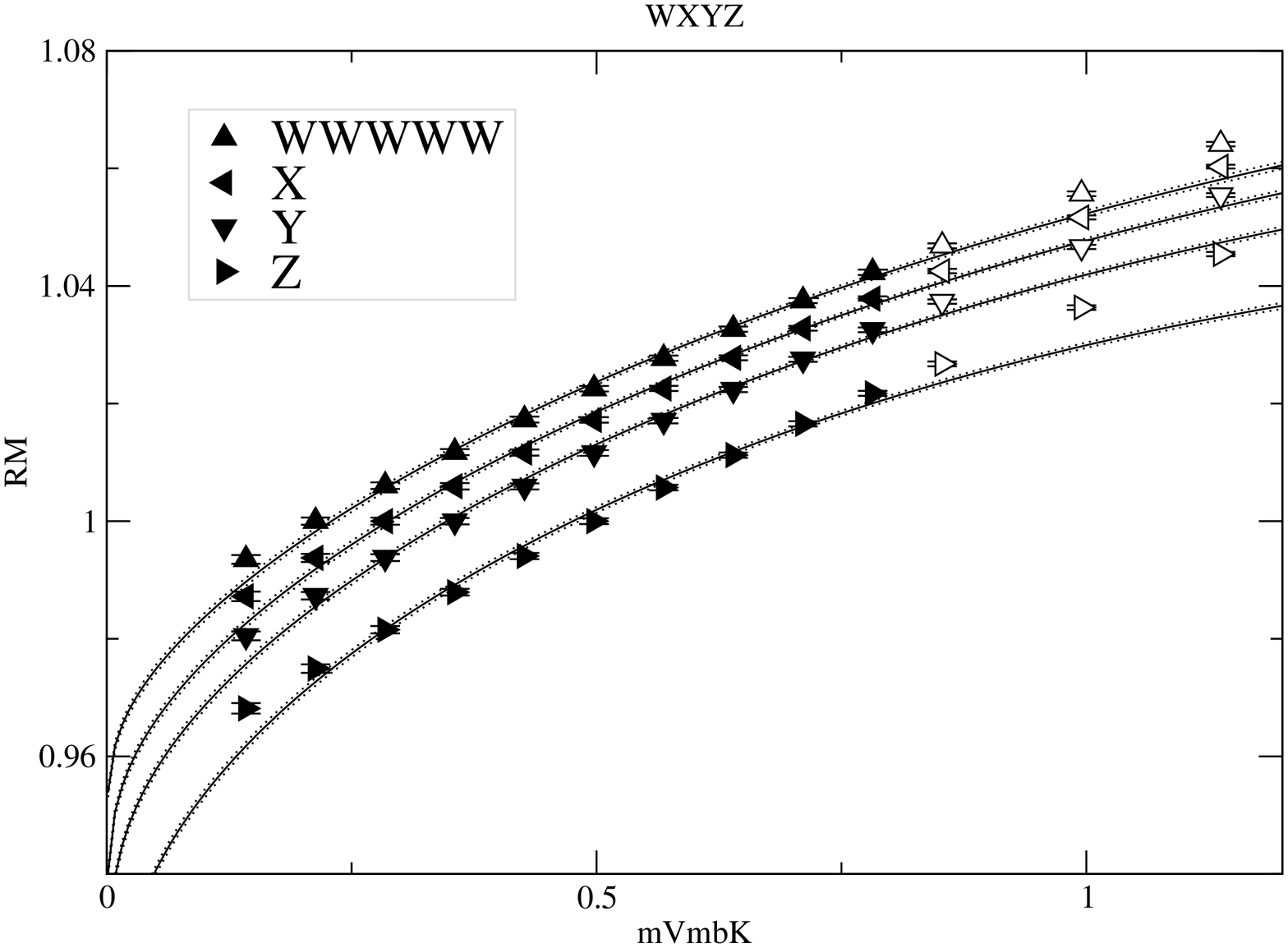}
\mycaption{$R_M$ versus valence quark mass from the four independent pqChPT
fits of ensembles \ensemble{W} through \ensemble{Z}.  Filled
diamonds correspond to valence-quark-mass values within the fit range, while
open diamonds correspond to those values beyond it.  An average of the
ensembles' values for $m_{Q_K}$ is used.}
\end{figure}

\begin{figure}
\centering
\psfrag{mVmbK}[t][t]{\Large $m_V / \avg{m}_{Q_K}$}
\psfrag{fpi}[b][B]{\Large $f_{\pi_5} \, (\text{MeV})$}
\psfrag{WXYZ}[b][B]{
\begin{tabular}{cc}
\Large \ensemble{W} - \ensemble{Z} \\
\large (independent)
\end{tabular}}
\psfrag{WWWWW}[l][l]{\large $m_S = 0.015$}
\psfrag{X}[l][l]{\large $m_S = 0.02$}
\psfrag{Y}[l][l]{\large $m_S = 0.025$}
\psfrag{Z}[l][l]{\large $m_S = 0.035$}
\includegraphics[width=\textwidth,clip=]{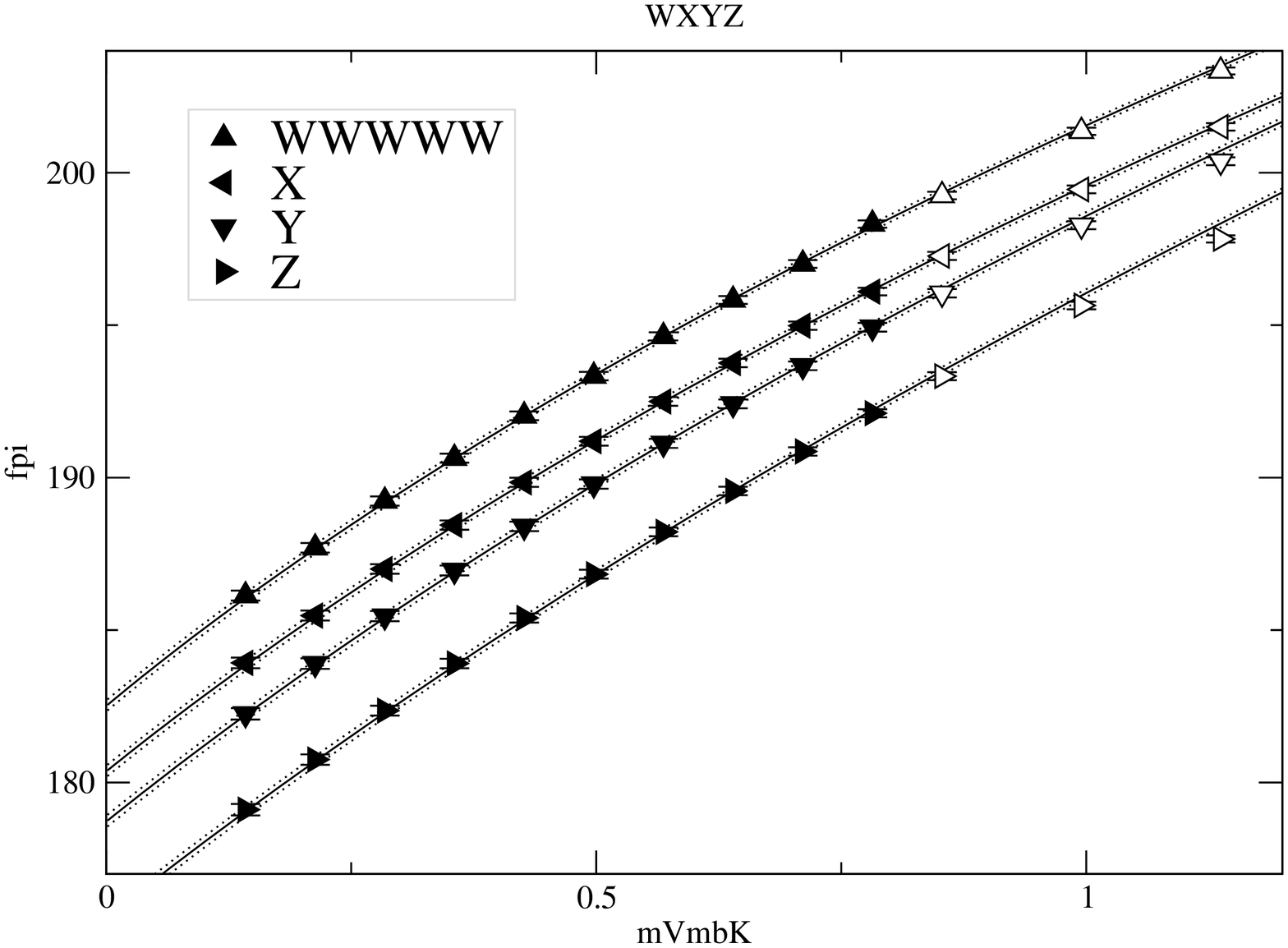}
\mycaption{Pion decay constant versus valence quark mass from the four
independent pqChPT fits of ensembles \ensemble{W} through
\ensemble{Z}.  Filled diamonds correspond to valence-quark-mass values within
the fit range, while open diamonds correspond to those values beyond it.  An
average of the ensembles' values for $m_{Q_K}$ and lattice spacing are used.}
\label{aq:figure}
\end{figure}

\clearpage

\begin{figure}
\centering
\psfrag{mQmbK}[t][t]{$m_Q / \avg{m}_{Q_K}$}
\psfrag{RMp}[b][B]{$R'_M$}
\psfrag{Mpi2}[b][B]{$M^2_{\pi_5} \, (\text{GeV}^2)$}
\psfrag{fpi}[b][B]{$f_{\pi_5} \, (\text{MeV})$}
\psfrag{WhXhYhZh}[b][B]{
\begin{tabular}{cc}
\Large \ensemble{W ~ hyp} - \ensemble{Z ~ hyp} \\
\large (simultaneous)
\end{tabular}}
\includegraphics[width=\textwidth,clip=]{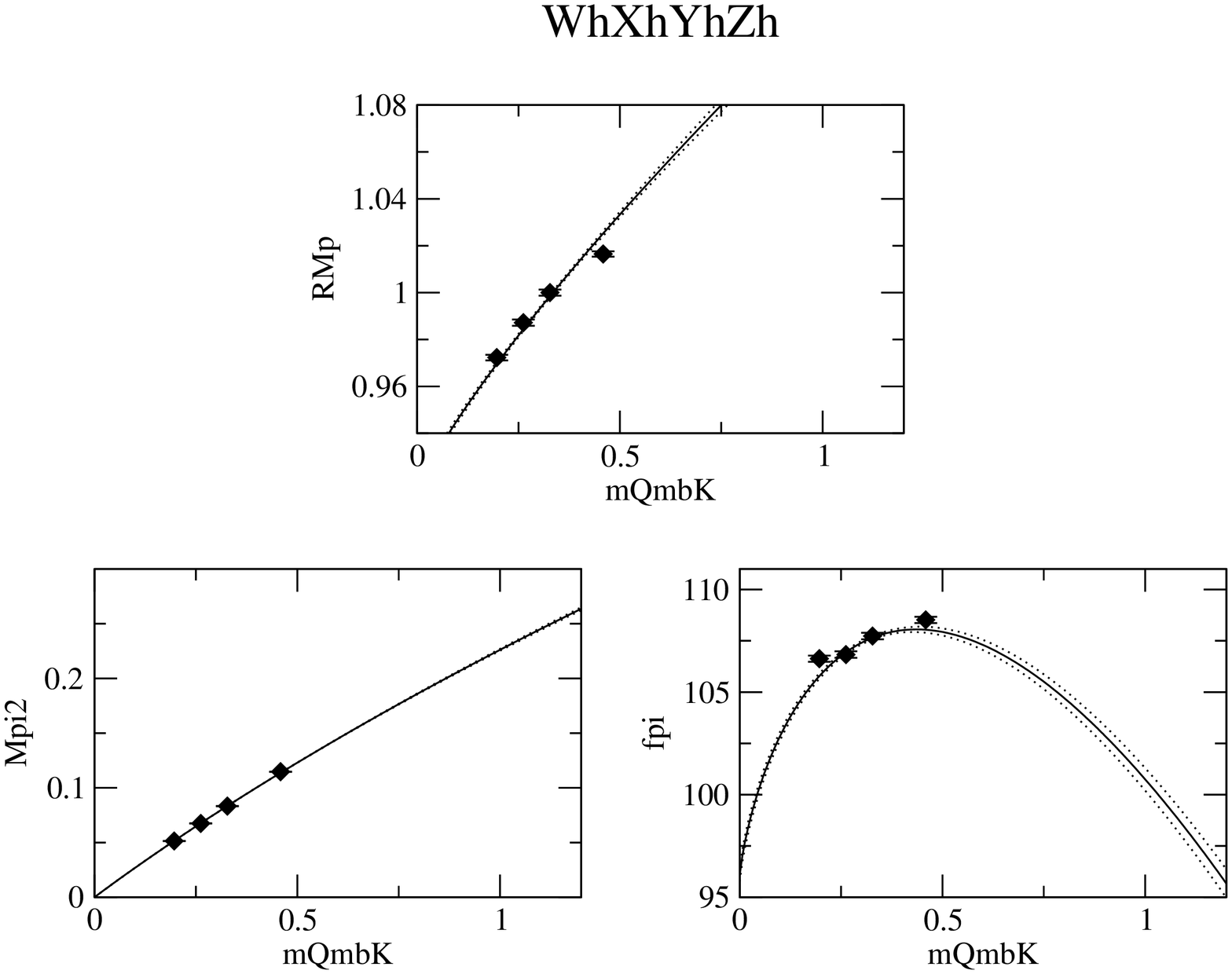}
\mycaption{Results from the simultaneous pqChPT fit of
ensembles \ensemble{W ~ hyp} through \ensemble{Z ~ hyp}, displayed along the
unquenched line of the quark-mass plane.  An average of the ensembles' values
for $m_{Q_K}$ and lattice spacing are used.}
\label{ak:figure}
\end{figure}

\begin{figure}
\centering
\psfrag{mQmbK}[t][t]{$m_Q / \avg{m}_{Q_K}$}
\psfrag{RMp}[b][B]{$R'_M$}
\psfrag{Mpi2}[b][B]{$M^2_{\pi_5} \, (\text{GeV}^2)$}
\psfrag{fpi}[b][B]{$f_{\pi_5} \, (\text{MeV})$}
\psfrag{WXYZ}[b][B]{
\begin{tabular}{cc}
\Large \ensemble{W} - \ensemble{Z} \\
\large (simultaneous)
\end{tabular}}
\includegraphics[width=\textwidth,clip=]{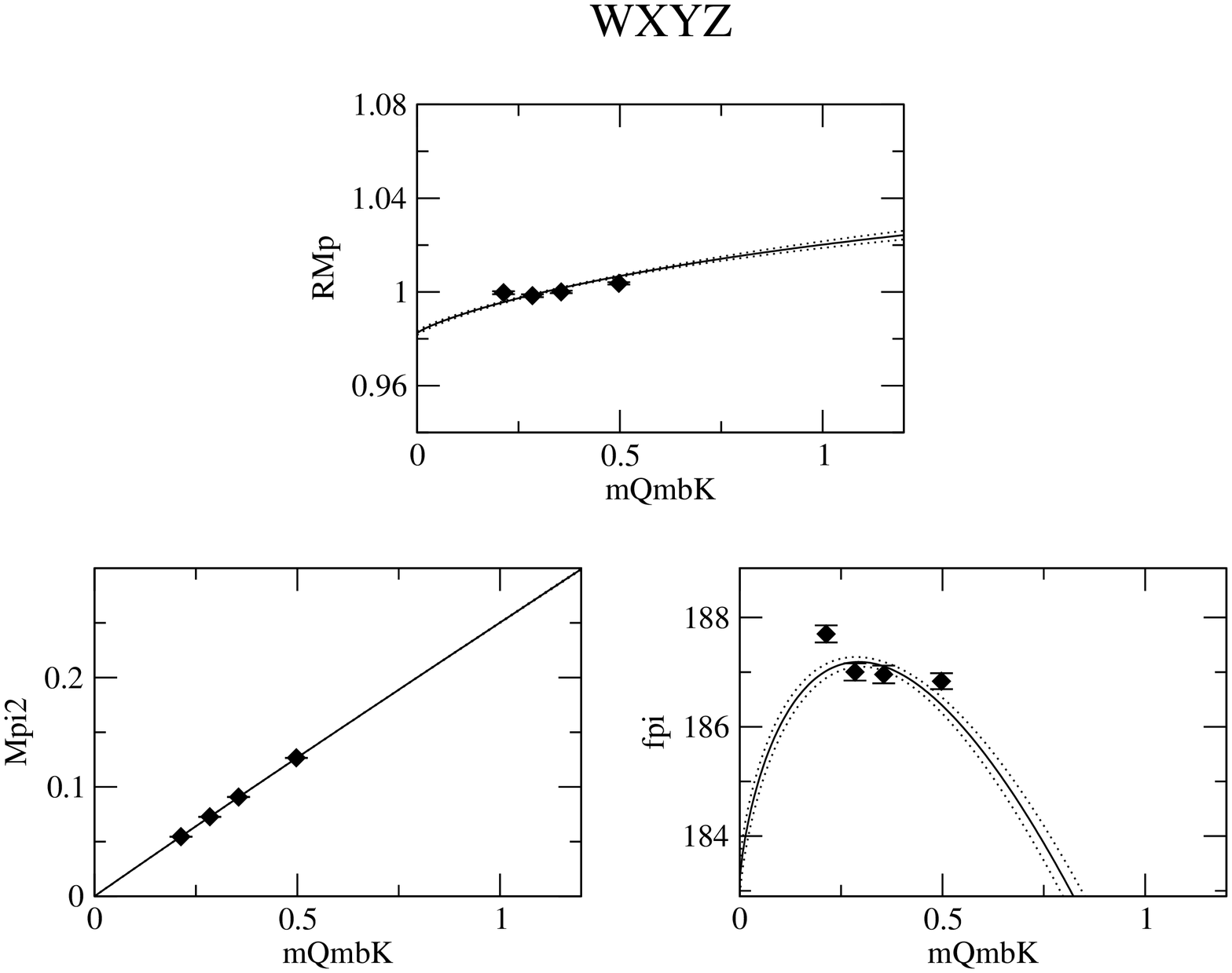}
\mycaption{Results from the simultaneous pqChPT fit of
ensembles \ensemble{W} through \ensemble{Z}, displayed along the unquenched
line of the quark-mass plane.  An average of the ensembles' values for
$m_{Q_K}$ and lattice spacing are used.}
\label{al:figure}
\end{figure}

\clearpage

\begin{figure}
\centering
\psfrag{LvmbK}[B][B]{$\Lambda_{m_V} / \avg{m}_{Q_K}$}
\psfrag{z}[b][B]{$\uz$}
\psfrag{f}[b][B]{$\uf$}
\psfrag{2a8a5}[b][B]{$2 \alpha_8 - \alpha_5$}
\psfrag{a5}[b][B]{$\alpha_5$}
\psfrag{2a6a4}[b][B]{$2 \alpha_6 - \alpha_4$}
\psfrag{a4}[b][B]{$\alpha_4$}
\psfrag{WhXhYhZh}[b][B]{
\begin{tabular}{cc}
\Large \ensemble{W ~ hyp} - \ensemble{Z ~ hyp} \\
\large (simultaneous)
\end{tabular}}
\includegraphics[width=\textwidth,clip=]{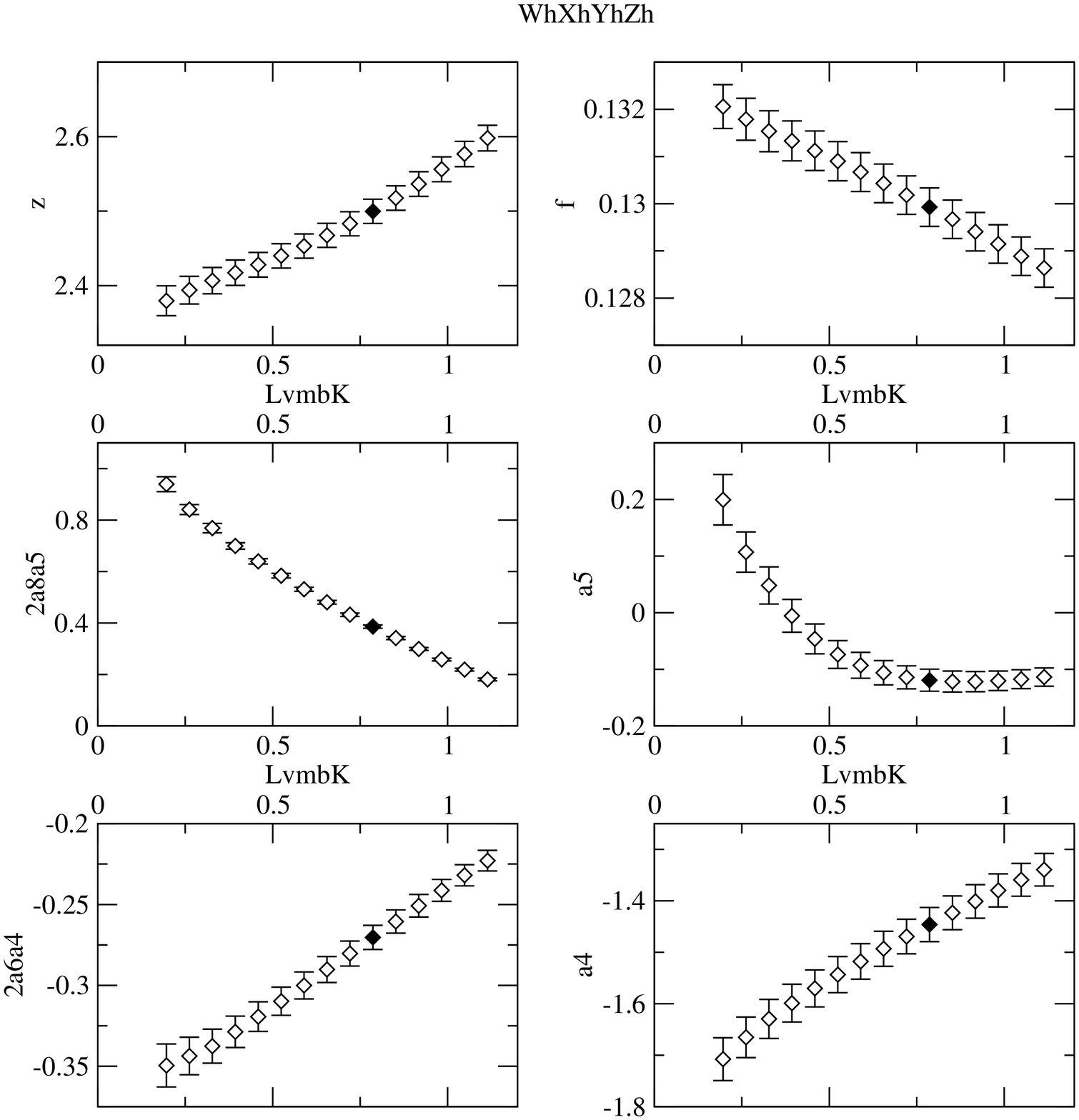}
\mycaption{Dependence of the results of the simultaneous pqChPT fit of
ensembles \ensemble{W ~ hyp} through \ensemble{Z ~ hyp} on the
valence-quark-mass cutoff.  The filled diamond corresponds to the cutoff used
in the final fit.  An average of the ensembles' values for $m_{Q_K}$ is used.}
\label{am:figure}
\end{figure}

\begin{figure}
\centering
\psfrag{LvmbK}[B][B]{$\Lambda_{m_V} / \avg{m}_{Q_K}$}
\psfrag{z}[b][B]{$\uz$}
\psfrag{f}[b][B]{$\uf$}
\psfrag{2a8a5}[b][B]{$2 \alpha_8 - \alpha_5$}
\psfrag{a5}[b][B]{$\alpha_5$}
\psfrag{2a6a4}[b][B]{$2 \alpha_6 - \alpha_4$}
\psfrag{a4}[b][B]{$\alpha_4$}
\psfrag{WXYZ}[b][B]{
\begin{tabular}{cc}
\Large \ensemble{W} - \ensemble{Z} \\
\large (simultaneous)
\end{tabular}}
\includegraphics[width=\textwidth,clip=]{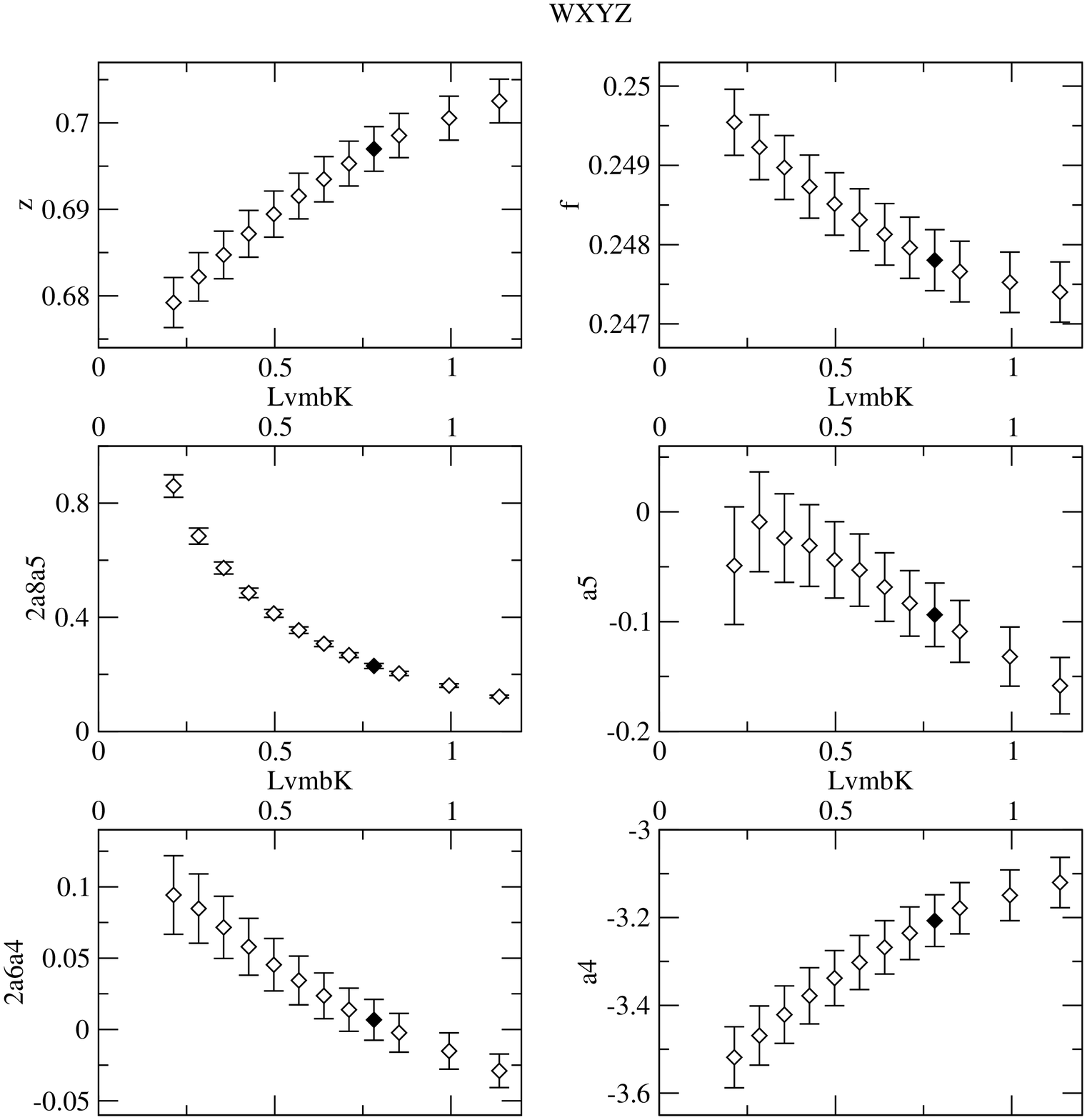}
\mycaption{Dependence of the results of the simultaneous pqChPT fit of
ensembles \ensemble{W} through \ensemble{Z} on the valence-quark-mass cutoff.
The filled diamond corresponds to the cutoff used in the final fit.  An average
of the ensembles' values for $m_{Q_K}$ is used.}
\label{an:figure}
\end{figure}

\clearpage

\nocite{Coleman:1985xx}
\nocite{DeGrand:1990ss}
\nocite{Ecker:1995gg}
\nocite{Gupta:1997nd}
\nocite{Georgi:1985kw}
\nocite{Kaku:1993ym}
\nocite{Kilcup:1995ww}
\nocite{Montvay:1994cy}
\nocite{Nelson:2001yq}
\nocite{Peskin:1995ev}
\nocite{Pich:1993uq}

\bibliographystyle{h-elsevier3}
\bibliography{thesis}

\end{document}